\documentclass[trackchanges, twocolumn,authoryear, resetfootnote]{aastex701}

\def\ra#1#2#3{#1$^{\rm h}$#2$^{\rm m}$#3$^{\rm s}$}
\def\dec#1#2#3{$#1^\circ #2' #3''$}

\usepackage[version=4]{mhchem}
\usepackage{graphicx,epsf}
\usepackage{subfigure}
\usepackage{array}
\usepackage{graphicx}	
\usepackage{amsmath}	
\usepackage{amssymb}	
\usepackage{newtxtext,newtxmath}
\usepackage{siunitx}
\usepackage{booktabs}
\usepackage{footnote}
\usepackage{natbib}
\usepackage{array}
\usepackage{multirow}

\newcommand{\MNi}{$M_{\rm Ni}$\,}
\newcommand{\MEj}{$M_{\rm ej}$\,}
\newcommand{\Msun}{${\rm M}_\odot$\,}

\newcommand{\kms}{km\,s$^{-1}$\,}
\newcommand{\niVI}{${}^{56}\textrm{Ni}$\,}
\newcommand{\coVI}{${}^{56}\textrm{Co}$\,}
\newcommand{\feVI}{${}^{56}\textrm{Fe}$\,}

\newcommand{\tzero}{$T_0$\,}
\newcommand{\tnity}{$T_{90}$\,}
\newcommand{\eiso}{$E_{\rm iso}$\,}
\newcommand{\ekiso}{$E_{\rm K, iso}$\,}
\newcommand{\ek}{$E_{\rm{K}}$\,}
\newcommand{\egiso}{$E_{\gamma,\rm iso}$\,}

\newcommand{\epeak}{$E_{\rm peak}$\,}

\newcommand{\thv}{\theta_{\rm v}}

\newcommand{\thj}{\theta_{\rm jet}}
\newcommand{\tpeak}{t_{\rm peak}}
\newcommand{\tdec}{t_{\rm dec}}
\newcommand{\gam}{$\gamma$}

\newcommand{\erg}{\mbox{$\rm erg$}\,}
\newcommand{\pcmsq}{\mbox{$\rm cm^{-2}$}\,}
\newcommand{\psec}{\mbox{$\rm s^{-1}$}\,}
\newcommand{\phz}{\mbox{$\rm Hz^{-1}$}\,}

\newcommand{\kev}{\mbox{$\rm keV$}\,}
\newcommand{\thisgrb}{GRB\,260310A\,}
\newcommand{\z}{$z$ }

\def\arcsec{\hbox{$^{\prime\prime}$}}

\usepackage[scale=0.94]{tgbonum}
\usepackage[mode=text]{siunitx}

\makeatletter
\DeclareTextCompositeCommand{\r}{OT1}{A}{%
  \leavevmode\vbox{%
    \offinterlineskip
    \ialign{\hfil##\hfil\cr\char23\cr\noalign{\kern-1.15ex}A\cr}%
  }%
}
\makeatother

\graphicspath{{./}{figures/}}

\begin{document}

\title{Early Multiwavelength Observations of AT\,2026fgk: \\ The Luminous Afterglow to Sub-luminous GRB\,260310A, Identified Independently of a Gamma-ray Trigger}

\author[orcid=0000-0002-0129-806X]{K. -R. Hinds}
\affiliation{Division of Physics, Mathematics and Astronomy, California Institute of Technology, Pasadena, CA 91125, USA}
\email{khinds@caltech.edu}

\author[orcid=0000-0002-9017-3567]{A. Y. Q. Ho}
\affiliation{Department of Astronomy, Cornell University, Ithaca, NY 14850, USA}
\email{ayh24@cornell.edu}

\author[orcid=0000-0002-5890-9298]{Y. Wagh}
\affiliation{Astrophysics Research Institute, Liverpool John Moores University, 146 Brownlow Hill, Liverpool, L3 5RF, UK}
\email{y.p.wagh@2026.ljmu.ac.uk}

\author[orcid=0000-0002-7778-3117]{R. Jayaraman}
\affiliation{Department of Astronomy, Cornell University, Ithaca, NY 14850, USA}
\email{rj438@cornell.edu}

\author[orcid=0000-0001-8472-1996]{D. A. Perley}
\affiliation{Astrophysics Research Institute, Liverpool John Moores University, 146 Brownlow Hill, Liverpool, L3 5RF, UK}
\email{d.a.perley@ljmu.ac.uk}

\author[orcid=0000-0003-3630-9440]{G. Waratkar}
\affiliation{Cahill Center for Astronomy \& Astrophysics, California Institute of Technology, 1216 East California Boulevard, Pasadena, CA 91125, USA}
\email{gauravw@caltech.edu}

\author[orcid=0009-0008-2714-2507]{A. Bochenek}
\affiliation{Astrophysics Research Institute, Liverpool John Moores University, 146 Brownlow Hill, Liverpool, L3 5RF, UK}
\email{a.m.bochenek@2023.ljmu.ac.uk}

\author[orcid=0000-0002-5826-0548]{B. P. Gompertz}
\affiliation{School of Physics and Astronomy \& Institute for Gravitational Wave Astronomy, University of Birmingham, Birmingham B15 2TT, United Kingdom}
\email{b.gompertz@bham.ac.uk}

\author[orcid=0000-0002-4223-103X]{C. Fremling}
\affiliation{Caltech Optical Observatories, California Institute of Technology, Pasadena, CA 91125, USA}
\affiliation{Division of Physics, Mathematics and Astronomy, California Institute of Technology, Pasadena, CA 91125, USA}
\email{fremling@caltech.edu}

\author[orcid=0000-0002-9267-6213]{J. Rastinejad}
\affiliation{HFP Einstein Fellow, Department of Astronomy, University of Maryland, College Park, MD~20742, USA}
\email{jcrastin@umd.edu}

% ...

\begin{abstract}

The origins of sub-luminous ($L_\mathrm{\gamma,\mathrm{iso}} < 10^{49.5}$\,erg\,s$^{-1}$) gamma-ray bursts (GRBs) associated with broad-lined Type~Ic supernovae (Ic-BL SNe) are poorly understood, in part due to the low discovery rate and faint afterglows. Here we present the identification of the optical afterglow of Fermi-GBM-detected GRB\,260310A (AT\,2026fgk) as a rapidly rising ($>1\,$mag\,d$^{-1}$), red ($g-r=0.4$\,mag) transient using the Gravitational-wave Optical Transient Observatory, Large Array Survey Telescope, and Zwicky Transient Facility (ZTF) data streams. We present multiwavelength follow-up observations from the first 50\,days, which reveal that GRB 260310A/AT\,2026fgk was sub-luminous ($L_\mathrm{\gamma,iso}=10^{48.8}\,$erg\,s$^{-1}$); it was the most nearby ($z=0.153$) afterglow identified blindly by an optical survey; and that it is one of the brightest afterglows ever observed at X-ray, optical, and radio (cm to mm) wavelengths. We spectroscopically confirm an underlying Ic-BL SN with properties typical of GRB-SNe ($M_\mathrm{ej}\approx3\,M_\odot$, $E_{\rm K}\approx 10^{52}\,$erg). With basic modeling of the afterglow, including the long optical rise ($\approx10^{3}\,$s), we infer either a low initial Lorentz factor ($\Gamma_0\approx40$) or a slightly off-axis viewing angle ($\lesssim3^\circ$). The host galaxy's mass and star formation rate are similar to the hosts of other sub-luminous GRBs. ZTF's flux-limited survey gives a volumetric rate of AT\,2026fgk-like events of $0.30^{+1.37}_{-0.29}\,$Gpc\,$^{-3}$\,yr$^{-1}$, which is consistent with the on-axis, high luminosity ($L_{\rm \gamma,iso}>10^{49.5}$\,erg\,s$^{-1}$) long-GRB rate. The similarity in the rates strongly constrains the prevalence of low-$\Gamma_0$ bursts and the beaming of the initial relativistic material in GRBs.

\end{abstract}

\keywords{\uat{Supernovae}{1668} --- \uat{High Energy astrophysics}{739} --- \uat{Circumstellar matter}{241} -- \uat{X-ray astronomy}{1810} } 


\section{Introduction}

The majority of long gamma-ray bursts (LGRBs) are powered by the core collapse of massive stars, in which an ultra-relativistic jet drives both the prompt gamma-ray emission and a subsequent broadband afterglow \citep[e.g.,][]{rees1992MNRAS.258P..41R, woosley1993ApJ...405..273W, piran2004RvMP...76.1143P, cano2017AdAst2017E...5C}\footnote{A small but growing number of long-duration events instead originate from compact-object mergers \citep[e.g.,][]{rastinejad2022Natur.612..223R,troja2022Natur.612..228T,yang2022Natur.612..232Y, gottlieb2023ApJ...958L..33G, gompertz2023NatAs...7...67G,levan2024Natur.626..737L, yang2024Natur.626..742Y}.}. The discovery of LGRB\,980425---associated with supernova (SN) 1998bw \citep{galama1998Natur.395..670G, iwamoto1998Natur.395..672I, kulkarni1998Natur.395..663K}---and over 30 spectroscopically confirmed LGRB-SNe since then \citep{cano2017AdAst2017E...5C, fiore2025Galax..13...57F, finneran2025A&C....5200954F} established a robust connection between collapsar LGRBs and broad-lined Type Ic supernovae (Ic-BL SNe). LGRB-SNe can have kinetic energies ($E_K$) over an order of magnitude greater than the canonical $10^{51}$\,erg of neutrino-driven supernovae \citep{cano2017AdAst2017E...5C}. 

Cosmological LGRBs are typically defined as having isotropic-equivalent gamma-ray energies \egiso$\sim10^{51}$--$10^{54}$\,\erg, with luminosities $L_{\gamma,\rm{iso}}\gtrsim10^{50}$--$10^{54}$\,\erg\psec and prompt emission peak energies \epeak$\sim$100--1000~keV \citep[e.g.,][]{frail2001ApJ...562L..55F, amati2002A&A...390...81A, cano2011ApJ...740...41C}. However, a number of nearby 
($z\leq0.3$) LGRBs have been discovered with low luminosities ($L_{\gamma,\rm{iso}}<10^{49.5}$\,\erg\psec; Table~\ref{tab:llgrb_sample}), initially by 
BeppoSAX and HETE-2 \citep[e.g., ][]{heise2001grba.conf...16H, sakamoto2005ApJ...629..311S,campana2006Natur.442.1008C, starling2011MNRAS.411.2792S}. Sub-luminous LGRBs were firmly tied to Ic-BL SNe through Swift-era events such as XRF\,060218/SN\,2006aj and GRB\,100316D/SN\,2010bh \citep[][]{campana2006Natur.442.1008C, pian2006Natur.442.1011P, starling2011MNRAS.411.2792S}. 
Sub-luminous LGRBs are challenging to detect due to their low $E_{\gamma,\mathrm{iso}}$, low $E_\mathrm{peak}$, and often longer prompt durations \citep[e.g.,][]{liang2007ApJ...662.1111L, virgili2009MNRAS.392...91V, bromberg2011ApJ...739L..55B, margutti2013ApJ...778...18M,  nakar2015ApJ...807..172N, cano2017AdAst2017E...5C, delia2018AA...619A..66D}. 
Modeling of the radio emission from several sub-luminous LGRBs implies quasi-isotropic emission (e.g., \citealt{soderberg2006Natur.442.1014S}). 
The volumetric rate of sub-luminous LGRBs, found to be $\mathcal{R} = 164^{+98}_{65}$\,Gpc$^{-3}$\,yr$^{-1}$ \citep[above $L_{\rm iso}=5\times10^{46}$\,\erg\psec, see][and references therein]{sun2015ApJ...812...33S} 
is comparable to the collimation-corrected high-luminosity LGRB rate, adopting an on-axis rate of $\mathcal{R} \sim$1~Gpc$^{-3}$\,yr$^{-1}$ \citep{virgili2009MNRAS.392...91V, pescalli2016A&A...587A..40P, ghirlanda2022ApJ...932...10G} and a beaming fraction of $\approx100$. 



\begin{deluxetable}{llcccc}[!ht]
\tablecaption{Properties of sub-luminous GRBs and their accompanying SNe.
$E_{\gamma,\mathrm{iso}}$ and $E_{\mathrm{p,i}}$ are from
\citet{minaev2020MNRAS.492.1919M} where available, and from
\citet{cano2017AdAst2017E...5C} otherwise ($\dagger$). GRBs are in the first section, EP events in the second, and SVOM events are in the third.
\label{tab:llgrb_sample}}
\tablewidth{0pt}
\tablehead{
  \colhead{GRB} &
  \colhead{SN}  &
  \colhead{$z$} &
  \colhead{$E_{\gamma,\mathrm{iso}}$ ($10^{50}$ erg)} &
  \colhead{$E_{\mathrm{p,i}}$ (keV)} & \colhead{Refs.}
}
\startdata
980425  & 1998bw  & 0.0085 & $0.010 \pm 0.002$           & $55 \pm 21$           & \tablenotemark{a} \\
020903  & \nodata     & 0.251  & $0.24 \pm 0.06$             & $3.4 \pm 1.8$         & \tablenotemark{b} \\
031203  & 2003lw  & 0.1055 & $1.0 \pm 0.4$               & $158 \pm 51$          & \tablenotemark{c} \\
060218  & 2006aj  & 0.0331 & $0.53 \pm 0.03$             & $4.9 \pm 0.3$         & \tablenotemark{d} \\
100316D & 2010bh  & 0.059  & $0.59^{\dagger}$            & $26 \pm 16^{\dagger}$ & \tablenotemark{e} \\
171205A & 2017iuk & 0.0368 & $0.218^{+0.063}_{-0.050}$  & $125^{+141}_{-37}$    & \tablenotemark{f} \\
190829A & 2019oyw   & 0.0785 & $1.8^{\dagger}$             & $130 \pm 20^{\dagger}$& \tablenotemark{g} \\
\textbf{260310A} & \textbf{2026fgk} & $\mathbf{0.153}$ & $\mathbf{3.44}$ & $\mathbf{190}$ & \tablenotemark{\textbf{h}} \\ \hline
240414A & 2024gsa & 0.401 & $0.53^{+0.08}_{-0.06}$  & $<1.3$ &\tablenotemark{i} \\
240801a & \ldots & 1.6734 & $55^{+5.4}_{-5.0}$ & $14.9^{+7.08}_{-4.71}$ & \tablenotemark{j}\\
250304A & 2025fhm & 0.2 & \ldots &$<$70 &\tablenotemark{k}\\
250108A & 2025kg & 0.176 & $0.08^{+0.22}_{-0.03}$ & $<1.8$ &\tablenotemark{l} \\ 
250827B & 2025wkm & 0.1194 & $0.01$ & $<1.5$&\tablenotemark{m} \\ 
260321A & 2026gzf & 0.0343&  $<0.01$ &\ldots & \tablenotemark{n} \\ \hline
241001A & 2024aiiq & 0.573 & 0.895 &$<$10 & \tablenotemark{o} \\
\hline
\enddata
\tablenotemark{a}{\citet{galama1998Natur.395..670G, kulkarni1998Natur.395..663K, pian2000ApJ...536..778P, patat2001ApJ...555..900P, nakamura2001ApJ...550..991N}}$\,\cdot$
\tablenotemark{b}
{\citet{soderberg2004ApJ...606..994S, ricker2002GCN..1530....1R, soderberg2005ApJ...627..877S, bersier2006ApJ...643..284B}}$\,\cdot$
\tablenotemark{c}{\citet{malesani2004ApJ...609L...5M, soderberg2004Natur.430..648S, galyam2004ApJ...609L..59G, mazzali2006ApJ...645.1323M, waxman2007ApJ...667..351W}}$\,\cdot$
\tablenotemark{d}{\citet{modjaz2006ApJ...645L..21M, sollerman2006AA...454..503S, cobb2006ApJ...645L.113C, soderberg2006Natur.442.1014S,pian2006Natur.442.1011P, mazzali2006Natur.442.1018M,ferror2006AA...457..857F, nakar2015ApJ...807..172N, irwin2025MNRAS.542.1269I}}$\,\cdot$
\tablenotemark{e}{\citet{starling2011MNRAS.411.2792S, cano2011ApJ...740...41C, fan2011ApJ...726...32F, bufano2012ApJ...753...67B, margutti2013ApJ...778...18M}}$\,\cdot$
\tablenotemark{f}{\citet{delia2018AA...619A..66D, wang2018ApJ...867..147W, izzo2019Natur.565..324I, urata2019ApJ...884L..58U}}$\,\cdot$
\tablenotemark{g}{\citet{chand2020ApJ...898...42C,hess2021Sci...372.1081H, fraija2021ApJ...918...12F, salafia2022ApJ...931L..19S, rhodes2020MNRAS.496.3326R}}$\,\cdot$
\tablenotemark{\textbf{h}}{\textbf{This work}} $\,\cdot$\tablenotemark{i}{\citet{sun2025NatAs...9.1073S}}$\,\cdot$
\tablenotemark{j}{\citet{jiang2025ApJ...988L..34J}}$\,\cdot$
\tablenotemark{k}{\citet{GCN.39594, GCN.39585, GCN.39600}}$\,\cdot$
\tablenotemark{l}{\citet{li2025arXiv250417034L, zhu2025MNRAS.544L.139Z}}$\,\cdot$
\tablenotemark{m}{\citet{srinivasaragavan2025arXiv251210239S}}$\,\cdot$
\tablenotemark{n}{\citet{GCN.44071}}$\,\cdot$
\tablenotemark{o}{\citet{schneider2026arXiv260420346S}.}

\end{deluxetable}


In part due to the small number of well-observed sub-luminous LGRBs, their underlying physical connection to classical high-luminosity LGRBs remains unresolved. Several scenarios have been proposed: choked jets, where a jet choked in the stellar envelope or the circumstellar material (CSM) transfers energy to a wide-angle cocoon \citep[][]{nakar2012ApJ...747...88N, nakar2015ApJ...807..172N}; baryon-loaded (``dirty") outflows, in which a successful jet is mass-loaded and thus has a low initial Lorentz factor \citep[$\Gamma\lesssim10$;][]{paczynski1998ApJ...494L..45P, dermer1999ApJ...513..656D,  mazzali2006Natur.442.1018M}; low-power jets from alternative central engines such as protomagnetars or low-angular-momentum accretion \citep{metzger2011MNRAS.413.2031M, irwin2016MNRAS.460.1680I}; and off-axis relativistic jets, where a standard LGRB jet is viewed significantly off-axis and thus dimmed by relativistic beaming \citep{granot2002ApJ...568..820G, vanerten2010ApJ...722..235V, sato2021MNRAS.504.5647S}. 

The study of sub-luminous LGRBs has been reinvigorated by the launch of soft wide-field X-ray missions such as the Einstein Probe Wide-field X-ray Telescope \citep[EP-WXT;][]{yuan_ep_mission,yuan_ep_science} and the ECLAIRs camera onboard the Space-based multi-band Variable Object Monitor \citep[SVOM;][]{wej2016arXiv161006892W, atteia2022IJMPD..3130008A}. 
EP and SVOM have already identified a number of nearby, sub-luminous, soft-spectrum GRB-SNe (Table~\ref{tab:llgrb_sample}; e.g., \citealt{rastinejad2025ApJ...988L..13R, srinivasaragavan2025ApJ...988L..60S, srinivasaragavan2025arXiv251210239S, jiang2025ApJ...988L..34J, li2025arXiv250417034L}). In parallel, wide-field optical surveys are now capable of finding afterglows independently of a gamma-ray trigger \citep{2013ApJ...769..130C, 2015ApJ...803L..24C, 2017ApJ...850..149S, 2017ApJ...845..152B, 2019A&A...621A..81M, 2020ApJ...904..155A, Ho2020ApJ...905...98H}, enabling the discovery of events whose prompt emission is below the trigger threshold for high-energy satellites \citep{2013ApJ...769..130C, Ho2020ApJ...905...98H, 2022ApJ...938...85H,lipunov2022MNRAS.516.4980L, xu2023A&A...679A.103X,2025MNRAS.537.2362P, 2025ApJ...985..124L, srinivasaragavan2025MNRAS.538..351S}. 

AT\,2026fgk is one of the most recent afterglows blindly identified in optical-survey data. It was confirmed to be associated with long-duration GRB\,260310A \citep{GCN.43974,oniell2026TNSTR1001....1O_tns}. Optical spectroscopy \citep[][]{GCN.43977, GCN.43984, GCN.43986} confirmed \z~=~0.153 ($d_{\rm L}$\,=~762~Mpc)\footnote{We assume $\Omega_M = 0.3$, $\Omega_\Lambda= 0.7$, $h$ = 0.7 \citep{planck2020A&A...641A...6P} throughout this work, implemented using \textsc{Astropy} \citep{astropy1}.} and established the presence of a Ic-BL SN, making this one of the closest sub-luminous LGRBs detected this decade, and by far the most nearby afterglow identified by an optical survey. 
In this paper, we present detailed, early (up to 50\,d post-trigger) multi-wavelength data of AT\,2026fgk. 
We present the data in Section~\ref{sec:observations}, and the inferred properties of the supernova in Section~\ref{sec:analysis-supernova-properties}. In Section~\ref{sec:comparison}, we compare the properties of AT\,2026fgk to other GRBs and Ic-BL SNe. In Section~\ref{sec:rates} we use the ZTF flux-limited experiment to infer the volumetric rate of similar events. We measure the host galaxy properties in Section~\ref{sec:host_properties}. We perform basic modeling and discuss possible physical origins of AT\,2026fgk in Section~\ref{sec:discussion}, and summarize in Section~\ref{sec:summary}. Throughout, magnitudes are reported in the AB system.

\section{Observations}
\label{sec:observations}

\subsection{Prompt Emission}
\label{sec:prompt}

\begin{figure}[!ht]
    \centering
    \includegraphics[width=1\linewidth]{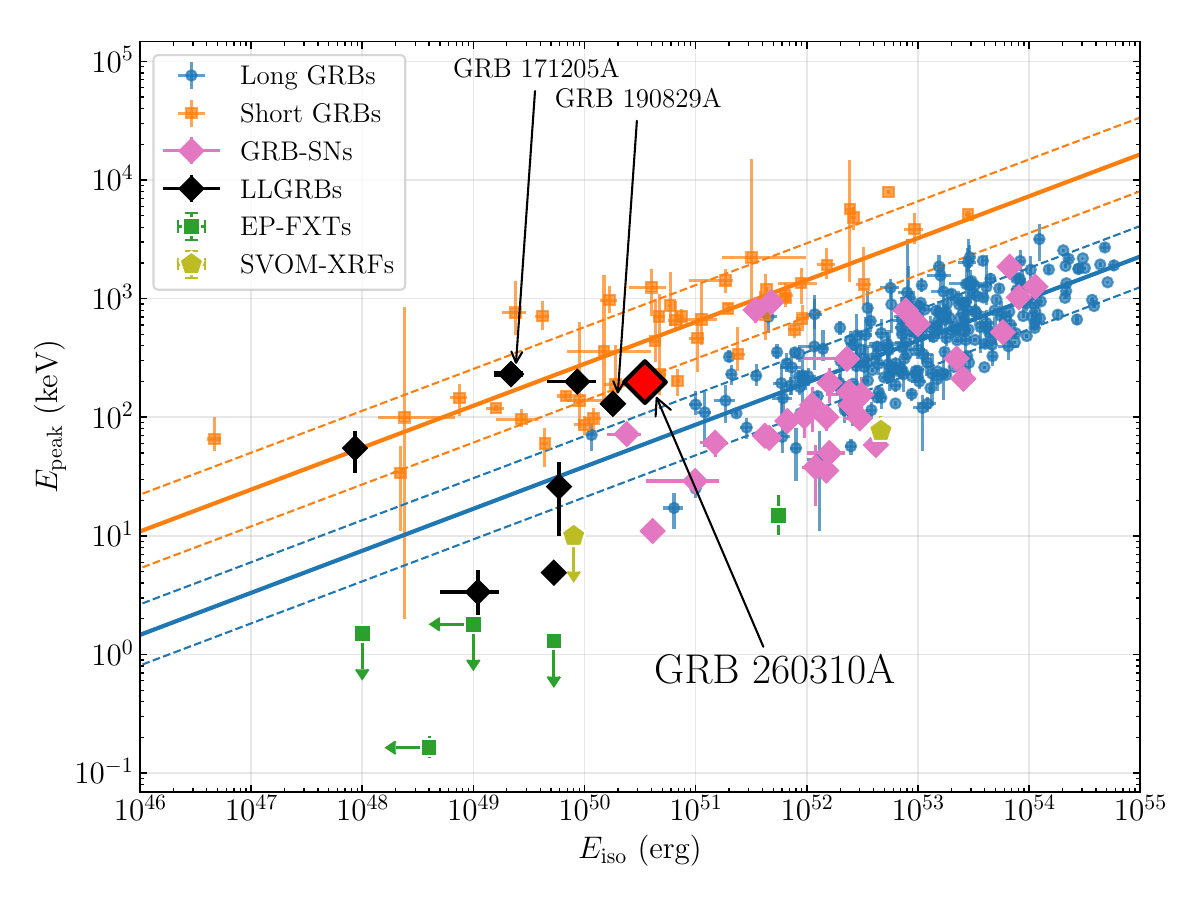}
    \caption{GRB 260310A (\emph{red diamond}) in the \epeak-\eiso plane (Amati relation). \emph{Blue} and \emph{orange} lines show the best-fit correlations and the 3$\sigma$ confidence intervals for long and short GRBs. GRB-SNe (\emph{magenta diamonds}; \citealt{cano2017AdAst2017E...5C} and \citealt{minaev2020MNRAS.492.1919M}), EP-FXT XRF transients (\emph{green squares}; \citealt{sun2025NatAs...9.1073S, jiang2025ApJ...988L..34J, GCN.39594, li2025arXiv250417034L, srinivasaragavan2025arXiv251210239S, GCN.44071}), SVOM-detected XRFs (\emph{yellow pentagons}; \citealt{schneider2026arXiv260420346S}), and sub-luminous GRBs (\emph{black diamonds}) are shown for comparison, with GRBs 171205A and 190829A labeled--see Table~\ref{tab:llgrb_sample} for a breakdown and references.}
    \label{fig:amati}
\end{figure}

GRB\,260310A was detected by the \textit{Fermi} Gamma-ray Burst Monitor \citep[GBM;][]{meegan2009ApJ...702..791M} at 04:56:38 UTC on 2026 March 10 \citep[hereafter $T_0$;][]{GCN.43951} and by \textit{AstroSat}/CZTI \citep{GCN.43958}. From \textit{Fermi}/GBM, the burst has a duration of \tnity$= 57.3 \pm 9.2$\,s and a fluence of $5.32 \pm 0.10 \times 10^{-6}$\,\erg\pcmsq\ in the 10--1000\,\kev band \citep{GCN.43975}. At $z=0.153$ (Section~\ref{sec:ztf}), the isotropic-equivalent energy is $E_{\rm \gamma,iso}\,=\,(3.44\,\pm\,0.24)\,\times\,10^{50}$\,\erg\, and the peak energy is $\text{\epeak}\sim190$\,\kev ($k$-corrected). The low \egiso\ and $L_\mathrm{\gamma,\mathrm{iso}}$ make \thisgrb\ sub-luminous compared to cosmological GRBs \citep[Figure~\ref{fig:amati}; e.g.,][]{liang2007ApJ...662.1111L, virgili2009MNRAS.392...91V, bromberg2011ApJ...739L..55B, cano2011ApJ...740...41C}. 

\subsection{Discovery of AT 2026fgk and Association with GRB 260310A}

\subsubsection{GOTO}

The Gravitational-wave Optical Transient Observer (\citealp[GOTO;][]{steeghs2022MNRAS.511.2405S, dryer2024SPIE13094E..1XD, lyman2026arXiv260302330L}, Steeghs et al., in prep.) is a robotic telescope network distributed across two observing sites: the Roque de los Muchachos Observatory on La Palma, Canary Islands, and Siding Spring Observatory in New South Wales, Australia. Each site comprises two mounts, each with eight 0.4\,m telescopes equipped with $\sim$50\,Megapixel CCDs, whose field of view slightly overlaps, providing a continuous sky footprint $\sim$45$^\circ$. GOTO responded autonomously to the GRB\,260310A \emph{Fermi}/GBM alert, with targeted imaging beginning $\sim$12\,minutes post-trigger. Difference imaging was performed against deeper pre-trigger site-specific template observations using a modified implementation of \textsc{hotpants}\footnote{\url{https://github.com/Lyalpha/hotpants}}\citep{becker2015ascl.soft04004B}. Following this process, AT\,2026fgk was detected in the difference images at $>5\sigma$ level. Forced photometry was subsequently performed using a forced photometry service (Jarvis et al. in prep.), and the resulting values are those presented here and shown in Figure~\ref{fig:optical_lc}.

AT\,2026fgk was discovered by GOTO on 2026 March 10 05:14 UTC (MJD~61109.22; \tzero+0.012\,d or 17 minutes), at a position (J2000) R.A. =  \ra{14}{37}{16.14}, Dec. = \dec{71}{50}{30.31}, corresponding to a Galactic latitude $b=+42.97$\,deg, and discovery magnitude of GOTO~$L = 18.84\pm0.06$\,mag \citep[$\lambda_{\rm eff}$=5325.68\AA, see][for details]{steeghs2022MNRAS.511.2405S}. The internal source name was GOTO26buh. The detection was reported to the Transient Name Server \citep{oniell2026TNSTR1001....1O_tns}.


\subsubsection{ATLAS}

Following GOTO, the next detection came from the Asteroid Terrestrial-impact Last Alert System \citep[ATLAS;][]{tonry2018PASP..130f4505T_atlasfp}. 
The ATLAS system comprises five wide-field optical telescopes. Four of the five telescopes (those situated in Haleakala and Mauna Loa in Hawaii, El Sauce in Chile, and Sutherland in South Africa) have identical designs, consisting of 0.5\,m aperture telescopes equipped with broad-band {\it cyan} ($c$) and {\it orange} ($o$) filters \citep{tonry2018PASP..130f4505T_atlasfp}. The fifth unit (located in Tenerife in Spain) is composed of four modules, each of which contains four commercial telescopes that combine to provide performance equivalent to a 56\,cm aperture telescope \citep{Licandro2023, Licandro2025}. All telescopes possess the same broad-band {\it wide} ($w$) filter \citep{Tonry2025}.

All ATLAS data is processed following the same standard pipeline. Target images are astrometrically and photometrically calibrated using RefCat2 \citep{tonry2018ApJ...867..105T} before template subtraction is performed utilizing deep reference sky images. PSF photometry is then performed on these difference images \citep[as outlined by][]{tonry2018ApJ...867..105T}. The ATLAS Forced Photometry Server\footnote{\url{https://fallingstar-data.com/forcedphot}} provides public access to ATLAS forced photometry for any sky location; however, it does not yet include $w$-band data.

Combined, the ATLAS telescopes robotically survey the night sky every $\approx 12 - 36$~hours in a dithered $4 \times 30$\,s exposure pattern (spaced across a 1\,hr period). 
ATLAS first detected the optical counterpart to GRB\,260310A on MJD~61109.50 ($T_0 + 7$\,hr) in the $o$-band, and continued to significantly detect the transient across $c$, $o$ and $w$ until MJDs~61124.5, 61125.5 and 61126.1, respectively (see Table~\ref{tab:opticalphotometry} and Figure~\ref{fig:optical_lc}).


\subsubsection{LAST}

AT\,2026fgk was first formally associated with \thisgrb by the Large Array Survey Telescope \citep[LAST;][]{ofek2023PASP..135f5001O, benami2023PASP..135h5002B, Konno2026}. LAST automatically triggered observations of the field of \thisgrb, initiating tiling observations at $T_0 + 0.486$\,d (11.7\,hr) and covering $\sim97\%$ of the localization probability area. The data was reduced using the LAST real-time pipeline \citep{2023PASP..135l4502O, Konno2026} and transients were searched using ZOGY and \textit{Translient} \citep{ofek2014ascl.soft07005O, 2024AJ....167..281S} algorithms and photometrically calibrated using the transmission fitting method \citep{garrappa2025A&A...699A..50G}. The optical counterpart was detected within the covered field over several epochs, and exhibited a rapid decay, with the last non-detection at $T_0-0.089$\,d ($\sim2.15$\,h; MJD~61109.12; $m_{\rm{lim}} = 20.74$\,mag). Based on observations at \tzero$+0.585$\,d (14\,hr), LAST reported the association of the optical transient with \thisgrb \citep{GCN.43974}. 

\subsubsection{ZTF}
\label{sec:ztf}

AT\,2026fgk was serendipitously detected by the Zwicky Transient Facility \citep[ZTF;][]{zbuilderbellm2019PASP..131a8002B, zbuilderdekany2020PASP..132c8001D, zbuildergraham2019PASP..131g8001G}, which operates on the Palomar 48-inch (P48) telescope with a 47~deg$^2$ field-of-view camera, surveying the entire northern sky on a cadence on $\sim$2~d \citep{zbuilderbellm2019PASP..131f8003B}. On any given night, ZTF produces a real time alert stream based on image differencing with \textsc{ZOGY} \citep{zackay2016ApJ...830...27Z} against coadded reference templates \citep[][]{zbuildermasci2019PASP..131a8003M}, releasing upwards of $10^{5}$ alerts \citep{patterson2019PASP..131a8001P}. 

AT\,2026fgk was flagged as a candidate of interest by a custom fast-transient pipeline (e.g., \citealt{Ho2020ApJ...905...98H}) for filtering alerts \citep[][]{zbuilderduev2019MNRAS.489.3582D, zbuilderdekany2020PASP..132c8001D, zbuilderduev2020AAS...23538608D, zbuilderduev2021arXiv211112142D}, with no knowledge of the GRB association at the time of flagging.  
The first ZTF detection of AT\,2026fgk (ZTF26aakjzdt) was on MJD~61110.405 ($T_0+$1.2\,d), yielding $g = 18.09 \pm 0.02$\,mag, followed 2.1~hr later by $r = 17.68 \pm 0.02$\,mag, and was preceded by a ZTF forced-photometry non-detection at $g>20.40$\,mag on MJD 61108.23 ($T_0-0.97$\,d). After correction for Galactic extinction \citep[$E(B-V) = 0.030$\,mag;][]{schlegel1998ApJ...500..525S} using the \citet{schlafly2011ApJ...737..103S} law ($A_g$\,=\,0.099\,mag, $A_r$\,=\,0.069\,mag), the intrinsic color at first detection in ZTF is $(g-r)_0 = 0.38 \pm 0.03$\,mag. The non-detection limit and the first detection implied a rise of $\gtrsim 1$\,mag\,d$^{-1}$ in $g$; together with its red color, this prompted high-priority flagging by the fast-transients filter. We present forced photometry on the ZTF images \citep{2023arXiv230516279M} in Table~\ref{tab:opticalphotometry} and Figure~\ref{fig:optical_lc}.

\subsection{Follow-up Observations: Optical Photometry}
\label{sec:optical-photometry}

\begin{figure*}
    \centering
    \includegraphics[width=1\linewidth]{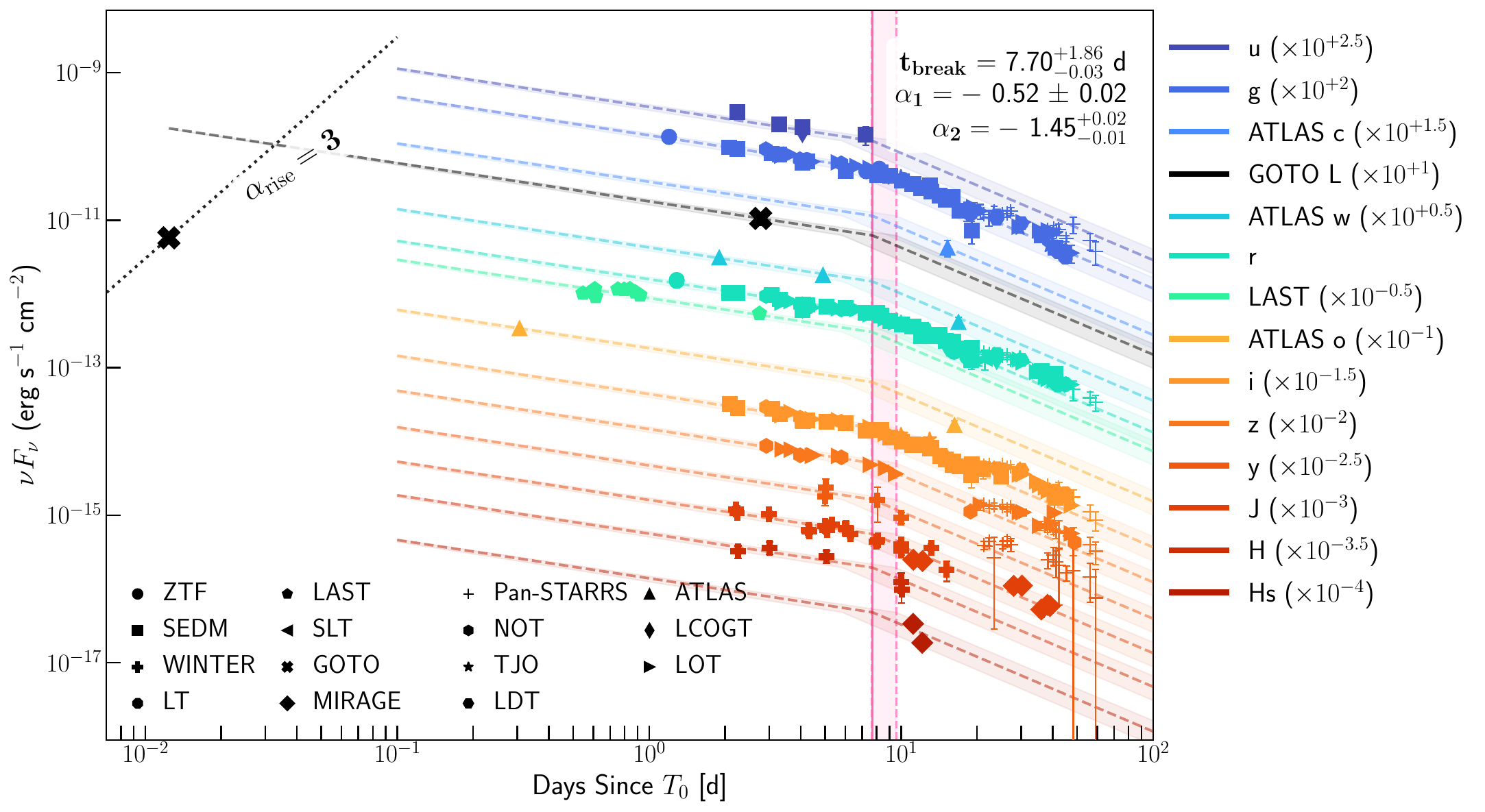}
    \caption{Multi-band optical and NIR light curves of GRB\,260310A / AT\,2026fgk, spanning from $\sim$17 minutes (first GOTO detection) to $\sim$+50~d post trigger. Filled symbols denote photometry from this work (ZTF, SEDM, WINTER, LT, LAST, GOTO, MIRAGE, NOT, TJO, ATLAS, LOT, LDT, LCOGT, and SLT). We extrapolate the pre-jet-break power-law in GOTO $L$ to the first GOTO epoch to demonstrate there is a brightening in GOTO $L$, indicating that the first epoch is pre-peak, and lies significantly below the extrapolated decay. We also plot a $t^{+3}$ rising power-law anchored to the GOTO-$L$ observation on the rise to predict the time of peak, $t_{\rm peak}$, based on \citet{sari1998ApJ...497L..17S, granot2002ApJ...568..820G}. Bands are labeled in the legend and offset vertically by factors indicated. Dashed lines show the best-fit smoothly broken power-law (see Section~\ref{sec:analysis-supernova-properties} for details); $\alpha_1 = -0.52 \pm 0.02$ (std) is the average pre-break index calculated from individual bands, $\alpha_2 = -1.45^{+0.02}_{-0.01}$ is the post-break index, with a break time $t_{\rm b} = 7.70^{+1.86}_{-0.03}$\,d (vertical pink band). All data are corrected for Galactic extinction.}
    \label{fig:optical_lc}
\end{figure*}





\begin{figure*}
    \centering
    \includegraphics[width=1\linewidth]{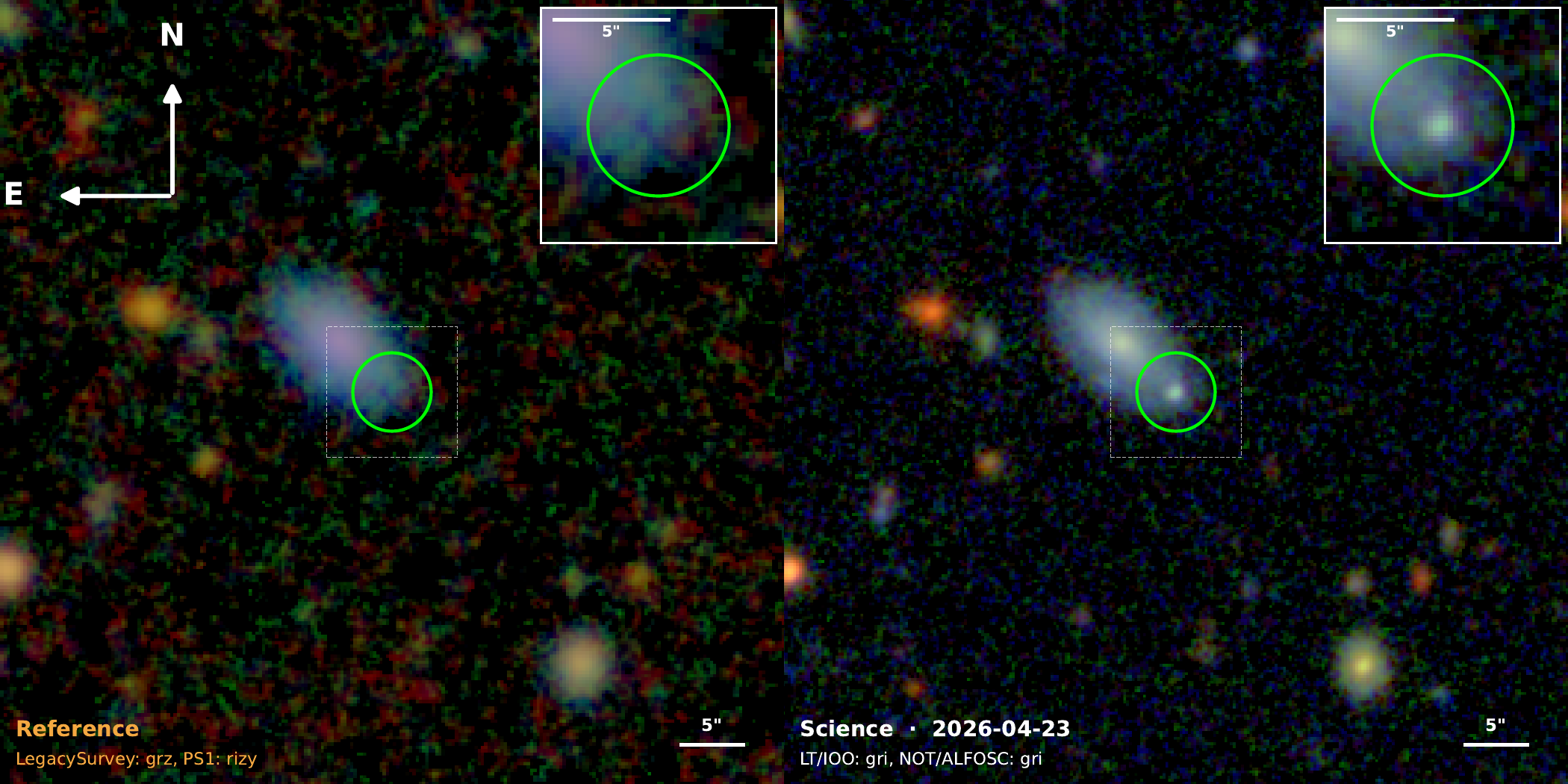}
    \caption{Color composite images of the field of \thisgrb/AT\,2026fgk. \emph{Left:} Pre-explosion reference image constructed from Legacy Survey ($grz$) and Pan-STARRS ($rizy$) imaging, combined into a single RGB frame. \emph{Right:} Science-epoch color composite assembled from $gri$ imaging from the Liverpool Telescope and Nordic Optical Telescope on 2026 April 23--24 ($\Delta t\approx44$\,d). Both panels are centered on the host, with a 10\arcsec inset; a green circle marks the transient position.}
    \label{fig:sedm_finder}
\end{figure*}

In this section we describe optical photometric follow-up observations obtained upon the identification of AT\,2026fgk as a strong afterglow candidate. A stacked image of \thisgrb\ is shown in Figure~\ref{fig:sedm_finder}, and the photometry can be found in Table~\ref{tab:opticalphotometry} and Figure~\ref{fig:optical_lc}.

Coordinated using the Fritz SkyPortal Marshal \citep[][]{fbuildercoughlin2023ApJS..267...31C, fbuildervanderwalt2019JOSS....4.1247V}, we obtained nightly $gri$ imaging with the Spectral Energy Distribution Machine \citep[SEDM;][]{sbuilderblagorodnova2018PASP..130c5003B, sbuilderrigault2019A&A...627A.115R, sbuilderkim2022PASP..134b4505K} on the 60-inch telescope at Palomar observatory (P60), spanning the full duration of our campaign, and SEDM $u$-band imaging for the first 5 nights. Image subtraction was performed using a custom pipeline, which uses \texttt{SExtractor} \citep{Bertin+1996_sourceextractor}, \texttt{SWarp} \citep{Bertin2010_swarp} \texttt{PSFex} \citep{bertin2011ASPC..442..435B} for source detection, image alignment, and cross-convolution of the science point spread function (PSF) with a reference image. For $r$ and $i$ band images we used Pan-STARRS \citep[PS;][]{chambers2019panstarrs1surveys}, for $g$-band images we used Legacy Survey Data Release 10 \citep[LS-DR10;][]{dey_legacysurveys_2019AJ....157..168D}, and for $u$-band images we used images from the Canada France Hawai'i Telescope Legacy Survey \citep[CFHT;][]{boulade2003SPIE.4841...72B, gwyn2012AJ....143...38G}. The convolved PSF is then fitted simultaneously to both images and subtraction performed after, following the methods of \citet{galyam2008ApJ...686..408G} and \citet{fremling2016A&A...593A..68F}.

Liverpool Telescope (LT; \citealt{Steele2004}) observations were taken with the IO:O camera in the SDSS $g$, $r$ and $i$ filters at multiple epochs---one at 6.77 days post GRB trigger, and then late-time observations after 40 days post trigger. Each frame was reduced by the automatic LT pipeline (providing bias subtraction, flat fielding, and astrometry). Same-filter frames at each epoch were stacked using \texttt{SWarp}. Image subtraction via PSF convolution, as well as final PSF photometry, were performed on each stack via a custom code utilizing \texttt{SExtractor} and \texttt{PSFex} . For $g$ and $r$ band, reference images for image subtraction were obtained from LS-DR10, and for $i$ band, the reference image used was taken from PS. The PS catalog was also used for the photometric calibration of the field. 

Imaging observations were carried out at the Nordic Optical Telescope (NOT), located at the Roque de los Muchachos Observatory (La Palma, Canary Islands, Spain). A total of seven epochs were secured, using the ALFOSC instrument equipped with the SDSS $ugriz$ filters, over the time period of approximately 3--50 days after the GRB. Typical seeing conditions were such that the transient was visible and well separated from the core of the host galaxy. The data were reduced following standard procedures using custom-built routines. Difference imaging was performed against PS1 and LS-DR10 references by convolving the sharper-seeing image with a Gaussian kernel to matched to the broader PSF.

We obtained observations with MEIA2 (2k\,x\,2k Andor iKon XL, pixel scale 0.6") camera on the 0.8\,m Joan Oro Telescope (TJO) located in the Catalan Pyrenees. In total, we obtained 5 epochs at 10, 13, 29.6, 38.7 and 40\,d post-trigger using SDSS $gri$ bands. 

We observed the target with the Large Monolithic Imager \citep[LMI;][]{ldt_12,ldt_14}, mounted on the 4.3m Lowell Discovery Telescope \citep[LDT;][]{ldt_12} at Lowell Observatory (PI: Rastinejad). Observations were conducted on multiple nights (see Table~\ref{tab:opticalphotometry} for a complete list), in $g$-, $r$- and $i$-band. Photometry was performed on images from TJO and LDT with the same methods as used on the NOT images.

AT\,2026fgk was observed with the Lulin One-meter Telescope (LOT) and the 40\,cm Seisei Lulin Telescope (SLT) at Lulin Observatory, Taiwan, as part of the Kinder collaboration \citep{chen2025ApJ...983...86C}. PSF photometry was performed at the transient’s position using \texttt{AutoPhOT} \citep{brennan2022A&A...667A..62B}. The photometry is calibrated against field stars from the ATLAS-RefCat2 catalog via MAST \citep{tonry2018ApJ...867..105T}. Details of the Lulin facilities and reduction procedures are given in \citet{aryan2025ApJS..281...20A}.

We obtained imaging of \thisgrb with the Sinistro cameras on the Las Cumbres Observatory Global Telescope Network \citep[LCOGT;][]{brown2013PASP..125.1031B} 1\,m telescopes at McDonald Observatory and Tenerife. We reduced the BANZAI-processed images and performed stacked photometry at the transient position. We adopted PSF-based photometry as our preferred measurement. The $r$ and $i$ calibration was done against PS, while the $u$-band point was calibrated using our synthetic/Gaia-based $u$-band approach. We also inspected one additional LCOGT epoch, but we do not include it in the attached table because it did not yield a reliable calibrated photometric measurement in our reduction.

We undertook multi-band $grizy$ \citep{Tonry2012} follow-up of AT\,2026fgk utilizing PS, commencing on MJD~61130.53 ($T_0 + 21.3$\,d). PS is a twin 1.8-m telescope system (dubbed PS1 and PS2), both situated atop Haleakala mountain on the Hawaiian island of Maui \citep{Chambers2016}. All observations of AT\,2026fgk utilized PS2, which possesses a pixel scale of 0.26" and a field of view of 7\,deg$^2$. We processed all images with the standard PS reduction pipeline, whereby target images are astrometrically and photometrically calibrated, before having a historical PS1 $3 \pi$ reference image subtracted \citep{Chambers2016, Magnier2020a, Magnier2020c, Waters2020}. This pipeline produces final, calibrated difference images, from which we measure PSF photometry of the transient \citep{Magnier2020b}. Our observations extend to MJD~61168.44, and the first 50\,d are summarized in Table~\ref{tab:opticalphotometry}.

\subsection{Nordic Optical Telescope - Polarimetry}
\label{sec:polarimetry}

\begin{figure}
    \centering
    \includegraphics[width=0.98\linewidth]{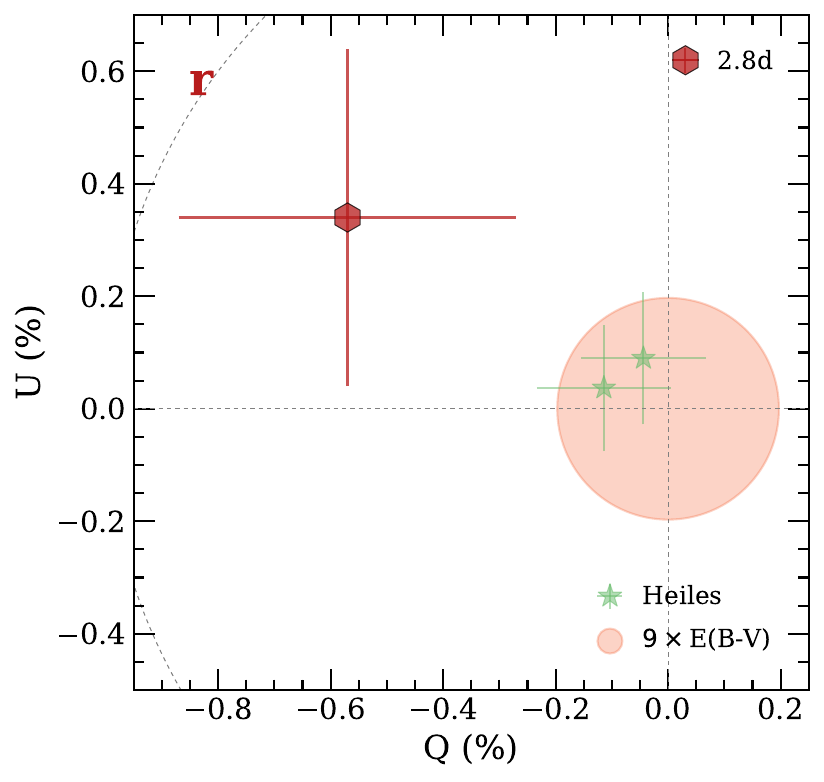}
    \caption{NOT/ALFOSC $r$-band polarimetry of GRB\,260310A / AT\,2026fgk. Stars in the Heiles catalog \citep{Heiles2000} within $5\degr$ and the maximum Galactic ISP \citep[$9\%\times E(B-V)$;][]{Serkowski1975} are shown. The event is consistent with MW ISP at $\sim2\sigma$.}
    \label{fig:impol}
\end{figure}

We also obtained $r$-band polarimetry starting on March 13, 2026 (MJD~61112.07; $T_0+2.8$\,d), over the half-wave plate angles $0.0$, $22.5$, $45.0$ and $67.5^\circ$. The data was reduced with a custom pipeline \citep{Pursiainen2025}, using a radius of $2\times\text{FWHM}$ to extract the extraordinary and ordinary beam photometry \citep{Pursiainen2023}. The pipeline also performs polarization bias correction following the canonical case of \citet{Plaszczynski2014}. Measured polarization is low at $P=0.60\pm0.30$\,\%, and consistent with $0\%$ at $2\sigma$ (see Figure \ref{fig:impol}). Milky Way interstellar polarization (ISP) appears low in the direction of the event. The maximum level of MW ISP \citep[$9\,\%\times E(B-V)$;][]{Serkowski1975} is $P_\mathrm{MW}=0.20$\,\%, supported by the two stars with $5^\circ$ in the Heiles catalog \citep{Heiles2000} that show low polarization consistent with $0\%$. As such the intrinsic polarization of the event is low, and consistent with Galactic ISP at $\sim2\sigma$.

\subsection{Infrared Photometry}

We observed the source in the near-infrared (NIR) with the Wide-Field Infrared Transient Explorer \citep[WINTER;][]{frostig_winter_2026}, mounted on a 1m telescope at Palomar Observatory. Images were taken nightly using both the J$-$, and Hs$-$band filters, and were reduced using the standard \texttt{mirar}\footnote{\url{https://github.com/winter-telescope/mirar}} \citep{mirar, mirar_zenodo}. Astrometric calibration was performed using Gaia position \citep{Gaia2021}, while photometric calibration was performed using 2MASS \citep{2mass}. No NIR reference images were available for subtraction, but we perform PSF photometry on the transient which remains spatially separated from the underlying host. 


We also observed the source as part of commissioning operations for the MDM Infrared Astronomy InGaAs Experiment (MIRAGE) instrument on the MDM\,1.3m telescope. MIRAGE is a new $YJH$-band imager using a Princeton 1280Scicam InGaAs device. Data were acquired on seven nights between UTC 2026-03-21 and 2026-04-18 (MJD~61120--61148). Each observation consisted of a sequence of dithered exposures in $J$ and $Hs$ bands. Dark-correction, flat-fielding, astrometric and photometric calibration against the 2MASS catalog \citep{2mass} were carried out using standard techniques \citep{De2020}. We report aperture photometry magnitudes matched to the seeing at the epoch of observation in Table~\ref{tab:opticalphotometry}. 

\subsection{Optical and NIR Spectroscopy}
\label{sec:spectroscopy}


We obtained spectroscopy using the Next Generation Palomar Spectrograph \citep[NGPS;][]{jiang2018SPIE10702E..2LJ, kasliwal2024TNSAN.340....1K} on the Palomar Observatory 200-inch telescope.
NGPS spectra were obtained on 2026-03-12, 2026-03-13, 2026-03-14 with a 1.5" wide slit and $2\times3$ spatial and spectral binning. NGPS operates with four arms and can obtain data from $\sim$310 nm -- 1040\,nm, with all observations taken using the 4 channels (U, G, R, and I). The data were reduced using standard methods with a custom pipeline developed for NGPS. 

\begin{figure}
    \centering
    \includegraphics[width=1\linewidth]{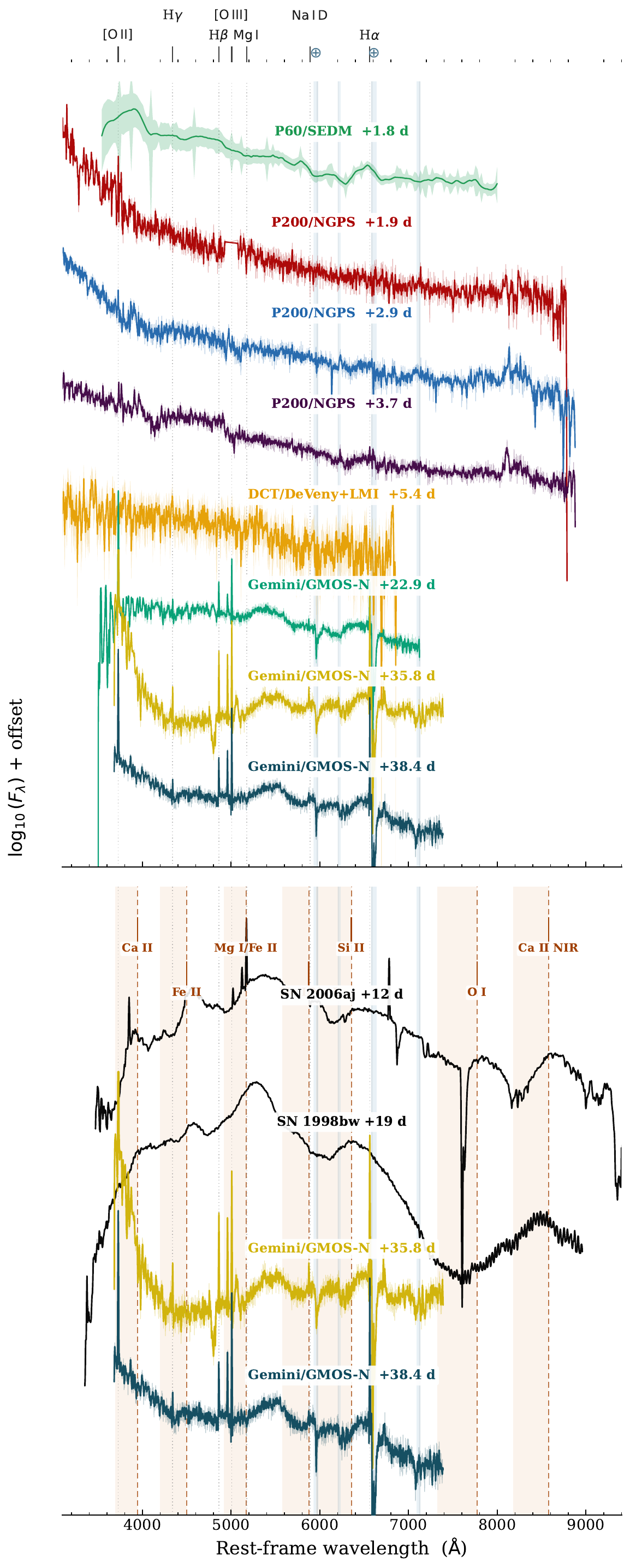}
    \caption{Optical spectral sequence of \thisgrb / AT\,2026fgk. Spectra are labeled by observing instrument and phase relative to \tzero. Spectra of SN\,1998bw and SN\,2006aj, well studied Type Ic-BL SNe, are included here for comparison in black. All spectra are smoothed and normalized for comparison. It is clear that no broad features appear until well into the decline phase, indicating the minimal contribution of the SN until late times (+23~d rest frame). Gemini/GMOS-N spectra at phases +22.8 and +35.8~d (rest frame) were observed at PAs to obtain host spectra, which suffered from differential slit loses that were corrected for in the final spectra.}
    \label{fig:specOpt}
\end{figure}


We observed the target with the DeVeney spectrograph \citep{ldt_12}, mounted on the 4.3\,m LDT at Lowell Observatory, on 2026-03-15 (PI: Stein). 
We observed with the 1.5" slit, 300/4000 grating, and took 3x600s exposures. The data was reduced using the standard DeVeney pipeline in \texttt{PypeIt} \citep{pypeit:joss_arXiv, pypeit:zenodo}. 


Three epochs of spectroscopy were obtained with the Gemini Multi-Object Spectrograph on the 8.1~m Gemini North Telescope \citep[GMOS-N;][]{hook2004PASP..116..425H} under a shared observing program (PI: Srinivasaragavan, O'Connor, and Tanaka). Observations were obtained at rest frame epochs $T_0+22.8$, +35.8, and +38.4\,d  (2026 April 5, 20, and 23 UTC; MJDs 61135.529, 61150.500, and 61153.503). All epochs used the B480+G5309 grating with a 1\arcsec\ wide longslit at two central wavelength settings (5900 and 6200\,\AA, $2 \times 1000$\,s each) to mitigate gaps between the three GMOS CCD detectors, yielding a combined wavelength range of approximately 3950--8215\,\AA\ ($\sim$3430--7130\,\AA\ rest frame at $z = 0.153$) at a resolution of $R \approx 1900$. A summary can be found in Table~\ref{tab:spectra}. Data was reduced using the Gemini \textsc{dragons} pipeline \citep{dragons2023RNAAS...7..214L, simpson_2026_19055103} with standard bias and flat-field corrections. For the two earlier epochs (April 5 and 20), the slit was oriented at PA = $47^{\circ}$ in order to place both the transient and host simultaneously on the longslit. This PA lay $\approx 120^{\circ}$ from the parallactic angle at the time of each observation (the $q \approx 169^{\circ}$ and $164^{\circ}$ respectively), so a significant component of atmospheric differential refraction (ADR) was directed perpendicular to the slit rather than along it. At airmasses of $\sim 1.7$, this produces differential slit losses that grow steeply towards blue wavelengths, substantially suppressing the blue continuum. The third epoch was oriented close to the parallactic angle (PA $\approx 142^{\circ}$, 16$^{\circ}$ from parallactic). We correct for the significant slit losses caused via a multiplicative throughput correction ---see Appendix~\ref{asec:gmos_adr_corr} for details\footnote{We note this did not improve the blue end of the spectrum for the spectrum at +35.8\,d rest frame.}. Figure~\ref{fig:specOpt} shows the optical spectral evolution.

We observed the position of GRB\,260310A with the Near Infra-Red Spectrograph (NIRSpec) on board the James Webb Space Telescope (JWST) under program GO 9254 (PI: Gompertz). Observations began at MJD 61133.931 ($T_0 + 24.7$\,d), using the 0.4" S400A1 fixed slit and the unfiltered prism for a total exposure time of 1.82\,hr in the range 0.6 -- 5.3\,$\mu$m (observed). Data was obtained from the Mikulski Archive for Space Telescopes (MAST) and were reprocessed with the NIRSpec pipeline using a offset of X~$= -4$ pixels and Y = +5 pixels due to the target not being properly centered in the slit.

The spectral evolution of AT2026fgk is shown in Figure~\ref{fig:specOpt}.

\subsection{X-ray Observations}
\label{sec:xray_obs}

We used several different X-ray observatories to follow the evolution of the
X-ray afterglow. In all cases, we fit the data to an absorbed power-law model, with Galactic absorption ($N_H$) set to $2.65\times10^{20}$\,cm$^{-2}$---calculated using the {\tt nh} utility in {\tt HEASoft}, which uses maps from \citet{hi4pi_maps} and abundances from \citet{wilms_abund}.

\subsubsection{Einstein Probe FXT}

\begin{figure}
    \centering
    \includegraphics[width=1\linewidth]{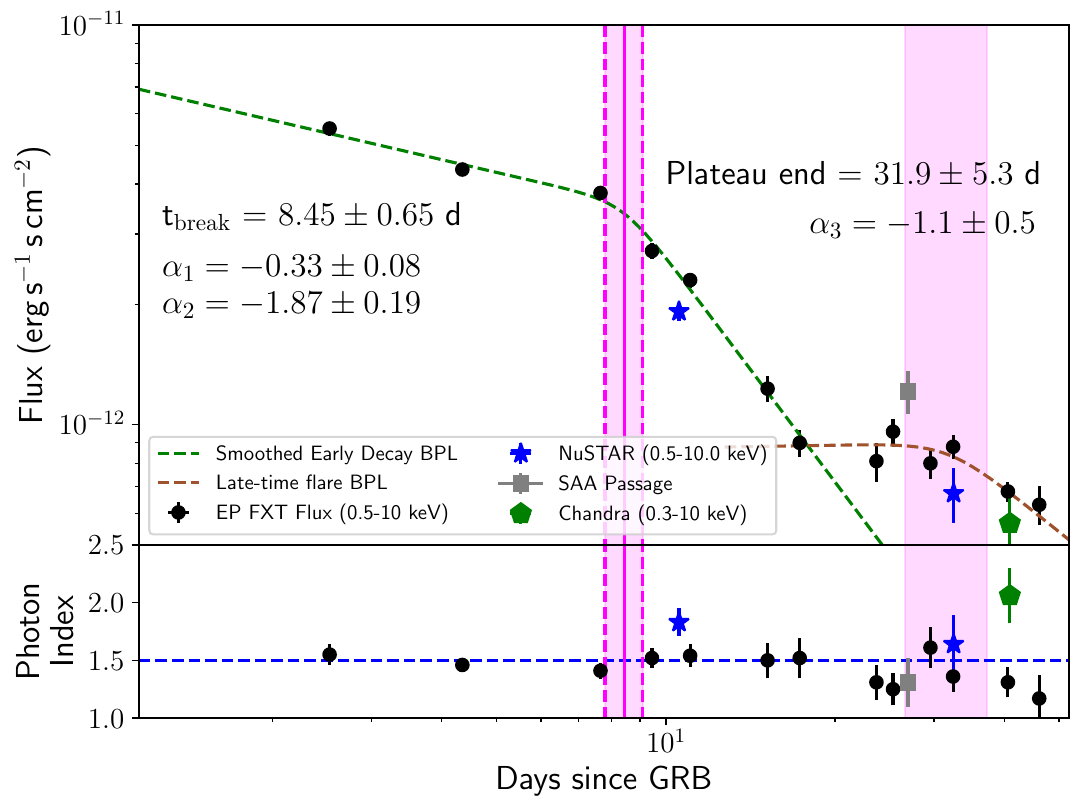}
    \caption{\emph{Top}: X-ray ($\sim0.5$--10\,keV) light curve of GRB\,260310A / AT\,2026fgk from EP/FXT (\emph{black circles}), NuSTAR (\emph{blue stars}), and Chandra (\emph{green pentagon}). The early decline is well described by a smoothly broken power law (\emph{green dashed}; $t_{\rm b} = 8.45 \pm 0.65$\,d, $\alpha_1 = -0.33 \pm 0.08\,, \alpha_2 = -1.87 \pm 0.19$). We observe a late time re-brightening/plateau until $31.9 \pm 5.3$\,d before a resumed decline with $\alpha_3 = -1.1 \pm 0.5$. \emph{Gray squares} mark South Atlantic Anomaly-affected epochs. \emph{Bottom}: The X-ray photon index $\Gamma$.} 
    \label{fig:xray_afterglow}
\end{figure}

We triggered the EP-FXT using the Target-of-Opportunity mode and observed the GRB at 13 total epochs, spaced between 2--5 days apart. Each observation lasted between 2 and 6 ks; this was suitable given the brightness of the X-ray afterglow.  Further follow-up was prevented by AT\,2026fgk moving into a region of the sky too close to the Sun for EP to observe safely. The observation logs are given in Table \ref{tab:xray}. We reduced the raw data from both the A and the B modules using the Einstein Probe {\tt fxttools} v1.20 \citep{zhao_epfxt_reduction}; this package also enabled the creation of response files ({\tt rmf} and {\tt arf}) using the EP CALDB. The observation with mid-time MJD 61124.3812 was cut short due to passage through the South Atlantic Anomaly, so we avoid using this observation in our analysis, although we present the results in Figure \ref{fig:xray_afterglow} and Table \ref{tab:xray} in the interest of completeness.

For each observation, we used {\tt XSPEC} \citep{arnaud_xspec} to analyze the count spectra. In order to ensure a statistically robust result, we binned the spectra to 5 counts per bin. We jointly fit the binned count spectra from both the A and B modules using an absorbed power-law model ({\tt tbabs*powerlaw}) and used Cash statistics \citep{cash_cstat,kaastra_cstat} to estimate the goodness-of-fit. This model gave good fits to the data, with reduced $\chi^2$ values of roughly 1. The fit statistics are reported in Table \ref{tab:xray}. To calculate the unabsorbed flux, we used the  {\tt cflux} convolution model in conjunction with an absorbed power-law; these fluxes are also reported in Table \ref{tab:xray}. 

\subsubsection{NuSTAR}
We triggered the Nuclear Spectroscopic Telescope Array \citep[NuSTAR; ][]{nustar_mission} as part of a Director's Discretionary Time (DDT; PI Waratkar) proposal. NuSTAR observed the afterglow of GRB\,260310A across two epochs (details on these observations are given in Table \ref{tab:xray}), for a total observational duration of roughly 55\,ks (23 + 12 + 20). We reduced the raw data using the NuSTAR Data Analysis Software ({\tt NuSTARDAS}), which applies calibration, screening, and attitude corrections using the NuSTAR CALDB \citep{nustar_calibration}. Source and background spectra, along with response files ({\tt rmf} and {\tt arf}), were extracted from both focal plane modules. The spectra were binned using the optimal binning scheme of \citet{kaastra_optimal_bin} to maximize spectral information while maintaining a minimum of one count per bin. We then used {\tt XSPEC} \citep{arnaud_xspec} to simultaneously analyze the FPMA and FPMB count spectra in the 3--79~keV energy range. We fit the count spectra using an absorbed power-law model ({\tt tbabs*ztbabs*powerlaw}) and used Cash statistics \citep{cash_cstat, kaastra_cstat} to estimate the goodness-of-fit. To calculate the unabsorbed flux in the 2--10\,keV band, we used the {\tt cflux} convolution model with the absorbed power-law. We report flux and photon index values in Table~\ref{tab:xray}---we restrict analysis to 2--10\,keV for comparison with EP\footnote{We note that a preliminary analysis of the first 7.5~ks of this observation was reported in \citet{GCN.44063}; here we report results from the full 23.4~ks dataset, which yields improved constraints on the spectral parameters. We also reported the results of the second observation in \citet{GCN.44278}.}.

\subsubsection{Chandra X-ray Observatory}
We analyzed late-time observations of this GRB through publicly-available Chandra data obtained as part of a DDT program (PI: Yang). This observation using ACIS-S started on 2026 April 20 at 01:47:42 UTC, corresponding to $T_0 + 40.87$\,d, and lasted roughly 5 ks. For the observation, chips S2 and S3 were turned on, and I3 was dropped. The target was centered on the S3 chip. We obtained the data from the Chandra data archive and reprocessed the data with {\tt CIAO v4.18.0} \citep{ciao_chandra} and {\tt CALDB v4.12.3}. We identify a source at the location of the afterglow, which matches the optical transient's location to within less than 1''. We extract the flux from this location using a circular aperture with radius 10'' (centered on the optical transient's coordinates) and a nearby source-free background region of radius 100''. We find a total of 107 net counts in our aperture. Using {\tt srcflux}, we find an unabsorbed flux of $5.66\pm0.93\times10^{-13}$\,erg\,cm$^{-2}$\,s$^{-1}$ (0.3--10\,keV).

We extract a spectrum for this source using {\tt specextract} and use {\tt XSPEC} \citep{arnaud_xspec} to fit this using an absorbed power-law (with a Galactic component, but not accounting for host galaxy absorption), after binning the spectrum to ensure a minimum of 2 counts per bin. We find a power-law index of $2.06\pm0.24$. This independent analysis of the data is consistent with the results reported in the GCN of \citet{yang2026arXiv260523818Y}. We are also able to identify two nearby sources reported therein that could be potential contaminants of the EP and NuSTAR analyses, although the flux obtained from Chandra is in good agreement with the EP flux from a near-contemporaneous observation, suggesting that any potential contamination in other X-ray data is negligible. 

The X-ray light curve of AT\,2026fgk is shown in Figure~\ref{fig:xray_afterglow}. 
We note that the photon indices and fluxes measured by EP/FXT, NuSTAR, and Chandra show differences that are larger than their statistical uncertainties, see Appendix~\ref{asec:xray_calib} for further details.

\subsection{Radio (cm- to mm-wavelength) Observations}
\begin{figure}[!ht]
    \centering
        \centering
        \includegraphics[width=\linewidth]{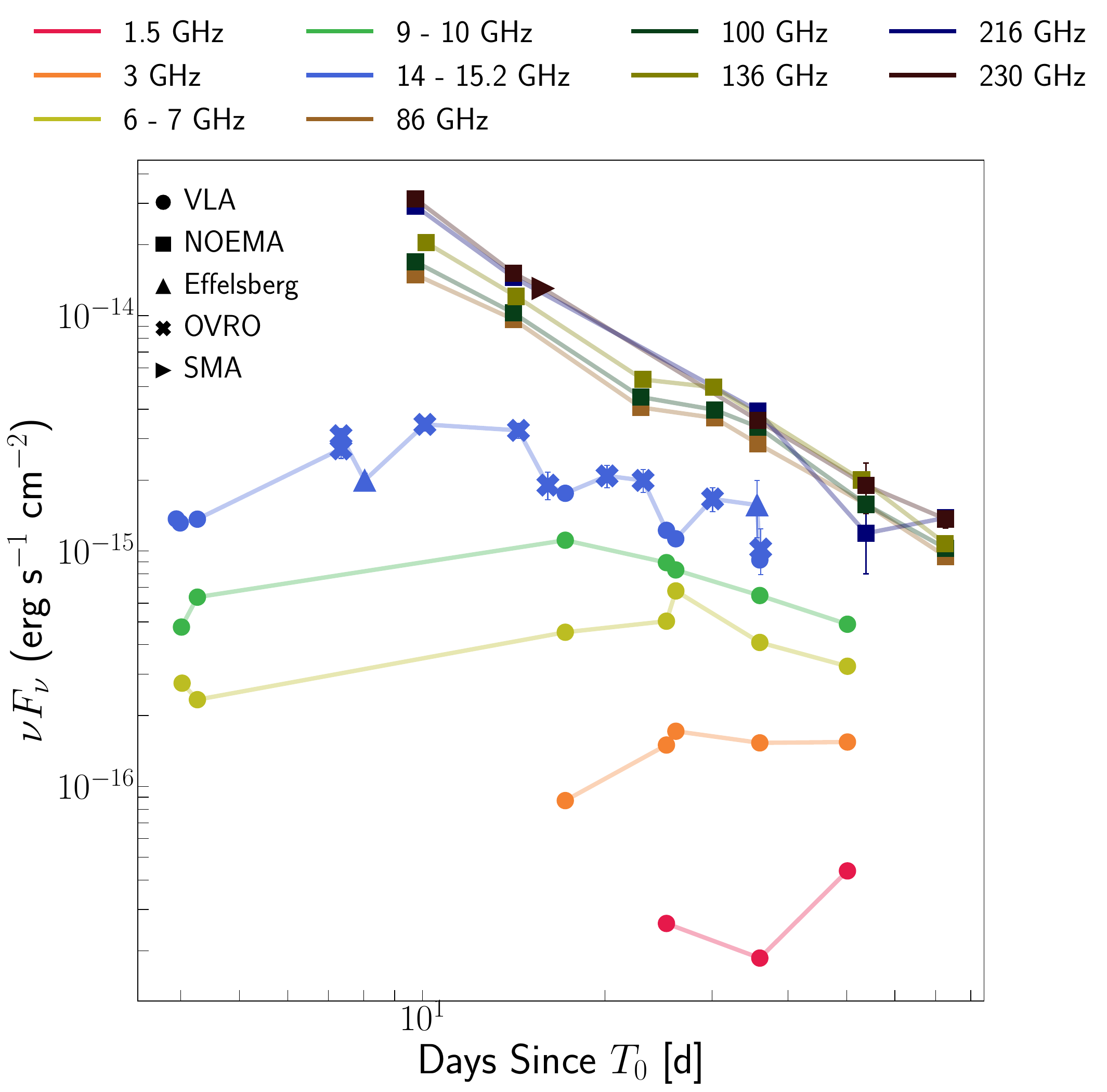}
        \caption{Radio (cm-to-mm) light curve of GRB\,260310A / AT\,2026fgk, color coded by frequency and with different symbols for different data sources.}
        \label{fig:radio_lc}
\end{figure}

The radio light curve of AT\,2026fgk is shown in Figure~\ref{fig:radio_lc}, and the data is provided in Table~\ref{tab:radio}. The radio counterpart of \thisgrb was first detected by the Arcminute Microkelvin Imager -- Large Array \citep[AMI-LA;][]{ami2008MNRAS.391.1545Z} at 15.2~GHz at \tzero$+\sim4$\,d \citep{GCN.44005}. Follow-up observations with the Karl G. Jansky Very Large Array (VLA) were obtained at \tzero$+4.26$\,d across 6, 10 and 15~GHz \citep{GCN.44045}, confirming a bright, spectrally evolving radio counterpart. 

\subsubsection{Very Large Array (VLA)}

Through a VLA DDT (Program ID 26A-608, PI Perley) and a regular cycle VLA proposal (Program ID 26A-177, PI Gompertz), we obtained several epochs at \tzero$+3.9$ to $+50$\,d with coverage from 1 -- 50~GHz \citep{2026GCN.44160....1P, 2026GCN.44235....1S}. All observations were taken with the array in the A-configuration, with 3C286 used as a flux and bandpass calibrator and a nearby source as a phase calibrator. Low-frequency observations were reduced without self-calibration using the Common Astronomy Software Applications \citep[\texttt{CASA};][]{2007ASPC..376..127M, 2022PASP..134k4501C} VLA Calibration\footnote{\url{https://science.nrao.edu/facilities/vla/data-processing/pipeline}} and Imaging\footnote{\url{https://science.nrao.edu/facilities/vla/data-processing/pipeline/vipl}} Pipelines. The \texttt{pwkit}/\texttt{imtool} package \citep{2017ascl.soft04001W} was used to determine the flux densities. For high-frequency observations (Ka-band and above on 2026-03-26, K-band and above on 2026-04-04, and Ku-band and above on 2026-04-15) there are significant phase errors evident in the basic calibrated data due to atmospheric variations; for these epochs we used Astronomical Image Processing System \citep[\texttt{AIPS};][]{2003ASSL..285..109G} to perform self-calibration (and flux calibration and bandpass calibration, and to determine the flux density). The results are summarized in Table~\ref{tab:radio}.

\subsubsection{Owens Valley Radio Observatory 40 m Telescope}


The California Institute of Technology’s Owens Valley Radio Observatory \citep[OVRO;][]{richards2011ApJS..194...29R} observed \thisgrb at 15~GHz on the 40~m Telescope on 9 occasions between March 17th and May 15th, 2026. A correction had to be made for a steep-spectrum, non-varying  confusing source with a 15~GHz flux density of $\sim$3.8~mJy located close to the half-power point in the 40~m Telescope beam. At this position in the 40~m Telescope beam, small offsets in pointing can introduce large fractional variations in the detected flux density of the confusing source. For this reason an uncertainty of $\pm$1~mJy has been added in quadrature to the measured uncertainties in the flux density of \thisgrb at 15~GHz.  

\subsubsection{Effelsberg 100 m Telescope}

Observations with the Effelsberg 100\,m radio telescope have been performed with the secondary focus receivers S28mm (10.45\,GHz), S20mm (14.25 and 16.75\,GHz) and S14mm (19.25, 24.75\,GHz on March 18, 19, 26, and April 14). The measurements were done with cross-scans in azimuth and elevation over the source position. Gaussian profiles were fitted to the scans, and the amplitudes taken as measure for the flux density of the sources. After correcting for the atmospheric attenuation and the gain-elevation effect of the telescope, the absolute calibration was done by comparison with observations of suitable flux density calibrators \citep[for details see e.g.,][]{kraus2003A&A...401..161K}.

\subsubsection{NOEMA}

The NOrthern Extended Millimeter Array (NOEMA) monitored \thisgrb under the project code E25AD in three spectral bands; 3~mm and 1.3~mm were monitored simultaneously with the recently implemented dual band mode, while~2 mm observations were performed in separate sub-projects. The young massive stars MWC349 and LKHA101 served as primary calibration sources. Monitoring started on March 19th 2026, on 22:41 UTC and is still ongoing. 

Calibration of the interferometric data was done with the CLIC software that is part of the GILDAS package\footnote{\url{https://www.iram.fr/IRAMFR/GILDAS}} in dual band mode; datasets from the PolyFiX I + II correlators were reduced independently. The calibrated interferometric visibilities from CLIC were analyzed in the complex UV plane with the MAPPING software (also part of GILDAS).

\subsubsection{SMA} 

We obtained two epochs of observations as part of the Submillimeter Array (SMA) POETS (Pursuit of Extragalactic Transients with the SMA; PI Berger) program on March 26, 2026 (MJD 61125.79; $\Delta t =13.69$\,d rest frame) which resulted in a detection and April 30, 2026 (MJD~61160.19; $\Delta t =44.22$\,d rest frame) which resulted in a non-detection.



Observations using the SMA were conducted on March 26, 2026 spanning UTC 12:53--19:42 (including calibration), with observations of the target AT\,2026fgk totaling 3.4 hours, spanning UTC 13:10--18:30. Six antennas were operational, tuned to a local oscillator frequency of 225.5\,GHz, with dual polarization spectral coverage from 209.5\,GHz to 221.5\,GHz in the lsb, and 229.5\,GHz to 241.5\,GHz in the usb. Weather conditions were very good with $\sim$1\,mm precipitable water vapor (median $\tau_{225}$ = 0.06), with good atmospheric phase stability. Complex gain calibrators were 1800+784 and 1459+716, the passband calibrator was BLLac, and MWC349A was used as the flux calibration standard. We found a strong detection of mm emission at the optically defined position for AT\,2026fgk, with a flux density of $5.67\pm0.23$\,mJy from the combined continuum data.

Followup SMA observations were conducted on April 30, 2026 spanning UTC 03:31--14:43
(including calibration). Due to early difficulty with atmospheric phase instability, usable observations of the target AT\,2026fgk only spanned UTC 08:43--14:07, with 3.41 usable hours on source. Four antennas were operational, with the same spectral coverage as above. Weather conditions were moderate with $\sim$2--3\,mm precipitable water vapor (median $\tau_{225}$ = 0.13), with acceptable atmospheric phase stability after UTC $\sim$08:38. Complex gain calibrators were 1800+784 and 1459+716, the passband calibrators were 3C84, 3C273, and 3C454.3, and MWC349A was used as the flux calibration standard. The observations resulted in no detection, with an rms sensitivity of 500\,$\mu$Jy/beam in the combined continuum.

\section{Supernova modeling}
\label{sec:analysis-supernova-properties}

\subsection{Light-curve Fitting}
\label{sec:lc_fit}
As shown in Figure~\ref{fig:optical_lc}, the UVOIR light curve of AT\,2026fgk is dominated by the afterglow. Nonetheless, we robustly detect a contribution to the emission from the underlying Ic-BL SN. In this section, we jointly model the optical light curve as a superposition of a broken power-law \citep[BPL;][]{beuermann1999A&A...352L..26B} afterglow and a SN component, using SN\,1998bw as a template, in order to infer the SN properties.

The fluxes of SN\,1998bw \citep[retrieved from][]{clocchiatti2011AJ....141..163C} were first redshift-corrected to the distance of \thisgrb and then combined with the afterglow model. To optimize the fit of this combined afterglow+SN model to the observed data, we introduced two SN transformation parameters: a timescale stretch factor ($s$) and a luminosity scaling factor ($k$). The BPL is formulated as
\begin{equation}
    F(t) = F_b \left[ \frac{1}{2} \left( \left( \frac{t}{t_b} \right)^{-y\alpha_1} + \left( \frac{t}{t_b} \right)^{-y\alpha_2} \right) \right]^{-\frac{1}{y}}.
\end{equation}

We performed a $\chi^2$ minimization across a fifteen-dimensional parameter space. The post-break decay index ($\alpha_2$), the break time ($t_b$), and the stretch factor ($s$) are shared across all bands. The pre-break decay index ($\alpha_1$), the normalization flux ($F_b$), and the luminosity scaling factor ($k$) are allowed to vary independently for each of the four optical bands. The analysis used photometry in the SDSS-$g$, $r$, $i$, and $z$ bands from ZTF, SEDM, LT, NOT, TJO, LOT, LDT, LCOGT, and SLT $g$- and $r$-bands due to their high temporal cadence. The resulting best-fit shared parameters --- post-break decay index $\alpha_2 = 1.45^{+0.02}_{-0.01}$, break time $t_b = 7.70^{+1.86}_{-0.03}$~days, and stretch factor $s = 1.23^{+0.37}_{-0.12}$--- are obtained from $\chi^2$ minimization of the afterglow+SN model. The per-band best-fit parameters are summarized in Table~\ref{tab:fitted_params_perband}.

\begin{table}[!h]
    \centering
    \caption{Best-fit per-band parameters from $\chi^2$ minimization of the afterglow+SN model to photometry (MW extinction-corrected, but not corrected for host-extinction): the pre-break decay index $\alpha_1$, the normalization flux $F_b$, the luminosity scaling factor $k$, and the $2\sigma$ confidence interval (CI) on $k$ from profile likelihood analysis.}
    \label{tab:fitted_params_perband}
    \begin{tabular}{lcccc}
        \toprule
        Band & $\alpha_1$ & $F_b$ (mJy) & $k$ & $2\sigma$ CI on $k$ \\
        \midrule
        $g$ & $0.510$ & $0.0836$ & $0.120$ & $[0.092,\ 0.152]$ \\
        $r$ & $0.518$ & $0.1110$ & $0.271$ & $[0.258,\ 0.289]$ \\
        $i$ & $0.547$ & $0.1218$ & $0.369$ & $[0.343,\ 0.380]$ \\
        $z$ & $0.524$ & $0.1514$ & $0.267$ & $[0.228,\ 0.486]$ \\
        \bottomrule
    \end{tabular}
\end{table}

\begin{figure*}[!ht]
    \centering
    \includegraphics[width=1\linewidth]{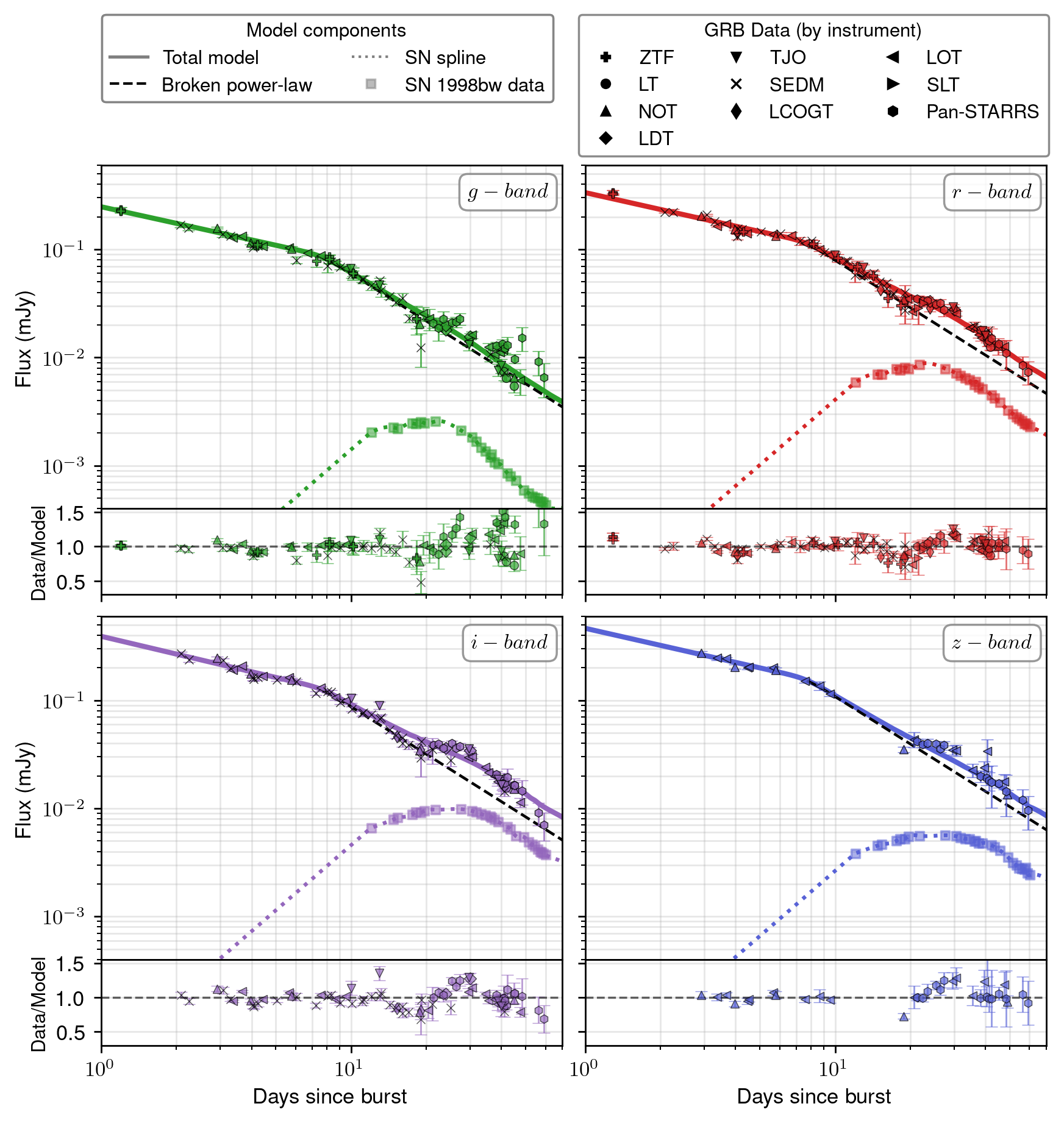}
    \caption{Optical light curves of \thisgrb / AT\,2026fgk in $g$, $r$, $i$, and $z$ bands, with photometry color-coded by band and marker shapes by instrument. In each band the best-fit model (\emph{solid}) is the sum of a BPL afterglow (\emph{dashed}) and a stretched+scaled SN\,1998bw template \citep[\emph{dotted}][]{clocchiatti2011AJ....141..163C}, obtained from a joint $\chi^2$ minimization across all four bands; scaled SN\,1998bw data points are shown as \emph{squares}. The emergence of the SN as a distinct bump above the fading afterglow is clearest beyond $\sim15$\,d. Sub-panels below each light curve show the residuals as the ratio of observed model flux, scattering about unit (\emph{dashed} line).}
    \label{fig:ag_sn_model}
\end{figure*}

To quantify the supernova contribution, we derived confidence intervals on the scaling parameter $k$ using a profile likelihood method. For each band, $\chi^2$  was evaluated over a grid of fixed $k$ values spanning $0.001$ to $0.8$, with  all other parameters freely minimized at each grid point. The profile $\Delta\chi^2$ was then computed relative to the global best-fit $\chi^2$, and $2\sigma$ confidence intervals were defined by the threshold $\Delta\chi^2 = 4.0$, corresponding to the 95.4\% confidence level for a single parameter of interest under Wilks' theorem \citep{wilksd543aecb-cd73-36d5-9101-f08a74f8e8c6}. The resulting intervals are listed alongside the best-fit parameters in Table~\ref{tab:fitted_params_perband}. In all four bands, the $2\sigma$ confidence intervals are strictly positive and exclude zero, ruling out the absence of a supernova component and providing robust evidence for a SN\,1998bw-like contribution to the late-time optical emission. The joint afterglow and supernova model fit is shown in Figure~\ref{fig:ag_sn_model}.

The SN associated with \thisgrb contributes negligibly to the afterglow in the optical light curve until $\gtrsim20$\,d (observer frame; $\sim$17~d rest frame), with the characteristic bump first emerging at this epoch. The extended afterglow dominance, compared to other GRB-SNe, is likely a consequence of the bright, slowly-declining afterglow---with a post-break decay index $\alpha = 1.45$---rather than an intrinsically slow SN. This is analogous to GRB\,221009A, where the extraordinarily luminous afterglow kept SN\,2022xiw subdominant in ground-based optical imaging well beyond 20~d \citep[rest frame;][]{fulton2023ApJ...946L..22F,levan2023ApJ...946L..28L, blanchard2024NatAs...8..774B, srinivasaragavan2023ApJ...949L..39S}. \citet{GCN.44124} first reported the spectroscopic confirmation of broad Type Ic-BL absorption features for AT\,2026fgk at $\sim T_0 + 17$\,d using GTC/OSIRIS; this observation was consistent with the first reported optical light curve evidence (a deviation from the afterglow power law due to an increase in brightness)  at the same epoch \citep[][]{GCN.44125}. This can be seen in Figure~\ref{fig:optical_lc} at $\sim20$\,d, where the $gri$ light curves begin to notably deviate from the $1.45$ afterglow power-law index.

The best-fit stretch factor $s = 1.23^{+0.37}_{-0.12}$ places the estimated rest frame SN peak at $\sim23$\,d ($\sim$27~d observer-frame), while the $r$-band luminosity scaling $k_r = 0.271^{+0.018}_{-0.013}$ corresponds to $\sim30$\% of the SN\,1998bw peak flux, yielding a peak absolute magnitude $M_r \approx -18.3$\,mag, derived from the SN\,1998bw reference $M_R = -19.07$\,mag \citep[][]{clocchiatti2011AJ....141..163C}.

\subsection{Spectral Properties}
\label{sec:nirspec}
\begin{figure*}
    \centering
    \includegraphics[width=0.8\linewidth]{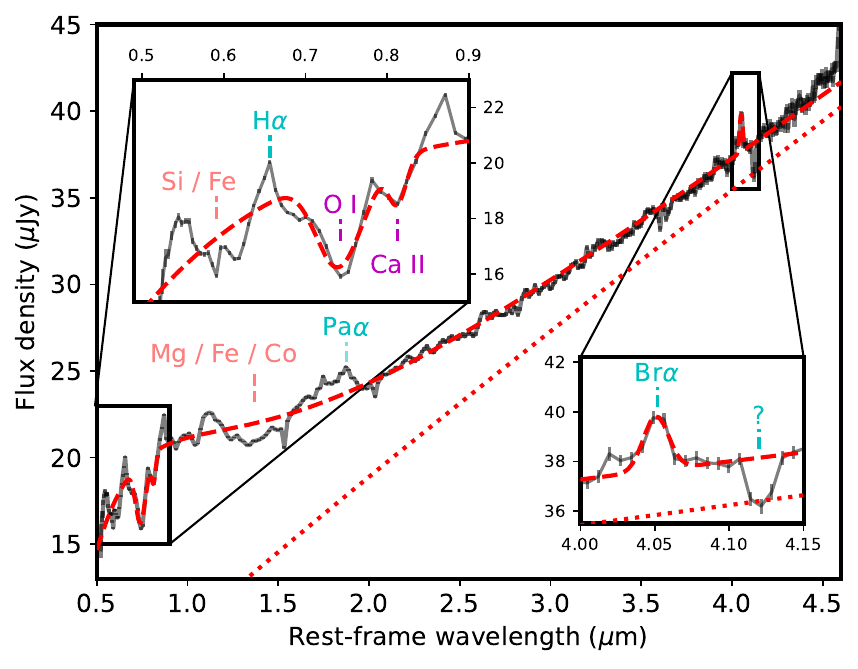}
    \caption{The JWST/NIRSpec spectrum of \thisgrb / AT\,2026fgk. The red end is dominated by a power-law from the GRB afterglow (red dotted line), while the blue end exhibits broad absorption and emission features characteristic of Ic-BL SNe. The continuum is reasonably approximated by a blackbody plus power-law model (red dashed line) which facilitates the identification of broad absorption features attributed to \ion{O}{1} and \ion{Ca}{2} (purple). Several other absorption complexes, likely associated with blended Mg-, Fe-, and Si-group transitions, are marked in red. Narrow H emission lines from the host galaxy are detected or tentatively identified in cyan. No obvious identification is found for the possible narrow absorption feature centered at $4.12$\,$\mu$m.}
    \label{fig:nirspec}
\end{figure*}

The NIRSpec spectrum (Figure~\ref{fig:nirspec}) was taken at 24.7\,d in the observer frame, corresponding to the peak of the SN light curve and equivalent to 21.4\,d in the rest frame. It is broadly characterized by a power-law-like profile in the red end (beyond $\sim$ 3\,$\mu$m) from the GRB afterglow with broad excesses at bluer wavelengths, consistent with a thermal SN component. We fit the data with a multi-component model comprised of a power-law + blackbody + Gaussians to represent broad and narrow absorption and emission features. We find that parameter uncertainties are implausibly low in many cases despite the intentionally approximate nature of the model. This is likely due to the high quality of the spectrum, where the mean signal-to-noise ratio per resolution element is 228. In order to provide a more reliable estimate of parameter uncertainties, we rescale them by a factor of $\sigma_r = \sigma_m \sqrt{\chi^2_{\rm red}}$, where $\sigma_r$ is the reported 1$\sigma$ uncertainty, $\sigma_m$ is the measured one, and $\sqrt{\chi^2_{\rm red}} = 6.8$ is the square root of $\chi^2$/dof, accounting for the underlying assumption when calculating covariance that $\chi^2_{\rm red} \approx 1$. The exception to this is the narrow line fits, which are far less sensitive to changes in the global model.

The continuum is well described by a power-law with an index of $F \propto \lambda^{0.907\pm0.014}$ and a blackbody with a rest frame temperature of $T = 6720 \pm 68$\,K and a radius of $r = (3.14 \pm 0.07) \times 10^{15}$\,cm. However, we caution that degeneracies exist between these fit parameters, and the blackbody is only an approximate description of the SN continuum, which is composed of multiple broad absorption and emission components. Taken at face value, these findings suggest an average expansion velocity of $v_{av} \approx 16,750$\,km\,s$^{-1}$ over the first 21.4 rest frame days.

Superimposed over the blackbody continuum are a number of broad absorption features, characteristic of Ic-BL SNe. Notably, we are able to fit Gaussians to broad features centered at $0.741 \pm 0.007$\,$\mu$m and $0.813 \pm 0.007$\,$\mu$m in the rest frame. We identify these as partially blended \ion{O}{1} and \ion{Ca}{2} absorption troughs, which are expected at rest frame minima of $0.777$\,$\mu$m and $0.850$\,$\mu$m, respectively. Their measured centroid blueshifts correspond to velocities of $v_{\mathrm{OI}} = 14,570 \pm 140$\,km\,s$^{-1}$ and $v_{\mathrm{CaII}} = 13,650 \pm 140$\,km\,s$^{-1}$, consistent with typical line velocities measured in Ic-BL SNe around peak \citep[e.g.][]{taddia2019A&A...621A..71T,finneran2025A&C....5200954F}. The Gaussian widths of the absorption profiles show $1\sigma$ velocity dispersions of $\Delta v_{\mathrm{OI}} = 11,600 \pm 1,400$\,km\,s$^{-1}$ and $\Delta v_{\mathrm{CaII}} = 4,020 \pm 1,700$\,km\,s$^{-1}$. The low value of the latter is likely due to degeneracy with the \ion{O}{1} feature as a result of their proximity. We therefore do not consider the velocity dispersions robust, but the blueshift of the line centroids and broadness of the features are sufficient to confirm expansion velocities typical of Ic-BL SNe with and without GRBs. Further absorption complexes and their possible contributing transitions are marked in Figure~\ref{fig:nirspec}.

We identify a number of narrow H emission lines from the host galaxy. The most robust of these is Br$\alpha$, at a measured rest frame wavelength of $4.051 \pm 0.001$\,$\mu$m, consistent with the expected rest frame wavelength of $4.052$\,$\mu$m. The measured significance of Br$\alpha$ is $6.8\sigma$ above the local continuum. We also note possible excesses at the positions of H$\alpha$ ($0.646$\,$\mu$m) and Pa$\alpha$ ($1.875$\,$\mu$m), but are unable to separate them from the much broader SN features in these regions (Figure~\ref{fig:nirspec}). The recovery of the H$\alpha$ line  would enable us to measure the optical extinction by taking the ratio of its flux with Br$\alpha$, assuming standard Case B recombination. However, fitting the prominent H$\alpha$ line in the GMOS 41.3\,d spectrum indicates a peak flux density of $\sim 0.9$\,$\mu$Jy, suggesting a contribution to the observed excess of at most $\sim 50$\%. This is consistent with our inability to isolate a narrow line component, compounded by the uncertain contribution of the complex continuum in this region. Comparing Br$\alpha$ from NIRSpec with H$\alpha$ from GMOS is not possible due to the different slit positions, aperture sizes and sensitivity curves, which would render any derived $A_V$ highly unreliable.

\subsection{Explosion Properties}
\label{sec:nimass}
The peak bolometric luminosity of stripped-envelope SNe is set, at first order, by the mass of the radioactive \niVI synthesized in the explosion \citep{arnett1982ApJ...253..785A, valenti2008MNRAS.383.1485V}. The rate of post-peak decline---and in particular the extended tail at late times---reflects the transparency of the ejecta to \gam-rays from the decay chain of \niVI to \coVI and \feVI, itself sensitive to \MEj and the density profile \citep[e.g.,][]{clocchiatti1997ApJ...491..375C, wheeler2015MNRAS.450.1295W}.

As the photometric coverage spans the GRB afterglow and the SN bump, we infer SN properties from a joint fit that models both the components together rather than from a SN-only fit to an afterglow-subtracted light curve. We use the modified Arnett implementation \citep[][see also \citealt{valenti2008MNRAS.383.1485V}]{arnett1982ApJ...253..785A} provided by \textsc{Redback} \citep{sarin2024MNRAS.531.1203S}, which couples the radioactive heating from \niVI\,$\to$\,\coVI\,$\to$\,\feVI decay chain to a one-zone diffusion calculation with the late time $\gamma$-ray leakage prescription of \citet{valenti2008MNRAS.383.1485V} and a temperature-floor photosphere \citep{nicholl2017ApJ...850...55N} radiating a blackbody SED. The free Arnett parameters are the nickel mass fraction, $f_{\rm Ni}$, the ejecta mass, \MEj, and a characteristic ejecta velocity, $v_{\rm ej}$; the optical opacity is fixed at $\kappa = 0.07$\,cm$^{-2}$g$^{-1}$ \citep[e.g.,][]{chugai2000AstL...26..797C,taddia2018A&A...609A.136T, tartaglia2021A&A...650A.174T, barbarino2021A&A...651A..81B}, and the $\gamma$-ray opacity $\kappa_\gamma=$0.03\,cm$^{2}$g$^{-1}$. The afterglow is modeled, as before, by a smoothly broken power-law parameterized by $\alpha_1, \alpha_2$ and break time $t_b$.

The afterglow parameters ($\alpha_1$, $\alpha_2$ and $t_b$) are anchored by the broken power-law fit to the same photometry presented in Section~\ref{sec:lc_fit}: we adopt Gaussian priors on $\alpha_1 \sim \mathcal{N}(-0.52,0.10)$, $\alpha_2 \sim \mathcal{N}(-1.45,0.20)$, and $t_b \sim \mathcal{N}(7.70,2.00)$, and the SN parameters are sampled from broad uniform priors. We sample the joint posterior with \textsc{bilby} \citep{ashton2019ApJS..241...27A} using the nested sampler \textsc{nessai} \citep{williams2021PhRvD.103j3006W} under a Gaussian likelihood evaluated in magnitude space.

The posterior medians and 68\% credible intervals on the SN parameters are shown in Table~\ref{tab:bpl_arnett_params} (fit is shown in Figure~\ref{fig:bpl_arnett}). $M_{\rm Ni}$ is a derived quantity, $M_{\rm Ni} = f_{\rm Ni}\,M_{\rm ej}$.

\begin{table}
    \centering
    \caption{Posterior medians and 68\% credible intervals for the joint Arnett SN
    and broken power-law afterglow fit.}
    \label{tab:bpl_arnett_params}
    \begin{tabular}{llc}
        \hline\hline
        Component & Parameter & Value \\
        \hline
        \multirow{4}{*}{SN (Arnett)}
            & $f_{\rm Ni}$  & $0.099 \pm 0.009$ \\
            & $M_{\rm ej}$  & $3.47^{+0.28}_{-0.27}\ \mathrm{M}_\odot$ \\
            & $v_{\rm ej}$  & $10{,}020^{+380}_{-330}\ \mathrm{km\,s}^{-1}$ \\
            & $M_{\rm Ni}$  & $0.344 \pm 0.016\ \mathrm{M}_\odot$ \\
        \hline
        \multirow{3}{*}{Afterglow (BPL)}
            & $\alpha_1$ & $-0.58 \pm 0.01$ \\
            & $t_b$      & $7.13 \pm 0.07\ \mathrm{d}$ \\
            & $\alpha_2$ & $-1.68 \pm 0.04$ \\
        \hline
    \end{tabular}
\end{table}


The fitted $v_{\rm ej} =10,020 \pm 380$\,\kms is $\sim$\,6700\,\kms lower than the SED-averaged velocity directly measured from our JWST NIRSpec observation ($v_{\rm av} \approx 16,750$\,\kms at $T_0+21.4$\,d). The Arnett $v_{\rm ej}$ is a one-zone effective velocity that sets the diffusion timescale, and is known to depart from the photospheric line velocity when the ejecta are stratified \citep[e.g.][]{dessart2016MNRAS.458.1618D, khatami2019ApJ...878...56K}; we therefore adopt the JWST line velocity as our photospheric velocity for the kinetic energy (\ek) budget, which is consistent with the population mean of peak-epoch line velocities for Type Ic-BL populations \citep[16,000 $\pm$ 1,100\,\kms;][]{srinivasaragavan2024ApJ...976...71S}. 

Adopting $v_{\rm ph} = v_{\rm av}= 16,750$\,\kms, the kinetic energy under the thin-shell prescription of \citet{valenti2008MNRAS.383.1485V} ($E_{\rm K} = \tfrac{1}{2}\,M_{\rm ej}\,v_{\rm ph}^2$) is \ek=$(9.7\pm0.8)\times10^{51}$\,\erg, while the homologous uniform-density model of \citet{lyman2016MNRAS.457..328L} ($E_{\rm K} = \tfrac{3}{10}\,M_{\rm ej}\,v_{\rm ph}^2$) gives \ek=$(5.8\pm0.5)\times10^{51}$\,\erg, where the uncertainty is propagated from the statistical uncertainty on \MEj.

We characterize the temporal evolution of the SN component using flux-threshold timescales measured from the scaled SN\,1998bw template (Figures~\ref{fig:ag_sn_model} and \ref{fig:comp_to_others}). The rise time $t_{\rm rise\,, 50}$ is defined as the time to peak from 50\% of the peak flux, with the fade time $t_{\rm fade\,, 50}$ similarly defined. The full width at half maximum is defined as FWHM $= t_{\rm rise\,, 50} + t_{\rm fade\,, 50}$; this parameter is primarily governed by the photon diffusion timescale and therefore encodes the ejecta mass and kinetic energy. 

The SN component peaks at $t_{\rm peak} \approx 22$--28\,d across $g$, $r$, $i$, and $z$ (observer frame), with rise times from 50\% peak flux to peak spanning $t_{\rm rise\,,50} \approx 11$--17\,d. These timescales are summarized in Table~\ref{tab:sn_timescales}. 

\begin{table}[!ht]
  \centering
  \caption{Rise and fade timescales of the \thisgrb\ SN component.
    All values are in rest frame days.}
  \label{tab:sn_timescales}
  \begin{tabular}{lccc}
    \hline
    Band &
      $t_{\rm rise\,,50}$\,[d] &
      $t_{\rm fade\,,50}$\,[d] &
      FWHM~[d]\\
    \hline\hline
    $g$ & 12.0 & 11.6 & 23.6 \\
    $r$ & 12.0 & 18.0 & 39.8 \\
    $i$ & 14.9 & 23.3 & 48.3 \\
    $z$ & 9.7 & 31.2 & 40.8 \\
    \hline
  \end{tabular}
\end{table}


\section{Comparison}
\label{sec:comparison}

\subsection{Prompt Emission}
\label{sec:comparison-prompt}

As shown in Figure~\ref{fig:amati}, sub-luminous LGRBs exhibit large heterogeneity in their positioning in the Amati \epeak--\eiso relation \citep{amati2002A&A...390...81A}. 
XRFs~020903 and 060218 are much softer than would be expected given their \eiso\ values, while GRB\,100316D follows the typical trend for LGRBs. By contrast, GRBs 980425, 031203, 171205A, 190829A, and the new event 260310A are all harder than classical LGRBs in the sense that their \epeak\ is higher than a classical LGRB of the same \eiso\ value. We discuss possible physical explanations for the diversity in Section~\ref{sec:discussion}. 

\subsection{Afterglow}


In addition to exhibiting large heterogeneity in their prompt emission properties (Section~\ref{sec:comparison-prompt}), sub-luminous GRBs also show diverse multiwavelength afterglow properties. In this section we compare the afterglow properties in each waveband to other LGRBs in the literature. Basic modeling of the afterglow+SN is presented in Section~\ref{sec:discussion}; more thorough modeling will be presented in a companion paper, along with our late-time multiwavelength data. 

\subsubsection{Optical}

As shown in Figure~\ref{fig:optical_lc}, the optical light curve of AT\,2026fgk shows a clear rise, then a power-law decay with a prominent break at $\sim7.70$\,d. The only other sub-luminous GRBs with early afterglow-dominated light curves are 020903 and 190829A. The optical light curves of the other events---GRB\,980425, 060218, 100316D, 171205A, 031203---are dominated by the SN and/or an early (first few days) thermal shock-cooling peak. 

\begin{figure}
    \centering
    \includegraphics[width=\linewidth]{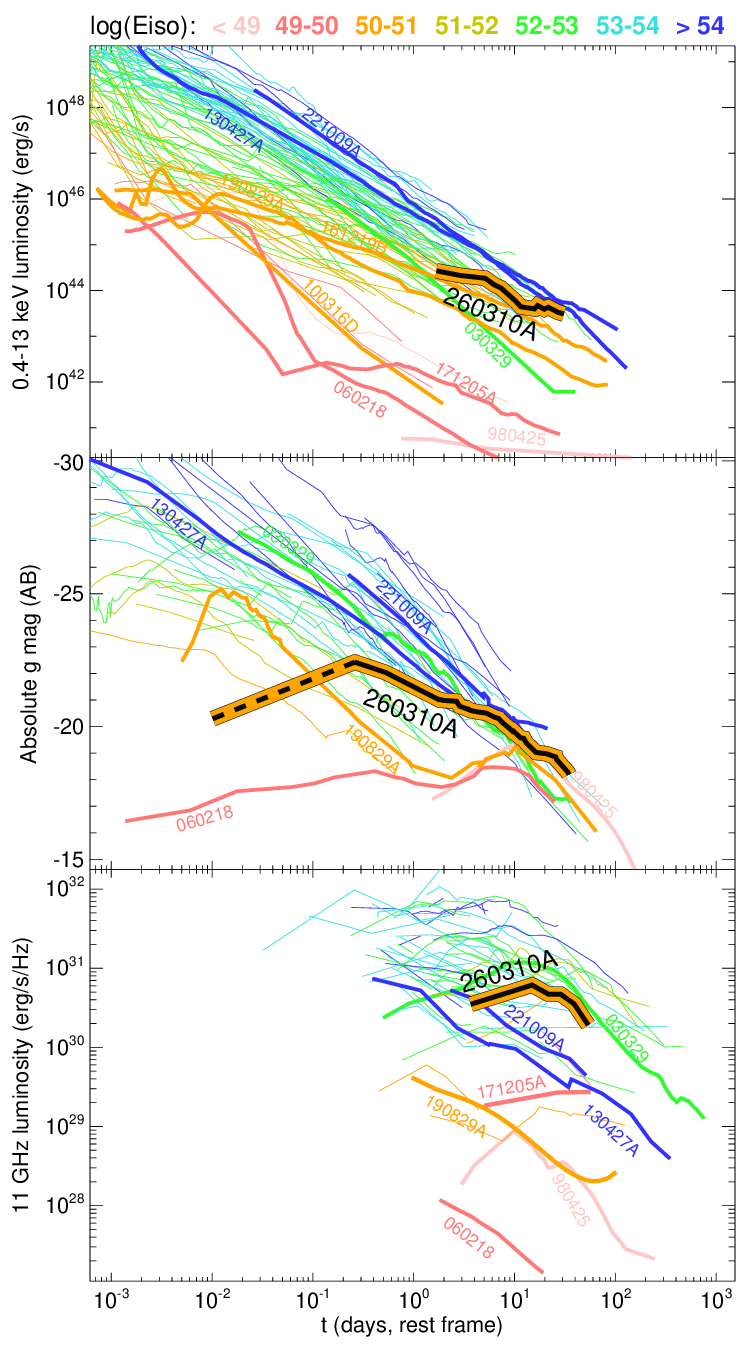}
    \caption{Comparison of the light curve of GRB\,260310A / AT\,2026fgk to other GRB afterglows at X-ray (\emph{top}), optical (\emph{middle}), and radio (\emph{bottom}) wavelengths. 
    X-ray light curves are drawn from the \emph{Swift}/XRT repository \citep{swiftrepo1_2007A&A...469..379E}, supplemented for individual events by GRB\,980425 \citep{pian2000ApJ...536..778P} and
    GRB\,030329 \citep{tiengo_grb030329_xray}. Optical light curves are taken from the compilation of \citet{kann2010ApJ...720.1513K}, supplemented by GRB\,980425 \citep{clocchiatti2011AJ....141..163C}, GRB\,060218 \citep{campana2006Natur.442.1008C}, GRB\,130427A \citep{perley2014ApJ...781...37P}, GRB\,171205A \citep{delia2018AA...619A..66D}, GRB\,190829A \citep{dichiara2022MNRAS.512.2337D, medlerthesis}, and GRB\,221009A \citep{laskar2023ApJ...946L..23L}. Radio light curves are taken from the compilation of \citet{chandra2012ApJ...746..156C}, supplemented by GRB\,130427A \citep{perley2014ApJ...781...37P}, GRB\,161219B \citep{alexander2019ApJ...870...67A}, GRB\,171205A \citep{maity2021ApJ...907...60M}, GRB\,190829A \citep{rhodes2020MNRAS.496.3326R}, and GRB\,221009A \citep{laskar2023ApJ...946L..23L}. 
    All light curves have been shifted to the rest frame of GRB\,260310A ($z = 0.153$) for direct comparison.    }
    \label{fig:grb_compare}
\end{figure}

\thisgrb\ is sub-luminous near peak (Figure~\ref{fig:grb_compare}): the optical peak ($M_r\lesssim-21.81$\,mag) lies $\sim1$--2\,mag below the median peak luminosity of Swift-era on-axis LGRB afterglows \citep[e.g.,][]{kann2010ApJ...720.1513K}. 
Both pre- and post-decay slopes are inconsistent at $>3\sigma$ with the population means from \citet{zeh2006ApJ...637..889Z} and \citet{kann2010ApJ...720.1513K}. The pre-break slope ($\alpha_1 = -0.52\pm0.02$) is significantly shallower than the sample means ($\langle\alpha_1\rangle \approx -1.05$ to $-1.12$), while the post-break slope ($\alpha_2 = -1.45^{+0.02}_{-0.01}$) is similarly shallower than the means ($\langle\alpha_2\rangle \approx -2.12$ to $-2.34$). The break time $t_b\sim$7.70\,d is also substantially later than either sample range, which extends to a few days at most, although this could be a selection effect (that monitoring cannot continue this late for more distant, fainter bursts). 



As shown in Figure~\ref{fig:optical_lc}, the rising phase of \thisgrb\ is bounded between the GOTO-$L$ detection at $T_0+17.9$\,min (1074\,s) and the first ATLAS $o$ detection at $T_0+7.15$\,hr ($2.57\times10^4$\,s). The brightening between GOTO-$L$ and ATLAS-$o$ detections, combined with the post-peak GOTO-$L$ detection at $T_0+2.76$\,d being $\sim0.66$\,mag brighter than the pre-peak detection and lying significantly below the extrapolation of the $t^{-0.52}$ decay to that first epoch (Figure~\ref{fig:optical_lc}), confirms that the rise time was between 1074\,s and $2.57\times10^4$\,s. For context, a systematic analysis of P60 automatic follow-up of optical afterglows by Bochenek et al. (in prep.) finds that of the early observed bursts, $\sim82\%$ universally fade, and the rest of the afterglows that show flaring/re-brightening start power law decay by $10^4$ s in observer frame; additionally, all of those features are observed to differ up to around one order of magnitude in flux space. GRB 190829A also exhibited a long rise (1500\,s), attributed to reverse shock emission \citep{salafia2022ApJ...931L..19S} or to the rise of an off-axis afterglow \citep{sato2021MNRAS.504.5647S}. In Section~\ref{sec:discussion} we model the afterglow under different assumptions about the origin of the slow rise. 

\subsubsection{X-ray}
\label{sec:comparison-xray}

As shown in Figure~\ref{fig:xray_afterglow}, the X-ray afterglow of AT\,2026fgk evolves very similarly to the optical afterglow, with consistent pre- and post-break temporal indices, as well as a consistent break time of $t_{\rm b}=8.45\pm0.65\,$d. As shown in Figure~\ref{fig:grb_compare}, the X-ray luminosity at late times (tens of days) is very high for a GRB with this $E_\mathrm{iso}$, and greatly exceeds the X-ray afterglow luminosity of other sub-luminous GRBs. 


Between roughly 15 and 30 days after \thisgrb, there is evidence for a plateau in the X-ray flux (see Figure~\ref{fig:xray_afterglow})\footnote{There is a flux measurement that is significantly above the plateau, though this occurred close to a passage through the South Atlantic Anomaly. We base any conclusions on the non-SAA data.}. The rebrightening was not accompanied by any noticeable change in the X-ray photon index. To our knowledge, such a late-time feature in the X-ray afterglow has never been observed before: while the most delayed optical flares were observed in GRB\,210204A/AT\,2021buv \citep{kumar_flares_210204a}, at roughly 10--15 days post-burst, it was not possible to obtain simultaneous X-ray measurements at these times, as the X-ray afterglow faded below the sensitivity of Swift-XRT. \citet{kumar_flares_210204a} suggest that these flares arise from refreshed shocks in the jet. An X-ray plateau was observed between roughly $10^4$--$10^5$\,s post-burst in GRB 171205A \citep{delia2018AA...619A..66D}, attributed to a magnetar central engine \citep{kong_171205_magnetar}. In GRB\,030329 \citep{granot_nakar_piran_refreshed_shocks}, the optical light curve appeared ``bumpy'' on timescales of $\sim0.5$\,d (before $+10$\,d). There was also a slight observed flattening of the X-ray light curve of GRB\,030329 (to a shallower power-law index) that was observed at  $+40$\,d, though it was not as pronounced as the plateau seen in AT\,2026fgk. It is worth noting that such late-time X-ray afterglow observations are rare and, in this case, were only enabled by this burst's cosmological proximity. The latest-time X-ray detections of afterglows have been for similarly cosmological nearby  bursts ($z\lesssim0.5$), including GRB\,030329  ($z\sim0.17$; \citealt{tiengo_grb030329_xray}) and GRB\,060729 ($z=0.54$; \citealt{grupe_2007_late_afterglow,grupe_2010_late_afterglow}).


\subsubsection{Radio}

\begin{figure*}
    \centering
    \includegraphics[width=1\linewidth]{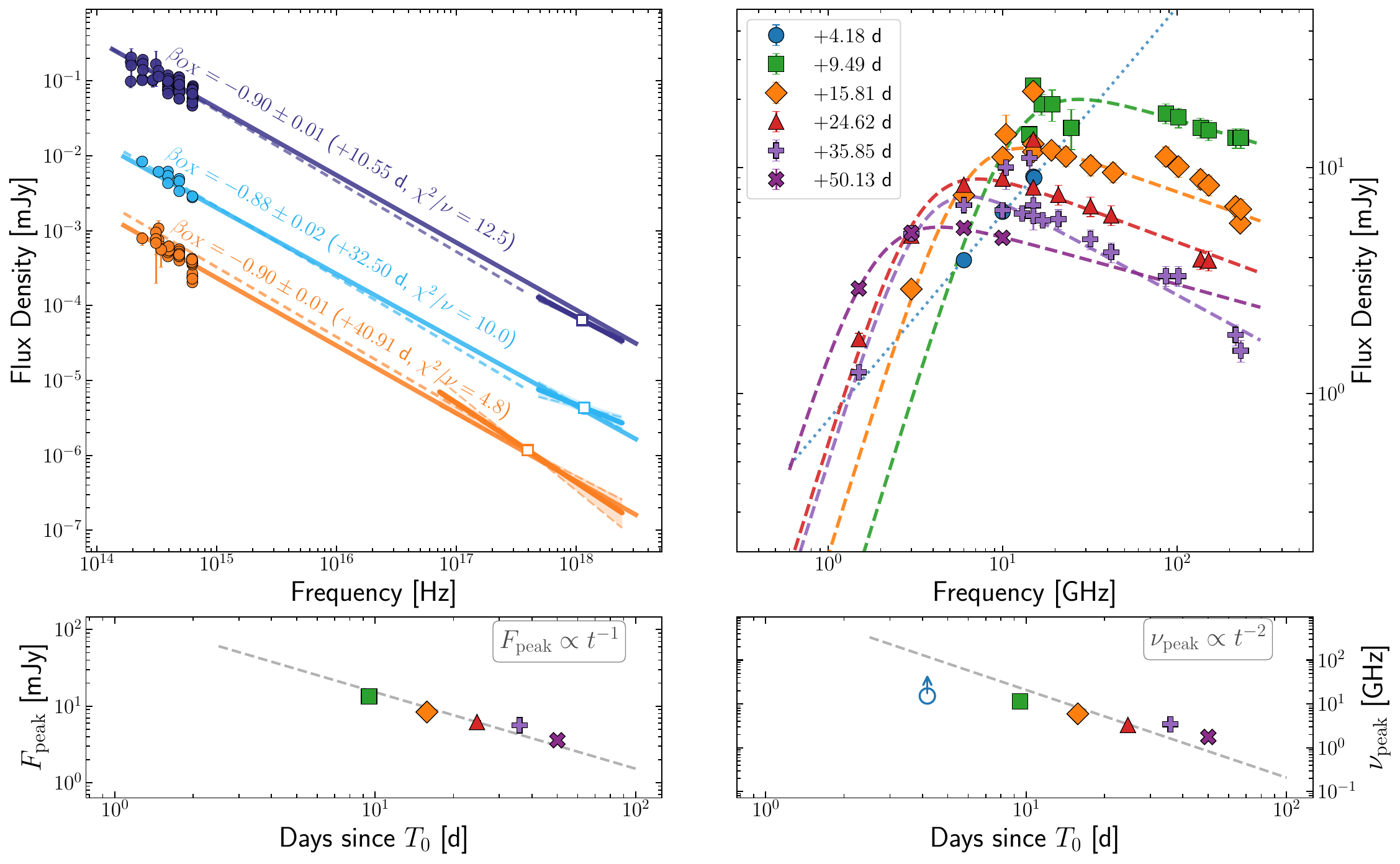}
    \caption{\emph{Top left}: Optical + X-ray SED at three epochs between (+10.55\,d, +32.50\,d, and +40.91\,d observer frame), offset vertically by -0.75\,dex per epoch for clarify. Circles show optical/NIR photometry MW de-reddened. Colored power-law segments render the X-ray $F_\nu(\nu)$ across each instruments quoted band (0.3--10\,keV for Chandra, 3--79\,keV for NuSTAR) using the measured photon index ($\Gamma$; see Table~\ref{tab:xray}). Translucent solid lines are the observed two-point spectral index $\beta_{OX}$ from mean optical to X-ray flux density at the lower band edge ($\beta_{OX}$ = $-0.93\pm0.01$, $-0.89\pm0.03$, $-0.80\pm0.02$ in time order). Dotted lines show the intrinsic optical-X-ray slopes $\beta\approx0.93$ from a joint optical+X-ray fit. \emph{Top right}: Radio SEDs at five epochs fitted with a smoothly broken power-law expected for SSA (dashed); dotted lines mark epochs with no peak in band. \emph{Bottom left, right}: Peak radio flux density ($F_\mathrm{peak}$) and peak frequency ($\nu_{\mathrm{peak}}$) vs. time, with $t^{-1}$ and $ t^{-2}$ references slopes (grey dashed). Open arrows denote upper/lower limits.
    }
    \label{fig:xrayradiosed}
\end{figure*}

The radio panel in Figure~\ref{fig:grb_compare} shows the cm-band radio light curves of \thisgrb\ alongside several other sub-luminous GRBs associated with SNe---GRBs 980425, 060218, 171205A and 190829A---and the energetic GRB\,030329/SN\,2003dh \citep[][]{matheson2003ApJ...599..394M, stanek2003ApJ...591L..17S, tiengo_grb030329_xray, vanderhorst2008A&A...480...35V, hjorth2003Natur.423..847H}. Radio emission arises from synchrotron radiation; electrons are accelerated to relativistic energies at the expanding blast-wave front \citep{chevalier1998ApJ...499..810C}. At the earliest epochs ($\lesssim$few days), a reverse shock propagating back through the decelerating ejecta can produce a short-lived radio flare as the shock shell rapidly cools \citep[e.g.,][]{sari1999ApJ...519L..17S, kobayashi000ApJ...542..819K, kobayashi2003ApJ...582L..75K}; this component fades steeply and becomes negligible once the reverse shock has crossed the ejecta shell, leaving the forward shock to dominate thereafter \citep{granot2014PASA...31....8G}.

\thisgrb peaks at $L_\nu \approx 7\times10^{30}$\,\erg\psec\phz at 11~GHz ($\sim$10~d rest frame), roughly 100 times more luminous than GRB\,980425/SN\,1998bw and consistent with the median peak luminosity $\sim10^{31}$\,\erg\psec\phz measured for LGRBs across a large redshift distribution \citep[][]{chandra2012ApJ...746..156C}. Among sub-luminous GRBs, it is the most radio-luminous event at comparable epochs, exceeding GRB\,031203/SN\,2003lw and GRB\,190829A.

The broadband radio SED of \thisgrb (Figure~\ref{fig:xrayradiosed}) exhibits a typical broken power law shape \citep[e.g.,][]{sari1998ApJ...497L..17S, chevalier1998ApJ...499..810C, chevalier2000ApJ...536..195C, granot2002ApJ...568..820G}: a steep low-frequency rise, a single peak, and an optically-thin decline at high frequency. 

\subsection{Supernova Properties}

\begin{figure*}[!ht]
    \centering
    \includegraphics[width=0.9\linewidth]{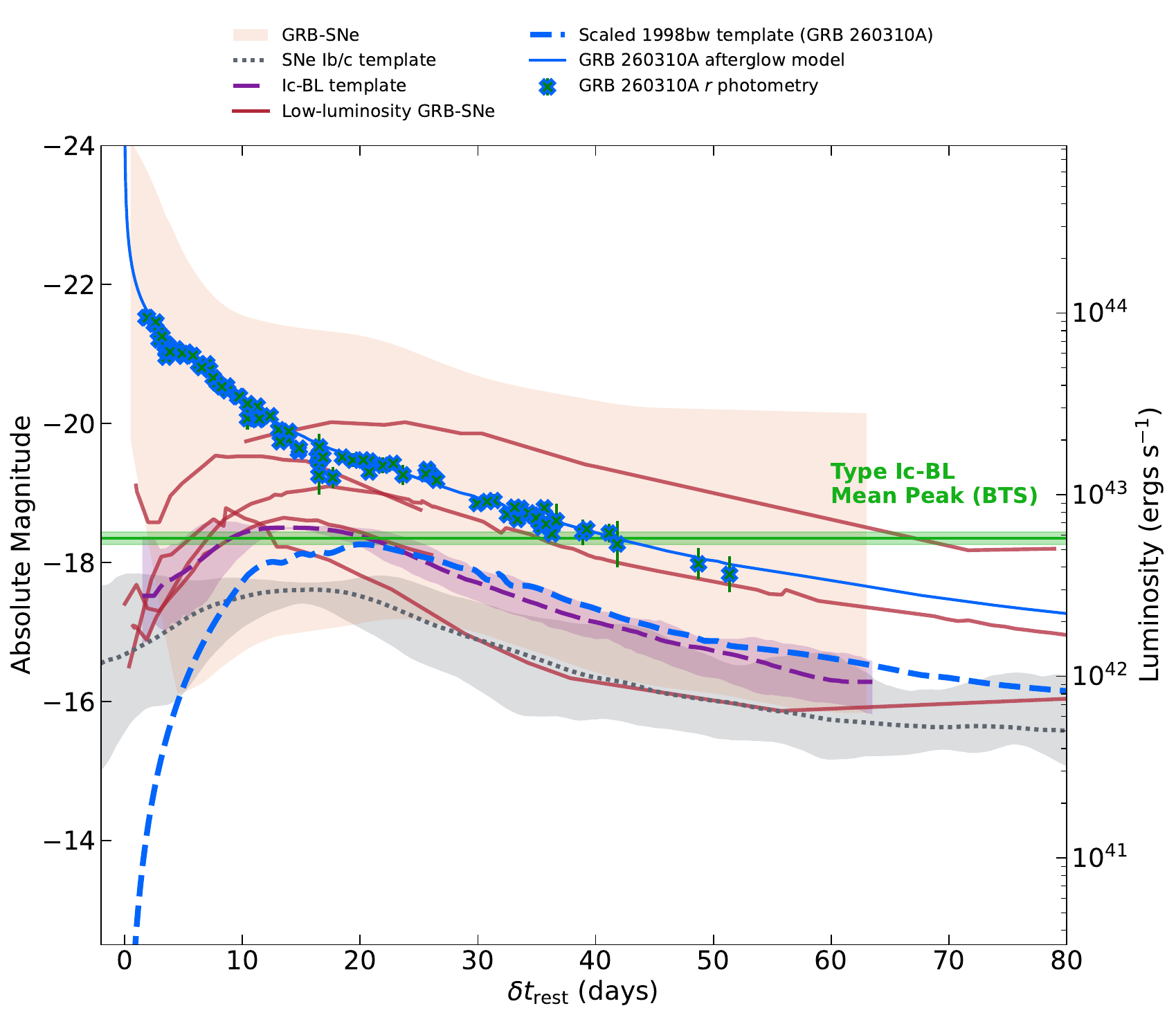}
    \caption{Rest frame $r$-band absolute magnitude light curve of \thisgrb (\emph{blue crosses}) compared to GRB-SNe, Type Ic-BL SNe and ordinary Type Ib/c SNe. The \emph{solid blue} curve is our best fit afterglow + SN model (see Section~\ref{sec:analysis-supernova-properties}), and the \emph{dashed blue} curve is the SN component (SN\,1998bw template scaled; $s=1.23^{+0.37}_{-0.12}$ and $k_r=0.271^{+0.018}_{-0.013}$). In \emph{red} are  other SNe with sub-luminous GRBs (SN\,1998bw, SN\,2003lw, SN\,2006aj, SN\,2010bh, and SN\,2017iuk). The horizontal \emph{green line} marks the Type Ic-BL mean peak from the ZTF Bright Transient Survey \citep{perley2020ApJ...904...35P, srinivasaragavan2024ApJ...960L..18S}. We plot Type Ib/c and Ic-BL SNe templates \citep[collected from][]{khakpash2024ApJS..275...37K} in \emph{grey} and \emph{purple}. All times are rest frame days since explosion/trigger, all magnitudes are k-corrected and corrected for Galactic extinction---LLGRBs are corrected for galactic extinction based on their original publications.}
    
    \label{fig:comp_to_others}
\end{figure*}

In Figure~\ref{fig:comp_to_others} we compare this SN to the Type Ic-BL population from the ZTF BTS \citep[][]{fremling2016A&A...593A..68F, perley2014ApJ...781...37P, rehemtulla2024ApJ...972....7R, anand2024ApJ...962...68A, srinivasaragavan2024ApJ...976...71S}, a spectroscopically complete, magnitude-limited survey classifying all extragalactic transients peaking brighter than 18.5~mag\footnote{The BTS is the largest, most complete magnitude-limited survey to date, having $>95\%$ completeness.}. The 104 confirmed Type Ic-BL SNe from the BTS span $M_r \approx -16$ to $-21$\,mag, with a mean of $\langle M_r \rangle = -18.32 \pm 0.08$\,mag (MSE). At $M_r \approx -18.3$\,mag, \thisgrb's SN falls $\sim 0.14$\,mag ($\sim1.8\times$\,MSE) below the BTS mean, placing it marginally towards the less luminous end of the SN Ic-BL distribution, albeit well within the intrinsic population scatter (see Figure~\ref{fig:comp_to_others}). 

Using the \MNi, we further contextualize \thisgrb's SN both within the GRB-SN population \citep{cano2017AdAst2017E...5C} and the wider Type Ic-BL population not associated with GRBs \citep{taddia2019A&A...621A..71T, srinivasaragavan2024ApJ...960L..18S}. The population-averaged \MNi for GRB-SNe is $\langle M_{\rm Ni}\rangle _{\rm GRB} = 0.37 \pm 0.20$\,\Msun \citep[][]{cano2017AdAst2017E...5C}, while for Ic-BL SNe not associated with GRBs it is $\langle M_{\rm Ni}\rangle _{\rm Ic-BL} = 0.37^{+0.08}_{-0.06}$\,\Msun \citep{srinivasaragavan2024ApJ...960L..18S}. For low luminosity (LL)GRB-SNe with $L_\mathrm{\gamma,\mathrm{iso}}<10^{48.5}\,$erg\,s$^{-1}$, \citet{cano2017AdAst2017E...5C} report average bolometric parameters of $\langle M_{\rm Ni}\rangle _{\rm LL} = 0.35 \pm 0.19$\,\Msun, $\langle E_{\rm{K}}\rangle _{\rm LL} = 27.8 \pm 19.6\times10^{51}$\,\erg \citep[][]{cano2017AdAst2017E...5C} and a filter-averaged luminosity scaling factor $\langle k\rangle_{\rm LL} = 0.94 \pm 0.41$ relative to SN\,1998bw. From Table~\ref{tab:fitted_params_perband}, we see a filter averaged $\langle k\rangle_{griz} = 0.26 \pm 0.10$, $\sim1.6\sigma$ below the LLGRB-SN mean, confirming it is sub-luminous to SN\,1998bw. Our value is fully consistent with the independent multi-band fits of \citet{oconnor2026grb260310asn2026fgkphotometric} and \citet{gill2026arXiv260526091G}, who infer $k \approx 0.4$--0.6 and $k \approx 0.3$ respectively. 

We find \MNi = $0.344 \pm 0.016$\,\Msun, placing \thisgrb's SN within the LLGRB-SN population mean ($\langle M_{\rm Ni}\rangle_{\rm LL} = 0.35 \pm 0.19$\,\Msun) and within $1\sigma$ of both the broader GRB-SN ($0.37 \pm 0.20$\,\Msun) and non-GRB Type Ic-BL ($0.37^{+0.08}_{-0.06}$\,\Msun) population means \citep{cano2017AdAst2017E...5C, taddia2019A&A...621A..71T, srinivasaragavan2024ApJ...960L..18S}. Our value sits at the lower end of the estimate $M_{\rm Ni} \approx 0.4$--0.5\,\Msun reported by \citet{oconnor2026grb260310asn2026fgkphotometric}.

For the \MEj, the joint broken power-law+Arnett fit directly yields \MEj=$3.47^{+0.28}_{-0.27}$\,\Msun. Adopting our JWST-anchored photospheric velocity $v_{\rm ph} = 16,750$\,\kms (Section~\ref{sec:nirspec}), we find the kinetic energy is \ek$= (5.8\pm0.5)\times10^{51}$\,\erg using the homologous uniform-density prescription of \citet{lyman2016MNRAS.457..328L}, and $E_{\rm K} = (9.7 \pm 0.8)\times10^{51}$\,\erg using the thin-shell prescription of \citet{valenti2008MNRAS.383.1485V}. The measured \MEj and \ek values agree closely with \citet{oconnor2026grb260310asn2026fgkphotometric}, who derive $M_{\rm ej} \approx 4$--$6$\,\Msun\ and $E_{\rm K} = 3$--$8\times 10^{51}$\,\erg. Our \MEj lies within $1\sigma$ of the GRB-SN population mean $\langle M_{\rm ej}\rangle_{\rm GRB} = 5.8\pm4.0$\,\Msun, but above the non-GRB Ic-BL mean $\langle M_{\rm ej}\rangle_{\rm Ic-BL} = 2.45^{+0.47}_{-0.41}$\,\Msun \citep{srinivasaragavan2024ApJ...960L..18S}. The kinetic energy is consistent with the non-GRB Ic-BL mean $\langle E_{\rm K}\rangle_{\rm Ic-BL} = (7.1 \pm 5.7)\times10^{51}$\,\erg \citep[][]{srinivasaragavan2024ApJ...960L..18S} and below the GRB-SN average $\langle E_{\rm K}\rangle_{\rm GRB} = (26\pm 18.3)\times10^{51}$\,\erg \citep[][]{cano2017AdAst2017E...5C} by $\sim1\sigma$.

Taken together, \thisgrb's SN is characterized by an average nickel mass relative to both the LLGRB-SN and SN Ic-BL populations, an ejecta mass consistent with LLGRB-SNe but marginally above the non-GRB Ic-BL, and a kinetic energy at the lower end of the GRB-SN distribution. The combination places SN as a sub-luminous, moderate-ejecta, low-energy GRB-SN.

\section{Volumetric Rate}
\label{sec:rates}

In this section, we take advantage of the proximity and optical brightness of \thisgrb\ to estimate the volumetric rate of optically selected afterglows. By comparing this rate measurement to the $\gamma$-ray-selected LGRB rate, we can constrain the angular structure of the initial relativistic material, as well as the rate of any other phenomena that might be preferentially selected in the optical band but missed by $\gamma$-ray satellites. 

Over the seven years of ZTF operations (2018--2024), fast transient searches in the ZTF alert stream have resulted in the blind identification of close to 20 afterglows \citep{2020ApJ...904..155A, Ho2020ApJ...905...98H, 2021ApJ...918...63A, 2022ApJ...938...85H, 2025ApJ...985..124L, 2025MNRAS.537.2362P, srinivasaragavan2025MNRAS.538..351S}. Almost all of the events with redshift measurements have been at cosmological distances ($z>1$). \thisgrb is the closest ($z=0.153$) and optically brightest of the ZTF afterglow sample: at $<18\,$mag it even passes the criteria of the ZTF BTS. 

We use the BTS to estimate the rate, following \citet{perley2026arXiv260103337P}. For rapidly fading events, a transient must both survive ZTF automated filtering ($\sim$hours--days cadences) and receive spectroscopic follow-up within $\sim$1--3\,d---the latter imposed to preserve classification completeness before the source drops below the nominal BTS follow-up limit. We therefore use the magnitude at $+2.5$\,d post trigger ($r = 18.40$\,mag; $M_r \leq -20.88$\,mag) as the effective brightness threshold, close to the BTS spectroscopic classification limit of 18.5~mag \citep{perley2020ApJ...904...35P} and conservative with respect to the peak. A source of this luminosity would satisfy $m\leq18.5$\,mag out to $z_{\rm max} \approx 0.16$ ($D_L \approx 764$\,Mpc), enclosing a co-moving volume of 1.2~Gpc$^3$ (all-sky). 

Over the course of the survey, two afterglows have been detected at $<18.5\,$mag in the ZTF BTS: AT\,2026fgk and AT\,2023sva (which was at $z=2.2810$ and did not have an associated detected GRB; \citealt{srinivasaragavan2025MNRAS.538..351S}). Therefore, only AT\,2026fgk has been discovered within the $z=0.16$ volume. Applying ZTF's sky coverage ($\sim35\%$)\footnote{This is based on the area approximately covered by ZTF on a 2 night cycle.} and an 8-year baseline, two-sided Poisson statistics \citep{gehrels1986ApJ...303..336G} at 95\% confidence yield $\mathcal{R} = 0.30^{+1.37}_{-0.29}~\text{Gpc}^{-3}\,\text{yr}^{-1}$, broadly consistent with the classical (high luminosity) LGRB rate \citep[$\sim1~\text{Gpc}^{-3}\,\text{yr}^{-1}$;][]{wanderman2010MNRAS.406.1944W}, albeit driven by a single event.

\section{Host Galaxy Properties}
\label{sec:host_properties}
We estimate host properties using \textsc{FrankenBlast} \citep[][]{nugent2026ApJ...997...38N}, a scalable transient host characterization package built on the Blast web app \citep{jones2024arXiv241017322J} that performs host association, aperture photometry on archival multiwavelength imaging, and Prospector SED-fitting accelerated by simulation-based inference \citep[SBI++;][]{Wang_2023}. The photometry spanning GALEX FUV/NUV through PS1 $grizy$, 2MASS $JHK$, and WISE $W1$--$W4$ (Table~\ref{tab:host_phot}, Figures~\ref{fig:sed1} and \ref{fig:sed2}---we omitted 2MASS $K$ from the fit as we retrieved an upper-limit) yield a stellar mass of $\log(M_*/\rm{M}_\odot) = 10.25^{+0.08}_{-0.09}$ and a star formation rate of $\log(\rm{SFR}) = 0.17^{+0.22}_{-0.21}$\,\Msun\,yr$^{-1}$, with the resulting $\log(\rm{sSFR}) \approx - 10.1$\,yr$^{-1}$. We place limits on host extinction in Appendix~\ref{asec:sed_host} using the ($grizyHJ$) transient photometry.

The host is atypical for a LGRB. \citet{taggart2021MNRAS.503.3931T}---who performed a comprehensive study of transient host properties that includes LGRB hosts at $z<0.3$---find a median stellar mass of $\log(M_*/\rm{M}_\odot) = 8.7 \pm 0.2$ and median $\log(\rm{sSFR}) = -9.1$\,yr$^{-1}$ for their 12 LGRB-SN hosts. \thisgrb's host sits $\sim1.5$\,dex above the median in mass, placing it among the most massive confirmed LGRB host galaxies at $z<0.3$. \thisgrb's sSFR ($\log(\rm{sSFR}) \approx -10.1$\,yr$^{-1}$) is correspondingly lower than the LGRB-SN median. 

Among sub-luminous GRBs specifically, the hosts of GRB\,980425, 031203, and 060218 span $\log(M_*/\rm{M}_\odot) = 7.5$--9.0 with $\log(\rm{sSFR})\geq -9.5$\,yr$^{-1}$ \citep[e.g.,][]{sollerman2005NewA...11..103S}. The host of \thisgrb is most similar to the hosts of GRB\,171205A, whose galaxy has $\log(M_*/\rm{M}_\odot) = 10.11^{+0.02}_{-0.08}$ \citep[][]{taggart2021MNRAS.503.3931T} and GRB\,190829A with $\log(M_*/\rm{M}_\odot) = 12.04^{+0.09}_{-0.10}$ \citep[][]{gupta2022JApA...43...82G}, whose host is one of the most massive, showing notably lower $\log(\rm{sSFR}) = -11.20$\,yr$^{-1}$.

\section{Discussion}
\label{sec:discussion}

There are several proposed models for sub-luminous GRBs: that they are classical LGRBs viewed off-axis \citep{izzo2019Natur.565..324I} or stifled in an extended stellar envelope or wind \citep[][]{campana2006Natur.442.1008C, nakar2015ApJ...807..172N, irwin2025MNRAS.542.1269I}, or low-power jets from alternative central engines \citep[][]{metzger2011MNRAS.413.2031M, irwin2016MNRAS.460.1680I}. Unlike in most previous sub-luminous GRBs, the strong afterglow emission from AT\,2026fgk disfavors a choked jet or low-power jet origin. In this section we perform basic modeling of AT\,2026fgk and find that it is consistent with being an LGRB viewed slightly off-axis, or a low-Lorentz factor jet viewed on-axis. 
We adopt fiducial values of \ekiso$=10^{53}$\,\erg and $n_{\rm ism} =1$\,cm$^{-3}$, typical of cosmological LGRB afterglow modeling \citep[e.g.,][]{frail2001ApJ...562L..55F, panaitescu2002ApJ...571..779P}. 


\subsection{A Slow Optical Rise}
\label{sec:optical-rise}

We bracket the time of optical peak, $\tpeak$, with three values: 1) a hard lower limit of $\tpeak\gtrsim1.1\times10^3$\,s set by the rising GOTO-$L$ detection; 2) a fiducial estimate of $\tpeak\approx3\times10^3$\,s obtained by anchoring a $F\propto(t-T_0)^3$ rising power-law \citep{granot2002ApJ...568..820G} to the GOTO-$L$ pre-peak observation and intersecting it with the post-peak (pre-jet-break) decay $F\propto t^{-0.52}$; and 3) a hard upper limit of $\tpeak\lesssim2.6\times10^4$\,s set by the first ATLAS $o$ detection, which we treat as already past peak given it is consistent with the post-peak decline index. 

If the rise reflects the deceleration of the blast wave to the observer's line of sight, the optical light curve peaks at the deceleration timescale, given by Equation~16 of \citet{2006RPPh...69.2259M}: 

\begin{equation}
    t_{\rm dec} = 10\,E_{53}^{1/3}\,n^{-1/3}\,\Gamma_{0,2.5}^{-8/3}{\,\rm s},
    \label{eq:tdec_onaxis}
\end{equation}

\noindent where $\Gamma_{0,2.5} = \Gamma_0/10^{2.5}$ and $\tpeak \equiv \tdec$. Inverting Equation~\ref{eq:tdec_onaxis} with $E_{53} = n_{\rm ism}=1$, the three peak-time bounds $\tpeak\simeq 1.1\times10^3$, $3\times10^3$, and $2.6\times10^4$\,s correspond to $\Gamma_0\simeq 60$, 40, and 16 respectively. Our fiducial bulk Lorentz factor is therefore $\Gamma_0\approx 40$, with an allowed range of $16\lesssim\Gamma_0\lesssim60$. This is notably lower than the $\Gamma_0\sim100$--$1000$ typical of cosmological LGRBs \citep[median$\sim300$;][]{ghirlanda2018A&A...609A.112G}, but consistent with values inferred for other sub-luminous GRBs such as GRB\,190829A \citep[$\Gamma_0\sim20$--60;][]{sato2021MNRAS.504.5647S,salafia2022ApJ...931L..19S}.

Alternatively, the optical rise may reflect the deceleration of a jet whose beaming cone first sweeps over an observer at $\thv > \thj$, with the flux rising as $\Gamma(t)$ drops until $\Gamma(\tpeak) = 1/(\thv-\thj)$ \citep[][]{granot2002ApJ...568..820G, ryan2020ApJ...896..166R, sato2021MNRAS.504.5647S}. From the Blanford-McKee result \citep[Equation~7 of][]{sato2021MNRAS.504.5647S} or the \textsc{afterglowpy}-calibrated equivalent \citep[Equation~36--37 of][with $E(\theta_w)/E_0\approx 1$]{ryan2020ApJ...896..166R}, the optical peak of the off-axis afterglow, in the top-hat limit $E(\theta_w)/E_0=1$, $\theta_w\equiv\thj$, is

\begin{equation}
    t_{\rm peak} = 12.1\,(1+z)\,E_{53}^{1/3}\,n_0^{-1/3}\,\left(\frac{\theta_v-\theta_j}{0.2\,\rm{rad}}\right)^{8/3}\;\rm{days}.
    \label{eq:tpeak_offaxis}
\end{equation}

Inverting Equation~\ref{eq:tpeak_offaxis} with $E_{53}=n_0=1$, the three peak-time bounds $\tpeak\simeq 1.1\times10^3$, $3\times10^3$, and $2.6\times10^4$\,s correspond to $\thv-\thj\simeq 0.8^\circ$, $1.2^\circ$, and $2.7^\circ$ respectively, with $\Gamma(t_{\rm peak})\simeq 70$, 50, and $20$. The fiducial $t^3$ estimate therefore implies a viewing-angle offset of $\thv-\thj\approx3^\circ$, with the bounds permitting at most a $\sim2^\circ$ offset. The implication is that \thisgrb is at most a few degrees off-axis---broadly consistent with the off-axis interpretation of GRB\,190829A \citep[e.g.,][]{sato2021MNRAS.504.5647S}. 

\citet{yang2026arXiv260523818Y} set the upper bound for the peak at $\approx0.1$\,d. \citet{gill2026arXiv260526091G} set the peak at $<0.5$\,d, with a deceleration time of $t_{\rm dec}\approx 0.23$\,d.

\subsection{A Jet Break in the X-ray and Optical Bands}
\label{sec:jetbreak}

A simultaneous, achromatic break was observed in the optical and X-ray light curves at $t_b \approx 8$\,d \citep[also seen in][]{oconnor2026grb260310asn2026fgkphotometric, gill2026arXiv260526091G, yang2026arXiv260523818Y}. A jet break is predicted to occur achromatically as the bulk Lorentz factor decelerates to $\Gamma \sim \theta_{\rm jet}^{-1}$ and the beaming cone encompasses the full jet \citep[e.g.,][]{sari1998ApJ...497L..17S, rhoads1999ApJ...525..737R}. This distinguishes the jet break from chromatic spectral breaks, such as the synchrotron cooling frequency passing though the observing band, which occur at different times in different wavebands. The optical and X-ray break times of \thisgrb are consistent within measurement uncertainties, providing strong evidence for a physical jet break \citep[e.g.,][]{harrison1999ApJ...523L.121H, zeh2006ApJ...637..889Z}. 

Under the interpretation that the break is a jet break, and that the jet was viewed on-axis, we can use the relation from \citet{sari1999ApJ...519L..17S}, converted from the rest- to observer frame:

\begin{equation}
    t_{\rm jet} \approx \frac{0.26}{(1+z)}\left(\frac{E_{\rm K, iso}}{10^{52}~{\rm erg}} \frac{n_0}{1~{\rm cm}^{-3}}\right)^{1/3}\left(\frac{\theta_j}{0.1~{\rm rad}}\right)^{8/3}~{\rm day}.\\
    \label{eq:tjet_on_sari}
\end{equation}

Using our fiducial values of $E_{53} =  n_{\rm ism}=1$, and inverting Equation~\ref{eq:tjet_on_sari} with our measured $t_b\approx8$\,d, we find $\thj\approx17^\circ$.

\citet{ryan2020ApJ...896..166R} provide an \textsc{afterglowpy}-calibrated fit for the case where the jet is viewed off axis \citep[their Equation~38, which applies for $\thv>1.01\,\thj$;][]{ryan2020ApJ...896..166R}:

\begin{equation}
    t_{\rm jet} = \frac{24.9}{(1+z)}\, E_{53}^{1/3}\,n_0^{-1/3}\left(\frac{\theta_{\rm obs} + 1.23\,\theta_{\rm j}}{0.5\, \rm{rad}}\right)^{8/3}\,\rm{days}.
    \label{eq:tjet_off_ryan}
\end{equation}

The rise-time analysis from the previous section indicates a viewing-angle offset of $\thv-\thj\approx0.8$--3$^\circ$. Substituting an offset of 1$^\circ$ and the same fiducial values used before into Equation~\ref{eq:tjet_off_ryan}, we find $\thj\approx6^\circ$. The factor of 2 difference is due to an observer outside the jet edge seeing a longer effective rise to $\Gamma\sim\thj^{-1}$ than an on-axis observer would. 
If the observed break is a jet break, then the steepening observed in the light curve must arise from the observer detecting the edge of the jet rather than lateral spreading \citep[e.g.,][]{rhoads1999ApJ...525..737R, granot2012MNRAS.421..570G}. 


\subsection{Spectral Energy Distribution}
\label{sec:sedox}

In the top left panel of Figure~\ref{fig:xrayradiosed}, we plot the optical to X-ray SED at the three epochs that have NuSTAR or Chandra coverage. These facilities provide the broadest spectral range for the X-ray afterglow (3--79\,keV for NuSTAR) or the highest sensitivity (Chandra). 
The measured photon indices (see Table~\ref{tab:xray}) 
correspond to X-ray spectral indices $\beta_X\,= \Gamma\,-1\,=\,0.83\pm0.12$, $0.64\pm0.25$, and $1.06\pm0.24$, all consistent within $\lesssim2.5\sigma$ with the magnitude of the optical-to-X-ray index $\beta_{OX}\,\approx\,0.9$. This is consistent with a single power-law spanning the optical to the X-ray throughout the first $\sim40$\,d. 
We therefore conclude that no cooling break lies between the optical and the X-ray, such that the UVOIR and X-ray observations lie in the regime 
$\nu_m < \nu < \nu_c$.

With $\nu_c$ above the X-ray band, we can use 
$\nu_c\geq10^{19}$\,Hz (from NuSTAR) to place an upper limit on the circumburst density. Inverting the synchrotron cooling-frequency expression for an adiabatic blast wave in a uniform medium \citep[Equation~11;][]{sari1998ApJ...497L..17S}:

\begin{equation}
    \nu_c = 2.7\times10^{12} \epsilon_B^{-3/2}\,E^{-1/2}_{52}\,n_1^{-1}\,t_d^{-1/2}\,\rm{Hz},
    \label{eq:nu_c}
\end{equation}

\noindent where $t_d$ is the time (in days) after explosion, $n_1=n_0/(1\;\rm{cm}^{-3})$, and $E_{52} = E_{\rm K,iso}/10^{52}$\,\erg. Requiring $\nu_c>10^{19}$\,Hz and $t_d = 9.15$\,d (rest frame phase of first NuSTAR epoch), and adopting our fiducial $E_{52} = 10$ together with the canonical upper bound $\epsilon_B<3\times10^{-4}$ \citep{panaitescu2001ApJ...560L..49P, chandra2012ApJ...746..156C, santana2014ApJ...785...29S} gives

\begin{equation}
    n_0 \lesssim0.1\rm{cm}^{-3}.
\end{equation}

This is consistent with the low circumburst densities inferred for other LGRB afterglows \citep[][]{panaitescu2001ApJ...560L..49P}.

There is also agreement on the spectral indices between \citet{oconnor2026grb260310asn2026fgkphotometric}, \citet{yang2026arXiv260523818Y}, and \citet{gill2026arXiv260526091G}, with all studies converging to $\beta_X\approx 0.8$ and $\beta_O\approx0.9$. Both in this work, and the uniform fit from \citet{gill2026arXiv260526091G} yield an electron power index of $p\approx2.8$, compared to the structured fit from \citet{gill2026arXiv260526091G} and radio constraint from \citet{yang2026arXiv260523818Y} giving $p=2.4$ and 2.2 respectively. This thus determines the location of $\nu_c$, which varies from above the X-ray frequency (here) to between X-ray and optical \citep[][]{ gill2026arXiv260526091G}.

From the radio spectral energy distribution (Figure~\ref{fig:xrayradiosed}), we find that the evolution of the peak radio frequency and peak radio flux density over time is consistent with expectations for $F_{\nu,m}$ and $\nu_m$ after the jet break \citep{sari1999ApJ...519L..17S}. 

\subsection{Physical Origin of the X-ray Rebrightening}


As discussed in Section~\ref{sec:comparison-xray}, a prominent X-ray rebrightening at such a late phase (tens of days) has, to our knowledge, not been observed in any past LGRB afterglow, although such late-time observations have only been obtained for a relatively small number of events. The plateau/flare reaches $L_{X} \simeq (5\text{--}7)\times10^{43}\,\text{\erg s}^{-1}$ at $\Delta t\sim$20--30\,d, with no evolution of the photon index ($\Gamma \approx$1.3--1.6) that is notably beyond calibration differences in instruments (see Appendix~\ref{asec:xray_calib}).

Under the assumption of a flare (taking into account the SAA point), the  variability timescale is roughly $\Delta t\sim6$\,d. A useful metric for discriminating between models of late-time variability is the ratio $\Delta t/t$, where $t$ is the duration after the burst at which this variability occurs. In this case, $t\sim$\,25\,d, so $\Delta t/t\sim 1/4$. Assuming a plateau, the timescale of this variability (before it begins to decay again) is roughly 15\,d, making $\Delta t/t$ roughly 0.6. \citet{ioka_grb_variability} discuss three physical causes for afterglow variability.  The first possibility arises from angular structure on the emitting surface,  which can result in variability on timescales $\Delta t \gtrsim t$; however, this is unlikely here, given our constraint that $\Delta t/t < 1$. The second arises from refreshed shocks, which should have a variability timescale of $\Delta t/t > 1/4$, which could be plausible here. In this scenario, an ``inner shell'' is ejected with a slower velocity than an ``outer shell''; once the Lorentz factor of the  outer shell drops to a value comparable to that of the inner shell, the two shocks interact and lead to a deviation from the typical afterglow decay. Another  consequence of this theory is that the afterglow energy should not return to the original decay (as now the afterglow has been ``refreshed'' with energy)---which we observe in this GRB's X-ray afterglow, after the plateau. The third possibility is a long-acting engine, whether black-hole accretion or a millisecond magnetar. 

Another possibility is that the X-rays originate from CSM shock-interaction. 
However, the luminosity is $\sim$50--100$\times$ brighter than the most X-ray luminous CSM-interacting SN known \citep[SN\,2010jl, $L_X \sim10^{42}$\,\erg\psec;][]{ofek_xray_sn_csm, Ofek2014ApJ...781...42O, fransson2014ApJ...797..118F, chandra2015ApJ...810...32C}, and 2--3 orders of magnitude above Type IIn/Ib/c interacting populations at comparable phases \citep[e.g., SN\,1998Z, SN\,2010jl, SN\,2014C;][]{schlegel2006ApJ...646..378S, brethauer2022ApJ...939..105B, Ferdinand2026ApJ..1001...26F}. 

A final possibility is that the plateau arises from  the slower wings of a structured jet entering the line of sight \citep[e.g.,][]{granot2002ApJ...570L..61G, troja2022Natur.612..228T, srinivasaragavan2025MNRAS.538..351S}. We consider this the most likely explanation in light of other properties of this event. A structured-jet fit to the full multiwavelength data out to late times will be performed in a follow-up paper.


\subsection{GRB 260310A/AT 2026fgk as an Off-axis GRB?} 

We showed that the slow optical rise could be attributed to an LGRB viewed slightly off-axis. A slightly off-axis viewing angle could also help explain the relatively strong break between $\alpha_1 = -0.52\pm0.02$ and $\alpha_2 = -1.45^{+0.02}_{-0.01}$. A similar explanation was invoked for both GRB\,171205A \citep[][]{delia2018AA...619A..66D, urata2019ApJ...884L..58U, kumar2022NewA...9701889K} and GRB\,190829A \citep[e.g.,][]{chand2020ApJ...898...42C, rhodes2020MNRAS.496.3326R, hess2021Sci...372.1081H, fraija2021ApJ...918...12F, sato2021MNRAS.504.5647S, hu2021AA...646A..50H, salafia2022ApJ...931L..19S, dichiara2022MNRAS.512.2337D}. The X-ray rebrightening could potentially be interpreted as arising from additional material entering the line of sight, although this would need to be tested in detail.

The placement of \thisgrb on the Amati diagram and its comparison to other events are of note in the context of off-axis models. In an off-axis geometry, the relativistic Doppler factor $\delta = [\Gamma(1-\beta\cos(\theta_{\rm obs} - \theta_{\rm jet}))]^{-1}$ decreases as the viewing angle $\theta_{\rm obs}  - \theta_{\rm jet}$ increases, leading to \epeak$\propto \delta$ and \eiso$\propto \delta^{1-\alpha}$, where $\alpha$ is the low-energy photon index, as shown by \citet{yamazaki2003ApJ...594L..79Y, amati2007A&A...463..913A}, and discussed by \citet{delia2018AA...619A..66D} in the context of GRB\,171205A. For the spectral indices typical of GRBs ($\alpha<0$), $1 - \alpha>1$, so \eiso is suppressed more steeply than \epeak as $\delta$ decreases, displacing off-axis events to a region of the Amati plane characterized by higher \epeak/\eiso ratios---to above the canonical \epeak-\eiso relation. This is an interpretation for the subset of sub-luminous GRBs that occupy this region of the Amati diagram, including GRB\,171205A \citep{delia2018AA...619A..66D, wang2018ApJ...867..147W, suzuki2019ApJ...870...38S, maity2021ApJ...907...60M}. 
Canonical shock breakout sub-luminous GRBs---likely GRB\,980425, XRF\,060218, and XRF\,100316B---instead lie below the Amati relation, typically being interpreted as intrinsically less collimated or quasi-spherical outflows rather than off-axis events \citep[e.g.][]{campana2006Natur.442.1008C, nakar2012ApJ...747...88N, margutti2013ApJ...778...18M}.


Other studies on \thisgrb \citep{oconnor2026grb260310asn2026fgkphotometric, yang2026arXiv260523818Y, gill2026arXiv260526091G} have also suggested that we are viewing the event slightly off-axis ($\thv-\thj\lesssim10^\circ$).

\subsection{Implication of the Volumetric Rate for Jet Structure}

In Section~\ref{sec:rates}, we used the ZTF flux-limited survey to infer that the volumetric rate for fast (and therefore observed close to the jet axis) optical afterglows is consistent with the local on-axis GRB rate. Since the ZTF optical afterglows are typically detectable for hours up to $\approx1\,$d, this suggests that the beaming factor of the material at 1\,d is comparable to the beaming factor of the initial relativistic material producing the gamma-rays. This is consistent with the fact that approximately half of the optically discovered afterglows are found to have an associated detected GRB \citep{2022ApJ...938...85H}. It also strongly limits the existence of other phenomena with similarly energetic and collimated relativistic jets. 

\subsection{Polarimetric Properties}

\thisgrb has been observed polarimetrically at both optical and radio wavelengths. Our NOT/ALFOSC $r$-band measurement at (+2.9\,d observer frame) yielded $P = 0.60 \pm 0.3$\% (Section~\ref{sec:polarimetry}), consistent with no polarization at $2\sigma$ (Figure~\ref{fig:impol}), consistent with the 3$\sigma$ limit from \citet{GCN.43990}. At cm wavelengths, \citet{christy2026arXiv260427480C} reported the first multi-frequency linear polarization detection in a GRB afterglow at $T_0+19.2$\,d, with polarization decreasing from $3.18\pm0.18$\% at 25\,GHz to $0.69\pm0.22$\% at 11\,GHz. The same observations yielded the first detection of Faraday rotation in a GRB afterglow \citep[RM$\approx8300$\,rad\,m$^{-2}$ at $z=0.153$;][]{christy2026arXiv260427480C}, consistent with propagation through a dense, magnetized progenitor environment. \citet{yang2026arXiv260523818Y} later measured $P =  1.7\pm0.4$\% at 15\,GHz at $T_0$+55\,d, by which time both 6 and 10\,GHz lie on the optically thin segment of the SED (see \citealt{christy2026arXiv260427480C} and \citealt{yang2026arXiv260523818Y} for more details). In the VLA data presented here, we have also confirmed strong polarization---analysis and results will be shown in a future publication.


\section{Summary and Conclusions}
\label{sec:summary}

AT\,2026fgk is by far the closest LGRB afterglow discovered blindly by wide-field optical surveys, and the first confirmed to be associated with a sub-luminous GRB. This paper presents detailed multiwavelength data of AT\,2026fgk from the first 50 days post-burst. The event is clearly in a rarely-studied regime of parameter space. GRB\,260310A had a high peak gamma-ray energy given its low $E_\mathrm{\gamma,iso}$. 
The optical light curve has a slow rise to peak, a shallow decay, a late achromatic break observed in the X-ray and optical bands, and a shallow post-break decay slope. The X-ray afterglow exhibits a very late ($>20\,$d) rebrightening or plateau. One possible explanation for these characteristics is that the event was observed marginally off-axis, but late-time afterglow monitoring will be needed to distinguish between possible models. Despite the unusual afterglow, the properties of the underlying broad-lined Type Ic SN are fairly typical for GRB-SNe. 
We infer a peak $M_r\approx-18.3\,$mag, an ejecta mass $M_\mathrm{ej}\approx3.5$,\Msun, and a kinetic energy $E_{\rm K}\approx5$--$9\times10^{51}\,$erg, depending on  modeling assumptions. 

Almost two dozen afterglows have been discovered blindly by wide-field optical surveys, but AT\,2026fgk is only the second afterglow bright enough to be detected in ZTF's flux-limited ($m<18.5\,$mag) survey. We use the results of the flux-limited experiment to infer a volumetric rate for AT\,2026fgk-like events of $0.30^{+1.37}_{-0.29\,}$\,Gpc\,$^{-3}$\,yr$^{-1}$, consistent with the local on-axis LGRB rate of $\sim 1\,$Gpc\,$^{-3}$\,yr$^{-1}$. Therefore, we conclude that the opening angle of the material producing the optical afterglow at $\approx1\,$d is similar to the opening angle of the initial relativistic material producing the gamma-rays, consistent with conclusions from the ratio of optical afterglows with and without associated detected GRBs. 

Clearly, sub-luminous GRBs are highly diverse; some do not show any luminous afterglow (e.g., GRB~60218/SN\,2006aj), and modeling the multiwavelength emission suggests a quasi-isotropic outflow. Others, such as GRB\,260310A/AT\,2026fgk, clearly harbor a successful and energetic collimated jet. The blind identification of \thisgrb/AT\,2026fgk demonstrates that optical wide-field surveys can access the population independently of satellite coverage, providing a complementary route to measuring the true volumetric rates of long- and low-luminosity GRBs. 

\begin{acknowledgments}

We are grateful to T. Laskar for valuable comments on earlier drafts on this manuscript and D. Malesani for assistance in coordinating observations during the early phases of this campaign.

Oschin Telescope 48-inch and the 60-inch Telescope at the Palomar Observatory as part of the Zwicky Transient Facility project. ZTF is supported by the National Science Foundation under Grants No.\,AST-1440341, AST-2034437, and currently Award 2407588. ZTF receives additional funding from the ZTF partnership. Current members include Caltech, USA; Caltech/IPAC, USA; University of Maryland, USA; University of California, Berkeley, USA; University of Wisconsin at Milwaukee, USA; Cornell University, USA; Drexel University, USA; University of North Carolina at Chapel Hill, USA; Institute of Science and Technology, Austria; National Central University, Taiwan, and OKC, University of Stockholm, Sweden. Operations are conducted by Caltech’s Optical Observatory (COO), Caltech/IPAC and the University of Washington at Seattle, USA.

Zwicky Transient Facility access was supported by Northwestern University and the Center for Interdisciplinary Exploration and Research in Astrophysics (CIERA).

The Gordon and Betty Moore Foundation, through both the Data-Driven Investigator Program and a dedicated grant, provided critical funding for SkyPortal. This research has made use of the NASA/IPAC Extragalactic Database (NED), which is funded by the National Aeronautics and Space Administration and operated by the California Institute of Technology. 

SED Machine is based upon work supported by the National Science Foundation under Grant No.\ 1106171. The ZTF forced-photometry service was funded under the Heising-Simons Foundation grant \#12540303 (PI: Graham). The Gordon and Betty Moore Foundation, through both the Data-Driven Investigator Program and a dedicated grant, provided critical funding for SkyPortal.

This work has made use of data from the NuSTAR mission, a project led by the California Institute of Technology, managed by the Jet Propulsion Laboratory, and funded by the National Aeronautics and Space Administration. This research has made use of the NuSTAR Data Analysis Software (NuSTARDAS) jointly developed by the ASI Science Data Center (ASDC, Italy) and the California Institute of Technology (USA). We thank the entire NuSTAR Science \& Mission operations teams for the rapid approval and execution of our DDT observation request. 

This work is partially based on data obtained with Einstein Probe, a space mission supported by the Strategic Priority Program on Space Science of Chinese Academy of Sciences, in collaboration with the European Space Agency, the Max-Planck-Institute for extraterrestrial Physics (Germany), and the Centre National d'Études Spatiales (France). We would like to thank Hui Sun and the ToO duty scientists for rapidly approving and executing all of our requested observations.

The National Radio Astronomy Observatory and Green Bank Observatory are facilities of the U.S. National Science Foundation operated under cooperative agreement by Associated Universities, Inc.

The Liverpool Telescope is operated on the island of La Palma by Liverpool John Moores University in the Spanish
Observatorio del Roque de los Muchachos of the Instituto de Astrofisica de Canarias with financial support from the UK Science and Technology Facilities Council.

Partly based on observations made with the Nordic Optical Telescope, owned in collaboration by the University of Turku and Aarhus University, and operated jointly by Aarhus University, the University of Turku and the University of Oslo, representing Denmark, Finland and Norway, the University of Iceland and Stockholm University at the Observatorio del Roque de los Muchachos, La Palma, Spain, of the Instituto de Astrofisica de Canarias. The NOT data were obtained under program ID P71-506 (PI Malesani, Fynbo, Xu).

The Gravitational-wave Optical Transient Observer (GOTO) project acknowledges support from the Science and Technology Facilities Council (STFC, grant nos ST T007184/1, ST/T003103/1, ST/T000406/1, ST/X001121/1, and ST/Z000165/1) and the GOTO consortium institutions; the University of Warwick; Monash University; the University of Sheffield; the University of Leicester; Armagh Observatory \& Planetarium; the National Astronomical Research Institute of Thailand (NARIT); the University of Manchester; the University of Birmingham; Instituto de Astrofisica de Canarias (IAC); the University of Portsmouth and the University of Turku.

This work makes use of observations from the Las Cumbres Observatory global telescope network.

Pan-STARRS is primarily funded to search for near-Earth asteroids through NASA grants NNX08AR22G and NNX14AM74G. The Pan-STARRS science products were made possible through the contributions of the University of Hawai’i Institute for Astronomy, the Queen’s University Belfast and the University of Oxford. 

This work has made use of data from the Asteroid Terrestrial-impact Last Alert System (ATLAS) project. The Asteroid Terrestrial-impact Last Alert System (ATLAS) project is primarily funded to search for near earth asteroids through NASA grants NN12AR55G, 80NSSC18K0284, and 80NSSC18K1575; byproducts of the NEO search include images and catalogs from the survey area. This work was partially funded by Kepler/K2 grant J1944/80NSSC19K0112 and HST GO-15889, and STFC grants ST/T000198/1 and ST/S006109/1. The ATLAS science products have been made possible through the contributions of the University of Hawaii Institute for Astronomy, the Queen’s University Belfast, the Space Telescope Science Institute, the South African Astronomical Observatory, and The Millennium Institute of Astrophysics (MAS), Chile and the University of Oxford. 

This work is based in part on observations made with the NASA/ESA/CSA James Webb Space Telescope. The data were obtained from the Mikulski Archive for Space Telescopes at the Space Telescope Science Institute, which is operated by the Association of Universities for Research in Astronomy, Inc., under NASA contract NAS 5–03127 for JWST. These observations are associated with program GO 9254.

We thank the Mt Cuba Astronomical Foundation for supporting the development of MIRAGE. This work is based on observations obtained at the MDM Observatory, operated by Dartmouth College, Columbia University, Ohio State University, Ohio University, and the University of Michigan.

The Joan Oró Telescope (TJO) of the Montsec Observatory (OdM) is owned by the Generalitat de Catalunya and operated by the Institute for Space Studies of Catalonia (IEEC). We acknowledge the support provided by telescope operators from Observatori del Montsec.

This publication has made use of data collected at Lulin Observatory, partly supported by the TAOvA with the NSTC grant 114-2740-M-008-002.

Based on observations obtained at the Canada-France-Hawai'i Telescope (CFHT) which is operated by the National Research Council of Canada, the Institut National des Sciences de l'Univers of the Centre National de la Recherche Scientifique of France, and the University of Hawai'i. CFHT is located on Maunakea on Hawai'i Island, a mountain of considerable cultural, natural, and ecological significance. Maunakea is a sacred site to Native Hawaiians, also known as Kānaka 'Ōiwi. Quality observations are made possible by relentless effort of the entire staff at Canada-France-Hawai'i Telescope. Based on observations obtained with MegaPrime/MegaCam, a joint project of CFHT and CEA/DAPNIA.

The Caltech OVRO 40 m Telescope monitoring program is supported by NSF grants AST 2407603 and AST 2407604. P.V.d.l.P. acknowledges support from ANID Basal AFB-170002, Núcleo Milenio TITANs (NCN2023\_002), CATA BASAL FB210003 and UdeC-VRID 2025001479INV.

The Submillimeter Array is a joint project between the Smithsonian Astrophysical Observatory and the Academia Sinica Institute of Astronomy and Astrophysics and is funded by the Smithsonian Institution and the Academia Sinica. We recognize that Maunakea, location of the SMA, is a culturally important site for the indigenous Hawaiian people; we are privileged to study the cosmos from its summit.

Partly based on observations with the 100\,m telescope of the MPIfR (Max-Planck-Institut für Radioastronomie) at Effelsberg.

This work is based on observations carried out under project number E25AD with the IRAM NOEMA Interferometer. IRAM is supported by INSU/CNRS (France), MPG (Germany) and IGN (Spain). 

This research has made use of data and software provided by the High Energy Astrophysics Science Archive Research Center (HEASARC), which is a service of the Astrophysics Science Division at NASA/GSFC and the High Energy Astrophysics Division of the Smithsonian Astrophysical Observatory.

Blast is based upon work originally supported by the National Aeronautics and Space Administration under Grant No. 80NSSC21K0834 issued through the Astrophysics Data Analysis Program.

KH would like to thank the Caltech Presidential Postdoctoral Fellowship program and President Rosenbaum.

JCR was supported by NASA through the NASA Hubble Fellowship grant \#HST-HF2-51587.001-A awarded by the Space Telescope Science Institute, which is operated by the Association of Universities for Research in Astronomy, Inc., for NASA, under contract NAS5-26555.

AMC and LC acknowledge support from the Irish Research Council Postgraduate Scholarship No. GOIPG/2022/1008

F. Cuadra acknowledges funding from the European Research Council (ERC) under the European Union’s Horizon 2020 research and innovation programme (grant agreement No. 101117455).

Dimple acknowledges support from STFC grant no. ST/Y002253/1.

BPG acknowledges support from STFC grant no. ST/Y002253/1 and the Leverhulme Trust grant no. RPG-2024-117.

MP acknowledges support from a UK Research and Innovation Future Leaders Fellowship (grant references MR/T020784/1 and UKRI1062).

A. Ruiz del Pozo acknowledges funding from the European Research Council (ERC) under the European Union’s Horizon Europe research and innovation programme (grant agreement No. 101221278, project OUTLIERS).

M.W.C. acknowledges support from the National Science Foundation with grant numbers PHY-2117997, PHY-2308862 and PHY-2409481.

DO acknowledges support from the Levrehulme Trust grant no. RPG-2024-117.

T.-W.C. and A.A. acknowledge financial support from the Yushan Fellow Program of the Ministry of Education, Taiwan (MOE-111-YSFMS-0008-001-P1), and from the National Science and Technology Council, Taiwan (NSTC 114-2112-M-008-021-MY3).

S.Y. acknowledges the funding from the National Natural Science Foundation of China under grant No. 12303046, the Startup Research Fund of Henan Academy of Sciences No. 242041217, the Joint Fund of Henan Province Science and Technology R\&D Program No. 235200810057 and the Henan Province High-Level Talent International Training Program.

KWS acknowledges funding from the Royal Society.

MN is supported by the European Research Council (ERC) under the European Union’s Horizon 2020 research and innovation programme (grant agreement No.~948381).

G.S.H.P. acknowledges support from the Pan-STARRS project, which is a project of the Institute for Astronomy of the University of Hawai'i, and is supported by the NASA SSO Near Earth Observation Program under grants 80NSSC18K0971, NNX14AM74G, NNX12AR65G, NNX13AQ47G, NNX08AR22G, 80NSSC21K1572, and by the State of Hawai'i.

SJS acknowledges funding from STFC Grants ST/Y001605/1, ST/X001253/1, a Royal Society Research Professorship and the Hintze Family Charitable Foundation. 

TLK acknowledges support from a Warwick Astrophysics prize post-doctoral fellowship made possible thanks to a generous philanthropic donation.

RJ would like to acknowledge the support of the Klarman Fellowship.

IA is supported by the National Science Foundation award AST 2505775 and NASA grant 24-ADAP24-0159.

VSD is supported by a Leverhulme Research Fellowship (RF-2025-297/9) and the Spanish Ministry of Science, Innovation and Universities (PID2023-151588NB-I00).

\end{acknowledgments}

\software{\textsc{numpy} \citep{numpy},
          \textsc{astropy} \citep{astropy1, astropy2, astropy3},
          \textsc{photutils} \citep{photutils},
          \textsc{astroquery} \citep{astroquery},
          \textsc{dynesty} \citep{dynesty1, Speagle_2020},
          \textsc{nessai} \citep{nessai}
          \textsc{prospector} \citep{prospector1, prospector2},
          \textsc{frankenblast-host} \citep{nugent2026ApJ...997...38N},
          \textsc{sedpy} \citep{sedpy},
          Python-FSPS \citep{fsps},
          \textsc{sbi} \citep{sbi1},
          \textsc{sbi++} \citep{sbi2}}

\facilities{Einstein Probe/FXT (EP), Nuclear Spectroscopic Telescope Array (NuSTAR), Liverpool Telescope (IO:O), P60 (SEDM), Hale Telescope (NGPS), Zwicky Transient Facility (ZTF), Chandra, Karl G. Jansky Very Large Array (VLA), Owens Valley Radio Observatory (OVRO), Effelsberg, NOrthern Extended Millimeter Array (NOEMA), Sub-Millimeter Array (SMA), Large Array Survey Telescope (LAST), Gravitational-wave Optical Transient Observatory (GOTO), Panoramic Survey Telescope and Rapid Response System (Pan-STARRS), Asteroid Terrestrial-impact Last Alert System (ATLAS), Lowell Discovery Telescope (LDT), Nordic Optical Telescope (NOT), Joan Oro Telescope (TJO), Wide Infrared Field Transient Explorer (WINTER),  MDM Infrared Astronomy InGaAs Explorer (MIRAGE), Gemini Multi-Object Spectrograph (GMOS), Las Cumbres Observatory Global Network (LCOGT), Lulin Optical Telescope (LOT), Nuclear Spectroscopic Telkescope Array (NuSTAR), }

\clearpage
\newpage
\bibliography{refs}{}

@ARTICLE{ishida_iachec,
       author = {{Ishida}, Manabu and {Tsujimoto}, Masahiro and {Kohmura}, Takayoshi and {Stuhlinger}, Martin and {Smith}, Michael and {Marshall}, Herman L. and {Guainazzi}, Matteo and {Kawai}, Kohei and {Ogawa}, Taiki},
        title = "{Cross Spectral Calibration of Suzaku, XMM-Newton, and Chandra with PKS 2155304 as an Activity of IACHEC}",
      journal = {\pasj},
         year = 2011,
        month = nov,
       volume = {63},
        pages = {S657-S668},
          doi = {10.1093/pasj/63.sp3.S657},
       adsurl = {https://ui.adsabs.harvard.edu/abs/2011PASJ...63S.657I}
}

@ARTICLE{madsen_iachec,
       author = {{Madsen}, Kristin K. and {Beardmore}, Andrew P. and {Forster}, Karl and {Guainazzi}, Matteo and {Marshall}, Herman L. and {Miller}, Eric D. and {Page}, Kim L. and {Stuhlinger}, Martin},
        title = "{IACHEC Cross-calibration of Chandra, NuSTAR, Swift, Suzaku, XMM-Newton with 3C 273 and PKS 2155-304}",
      journal = {\aj},
         year = 2017,
        month = jan,
       volume = {153},
       number = {1},
          eid = {2},
        pages = {2},
          doi = {10.3847/1538-3881/153/1/2},
archivePrefix = {arXiv},
       eprint = {1609.09032},
 primaryClass = {astro-ph.IM},
       adsurl = {https://ui.adsabs.harvard.edu/abs/2017AJ....153....2M}
}

@ARTICLE{nevalainen_calibration,
       author = {{Nevalainen}, J. and {David}, L. and {Guainazzi}, M.},
        title = "{Cross-calibrating X-ray detectors with clusters of galaxies: an IACHEC study}",
      journal = {\aap},
         year = 2010,
        month = nov,
       volume = {523},
          eid = {A22},
        pages = {A22},
          doi = {10.1051/0004-6361/201015176},
archivePrefix = {arXiv},
       eprint = {1008.2102},
 primaryClass = {astro-ph.IM},
       adsurl = {https://ui.adsabs.harvard.edu/abs/2010A&A...523A..22N}
}

@ARTICLE{GCN.44278,
       author = {{Waratkar}, Gaurav and {Kammoun}, Elias and {Jayaraman}, Rahul and {Hinds}, K.-Ryan and {Ho}, Anna Y.~Q.},
        title = "{GRB 260310A / AT2026fgk: NuSTAR detection of the rebrightening}",
      journal = {GRB Coordinates Network},
         year = 2026,
        month = apr,
       volume = {44278},
        pages = {1},
       adsurl = {https://ui.adsabs.harvard.edu/abs/2026GCN.44278....1W}
}

@ARTICLE{2022ApJ...938...85H,
       author = {{Ho}, Anna Y.~Q. and {Perley}, Daniel A. and {Yao}, Yuhan and {Svinkin}, Dmitry and {de Ugarte Postigo}, A. and {Perley}, R.~A. and {Kann}, D. Alexander and {Burns}, Eric and {Andreoni}, Igor and {Bellm}, Eric C. and {Bissaldi}, Elisabetta and {Bloom}, Joshua S. and {Brink}, Thomas G. and {Dekany}, Richard and {Drake}, Andrew J. and {Ag{\"u}{\'\i} Fern{\'a}ndez}, Jos{\'e} Feliciano and {Filippenko}, Alexei V. and {Frederiks}, Dmitry and {Graham}, Matthew J. and {Hristov}, Boyan A. and {Kasliwal}, Mansi M. and {Kulkarni}, S.~R. and {Kumar}, Harsh and {Laher}, Russ R. and {Lysenko}, Alexandra L. and {Mailyan}, Bagrat and {Malacaria}, Christian and {Miller}, A.~A. and {Poolakkil}, S. and {Riddle}, Reed and {Ridnaia}, Anna and {Rusholme}, Ben and {Savchenko}, Volodymyr and {Sollerman}, Jesper and {Th{\"o}ne}, Christina and {Tsvetkova}, Anastasia and {Ulanov}, Mikhail and {von Kienlin}, Andreas},
        title = "{Cosmological Fast Optical Transients with the Zwicky Transient Facility: A Search for Dirty Fireballs}",
      journal = {\apj},
         year = 2022,
        month = oct,
       volume = {938},
       number = {1},
          eid = {85},
        pages = {85},
          doi = {10.3847/1538-4357/ac8bd0},
archivePrefix = {arXiv},
       eprint = {2201.12366},
 primaryClass = {astro-ph.HE},
       adsurl = {https://ui.adsabs.harvard.edu/abs/2022ApJ...938...85H}
}

@ARTICLE{2025ApJ...985..124L,
       author = {{Li}, Maggie L. and {Ho}, Anna Y.~Q. and {Ryan}, Geoffrey and {Perley}, Daniel A. and {Lamb}, Gavin P. and {Nayana}, A.~J. and {Andreoni}, Igor and {Anupama}, G.~C. and {Bellm}, Eric C. and {Berger}, Edo and {Bloom}, Joshua S. and {Burns}, Eric and {Caiazzo}, Ilaria and {Chandra}, Poonam and {Coughlin}, Michael W. and {El-Badry}, Kareem and {Graham}, Matthew J. and {Kasliwal}, Mansi and {Keating}, Garrett K. and {Kulkarni}, S.~R. and {Kumar}, Harsh and {Masci}, Frank J. and {Perley}, Richard A. and {Purdum}, Josiah and {Rao}, Ramprasad and {Rodriguez}, Antonio C. and {Rusholme}, Ben and {Sarin}, Nikhil and {Sollerman}, Jesper and {Srinivasaragavan}, Gokul P. and {Swain}, Vishwajeet and {Vanderbosch}, Zachary},
        title = "{The Nature of Optical Afterglows without Gamma-Ray Bursts: Identification of AT2023lcr and Multiwavelength Modeling}",
      journal = {\apj},
         year = 2025,
        month = may,
       volume = {985},
       number = {1},
          eid = {124},
        pages = {124},
          doi = {10.3847/1538-4357/adc800},
archivePrefix = {arXiv},
       eprint = {2411.07973},
 primaryClass = {astro-ph.HE},
       adsurl = {https://ui.adsabs.harvard.edu/abs/2025ApJ...985..124L}
}

@ARTICLE{2020ApJ...904..155A,
       author = {{Andreoni}, Igor and {Kool}, Erik C. and {Sagu{\'e}s Carracedo}, Ana and {Kasliwal}, Mansi M. and {Bulla}, Mattia and {Ahumada}, Tom{\'a}s and {Coughlin}, Michael W. and {Anand}, Shreya and {Sollerman}, Jesper and {Goobar}, Ariel and {Kaplan}, David L. and {Loveridge}, Tegan T. and {Karambelkar}, Viraj and {Cooke}, Jeff and {Bagdasaryan}, Ashot and {Bellm}, Eric C. and {Cenko}, S. Bradley and {Cook}, David O. and {De}, Kishalay and {Dekany}, Richard and {Delacroix}, Alexandre and {Drake}, Andrew and {Duev}, Dmitry A. and {Fremling}, Christoffer and {Golkhou}, V. Zach and {Graham}, Matthew J. and {Hale}, David and {Kulkarni}, S.~R. and {Kupfer}, Thomas and {Laher}, Russ R. and {Mahabal}, Ashish A. and {Masci}, Frank J. and {Rusholme}, Ben and {Smith}, Roger M. and {Tzanidakis}, Anastasios and {Van Sistine}, Angela and {Yao}, Yuhan},
        title = "{Constraining the Kilonova Rate with Zwicky Transient Facility Searches Independent of Gravitational Wave and Short Gamma-Ray Burst Triggers}",
      journal = {\apj},
         year = 2020,
        month = dec,
       volume = {904},
       number = {2},
          eid = {155},
        pages = {155},
          doi = {10.3847/1538-4357/abbf4c},
archivePrefix = {arXiv},
       eprint = {2008.00008},
 primaryClass = {astro-ph.HE},
       adsurl = {https://ui.adsabs.harvard.edu/abs/2020ApJ...904..155A}
}

@ARTICLE{2021ApJ...918...63A,
       author = {{Andreoni}, Igor and {Coughlin}, Michael W. and {Kool}, Erik C. and {Kasliwal}, Mansi M. and {Kumar}, Harsh and {Bhalerao}, Varun and {Carracedo}, Ana Sagu{\'e}s and {Ho}, Anna Y.~Q. and {Pang}, Peter T.~H. and {Saraogi}, Divita and {Sharma}, Kritti and {Shenoy}, Vedant and {Burns}, Eric and {Ahumada}, Tom{\'a}s and {Anand}, Shreya and {Singer}, Leo P. and {Perley}, Daniel A. and {De}, Kishalay and {Fremling}, U.~C. and {Bellm}, Eric C. and {Bulla}, Mattia and {Crellin-Quick}, Arien and {Dietrich}, Tim and {Drake}, Andrew and {Duev}, Dmitry A. and {Goobar}, Ariel and {Graham}, Matthew J. and {Kaplan}, David L. and {Kulkarni}, S.~R. and {Laher}, Russ R. and {Mahabal}, Ashish A. and {Shupe}, David L. and {Sollerman}, Jesper and {Walters}, Richard and {Yao}, Yuhan},
        title = "{Fast-transient Searches in Real Time with ZTFReST: Identification of Three Optically Discovered Gamma-Ray Burst Afterglows and New Constraints on the Kilonova Rate}",
      journal = {\apj},
         year = 2021,
        month = sep,
       volume = {918},
       number = {2},
          eid = {63},
        pages = {63},
          doi = {10.3847/1538-4357/ac0bc7},
archivePrefix = {arXiv},
       eprint = {2104.06352},
 primaryClass = {astro-ph.HE},
       adsurl = {https://ui.adsabs.harvard.edu/abs/2021ApJ...918...63A}
}

@ARTICLE{2025MNRAS.537.2362P,
       author = {{Perley}, Daniel A. and {Ho}, Anna Y.~Q. and {Fausnaugh}, Michael and {Lamb}, Gavin P. and {Kasliwal}, Mansi M. and {Ahumada}, Tomas and {Anand}, Shreya and {Andreoni}, Igor and {Bellm}, Eric and {Bhalerao}, Varun and {Bolin}, Bryce and {Brink}, Thomas G. and {Burns}, Eric and {Cenko}, S. Bradley and {Corsi}, Alessandra and {Filippenko}, Alexei V. and {Frederiks}, Dmitry and {Goldstein}, Adam and {Hamburg}, Rachel and {Jayaraman}, Rahul and {Jonker}, Peter G. and {Kool}, Erik C. and {Kulkarni}, Shrinivas R. and {Kumar}, Harsh and {Laher}, Russ and {Levan}, Andrew and {Lysenko}, Alexandra and {Perley}, Richard A. and {Ricker}, George R. and {Riddle}, Reed and {Ridnaia}, Anna and {Rusholme}, Ben and {Smith}, Roger and {Svinkin}, Dmitry and {Ulanov}, Mikhail and {Vanderspek}, Roland and {Waratkar}, Gaurav and {Yao}, Yuhan},
        title = "{The luminous, slow-rising orphan afterglow AT2019pim as a candidate moderately relativistic outflow}",
      journal = {\mnras},
         year = 2025,
        month = mar,
       volume = {537},
       number = {3},
        pages = {2362-2379},
          doi = {10.1093/mnras/staf125},
archivePrefix = {arXiv},
       eprint = {2401.16470},
 primaryClass = {astro-ph.HE},
       adsurl = {https://ui.adsabs.harvard.edu/abs/2025MNRAS.537.2362P}
}

@ARTICLE{2013ApJ...769..130C,
       author = {{Cenko}, S. Bradley and {Kulkarni}, S.~R. and {Horesh}, Assaf and {Corsi}, Alessandra and {Fox}, Derek B. and {Carpenter}, John and {Frail}, Dale A. and {Nugent}, Peter E. and {Perley}, Daniel A. and {Gruber}, D. and {Gal-Yam}, Avishay and {Groot}, Paul J. and {Hallinan}, G. and {Ofek}, Eran O. and {Rau}, Arne and {MacLeod}, Chelsea L. and {Miller}, Adam A. and {Bloom}, Joshua S. and {Filippenko}, Alexei V. and {Kasliwal}, Mansi M. and {Law}, Nicholas M. and {Morgan}, Adam N. and {Polishook}, David and {Poznanski}, Dovi and {Quimby}, Robert M. and {Sesar}, Branimir and {Shen}, Ken J. and {Silverman}, Jeffrey M. and {Sternberg}, Assaf},
        title = "{Discovery of a Cosmological, Relativistic Outburst via its Rapidly Fading Optical Emission}",
      journal = {\apj},
         year = 2013,
        month = jun,
       volume = {769},
       number = {2},
          eid = {130},
        pages = {130},
          doi = {10.1088/0004-637X/769/2/130},
archivePrefix = {arXiv},
       eprint = {1304.4236},
 primaryClass = {astro-ph.CO},
       adsurl = {https://ui.adsabs.harvard.edu/abs/2013ApJ...769..130C}
}

@ARTICLE{2015ApJ...803L..24C,
       author = {{Cenko}, S. Bradley and {Urban}, Alex L. and {Perley}, Daniel A. and {Horesh}, Assaf and {Corsi}, Alessandra and {Fox}, Derek B. and {Cao}, Yi and {Kasliwal}, Mansi M. and {Lien}, Amy and {Arcavi}, Iair and {Bloom}, Joshua S. and {Butler}, Nat R. and {Cucchiara}, Antonino and {de Diego}, Jos{\'e} A. and {Filippenko}, Alexei V. and {Gal-Yam}, Avishay and {Gehrels}, Neil and {Georgiev}, Leonid and {Jes{\'u}s Gonz{\'a}lez}, J. and {Graham}, John F. and {Greiner}, Jochen and {Kann}, D. Alexander and {Klein}, Christopher R. and {Knust}, Fabian and {Kulkarni}, S.~R. and {Kutyrev}, Alexander and {Laher}, Russ and {Lee}, William H. and {Nugent}, Peter E. and {Prochaska}, J. Xavier and {Ramirez-Ruiz}, Enrico and {Richer}, Michael G. and {Rubin}, Adam and {Urata}, Yuji and {Varela}, Karla and {Watson}, Alan M. and {Wozniak}, Przemek R.},
        title = "{iPTF14yb: The First Discovery of a Gamma-Ray Burst Afterglow Independent of a High-energy Trigger}",
      journal = {\apjl},
         year = 2015,
        month = apr,
       volume = {803},
       number = {2},
          eid = {L24},
        pages = {L24},
          doi = {10.1088/2041-8205/803/2/L24},
archivePrefix = {arXiv},
       eprint = {1504.00673},
 primaryClass = {astro-ph.HE},
       adsurl = {https://ui.adsabs.harvard.edu/abs/2015ApJ...803L..24C}
}

@ARTICLE{2017ApJ...850..149S,
       author = {{Stalder}, B. and {Tonry}, J. and {Smartt}, S.~J. and {Coughlin}, M. and {Chambers}, K.~C. and {Stubbs}, C.~W. and {Chen}, T.-W. and {Kankare}, E. and {Smith}, K.~W. and {Denneau}, L. and {Sherstyuk}, A. and {Heinze}, A. and {Weiland}, H. and {Rest}, A. and {Young}, D.~R. and {Huber}, M.~E. and {Flewelling}, H. and {Lowe}, T. and {Magnier}, E.~A. and {Schultz}, A.~S.~B. and {Waters}, C. and {Wainscoat}, R. and {Willman}, M. and {Wright}, D.~E. and {Chu}, J. and {Sanders}, D. and {Inserra}, C. and {Maguire}, K. and {Kotak}, R.},
        title = "{Observations of the GRB Afterglow ATLAS17aeu and Its Possible Association with GW 170104}",
      journal = {\apj},
         year = 2017,
        month = dec,
       volume = {850},
       number = {2},
          eid = {149},
        pages = {149},
          doi = {10.3847/1538-4357/aa95c1},
archivePrefix = {arXiv},
       eprint = {1706.00175},
 primaryClass = {astro-ph.HE},
       adsurl = {https://ui.adsabs.harvard.edu/abs/2017ApJ...850..149S}
}

@ARTICLE{2017ApJ...845..152B,
       author = {{Bhalerao}, V. and {Kasliwal}, M.~M. and {Bhattacharya}, D. and {Corsi}, A. and {Aarthy}, E. and {Adams}, S.~M. and {Blagorodnova}, N. and {Cantwell}, T. and {Cenko}, S.~B. and {Fender}, R. and {Frail}, D. and {Itoh}, R. and {Jencson}, J. and {Kawai}, N. and {Kong}, A.~K.~H. and {Kupfer}, T. and {Kutyrev}, A. and {Mao}, J. and {Mate}, S. and {Mithun}, N.~P.~S. and {Mooley}, K. and {Perley}, D.~A. and {Perrott}, Y.~C. and {Quimby}, R.~M. and {Rao}, A.~R. and {Singer}, L.~P. and {Sharma}, V. and {Titterington}, D.~J. and {Troja}, E. and {Vadawale}, S.~V. and {Vibhute}, A. and {Vedantham}, H. and {Veilleux}, S.},
        title = "{A Tale of Two Transients: GW 170104 and GRB 170105A}",
      journal = {\apj},
         year = 2017,
        month = aug,
       volume = {845},
       number = {2},
          eid = {152},
        pages = {152},
          doi = {10.3847/1538-4357/aa81d2},
archivePrefix = {arXiv},
       eprint = {1706.00024},
 primaryClass = {astro-ph.HE},
       adsurl = {https://ui.adsabs.harvard.edu/abs/2017ApJ...845..152B}
}

@ARTICLE{2019A&A...621A..81M,
       author = {{Melandri}, A. and {Rossi}, A. and {Benetti}, S. and {D'Elia}, V. and {Piranomonte}, S. and {Palazzi}, E. and {Levan}, A.~J. and {Branchesi}, M. and {Castro-Tirado}, A.~J. and {D'Avanzo}, P. and {Hu}, Y.-D. and {Raimondo}, G. and {Tanvir}, N.~R. and {Tomasella}, L. and {Amati}, L. and {Campana}, S. and {Carini}, R. and {Covino}, S. and {Cusano}, F. and {Dadina}, M. and {Della Valle}, M. and {Fan}, X. and {Garnavich}, P. and {Grado}, A. and {Greco}, G. and {Hjorth}, J. and {Lyman}, J.~D. and {Masetti}, N. and {O'Brien}, P. and {Pian}, E. and {Perego}, A. and {Salvaterra}, R. and {Stella}, L. and {Stratta}, G. and {Yang}, S. and {di Paola}, A. and {Caballero-Garc{\'\i}a}, M.~D. and {Fruchter}, A.~S. and {Giunta}, A. and {Longo}, F. and {Pinamonti}, M. and {Sokolov}, V.~V. and {Testa}, V. and {Valeev}, A.~F. and {Brocato}, E.},
        title = "{Unveiling the enigma of ATLAS17aeu}",
      journal = {\aap},
         year = 2019,
        month = jan,
       volume = {621},
          eid = {A81},
        pages = {A81},
          doi = {10.1051/0004-6361/201833814},
archivePrefix = {arXiv},
       eprint = {1807.03681},
 primaryClass = {astro-ph.HE},
       adsurl = {https://ui.adsabs.harvard.edu/abs/2019A&A...621A..81M}
}

@ARTICLE{2006RPPh...69.2259M,
       author = {{M{\'e}sz{\'a}ros}, P.},
        title = "{Gamma-ray bursts}",
      journal = {Reports on Progress in Physics},
         year = 2006,
        month = aug,
       volume = {69},
       number = {8},
        pages = {2259-2321},
          doi = {10.1088/0034-4885/69/8/R01},
archivePrefix = {arXiv},
       eprint = {astro-ph/0605208},
 primaryClass = {astro-ph},
       adsurl = {https://ui.adsabs.harvard.edu/abs/2006RPPh...69.2259M}
}

@ARTICLE{2023arXiv230516279M,
       author = {{Masci}, Frank J. and {Laher}, Russ R. and {Rusholme}, Benjamin and {Shupe}, David and {Paladini}, Roberta and {Groom}, Steve and {Wold}, Avery and {Miller}, Adam A. and {Drake}, Andrew},
        title = "{A New Forced Photometry Service for the Zwicky Transient Facility}",
      journal = {arXiv e-prints},
         year = 2023,
        month = may,
          eid = {arXiv:2305.16279},
        pages = {arXiv:2305.16279},
          doi = {10.48550/arXiv.2305.16279},
archivePrefix = {arXiv},
       eprint = {2305.16279},
 primaryClass = {astro-ph.IM},
       adsurl = {https://ui.adsabs.harvard.edu/abs/2023arXiv230516279M}
}

@ARTICLE{Ho2020ApJ...905...98H,
       author = {{Ho}, Anna Y.~Q. and {Perley}, Daniel A. and {Beniamini}, Paz and {Cenko}, S. Bradley and {Kulkarni}, S.~R. and {Andreoni}, Igor and {Singer}, Leo P. and {De}, Kishalay and {Kasliwal}, Mansi M. and {Fremling}, Christoffer and {Bellm}, Eric C. and {Dekany}, Richard and {Delacroix}, Alexandre and {Duev}, Dmitry A. and {Goldstein}, Daniel A. and {Golkhou}, V. Zach and {Goobar}, Ariel and {Graham}, Matthew J. and {Hale}, David and {Kupfer}, Thomas and {Laher}, Russ R. and {Masci}, Frank J. and {Miller}, Adam A. and {Neill}, James D. and {Riddle}, Reed and {Rusholme}, Ben and {Shupe}, David L. and {Smith}, Roger and {Sollerman}, Jesper and {van Roestel}, Jan},
        title = "{ZTF20aajnksq (AT 2020blt): A Fast Optical Transient at z ≍ 2.9 with No Detected Gamma-Ray Burst Counterpart}",
      journal = {\apj},
         year = 2020,
        month = dec,
       volume = {905},
       number = {2},
          eid = {98},
        pages = {98},
          doi = {10.3847/1538-4357/abc34d},
archivePrefix = {arXiv},
       eprint = {2006.10761},
 primaryClass = {astro-ph.HE},
       adsurl = {https://ui.adsabs.harvard.edu/abs/2020ApJ...905...98H}
}

@ARTICLE{ofek_xray_sn_csm,
       author = {{Ofek}, E.~O. and {Fox}, D. and {Cenko}, S.~B. and {Sullivan}, M. and {Gnat}, O. and {Frail}, D.~A. and {Horesh}, A. and {Corsi}, A. and {Quimby}, R.~M. and {Gehrels}, N. and {Kulkarni}, S.~R. and {Gal-Yam}, A. and {Nugent}, P.~E. and {Yaron}, O. and {Filippenko}, A.~V. and {Kasliwal}, M.~M. and {Bildsten}, L. and {Bloom}, J.~S. and {Poznanski}, D. and {Arcavi}, I. and {Laher}, R.~R. and {Levitan}, D. and {Sesar}, B. and {Surace}, J.},
        title = "{X-Ray Emission from Supernovae in Dense Circumstellar Matter Environments: A Search for Collisionless Shocks}",
      journal = {\apj},
         year = 2013,
        month = jan,
       volume = {763},
       number = {1},
          eid = {42},
        pages = {42},
          doi = {10.1088/0004-637X/763/1/42},
archivePrefix = {arXiv},
       eprint = {1206.0748},
 primaryClass = {astro-ph.HE},
       adsurl = {https://ui.adsabs.harvard.edu/abs/2013ApJ...763...42O}
}

@ARTICLE{ioka_grb_variability,
       author = {{Ioka}, Kunihito and {Kobayashi}, Shiho and {Zhang}, Bing},
        title = "{Variabilities of Gamma-Ray Burst Afterglows: Long-acting Engine, Anisotropic Jet, or Many Fluctuating Regions?}",
      journal = {\apj},
         year = 2005,
        month = sep,
       volume = {631},
       number = {1},
        pages = {429-434},
          doi = {10.1086/432567},
archivePrefix = {arXiv},
       eprint = {astro-ph/0409376},
 primaryClass = {astro-ph},
       adsurl = {https://ui.adsabs.harvard.edu/abs/2005ApJ...631..429I}
}

@ARTICLE{granot_nakar_piran_refreshed_shocks,
       author = {{Granot}, Jonathan and {Nakar}, Ehud and {Piran}, Tsvi},
        title = "{Astrophysics: refreshed shocks from a {\ensuremath{\gamma}}-ray burst}",
      journal = {\nat},
         year = 2003,
        month = nov,
       volume = {426},
       number = {6963},
        pages = {138-139},
          doi = {10.1038/426138a},
archivePrefix = {arXiv},
       eprint = {astro-ph/0304563},
 primaryClass = {astro-ph},
       adsurl = {https://ui.adsabs.harvard.edu/abs/2003Natur.426..138G}
}

@ARTICLE{tiengo_grb030329_xray,
       author = {{Tiengo}, A. and {Mereghetti}, S. and {Ghisellini}, G. and {Tavecchio}, F. and {Ghirlanda}, G.},
        title = "{Late evolution of the X-ray afterglow of GRB 030329}",
      journal = {\aap},
         year = 2004,
        month = sep,
       volume = {423},
        pages = {861-865},
          doi = {10.1051/0004-6361:20041027},
archivePrefix = {arXiv},
       eprint = {astro-ph/0402644},
 primaryClass = {astro-ph},
       adsurl = {https://ui.adsabs.harvard.edu/abs/2004A&A...423..861T}
}

@ARTICLE{hu2021AA...646A..50H,
       author = {{Hu}, Y. -D. and {Castro-Tirado}, A.~J. and {Kumar}, A. and {Gupta}, R. and {Valeev}, A.~F. and {Pandey}, S.~B. and {Kann}, D.~A. and {Castell{\'o}n}, A. and {Agudo}, I. and {Aryan}, A. and {Caballero-Garc{\'\i}a}, M.~D. and {Guziy}, S. and {Martin-Carrillo}, A. and {Oates}, S.~R. and {Pian}, E. and {S{\'a}nchez-Ram{\'\i}rez}, R. and {Sokolov}, V.~V. and {Zhang}, B. -B.},
        title = "{10.4 m GTC observations of the nearby VHE-detected GRB 190829A/SN 2019oyw}",
      journal = {\aap},
         year = 2021,
        month = feb,
       volume = {646},
          eid = {A50},
        pages = {A50},
          doi = {10.1051/0004-6361/202039349},
archivePrefix = {arXiv},
       eprint = {2009.04021},
 primaryClass = {astro-ph.HE},
       adsurl = {https://ui.adsabs.harvard.edu/abs/2021A&A...646A..50H}
}

@ARTICLE{kaastra_optimal_bin,
       author = {{Kaastra}, J.~S. and {Bleeker}, J.~A.~M.},
        title = "{Optimal binning of X-ray spectra and response matrix design}",
      journal = {\aap},
         year = 2016,
        month = mar,
       volume = {587},
          eid = {A151},
        pages = {A151},
          doi = {10.1051/0004-6361/201527395},
archivePrefix = {arXiv},
       eprint = {1601.05309},
 primaryClass = {astro-ph.IM},
       adsurl = {https://ui.adsabs.harvard.edu/abs/2016A&A...587A.151K}
}

@ARTICLE{2023PASP..135l4502O,
       author = {{Ofek}, E.~O. and {Shvartzvald}, Y. and {Sharon}, A. and {Tishler}, C. and {Elhanati}, D. and {Segev}, N. and {Ben-Ami}, S. and {Nir}, G. and {Segre}, E. and {Sofer-Rimalt}, Y. and {Blumenzweig}, A. and {Strotjohann}, N.~L. and {Polishook}, D. and {Krassilchtchikov}, A. and {Zenin}, A. and {Fallah Ramazani}, V. and {Weimann}, S. and {Garrappa}, S. and {Shanni}, Y. and {Chen}, P. and {Zimmerman}, E.},
        title = "{The Large Array Survey Telescope-Pipeline. I. Basic Image Reduction and Visit Coaddition}",
      journal = {\pasp},
         year = 2023,
        month = dec,
       volume = {135},
       number = {1054},
          eid = {124502},
        pages = {124502},
          doi = {10.1088/1538-3873/ad0977},
archivePrefix = {arXiv},
       eprint = {2310.13063},
 primaryClass = {astro-ph.IM},
       adsurl = {https://ui.adsabs.harvard.edu/abs/2023PASP..135l4502O}
}

@ARTICLE{2024AJ....167..281S,
       author = {{Springer}, Ofer and {Ofek}, Eran O. and {Zackay}, Barak and {Konno}, Ruslan and {Sharon}, Amir and {Nir}, Guy and {Rubin}, Adam and {Haddad}, Asaf and {Friedman}, Jonathan and {Schein-Lubomirsky}, Leora and {Aizenberg}, Iakov and {Krassilchtchikov}, Alexander and {Gal-Yam}, Avishay},
        title = "{TRANSLIENT: Detecting Transients Resulting from Point-source Motion or Astrometric Errors}",
      journal = {\aj},
         year = 2024,
        month = jun,
       volume = {167},
       number = {6},
          eid = {281},
        pages = {281},
          doi = {10.3847/1538-3881/ad408d},
archivePrefix = {arXiv},
       eprint = {2403.09771},
 primaryClass = {astro-ph.IM},
       adsurl = {https://ui.adsabs.harvard.edu/abs/2024AJ....167..281S}
}

@ARTICLE{Konno2026,
       author = {{Konno}, R. and {Ofek}, E.~O. and {Krassilchtchikov}, A. and {Shvartzvald}, Y. and {Ben-Ami}, S. and {Polishook}, D. and {Tishler}, C. and {Segre}, E. and {Garrappa}, S. and {Zimmermann}, E.~A. and {Horowicz}, A. and {Chen}, P. and {Gal-Yam}, A. and {Engel}, M. and {Shani}, Y.~M. and {Spitzer}, S.~A. and {Fainer}, S. and {Yaron}, O. and {Blumenzweig}, A.},
        title = "{The Large Array Survey Telescope-Pipeline. II. Image Subtraction and Transient Detection}",
      journal = {arXiv e-prints},
         year = 2026,
        month = apr,
          eid = {arXiv:2604.27921},
        pages = {arXiv:2604.27921},
          doi = {10.48550/arXiv.2604.27921},
archivePrefix = {arXiv},
       eprint = {2604.27921},
 primaryClass = {astro-ph.IM},
       adsurl = {https://ui.adsabs.harvard.edu/abs/2026arXiv260427921K}
}

@INCOLLECTION{yuan_ep_mission,
       author = {{Yuan}, Weimin and {Zhang}, Chen and {Chen}, Yong and {Ling}, Zhixing},
        title = "{The Einstein Probe Mission}",
    booktitle = {Handbook of X-ray and Gamma-ray Astrophysics},
         year = 2022,
       editor = {{Bambi}, Cosimo and {Sangangelo}, Andrea},
          eid = {86},
        pages = {86},
          doi = {10.1007/978-981-16-4544-0_151-1},
       adsurl = {https://ui.adsabs.harvard.edu/abs/2022hxga.book...86Y}
}

@ARTICLE{yuan_ep_science,
       author = {{Yuan}, Weimin and {Dai}, Lixin and {Feng}, Hua and {Jin}, Chichuan and {Jonker}, Peter and {Kuulkers}, Erik and {Liu}, Yuan and {Nandra}, Kirpal and {O'Brien}, Paul and {Piro}, Luigi and {Rau}, Arne and {Rea}, Nanda and {Sanders}, Jeremy and {Tao}, Lian and {Wang}, Junfeng and {Wu}, Xuefeng and {Zhang}, Bing and {Zhang}, Shuangnan and {Ai}, Shunke and {Buchner}, Johannes and {Bulbul}, Esra and {Chen}, Hechao and {Chen}, Minghua and {Chen}, Yong and {Chen}, Yu-Peng and {Coleiro}, Alexis and {Coti Zelati}, Francesco and {Dai}, Zigao and {Fan}, Xilong and {Fan}, Zhou and {Friedrich}, Susanne and {Gao}, He and {Ge}, Chong and {Ge}, Mingyu and {Geng}, Jinjun and {Ghirlanda}, Giancarlo and {Gianfagna}, Giulia and {Gou}, Lijun and {Guillot}, S{\'e}bastien and {Hou}, Xian and {Hu}, Jingwei and {Huang}, Yongfeng and {Ji}, Long and {Jia}, Shumei and {Komossa}, S. and {Kong}, Albert K.~H. and {Lan}, Lin and {Li}, An and {Li}, Ang and {Li}, Chengkui and {Li}, Dongyue and {Li}, Jian and {Li}, Zhaosheng and {Ling}, Zhixing and {Liu}, Ang and {Liu}, Jinzhong and {Liu}, Liangduan and {Liu}, Zhu and {Luo}, Jiawei and {Ma}, Ruican and {Maggi}, Pierre and {Maitra}, Chandreyee and {Marino}, Alessio and {Ng}, Stephen Chi-Yung and {Pan}, Haiwu and {Rukdee}, Surangkhana and {Soria}, Roberto and {Sun}, Hui and {Tam}, Pak-Hin Thomas and {Thakur}, Aishwarya Linesh and {Tian}, Hui and {Troja}, Eleonora and {Wang}, Wei and {Wang}, Xiangyu and {Wang}, Yanan and {Wei}, Junjie and {Wen}, Sixiang and {Wu}, Jianfeng and {Wu}, Ting and {Xiao}, Di and {Xu}, Dong and {Xu}, Renxin and {Xu}, Yanjun and {Xu}, Yu and {Yang}, Haonan and {You}, Bei and {Yu}, Heng and {Yu}, Yunwei and {Zhang}, Binbin and {Zhang}, Chen and {Zhang}, Guobao and {Zhang}, Liang and {Zhang}, Wenda and {Zhang}, Yu and {Zhou}, Ping and {Zou}, Zecheng},
        title = "{Science objectives of the Einstein Probe mission}",
      journal = {Science China Physics, Mechanics, and Astronomy},
         year = 2025,
        month = mar,
       volume = {68},
       number = {3},
          eid = {239501},
        pages = {239501},
          doi = {10.1007/s11433-024-2600-3},
archivePrefix = {arXiv},
       eprint = {2501.07362},
 primaryClass = {astro-ph.HE},
       adsurl = {https://ui.adsabs.harvard.edu/abs/2025SCPMA..6839501Y}
}

@ARTICLE{De2020,
       author = {{De}, Kishalay and {Hankins}, Matthew J. and {Kasliwal}, Mansi M. and {Moore}, Anna M. and {Ofek}, Eran O. and {Adams}, Scott M. and {Ashley}, Michael C.~B. and {Babul}, Aliya-Nur and {Bagdasaryan}, Ashot and {Burdge}, Kevin B. and {Burnham}, Jill and {Dekany}, Richard G. and {Declacroix}, Alexander and {Galla}, Antony and {Greffe}, Tim and {Hale}, David and {Jencson}, Jacob E. and {Lau}, Ryan M. and {Mahabal}, Ashish and {McKenna}, Daniel and {Sharma}, Manasi and {Shopbell}, Patrick L. and {Smith}, Roger M. and {Soon}, Jamie and {Sokoloski}, Jennifer and {Soria}, Roberto and {Travouillon}, Tony},
        title = "{Palomar Gattini-IR: Survey Overview, Data Processing System, On-sky Performance and First Results}",
      journal = {\pasp},
         year = 2020,
        month = feb,
       number = {1008},
          eid = {025001},
        pages = {025001},
          doi = {10.1088/1538-3873/ab6069},
archivePrefix = {arXiv},
       eprint = {1910.13319},
 primaryClass = {astro-ph.IM},
       adsurl = {https://ui.adsabs.harvard.edu/abs/2020PASP..132b5001D}
}

@ARTICLE{nustar_calibration,
       author = {{Madsen}, Kristin K. and {Harrison}, Fiona A. and {Markwardt}, Craig B. and {An}, Hongjun and {Grefenstette}, Brian W. and {Bachetti}, Matteo and {Miyasaka}, Hiromasa and {Kitaguchi}, Takao and {Bhalerao}, Varun and {Boggs}, Steve and {Christensen}, Finn E. and {Craig}, William W. and {Forster}, Karl and {Fuerst}, Felix and {Hailey}, Charles J. and {Perri}, Matteo and {Puccetti}, Simonetta and {Rana}, Vikram and {Stern}, Daniel and {Walton}, Dominic J. and {J{\o}rgen Westergaard}, Niels and {Zhang}, William W.},
        title = "{Calibration of the NuSTAR High-energy Focusing X-ray Telescope.}",
      journal = {\apjs},
         year = 2015,
        month = sep,
       volume = {220},
       number = {1},
          eid = {8},
        pages = {8},
          doi = {10.1088/0067-0049/220/1/8},
archivePrefix = {arXiv},
       eprint = {1504.01672},
 primaryClass = {astro-ph.IM},
       adsurl = {https://ui.adsabs.harvard.edu/abs/2015ApJS..220....8M}
}

@INPROCEEDINGS{ciao_chandra,
       author = {{Fruscione}, Antonella and {McDowell}, Jonathan C. and {Allen}, Glenn E. and {Brickhouse}, Nancy S. and {Burke}, Douglas J. and {Davis}, John E. and {Durham}, Nick and {Elvis}, Martin and {Galle}, Elizabeth C. and {Harris}, Daniel E. and {Huenemoerder}, David P. and {Houck}, John C. and {Ishibashi}, Bish and {Karovska}, Margarita and {Nicastro}, Fabrizio and {Noble}, Michael S. and {Nowak}, Michael A. and {Primini}, Frank A. and {Siemiginowska}, Aneta and {Smith}, Randall K. and {Wise}, Michael},
        title = "{CIAO: Chandra's data analysis system}",
    booktitle = {Observatory Operations: Strategies, Processes, and Systems},
         year = 2006,
       editor = {{Silva}, David R. and {Doxsey}, Rodger E.},
       series = {Society of Photo-Optical Instrumentation Engineers (SPIE) Conference Series},
       volume = {6270},
        month = jun,
          eid = {62701V},
        pages = {62701V},
          doi = {10.1117/12.671760},
       adsurl = {https://ui.adsabs.harvard.edu/abs/2006SPIE.6270E..1VF}
}

@ARTICLE{yang_chandra_gcn,
       author = {{Yang}, Yu-Han and {Troja}, Eleonora and {Yin}, Yi-Han Iris and {O'Connor}, Brendan and {Yadav}, Muskan and {Passaleva} Niccolo and {Ricci}, Roberto},
        title = "{GRB 260310A: Late-time Chandra X-ray Detection}",
      journal = {GRB Coordinates Network},
         year = 2026,
        month = apr,
       volume = {44375},
        pages = {1}
}

@ARTICLE{kumar_flares_210204a,
       author = {{Kumar}, Harsh and {Gupta}, Rahul and {Saraogi}, Divita and {Ahumada}, Tom{\'a}s and {Andreoni}, Igor and {Anupama}, G.~C. and {Aryan}, Amar and {Barway}, Sudhanshu and {Bhalerao}, Varun and {Chandra}, Poonam and {Coughlin}, Michael W. and {Dimple} and {Dutta}, Anirban and {ghosh}, Ankur and {Ho}, Anna Y.~Q. and {Kool}, E.~C. and {Kumar}, Amit and {Medford}, Michael S. and {Misra}, Kuntal and {Pandey}, Shashi B. and {Perley}, Daniel A. and {Riddle}, Reed and {Ror}, Amit Kumar and {Setiadi}, Jason M. and {Yao}, Yuhan},
        title = "{The long-active afterglow of GRB 210204A: detection of the most delayed flares in a gamma-ray burst}",
      journal = {\mnras},
         year = 2022,
        month = jun,
       volume = {513},
       number = {2},
        pages = {2777-2793},
          doi = {10.1093/mnras/stac1061},
archivePrefix = {arXiv},
       eprint = {2204.07587},
 primaryClass = {astro-ph.HE},
       adsurl = {https://ui.adsabs.harvard.edu/abs/2022MNRAS.513.2777K}
}

@ARTICLE{nustar_mission,
       author = {{Harrison}, Fiona A. and {Craig}, William W. and {Christensen}, Finn E. and {Hailey}, Charles J. and {Zhang}, William W. and {Boggs}, Steven E. and {Stern}, Daniel and {Cook}, W. Rick and {Forster}, Karl and {Giommi}, Paolo and {Grefenstette}, Brian W. and {Kim}, Yunjin and {Kitaguchi}, Takao and {Koglin}, Jason E. and {Madsen}, Kristin K. and {Mao}, Peter H. and {Miyasaka}, Hiromasa and {Mori}, Kaya and {Perri}, Matteo and {Pivovaroff}, Michael J. and {Puccetti}, Simonetta and {Rana}, Vikram R. and {Westergaard}, Niels J. and {Willis}, Jason and {Zoglauer}, Andreas and {An}, Hongjun and {Bachetti}, Matteo and {Barri{\`e}re}, Nicolas M. and {Bellm}, Eric C. and {Bhalerao}, Varun and {Brejnholt}, Nicolai F. and {Fuerst}, Felix and {Liebe}, Carl C. and {Markwardt}, Craig B. and {Nynka}, Melania and {Vogel}, Julia K. and {Walton}, Dominic J. and {Wik}, Daniel R. and {Alexander}, David M. and {Cominsky}, Lynn R. and {Hornschemeier}, Ann E. and {Hornstrup}, Allan and {Kaspi}, Victoria M. and {Madejski}, Greg M. and {Matt}, Giorgio and {Molendi}, Silvano and {Smith}, David M. and {Tomsick}, John A. and {Ajello}, Marco and {Ballantyne}, David R. and {Balokovi{\'c}}, Mislav and {Barret}, Didier and {Bauer}, Franz E. and {Blandford}, Roger D. and {Brandt}, W. Niel and {Brenneman}, Laura W. and {Chiang}, James and {Chakrabarty}, Deepto and {Chenevez}, Jerome and {Comastri}, Andrea and {Dufour}, Francois and {Elvis}, Martin and {Fabian}, Andrew C. and {Farrah}, Duncan and {Fryer}, Chris L. and {Gotthelf}, Eric V. and {Grindlay}, Jonathan E. and {Helfand}, David J. and {Krivonos}, Roman and {Meier}, David L. and {Miller}, Jon M. and {Natalucci}, Lorenzo and {Ogle}, Patrick and {Ofek}, Eran O. and {Ptak}, Andrew and {Reynolds}, Stephen P. and {Rigby}, Jane R. and {Tagliaferri}, Gianpiero and {Thorsett}, Stephen E. and {Treister}, Ezequiel and {Urry}, C. Megan},
        title = "{The Nuclear Spectroscopic Telescope Array (NuSTAR) High-energy X-Ray Mission}",
      journal = {\apj},
         year = 2013,
        month = jun,
       volume = {770},
       number = {2},
          eid = {103},
        pages = {103},
          doi = {10.1088/0004-637X/770/2/103},
archivePrefix = {arXiv},
       eprint = {1301.7307},
 primaryClass = {astro-ph.IM},
       adsurl = {https://ui.adsabs.harvard.edu/abs/2013ApJ...770..103H}
}

@ARTICLE{tonry2018PASP..130f4505T_atlasfp,
       author = {{Tonry}, J.~L. and {Denneau}, L. and {Heinze}, A.~N. and {Stalder}, B. and {Smith}, K.~W. and {Smartt}, S.~J. and {Stubbs}, C.~W. and {Weiland}, H.~J. and {Rest}, A.},
        title = "{ATLAS: A High-cadence All-sky Survey System}",
      journal = {\pasp},
         year = 2018,
        month = jun,
       volume = {130},
       number = {988},
        pages = {064505},
          doi = {10.1088/1538-3873/aabadf},
archivePrefix = {arXiv},
       eprint = {1802.00879},
 primaryClass = {astro-ph.IM},
       adsurl = {https://ui.adsabs.harvard.edu/abs/2018PASP..130f4505T}
}

@ARTICLE{zhao_epfxt_reduction,
       author = {{Zhao}, Hai-Sheng and {Li}, Cheng-Kui and {Wang}, Jin and {Zhang}, Juan and {Jia}, Shu-Mei and {Guan}, Ju and {Zhao}, Xiao-Fan and {Chen}, Yong and {Xu}, Jing-Jing and {Han}, Da-Wei and {Song}, Li-Ming and {Cui}, Wei-Wei},
        title = "{Data reduction and processing for the Follow-up X-ray Telescope onboard Einstein Probe}",
      journal = {Radiation Detection Technology and Methods},
         year = 2025,
        month = jun,
       volume = {9},
       number = {2},
        pages = {215-222},
          doi = {10.1007/s41605-025-00526-8},
       adsurl = {https://ui.adsabs.harvard.edu/abs/2025RDTM....9..215Z}
}

@INPROCEEDINGS{arnaud_xspec,
       author = {{Arnaud}, K.~A.},
        title = "{XSPEC: The First Ten Years}",
    booktitle = {Astronomical Data Analysis Software and Systems V},
         year = 1996,
       editor = {{Jacoby}, George H. and {Barnes}, Jeannette},
       series = {Astronomical Society of the Pacific Conference Series},
       volume = {101},
        month = jan,
        pages = {17},
       adsurl = {https://ui.adsabs.harvard.edu/abs/1996ASPC..101...17A}
}

@ARTICLE{cash_cstat,
       author = {{Cash}, W.},
        title = "{Parameter estimation in astronomy through application of the likelihood ratio.}",
      journal = {\apj},
         year = 1979,
        month = mar,
       volume = {228},
        pages = {939-947},
          doi = {10.1086/156922},
       adsurl = {https://ui.adsabs.harvard.edu/abs/1979ApJ...228..939C}
}

@ARTICLE{kaastra_cstat,
       author = {{Kaastra}, J.~S.},
        title = "{On the use of C-stat in testing models for X-ray spectra}",
      journal = {\aap},
         year = 2017,
        month = sep,
       volume = {605},
          eid = {A51},
        pages = {A51},
          doi = {10.1051/0004-6361/201629319},
archivePrefix = {arXiv},
       eprint = {1707.09202},
 primaryClass = {astro-ph.HE},
       adsurl = {https://ui.adsabs.harvard.edu/abs/2017A&A...605A..51K}
}

@ARTICLE{oniell2026TNSTR1001....1O_tns,
       author = {{O'Neill}, D. and {Ramsay}, G. and {Ackley}, K. and {Dyer}, M. and {Lyman}, J. and {Ulaczyk}, K. and {Steeghs}, D. and {Galloway}, D. and {Dhillon}, V. and {O'Brien}, P. and {Noysena}, K. and {Kotak}, R. and {Breton}, R. and {Casares}, J. and {Nuttall}, L. and {Starling}, R. and {Gompertz}, B. and {Godson}, B. and {Killestein}, T. and {Kumar}, A. and {Pursiainen}, M.},
        title = "{GOTO Transient Discovery Report for 2026-03-10}",
      journal = {Transient Name Server Discovery Report},
         year = 2026,
        month = mar,
       volume = {2026-1001},
        pages = {1},
       adsurl = {https://ui.adsabs.harvard.edu/abs/2026TNSTR1001....1O}
}

@ARTICLE{meegan2009ApJ...702..791M,
       author = {{Meegan}, Charles and {Lichti}, Giselher and {Bhat}, P.~N. and {Bissaldi}, Elisabetta and {Briggs}, Michael S. and {Connaughton}, Valerie and {Diehl}, Roland and {Fishman}, Gerald and {Greiner}, Jochen and {Hoover}, Andrew S. and {van der Horst}, Alexander J. and {von Kienlin}, Andreas and {Kippen}, R. Marc and {Kouveliotou}, Chryssa and {McBreen}, Sheila and {Paciesas}, W.~S. and {Preece}, Robert and {Steinle}, Helmut and {Wallace}, Mark S. and {Wilson}, Robert B. and {Wilson-Hodge}, Colleen},
        title = "{The Fermi Gamma-ray Burst Monitor}",
      journal = {\apj},
         year = 2009,
        month = sep,
       volume = {702},
       number = {1},
        pages = {791-804},
          doi = {10.1088/0004-637X/702/1/791},
archivePrefix = {arXiv},
       eprint = {0908.0450},
 primaryClass = {astro-ph.IM},
       adsurl = {https://ui.adsabs.harvard.edu/abs/2009ApJ...702..791M}
}

@ARTICLE{GCN.43951,
       author = {{Fermi GBM Team}},
        title = "{GRB 260310A: Fermi GBM Final Real-time Localization}",
      journal = {GRB Coordinates Network},
         year = 2026,
        month = mar,
       volume = {43951},
        pages = {1},
}

@ARTICLE{GCN.44125,
       author = {{Guelfand}, M. and {Moreno M{\'e}ndez}, E. and {Becerra}, R.~L. and
                 {de Ugarte Postigo}, A. and {Globus}, N. and {Watson}, A.~M. and
                 {Basa}, S. and {Lee}, W.~H. and {Aguilar-Ruiz}, E. and
                 {Atteia}, J.-L. and {Angulo}, C. and {Akl}, D. and {Antier}, S. and
                 {Butler}, N.~R. and {Dornic}, D. and {Ducoin}, J.-G. and
                 {Fortin}, F. and {Garc{\'i}a Garc{\'i}a}, L. and {Gill}, R. and
                 {Kuwata}, A. and {Lincetto}, M. and {Mandarakas}, N. and
                 {L{\'o}pez-C{\'a}mara}, D. and {Magnani}, F. and {Pereyra}, M. and
                 {Rakotondrainibe}, N.~A. and {S{\'a}nchez {\'A}lvarez}, F. and
                 {Schneider}, B.},
        title = "{GRB 260310A / AT 2026fgk: COLIBR{\'I} photometric evidence of the emerging supernova}",
      journal = {GRB Coordinates Network},
         year = 2026,
        month = mar,
       volume = {44125},
        pages = {1},
}

@ARTICLE{GCN.44124,
       author = {{de Ugarte Postigo}, A. and {Geier}, S. and {Izzo}, L. and
                 {Malesani}, D.~B. and {Martin-Carrillo}, A. and {Thoene}, C.~C. and
                 {Aloy}, M.~A. and {Fynbo}, J.~P.~U. and {Galbany}, L. and
                 {Lombardi}, G. and {Rakotondrainibe}, N.~A. and {Schneider}, B. and
                 {Tanvir}, N.~R. and {Gonz{\'a}lez Gonz{\'a}lez}, D.},
        title = "{GRB 260310A: OSIRIS+/GTC spectroscopic detection of the associated BL-Ic, SN 2026fgk}",
      journal = {GRB Coordinates Network},
         year = 2026,
        month = mar,
       volume = {44124},
        pages = {1},
}

@ARTICLE{GCN.43958,
       author = {{Salunke}, S. and {Harsha K. H.} and {Tembhurnikar}, M. and {Arya}, A. and {Goyal}, A. and {Waratkar}, G. and {Vibhute}, A. and {Bhalerao}, V. and {Bhattacharya}, D. and {Rao}, A.~R. and {Vadawale}, S.},
        title = "{GRB 260310A: AstroSat CZTI detection of a long burst}",
      journal = {GRB Coordinates Network},
         year = 2026,
        month = mar,
       volume = {43958},
        pages = {1},
}

@ARTICLE{GCN.43974,
       author = {{Konno}, R. and {Garrappa}, S. and {Zimmerman}, E.~A. and {Horowicz}, A. and {Ofek}, E.~O. and {Ben-Ami}, S. and {Polishook}, D. and {Yaron}, O. and {Fainer}, S. and {Krassilchtchikov}, A. and {Shani}, Y.~M. and {Segre}, E. and {Gal-Yam}, A. and {Spitzer}, S.},
        title = "{GRB 260310A: potential optical counterpart}",
      journal = {GRB Coordinates Network},
         year = 2026,
        month = mar,
       volume = {43974},
        pages = {1},
}

@ARTICLE{GCN.43975,
       author = {{Hamburg}, R. and {Meegan}, C.},
        title = "{GRB 260310A: Fermi GBM Observation}",
      journal = {GRB Coordinates Network},
         year = 2026,
        month = mar,
       volume = {43975},
        pages = {1},
}

@ARTICLE{GCN.43977,
       author = {{Hinds}, {K.-R.} and {Perley}, D.~A. and {Ho}, A.~Y.~Q. and {Rose}, S. and {Rastinejad}, J. and {Stein}, R. and {Mo}, G. and {Kasliwal}, M. and {Sollerman}, J. and {Hall}, X.~J. and {Wise}, J. and {Jayaraman}, R. and {Bochenek}, A.},
        title = "{GRB260310A / AT2026fgk: P200 Spectroscopy}",
      journal = {GRB Coordinates Network},
         year = 2026,
        month = mar,
       volume = {43977},
        pages = {1},
}

@ARTICLE{GCN.43984,
       author = {{de Ugarte Postigo}, A. and {Izzo}, L. and {Martin-Carrillo}, A. and {Malesani}, D.~B. and {Geier}, S. and {Thoene}, C.~C. and {Aloy}, M.~A. and {Fynbo}, J.~P.~U. and {Galbany}, L. and {Lombardi}, G. and {Rakotondrainibe}, N.~A. and {Schneider}, B. and {Tanvir}, N.~R. and {Perley}, D.~A. and {P{\'e}rez Toledo}, F.~M. and {P{\'e}rez Valladares}, D.},
        title = "{GRB 260310A / AT2026fgk: OSIRIS+/GTC spectroscopy confirms redshift z = 0.153}",
      journal = {GRB Coordinates Network},
         year = 2026,
        month = mar,
       volume = {43984},
        pages = {1},
}

@ARTICLE{GCN.43986,
       author = {{Hsu}, B. and {Shrestha}, M. and {Andrews}, J. and {Sand}, D.~J. and {Franz}, N. and {Pearson}, J. and {Christy}, C. and {Ransome}, C.~L. and {Subrayan}, B. and {Bostroem}, K. and {Hosseinzadeh}, G. and {Smith}, N.},
        title = "{GRB 260310A/AT 2026fgk: Bok spectroscopic observation}",
      journal = {GRB Coordinates Network},
         year = 2026,
        month = mar,
       volume = {43986},
        pages = {1},
}

@ARTICLE{GCN.43990,
       author = {{Pursiainen}, M. and {Malesani}, D.~B. and {Leloudas}, G. and {Dimple} and {Martin-Carrillo}, A. and {Gompertz}, B.~P. and {Perley}, D.~A. and {Pyykkinen}, N.},
        title = "{GRB 260310A: NOT optical polarimetry}",
      journal = {GRB Coordinates Network},
         year = 2026,
        month = mar,
       volume = {43990},
        pages = {1},
}

@ARTICLE{GCN.43994,
       author = {{Jayaraman}, R. and {Ho}, A.~Y.~Q. and {Hinds}, {K.-R.} and {Fu}, S.~Y. and {Liang}, R.~D. and {Liu}, M.~J. and {Ling}, Z.~X. and {Sun}, H. and {Yuan}, W. and {Schroeder}, G. and {Hall}, X.~J. and {Coughlin}, M.},
        title = "{GRB 260310A: Einstein Probe Follow-up X-ray Telescope (FXT) Detection}",
      journal = {GRB Coordinates Network},
         year = 2026,
        month = mar,
       volume = {43994},
        pages = {1},
}

@ARTICLE{GCN.44005,
       author = {{Rhodes}, L. and {Hughes}, A. and {Fender}, R. and {Bright}, J. and {Green}, D. and {Titterington}, D.},
        title = "{GRB 260310A/AT2026fgk: radio detection at 15GHz}",
      journal = {GRB Coordinates Network},
         year = 2026,
        month = mar,
       volume = {44005},
        pages = {1},
}

@ARTICLE{GCN.44045,
       author = {{Giarratana}, S. and {Giroletti}, M. and {Ghirlanda}, G. and {Di Lalla}, N. and {Omodei}, N. and {Salafia}, O.~S. and {Nava}, L.},
        title = "{GRB 260310A / AT 2026fgk: radio detection with the VLA}",
      journal = {GRB Coordinates Network},
         year = 2026,
        month = mar,
       volume = {44045},
        pages = {1},
}

@ARTICLE{GCN.44063,
       author = {{Kammoun}, Elias and {Waratkar}, Gaurav and {Jayaraman}, Rahul and {Hinds}, K.-Ryan and {Ho}, Anna Y.~Q.},
        title = "{GRB 260310A / AT2026fgk: NuSTAR follow-up detection}",
      journal = {GRB Coordinates Network},
         year = 2026,
        month = mar,
       volume = {44063},
        pages = {1},
       adsurl = {https://ui.adsabs.harvard.edu/abs/2026GCN.44063....1K}
}

@ARTICLE{steeghs2022MNRAS.511.2405S,
       author = {{Steeghs}, D. and {Galloway}, D.~K. and {Ackley}, K. and {Dyer}, M.~J. and {Lyman}, J. and {Ulaczyk}, K. and {Cutter}, R. and {Mong}, Y.-L. and {Dhillon}, V. and {O'Brien}, P. and {Ramsay}, G. and {Poshyachinda}, S. and {Kotak}, R. and {Nuttall}, L.~K. and {Pall{\'e}}, E. and {Breton}, R.~P. and {Pollacco}, D. and {Thrane}, E. and {Aukkaravittayapun}, S. and {Awiphan}, S. and {Burhanudin}, U. and {Chote}, P. and {Chrimes}, A. and {Daw}, E. and {Duffy}, C. and {Eyles-Ferris}, R. and {Gompertz}, B. and {Heikkil{\"a}}, T. and {Irawati}, P. and {Kennedy}, M.~R. and {Killestein}, T. and {Kuncarayakti}, H. and {Levan}, A.~J. and {Littlefair}, S. and {Makrygianni}, L. and {Marsh}, T. and {Mata-Sanchez}, D. and {Mattila}, S. and {Maund}, J. and {McCormac}, J. and {Mkrtichian}, D. and {Mullaney}, J. and {Noysena}, K. and {Patel}, M. and {Rol}, E. and {Sawangwit}, U. and {Stanway}, E.~R. and {Starling}, R. and {Str{\o}m}, P. and {Tooke}, S. and {West}, R. and {White}, D.~J. and {Wiersema}, K.},
        title = "{The Gravitational-wave Optical Transient Observer (GOTO): prototype performance and prospects for transient science}",
      journal = {\mnras},
         year = 2022,
        month = apr,
       volume = {511},
       number = {2},
        pages = {2405-2422},
          doi = {10.1093/mnras/stac013},
archivePrefix = {arXiv},
       eprint = {2110.05539},
 primaryClass = {astro-ph.IM},
       adsurl = {https://ui.adsabs.harvard.edu/abs/2022MNRAS.511.2405S}
}

@ARTICLE{ofek2023PASP..135f5001O,
       author = {{Ofek}, E.~O. and {Ben-Ami}, S. and {Polishook}, D. and {Segre}, E. and {Blumenzweig}, A. and {Strotjohann}, N.-L. and {Yaron}, O. and {Shani}, Y.~M. and {Nachshon}, S. and {Shvartzvald}, Y. and {Hershko}, O. and {Engel}, M. and {Segre}, M. and {Segev}, N. and {Zimmerman}, E. and {Nir}, G. and {Judkovsky}, Y. and {Gal-Yam}, A. and {Zackay}, B. and {Waxman}, E. and {Kushnir}, D. and {Chen}, P. and {Azaria}, R. and {Manulis}, I. and {Diner}, O. and {Vandeventer}, B. and {Franckowiak}, A. and {Weimann}, S. and {Borowska}, J. and {Garrappa}, S. and {Zenin}, A. and {Fallah Ramazani}, V. and {Konno}, R. and {K{\"u}sters}, D. and {Sadeh}, I. and {Parsons}, R.~D. and {Berge}, D. and {Kowalski}, M. and {Ohm}, S. and {Arcavi}, I. and {Bruch}, R.},
        title = "{The Large Array Survey Telescope-System Overview and Performances}",
      journal = {\pasp},
         year = 2023,
        month = jun,
       volume = {135},
       number = {1048},
          eid = {065001},
        pages = {065001},
          doi = {10.1088/1538-3873/acd8f0},
archivePrefix = {arXiv},
       eprint = {2304.04796},
 primaryClass = {astro-ph.IM},
       adsurl = {https://ui.adsabs.harvard.edu/abs/2023PASP..135f5001O}
}

@ARTICLE{benami2023PASP..135h5002B,
       author = {{Ben-Ami}, S. and {Ofek}, E.~O. and {Polishook}, D. and {Franckowiak}, A. and {Hallakoun}, N. and {Segre}, E. and {Shvartzvald}, Y. and {Strotjohann}, N.~L. and {Yaron}, O. and {Aharonson}, O. and {Arcavi}, I. and {Berge}, D. and {Ramazani}, V. Fallah and {Gal-Yam}, A. and {Garrappa}, S. and {Hershko}, O. and {Nir}, G. and {Ohm}, S. and {Rybicki}, K. and {Sadeh}, I. and {Segev}, N. and {Shani}, Y.~M. and {Sofer-Rimalt}, Y. and {Weimann}, S.},
        title = "{The Large Array Survey Telescope-Science Goals}",
      journal = {\pasp},
         year = 2023,
        month = aug,
       volume = {135},
       number = {1050},
          eid = {085002},
        pages = {085002},
          doi = {10.1088/1538-3873/aceb30},
archivePrefix = {arXiv},
       eprint = {2304.02719},
 primaryClass = {astro-ph.IM},
       adsurl = {https://ui.adsabs.harvard.edu/abs/2023PASP..135h5002B}
}

@ARTICLE{ami2008MNRAS.391.1545Z,
       author = {{AMI Consortium: Zwart}, J.~T.~L. and others},
        title = "{The Arcminute Microkelvin Imager}",
      journal = {\mnras},
         year = 2008,
        month = dec,
       volume = {391},
       number = {4},
        pages = {1545-1558},
          doi = {10.1111/j.1365-2966.2008.13953.x},
archivePrefix = {arXiv},
       eprint = {0807.2469},
 primaryClass = {astro-ph},
       adsurl = {https://ui.adsabs.harvard.edu/abs/2008MNRAS.391.1545Z}
}

@ARTICLE{piran2004RvMP...76.1143P,
       author = {{Piran}, Tsvi},
        title = "{The physics of gamma-ray bursts}",
      journal = {Reviews of Modern Physics},
         year = 2004,
        month = oct,
       volume = {76},
       number = {4},
        pages = {1143-1210},
          doi = {10.1103/RevModPhys.76.1143},
archivePrefix = {arXiv},
       eprint = {astro-ph/0405503},
 primaryClass = {astro-ph},
       adsurl = {https://ui.adsabs.harvard.edu/abs/2004RvMP...76.1143P}
}

@ARTICLE{chand2020ApJ...898...42C,
       author = {{Chand}, Vikas and {Banerjee}, Ankush and {Gupta}, Rahul and {Dimple} and {Pal}, Partha Sarathi and {Joshi}, Jagdish C. and {Zhang}, Bin-Bin and {Basak}, R. and {Tam}, P.~H.~T. and {Sharma}, Vidushi and {Pandey}, S.~B. and {Kumar}, Amit and {Yang}, Yi-Si},
        title = "{Peculiar Prompt Emission and Afterglow in the H.E.S.S.-detected GRB 190829A}",
      journal = {\apj},
         year = 2020,
        month = jul,
       volume = {898},
       number = {1},
          eid = {42},
        pages = {42},
          doi = {10.3847/1538-4357/ab9606},
archivePrefix = {arXiv},
       eprint = {2001.00648},
 primaryClass = {astro-ph.HE},
       adsurl = {https://ui.adsabs.harvard.edu/abs/2020ApJ...898...42C}
}

@ARTICLE{salafia2022ApJ...931L..19S,
       author = {{Salafia}, Om Sharan and {Ravasio}, Maria Edvige and {Yang}, Jun and {An}, Tao and {Orienti}, Monica and {Ghirlanda}, Giancarlo and {Nava}, Lara and {Giroletti}, Marcello and {Mohan}, Prashanth and {Spinelli}, Riccardo and {Zhang}, Yingkang and {Marcote}, Benito and {Cim{\`o}}, Giuseppe and {Wu}, Xuefeng and {Li}, Zhixuan},
        title = "{Multiwavelength View of the Close-by GRB 190829A Sheds Light on Gamma-Ray Burst Physics}",
      journal = {\apjl},
         year = 2022,
        month = jun,
       volume = {931},
       number = {2},
          eid = {L19},
        pages = {L19},
          doi = {10.3847/2041-8213/ac6c28},
archivePrefix = {arXiv},
       eprint = {2106.07169},
 primaryClass = {astro-ph.HE},
       adsurl = {https://ui.adsabs.harvard.edu/abs/2022ApJ...931L..19S}
}

@ARTICLE{delia2018AA...619A..66D,
       author = {{D'Elia}, V. and {Campana}, S. and {D'A{\`\i}}, A. and {De Pasquale}, M. and {Emery}, S.~W.~K. and {Frederiks}, D.~D. and {Lien}, A. and {Melandri}, A. and {Page}, K.~L. and {Starling}, R.~L.~C. and {Burrows}, D.~N. and {Breeveld}, A.~A. and {Oates}, S.~R. and {O'Brien}, P.~T. and {Osborne}, J.~P. and {Siegel}, M.~H. and {Tagliaferri}, G. and {Brown}, P.~J. and {Cenko}, S.~B. and {Svinkin}, D.~S. and {Tohuvavohu}, A. and {Tsvetkova}, A.~E.},
        title = "{GRB 171205A/SN 2017iuk: A local low-luminosity gamma-ray burst}",
      journal = {\aap},
         year = 2018,
        month = nov,
       volume = {619},
          eid = {A66},
        pages = {A66},
          doi = {10.1051/0004-6361/201833847},
archivePrefix = {arXiv},
       eprint = {1810.03339},
 primaryClass = {astro-ph.HE},
       adsurl = {https://ui.adsabs.harvard.edu/abs/2018A&A...619A..66D}
}

@ARTICLE{sun2015ApJ...812...33S,
       author = {{Sun}, Hui and {Zhang}, Bing and {Li}, Zhuo},
        title = "{Extragalactic High-energy Transients: Event Rate Densities and Luminosity Functions}",
      journal = {\apj},
         year = 2015,
        month = oct,
       volume = {812},
       number = {1},
          eid = {33},
        pages = {33},
          doi = {10.1088/0004-637X/812/1/33},
archivePrefix = {arXiv},
       eprint = {1509.01592},
 primaryClass = {astro-ph.HE},
       adsurl = {https://ui.adsabs.harvard.edu/abs/2015ApJ...812...33S}
}

@ARTICLE{liang2007ApJ...662.1111L,
       author = {{Liang}, Enwei and {Zhang}, Bing and {Virgili}, Francisco and {Dai}, Z.~G.},
        title = "{Low-Luminosity Gamma-Ray Bursts as a Unique Population: Luminosity Function, Local Rate, and Beaming Factor}",
      journal = {\apj},
         year = 2007,
        month = jun,
       volume = {662},
       number = {2},
        pages = {1111-1118},
          doi = {10.1086/517959},
archivePrefix = {arXiv},
       eprint = {astro-ph/0605200},
 primaryClass = {astro-ph},
       adsurl = {https://ui.adsabs.harvard.edu/abs/2007ApJ...662.1111L}
}

@ARTICLE{perley2014ApJ...781...37P,
       author = {{Perley}, D.~A. and {Cenko}, S.~B. and {Corsi}, A. and {Tanvir}, N.~R. and {Levan}, A.~J. and {Kann}, D.~A. and {Sonbas}, E. and {Wiersema}, K. and {Zheng}, W. and {Zhao}, X.-H. and {Bai}, J.-M. and {Bremer}, M. and {Castro-Tirado}, A.~J. and {Chang}, L. and {Clubb}, K.~I. and {Frail}, D. and {Fruchter}, A. and {G{\"o}{\u{g}}{\"u}{\textcommabelow s}}, E. and {Greiner}, J. and {G{\"u}ver}, T. and {Horesh}, A. and {Filippenko}, A.~V. and {Klose}, S. and {Mao}, J. and {Morgan}, A.~N. and {Pozanenko}, A.~S. and {Schmidl}, S. and {Stecklum}, B. and {Tanga}, M. and {Volnova}, A.~A. and {Volvach}, A.~E. and {Wang}, J.-G. and {Winters}, J.-M. and {Xin}, Y.-X.},
        title = "{The Afterglow of GRB 130427A from 1 to {}10$^{16}$ GHz}",
      journal = {\apj},
         year = 2014,
        month = jan,
       volume = {781},
       number = {1},
          eid = {37},
        pages = {37},
          doi = {10.1088/0004-637X/781/1/37},
archivePrefix = {arXiv},
       eprint = {1307.4401},
 primaryClass = {astro-ph.HE},
       adsurl = {https://ui.adsabs.harvard.edu/abs/2014ApJ...781...37P}
}

@ARTICLE{dermer1999ApJ...513..656D,
       author = {{Dermer}, Charles D. and {Chiang}, James and {B{\"o}ttcher}, Markus},
        title = "{Fireball Loading and the Blast-Wave Model of Gamma-Ray Bursts}",
      journal = {\apj},
         year = 1999,
        month = mar,
       volume = {513},
       number = {2},
        pages = {656-668},
          doi = {10.1086/306871},
archivePrefix = {arXiv},
       eprint = {astro-ph/9804174},
 primaryClass = {astro-ph},
       adsurl = {https://ui.adsabs.harvard.edu/abs/1999ApJ...513..656D}
}

@ARTICLE{gehrels1986ApJ...303..336G,
       author = {{Gehrels}, N.},
        title = "{Confidence Limits for Small Numbers of Events in Astrophysical Data}",
      journal = {\apj},
         year = 1986,
        month = apr,
       volume = {303},
        pages = {336},
          doi = {10.1086/164079},
       adsurl = {https://ui.adsabs.harvard.edu/abs/1986ApJ...303..336G}
}

@ARTICLE{vanerten2010ApJ...722..235V,
       author = {{van Eerten}, Hendrik and {Zhang}, Weiqun and {MacFadyen}, Andrew},
        title = "{Off-axis Gamma-ray Burst Afterglow Modeling Based on a Two-dimensional Axisymmetric Hydrodynamics Simulation}",
      journal = {\apj},
         year = 2010,
        month = oct,
       volume = {722},
       number = {1},
        pages = {235-247},
          doi = {10.1088/0004-637X/722/1/235},
archivePrefix = {arXiv},
       eprint = {1006.5125},
 primaryClass = {astro-ph.HE},
       adsurl = {https://ui.adsabs.harvard.edu/abs/2010ApJ...722..235V}
}

@ARTICLE{granot2002ApJ...570L..61G,
       author = {{Granot}, Jonathan and {Panaitescu}, Alin and {Kumar}, Pawan and
         {Woosley}, Stan E.},
        title = "{Off-Axis Afterglow Emission from Jetted Gamma-Ray Bursts}",
      journal = {\apjl},
         year = 2002,
        month = may,
       volume = {570},
       number = {2},
        pages = {L61-L64},
          doi = {10.1086/340991},
archivePrefix = {arXiv},
       eprint = {astro-ph/0201322},
 primaryClass = {astro-ph},
       adsurl = {https://ui.adsabs.harvard.edu/abs/2002ApJ...570L..61G}
}

@ARTICLE{granot2002ApJ...568..820G,
       author = {{Granot}, Jonathan and {Sari}, Re'em},
        title = "{The Shape of Spectral Breaks in Gamma-Ray Burst Afterglows}",
      journal = {\apj},
         year = 2002,
        month = apr,
       volume = {568},
       number = {2},
        pages = {820-829},
          doi = {10.1086/338966},
archivePrefix = {arXiv},
       eprint = {astro-ph/0108027},
 primaryClass = {astro-ph},
       adsurl = {https://ui.adsabs.harvard.edu/abs/2002ApJ...568..820G}
}

@ARTICLE{rees1992MNRAS.258P..41R,
       author = {{Rees}, M.~J. and {Meszaros}, P.},
        title = "{Relativistic fireballs - Energy conversion and time-scales.}",
      journal = {\mnras},
         year = 1992,
        month = sep,
       volume = {258},
        pages = {41},
          doi = {10.1093/mnras/258.1.41P},
       adsurl = {https://ui.adsabs.harvard.edu/abs/1992MNRAS.258P..41R}
}

@ARTICLE{amati2002A&A...390...81A,
       author = {{Amati}, L. and {Frontera}, F. and {Tavani}, M. and {in't Zand}, J.~J.~M. and {Antonelli}, A. and {Costa}, E. and {Feroci}, M. and {Guidorzi}, C. and {Heise}, J. and {Masetti}, N. and {Montanari}, E. and {Nicastro}, L. and {Palazzi}, E. and {Pian}, E. and {Piro}, L. and {Soffitta}, P.},
        title = "{Intrinsic spectra and energetics of BeppoSAX Gamma-Ray Bursts with known redshifts}",
      journal = {\aap},
         year = 2002,
        month = jul,
       volume = {390},
        pages = {81-89},
          doi = {10.1051/0004-6361:20020722},
archivePrefix = {arXiv},
       eprint = {astro-ph/0205230},
 primaryClass = {astro-ph},
       adsurl = {https://ui.adsabs.harvard.edu/abs/2002A&A...390...81A}
}

@ARTICLE{rhodes2020MNRAS.496.3326R,
       author = {{Rhodes}, L. and {van der Horst}, A.~J. and {Fender}, R. and {Monageng}, I.~M. and {Anderson}, G.~E. and {Antoniadis}, J. and {Bietenholz}, M.~F. and {B{\"o}ttcher}, M. and {Bright}, J.~S. and {Green}, D.~A. and {Kouveliotou}, C. and {Kramer}, M. and {Motta}, S.~E. and {Wijers}, R.~A.~M.~J. and {Williams}, D.~R.~A. and {Woudt}, P.~A.},
        title = "{Radio afterglows of very high-energy gamma-ray bursts 190829A and 180720B}",
      journal = {\mnras},
         year = 2020,
        month = aug,
       volume = {496},
       number = {3},
        pages = {3326-3335},
          doi = {10.1093/mnras/staa1715},
archivePrefix = {arXiv},
       eprint = {2004.01538},
 primaryClass = {astro-ph.HE},
       adsurl = {https://ui.adsabs.harvard.edu/abs/2020MNRAS.496.3326R}
}

@ARTICLE{hess2021Sci...372.1081H,
       author = {{H.~E.~S.~S. Collaboration} and {Abdalla}, H. and {Aharonian}, F. and {Ait Benkhali}, F. and {Ang{\"u}ner}, E.~O. and {Arcaro}, C. and {Armand}, C. and {Armstrong}, T. and {Ashkar}, H. and {Backes}, M. and {Baghmanyan}, V. and {Barbosa Martins}, V. and {Barnacka}, A. and {Barnard}, M. and {Becherini}, Y. and {Berge}, D. and {Bernl{\"o}hr}, K. and {Bi}, B. and {Bissaldi}, E. and {B{\"o}ttcher}, M. and {Boisson}, C. and {Bolmont}, J. and {de Bony de Lavergne}, M. and {Breuhaus}, M. and {Brun}, F. and {Brun}, P. and {Bryan}, M. and {B{\"u}chele}, M. and {Bulik}, T. and {Bylund}, T. and {Caroff}, S. and {Carosi}, A. and {Casanova}, S. and {Chand}, T. and {Chandra}, S. and {Chen}, A. and {Cotter}, G. and {Cury{\l}o}, M. and {Damascene Mbarubucyeye}, J. and {Davids}, I.~D. and {Davies}, J. and {Deil}, C. and {Devin}, J. and {Dirson}, L. and {Djannati-Ata{\"\i}}, A. and {Dmytriiev}, A. and {Donath}, A. and {Doroshenko}, V. and {Dreyer}, L. and {Duffy}, C. and {Dyks}, J. and {Egberts}, K. and {Eichhorn}, F. and {Einecke}, S. and {Emery}, G. and {Ernenwein}, J.-P. and {Feijen}, K. and {Fegan}, S. and {Fiasson}, A. and {Fichet de Clairfontaine}, G. and {Fontaine}, G. and {Funk}, S. and {F{\"u}{\ss}ling}, M. and {Gabici}, S. and {Gallant}, Y.~A. and {Giavitto}, G. and {Giunti}, L. and {Glawion}, D. and {Glicenstein}, J.~F. and {Grondin}, M.-H. and {Hahn}, J. and {Haupt}, M. and {Hermann}, G. and {Hinton}, J.~A. and {Hofmann}, W. and {Hoischen}, C. and {Holch}, T.~L. and {Holler}, M. and {H{\"o}rbe}, M. and {Horns}, D. and {Huber}, D. and {Jamrozy}, M. and {Jankowsky}, D. and {Jankowsky}, F. and {Jardin-Blicq}, A. and {Joshi}, V. and {Jung-Richardt}, I. and {Kasai}, E. and {Kastendieck}, M.~A. and {Katarzy{\'n}ski}, K. and {Katz}, U. and {Khangulyan}, D. and {Kh{\'e}lifi}, B. and {Klepser}, S. and {Klu{\'z}niak}, W. and {Komin}, Nu. and {Konno}, R. and {Kosack}, K. and {Kostunin}, D. and {Kreter}, M. and {Lamanna}, G. and {Lemi{\`e}re}, A. and {Lemoine-Goumard}, M. and {Lenain}, J.-P. and {Leuschner}, F. and {Levy}, C. and {Lohse}, T. and {Lypova}, I. and {Mackey}, J. and {Majumdar}, J. and {Malyshev}, D. and {Malyshev}, D. and {Marandon}, V. and {Marchegiani}, P. and {Marcowith}, A. and {Mares}, A. and {Mart{\'\i}-Devesa}, G. and {Marx}, R. and {Maurin}, G. and {Meintjes}, P.~J. and {Meyer}, M. and {Mitchell}, A. and {Moderski}, R. and {Mohrmann}, L. and {Montanari}, A. and {Moore}, C. and {Morris}, P. and {Moulin}, E. and {Muller}, J. and {Murach}, T. and {Nakashima}, K. and {Nayerhoda}, A. and {de Naurois}, M. and {Ndiyavala}, H. and {Niemiec}, J. and {Oakes}, L. and {O'Brien}, P. and {Odaka}, H. and {Ohm}, S. and {Olivera-Nieto}, L. and {de Ona Wilhelmi}, E. and {Ostrowski}, M. and {Panny}, S. and {Panter}, M. and {Parsons}, R.~D. and {Peron}, G. and {Peyaud}, B. and {Piel}, Q. and {Pita}, S. and {Poireau}, V. and {Priyana Noel}, A. and {Prokhorov}, D.~A. and {Prokoph}, H. and {P{\"u}hlhofer}, G. and {Punch}, M. and {Quirrenbach}, A. and {Raab}, S. and {Rauth}, R. and {Reichherzer}, P. and {Reimer}, A. and {Reimer}, O. and {Remy}, Q. and {Renaud}, M. and {Rieger}, F. and {Rinchiuso}, L. and {Romoli}, C. and {Rowell}, G. and {Rudak}, B. and {Ruiz-Velasco}, E. and {Sahakian}, V. and {Sailer}, S. and {Salzmann}, H. and {Sanchez}, D.~A. and {Santangelo}, A. and {Sasaki}, M. and {Scalici}, M. and {Sch{\"a}fer}, J. and {Sch{\"u}ssler}, F. and {Schutte}, H.~M. and {Schwanke}, U. and {Seglar-Arroyo}, M. and {Senniappan}, M. and {Seyffert}, A.~S. and {Shafi}, N. and {Shapopi}, J.~N.~S. and {Shiningayamwe}, K. and {Simoni}, R. and {Sinha}, A. and {Sol}, H. and {Specovius}, A. and {Spencer}, S. and {Spir-Jacob}, M. and {Stawarz}, {\L}. and {Sun}, L. and {Steenkamp}, R. and {Stegmann}, C. and {Steinmassl}, S. and {Steppa}, C. and {Takahashi}, T. and {Tam}, T.},
        title = "{Revealing x-ray and gamma ray temporal and spectral similarities in the GRB 190829A afterglow}",
      journal = {Science},
         year = 2021,
        month = jun,
       volume = {372},
       number = {6546},
        pages = {1081-1085},
          doi = {10.1126/science.abe8560},
archivePrefix = {arXiv},
       eprint = {2106.02510},
 primaryClass = {astro-ph.HE},
       adsurl = {https://ui.adsabs.harvard.edu/abs/2021Sci...372.1081H}
}

@ARTICLE{fraija2021ApJ...918...12F,
       author = {{Fraija}, N. and {Veres}, P. and {Beniamini}, P. and {Galvan-Gamez}, A. and {Metzger}, B.~D. and {Barniol Duran}, R. and {Becerra}, R.~L.},
        title = "{On the Origin of the Multi-GeV Photons from the Closest Burst with Intermediate Luminosity: GRB 190829A}",
      journal = {\apj},
         year = 2021,
        month = sep,
       volume = {918},
       number = {1},
          eid = {12},
        pages = {12},
          doi = {10.3847/1538-4357/ac0aed},
archivePrefix = {arXiv},
       eprint = {2003.11252},
 primaryClass = {astro-ph.HE},
       adsurl = {https://ui.adsabs.harvard.edu/abs/2021ApJ...918...12F}
}

@ARTICLE{sato2021MNRAS.504.5647S,
       author = {{Sato}, Yuri and {Obayashi}, Kaori and {Yamazaki}, Ryo and {Murase}, Kohta and {Ohira}, Yutaka},
        title = "{Off-axis jet scenario for early afterglow emission of low-luminosity gamma-ray burst GRB 190829A}",
      journal = {\mnras},
         year = 2021,
        month = jul,
       volume = {504},
       number = {4},
        pages = {5647-5655},
          doi = {10.1093/mnras/stab1273},
archivePrefix = {arXiv},
       eprint = {2101.10581},
 primaryClass = {astro-ph.HE},
       adsurl = {https://ui.adsabs.harvard.edu/abs/2021MNRAS.504.5647S}
}

@ARTICLE{matheson2003ApJ...599..394M,
       author = {{Matheson}, T. and {Garnavich}, P.~M. and {Stanek}, K.~Z. and {Bersier}, D. and {Holland}, S.~T. and {Krisciunas}, K. and {Caldwell}, N. and {Berlind}, P. and {Bloom}, J.~S. and {Bolte}, M. and {Bonanos}, A.~Z. and {Brown}, M.~J.~I. and {Brown}, W.~R. and {Calkins}, M.~L. and {Challis}, P. and {Chornock}, R. and {Echevarria}, L. and {Eisenstein}, D.~J. and {Everett}, M.~E. and {Filippenko}, A.~V. and {Flint}, K. and {Foley}, R.~J. and {Freedman}, D.~L. and {Hamuy}, Mario and {Harding}, P. and {Hathi}, N.~P. and {Hicken}, M. and {Hoopes}, C. and {Impey}, C. and {Jannuzi}, B.~T. and {Jansen}, R.~A. and {Jha}, S. and {Kaluzny}, J. and {Kannappan}, S. and {Kirshner}, R.~P. and {Latham}, D.~W. and {Lee}, J.~C. and {Leonard}, D.~C. and {Li}, W. and {Luhman}, K.~L. and {Martini}, P. and {Mathis}, H. and {Maza}, J. and {Megeath}, S.~T. and {Miller}, L.~R. and {Minniti}, D. and {Olszewski}, E.~W. and {Papenkova}, M. and {Phillips}, M.~M. and {Pindor}, B. and {Sasselov}, D.~D. and {Schild}, R. and {Schweiker}, H. and {Spahr}, T. and {Thomas-Osip}, J. and {Thompson}, I. and {Weisz}, D. and {Windhorst}, R. and {Zaritsky}, D.},
        title = "{Photometry and Spectroscopy of GRB 030329 and Its Associated Supernova 2003dh: The First Two Months}",
      journal = {\apj},
         year = 2003,
        month = dec,
       volume = {599},
       number = {1},
        pages = {394-407},
          doi = {10.1086/379228},
archivePrefix = {arXiv},
       eprint = {astro-ph/0307435},
 primaryClass = {astro-ph},
       adsurl = {https://ui.adsabs.harvard.edu/abs/2003ApJ...599..394M}
}

@ARTICLE{nakamura2001ApJ...550..991N,
       author = {{Nakamura}, Takayoshi and {Mazzali}, Paolo A. and {Nomoto}, Ken'ichi and {Iwamoto}, Koichi},
        title = "{Light Curve and Spectral Models for the Hypernova SN 1998BW Associated with GRB 980425}",
      journal = {\apj},
         year = 2001,
        month = apr,
       volume = {550},
       number = {2},
        pages = {991-999},
          doi = {10.1086/319784},
archivePrefix = {arXiv},
       eprint = {astro-ph/0007010},
 primaryClass = {astro-ph},
       adsurl = {https://ui.adsabs.harvard.edu/abs/2001ApJ...550..991N}
}

@ARTICLE{rhoads1999ApJ...525..737R,
       author = {{Rhoads}, James E.},
        title = "{The Dynamics and Light Curves of Beamed Gamma-Ray Burst Afterglows}",
      journal = {\apj},
         year = 1999,
        month = nov,
       volume = {525},
       number = {2},
        pages = {737-749},
          doi = {10.1086/307907},
archivePrefix = {arXiv},
       eprint = {astro-ph/9903399},
 primaryClass = {astro-ph},
       adsurl = {https://ui.adsabs.harvard.edu/abs/1999ApJ...525..737R}
}

@ARTICLE{sari1998ApJ...497L..17S,
       author = {{Sari}, Re'em and {Piran}, Tsvi and {Narayan}, Ramesh},
        title = "{Spectra and Light Curves of Gamma-Ray Burst Afterglows}",
      journal = {\apjl},
         year = 1998,
        month = apr,
       volume = {497},
       number = {1},
        pages = {L17-L20},
          doi = {10.1086/311269},
archivePrefix = {arXiv},
       eprint = {astro-ph/9712005},
 primaryClass = {astro-ph},
       adsurl = {https://ui.adsabs.harvard.edu/abs/1998ApJ...497L..17S}
}

@ARTICLE{sari1999ApJ...519L..17S,
       author = {{Sari}, Re'em and {Piran}, Tsvi and {Halpern}, J.~P.},
        title = "{Jets in Gamma-Ray Bursts}",
      journal = {\apjl},
         year = 1999,
        month = jul,
       volume = {519},
       number = {1},
        pages = {L17-L20},
          doi = {10.1086/312109},
archivePrefix = {arXiv},
       eprint = {astro-ph/9903339},
 primaryClass = {astro-ph},
       adsurl = {https://ui.adsabs.harvard.edu/abs/1999ApJ...519L..17S},
}

@ARTICLE{fbuildercoughlin2023ApJS..267...31C,
       author = {{Coughlin}, Michael W. and {Bloom}, Joshua S. and {Nir}, Guy and {Antier}, Sarah and {du Laz}, Theophile Jegou and {van der Walt}, St{\'e}fan and {Crellin-Quick}, Arien and {Culino}, Thomas and {Duev}, Dmitry A. and {Goldstein}, Daniel A. and {Healy}, Brian F. and {Karambelkar}, Viraj and {Lilleboe}, Jada and {Shin}, Kyung Min and {Singer}, Leo P. and {Ahumada}, Tom{\'a}s and {Anand}, Shreya and {Bellm}, Eric C. and {Dekany}, Richard and {Graham}, Matthew J. and {Kasliwal}, Mansi M. and {Kostadinova}, Ivona and {Kiendrebeogo}, R. Weizmann and {Kulkarni}, Shrinivas R. and {Jenkins}, Sydney and {LeBaron}, Natalie and {Mahabal}, Ashish A. and {Neill}, James D. and {Parazin}, B. and {Peloton}, Julien and {Perley}, Daniel A. and {Riddle}, Reed and {Rusholme}, Ben and {van Santen}, Jakob and {Sollerman}, Jesper and {Stein}, Robert and {Turpin}, D. and {Wold}, Avery and {Amat}, Carla and {Bonnefon}, Adrien and {Bonnefoy}, Adrien and {Flament}, Manon and {Kerkow}, Frank and {Kishore}, Sulekha and {Jani}, Shloke and {Mahanty}, Stephen K. and {Liu}, C{\'e}line and {Llinares}, Laura and {Makarison}, Jolyane and {Olli{\'e}ric}, Alix and {Perez}, In{\`e}s and {Pont}, Lydie and {Sharma}, Vyom},
        title = "{A Data Science Platform to Enable Time-domain Astronomy}",
      journal = {\apjs},
         year = 2023,
        month = aug,
       volume = {267},
       number = {2},
          eid = {31},
        pages = {31},
          doi = {10.3847/1538-4365/acdee1},
archivePrefix = {arXiv},
       eprint = {2305.00108},
 primaryClass = {astro-ph.IM},
       adsurl = {https://ui.adsabs.harvard.edu/abs/2023ApJS..267...31C}
}

@ARTICLE{sbuilderkim2022PASP..134b4505K,
       author = {{Kim}, Y.-L. and {Rigault}, M. and {Neill}, J.~D. and {Briday}, M. and {Copin}, Y. and {Lezmy}, J. and {Nicolas}, N. and {Riddle}, R. and {Sharma}, Y. and {Smith}, M. and {Sollerman}, J. and {Walters}, R.},
        title = "{New Modules for the SEDMachine to Remove Contaminations from Cosmic Rays and Non-target Light: BYECR and CONTSEP}",
      journal = {\pasp},
         year = 2022,
        month = feb,
       volume = {134},
       number = {1032},
          eid = {024505},
        pages = {024505},
          doi = {10.1088/1538-3873/ac50a0},
archivePrefix = {arXiv},
       eprint = {2203.01346},
 primaryClass = {astro-ph.IM},
       adsurl = {https://ui.adsabs.harvard.edu/abs/2022PASP..134b4505K}
}

@ARTICLE{zbuilderduev2021arXiv211112142D,
       author = {{Duev}, Dmitry A. and {van der Walt}, St{\'e}fan J.},
        title = "{Phenomenological classification of the Zwicky Transient Facility astronomical event alerts}",
      journal = {arXiv e-prints},
         year = 2021,
        month = nov,
          eid = {arXiv:2111.12142},
        pages = {arXiv:2111.12142},
          doi = {10.48550/arXiv.2111.12142},
archivePrefix = {arXiv},
       eprint = {2111.12142},
 primaryClass = {astro-ph.IM},
       adsurl = {https://ui.adsabs.harvard.edu/abs/2021arXiv211112142D}
}

@ARTICLE{zbuilderdekany2020PASP..132c8001D,
       author = {{Dekany}, Richard and {Smith}, Roger M. and {Riddle}, Reed and {Feeney}, Michael and {Porter}, Michael and {Hale}, David and {Zolkower}, Jeffry and {Belicki}, Justin and {Kaye}, Stephen and {Henning}, John and {Walters}, Richard and {Cromer}, John and {Delacroix}, Alex and {Rodriguez}, Hector and {Reiley}, Daniel J. and {Mao}, Peter and {Hover}, David and {Murphy}, Patrick and {Burruss}, Rick and {Baker}, John and {Kowalski}, Marek and {Reif}, Klaus and {Mueller}, Phillip and {Bellm}, Eric and {Graham}, Matthew and {Kulkarni}, Shrinivas R.},
        title = "{The Zwicky Transient Facility: Observing System}",
      journal = {\pasp},
         year = 2020,
        month = mar,
       volume = {132},
       number = {1009},
          eid = {038001},
        pages = {038001},
          doi = {10.1088/1538-3873/ab4ca2},
archivePrefix = {arXiv},
       eprint = {2008.04923},
 primaryClass = {astro-ph.IM},
       adsurl = {https://ui.adsabs.harvard.edu/abs/2020PASP..132c8001D}
}

@INPROCEEDINGS{zbuilderduev2020AAS...23538608D,
       author = {{Duev}, D.~A. and {Caltech ZTF Team}},
        title = "{Deep learning for the Zwicky Transient Facility: real/bogus classification and identification of fast-moving objects}",
    booktitle = {American Astronomical Society Meeting Abstracts \#235},
         year = 2020,
       series = {American Astronomical Society Meeting Abstracts},
       volume = {235},
        month = jan,
          eid = {386.08},
        pages = {386.08},
       adsurl = {https://ui.adsabs.harvard.edu/abs/2020AAS...23538608D}
}

@ARTICLE{zbuilderduev2019MNRAS.489.3582D,
       author = {{Duev}, Dmitry A. and {Mahabal}, Ashish and {Masci}, Frank J. and {Graham}, Matthew J. and {Rusholme}, Ben and {Walters}, Richard and {Karmarkar}, Ishani and {Frederick}, Sara and {Kasliwal}, Mansi M. and {Rebbapragada}, Umaa and {Ward}, Charlotte},
        title = "{Real-bogus classification for the Zwicky Transient Facility using deep learning}",
      journal = {\mnras},
         year = 2019,
        month = nov,
       volume = {489},
       number = {3},
        pages = {3582-3590},
          doi = {10.1093/mnras/stz2357},
archivePrefix = {arXiv},
       eprint = {1907.11259},
 primaryClass = {astro-ph.IM},
       adsurl = {https://ui.adsabs.harvard.edu/abs/2019MNRAS.489.3582D}
}

@ARTICLE{zbuildergraham2019PASP..131g8001G,
       author = {{Graham}, Matthew J. and {Kulkarni}, S.~R. and {Bellm}, Eric C. and {Adams}, Scott M. and {Barbarino}, Cristina and {Blagorodnova}, Nadejda and {Bodewits}, Dennis and {Bolin}, Bryce and {Brady}, Patrick R. and {Cenko}, S. Bradley and {Chang}, Chan-Kao and {Coughlin}, Michael W. and {De}, Kishalay and {Eadie}, Gwendolyn and {Farnham}, Tony L. and {Feindt}, Ulrich and {Franckowiak}, Anna and {Fremling}, Christoffer and {Gezari}, Suvi and {Ghosh}, Shaon and {Goldstein}, Daniel A. and {Golkhou}, V. Zach and {Goobar}, Ariel and {Ho}, Anna Y.~Q. and {Huppenkothen}, Daniela and {Ivezi{\'c}}, {\v{Z}}eljko and {Jones}, R. Lynne and {Juric}, Mario and {Kaplan}, David L. and {Kasliwal}, Mansi M. and {Kelley}, Michael S.~P. and {Kupfer}, Thomas and {Lee}, Chien-De and {Lin}, Hsing Wen and {Lunnan}, Ragnhild and {Mahabal}, Ashish A. and {Miller}, Adam A. and {Ngeow}, Chow-Choong and {Nugent}, Peter and {Ofek}, Eran O. and {Prince}, Thomas A. and {Rauch}, Ludwig and {van Roestel}, Jan and {Schulze}, Steve and {Singer}, Leo P. and {Sollerman}, Jesper and {Taddia}, Francesco and {Yan}, Lin and {Ye}, Quan-Zhi and {Yu}, Po-Chieh and {Barlow}, Tom and {Bauer}, James and {Beck}, Ron and {Belicki}, Justin and {Biswas}, Rahul and {Brinnel}, Valery and {Brooke}, Tim and {Bue}, Brian and {Bulla}, Mattia and {Burruss}, Rick and {Connolly}, Andrew and {Cromer}, John and {Cunningham}, Virginia and {Dekany}, Richard and {Delacroix}, Alex and {Desai}, Vandana and {Duev}, Dmitry A. and {Feeney}, Michael and {Flynn}, David and {Frederick}, Sara and {Gal-Yam}, Avishay and {Giomi}, Matteo and {Groom}, Steven and {Hacopians}, Eugean and {Hale}, David and {Helou}, George and {Henning}, John and {Hover}, David and {Hillenbrand}, Lynne A. and {Howell}, Justin and {Hung}, Tiara and {Imel}, David and {Ip}, Wing-Huen and {Jackson}, Edward and {Kaspi}, Shai and {Kaye}, Stephen and {Kowalski}, Marek and {Kramer}, Emily and {Kuhn}, Michael and {Landry}, Walter and {Laher}, Russ R. and {Mao}, Peter and {Masci}, Frank J. and {Monkewitz}, Serge and {Murphy}, Patrick and {Nordin}, Jakob and {Patterson}, Maria T. and {Penprase}, Bryan and {Porter}, Michael and {Rebbapragada}, Umaa and {Reiley}, Dan and {Riddle}, Reed and {Rigault}, Mickael and {Rodriguez}, Hector and {Rusholme}, Ben and {van Santen}, Jakob and {Shupe}, David L. and {Smith}, Roger M. and {Soumagnac}, Maayane T. and {Stein}, Robert and {Surace}, Jason and {Szkody}, Paula and {Terek}, Scott and {Van Sistine}, Angela and {van Velzen}, Sjoert and {Vestrand}, W. Thomas and {Walters}, Richard and {Ward}, Charlotte and {Zhang}, Chaoran and {Zolkower}, Jeffry},
        title = "{The Zwicky Transient Facility: Science Objectives}",
      journal = {\pasp},
         year = 2019,
        month = jul,
       volume = {131},
       number = {1001},
        pages = {078001},
          doi = {10.1088/1538-3873/ab006c},
archivePrefix = {arXiv},
       eprint = {1902.01945},
 primaryClass = {astro-ph.IM},
       adsurl = {https://ui.adsabs.harvard.edu/abs/2019PASP..131g8001G}
}

@ARTICLE{sbuilderrigault2019A&A...627A.115R,
       author = {{Rigault}, M. and {Neill}, J.~D. and {Blagorodnova}, N. and {Dugas}, A. and {Feeney}, M. and {Walters}, R. and {Brinnel}, V. and {Copin}, Y. and {Fremling}, C. and {Nordin}, J. and {Sollerman}, J.},
        title = "{Fully automated integral field spectrograph pipeline for the SEDMachine: pysedm}",
      journal = {\aap},
         year = 2019,
        month = jul,
       volume = {627},
          eid = {A115},
        pages = {A115},
          doi = {10.1051/0004-6361/201935344},
archivePrefix = {arXiv},
       eprint = {1902.08526},
 primaryClass = {astro-ph.IM},
       adsurl = {https://ui.adsabs.harvard.edu/abs/2019A&A...627A.115R}
}

@ARTICLE{zbuilderbellm2019PASP..131f8003B,
       author = {{Bellm}, Eric C. and {Kulkarni}, Shrinivas R. and {Barlow}, Tom and {Feindt}, Ulrich and {Graham}, Matthew J. and {Goobar}, Ariel and {Kupfer}, Thomas and {Ngeow}, Chow-Choong and {Nugent}, Peter and {Ofek}, Eran and {Prince}, Thomas A. and {Riddle}, Reed and {Walters}, Richard and {Ye}, Quan-Zhi},
        title = "{The Zwicky Transient Facility: Surveys and Scheduler}",
      journal = {\pasp},
         year = 2019,
        month = jun,
       volume = {131},
       number = {1000},
        pages = {068003},
          doi = {10.1088/1538-3873/ab0c2a},
archivePrefix = {arXiv},
       eprint = {1905.02209},
 primaryClass = {astro-ph.IM},
       adsurl = {https://ui.adsabs.harvard.edu/abs/2019PASP..131f8003B}
}

@ARTICLE{fbuildervanderwalt2019JOSS....4.1247V,
       author = {{van der Walt}, St{\'e}fan and {Crellin-Quick}, Arien and {Bloom}, Joshua},
        title = "{SkyPortal: An Astronomical Data Platform}",
      journal = {The Journal of Open Source Software},
         year = 2019,
        month = may,
       volume = {4},
       number = {37},
          eid = {1247},
        pages = {1247},
          doi = {10.21105/joss.01247},
       adsurl = {https://ui.adsabs.harvard.edu/abs/2019JOSS....4.1247V}
}

@ARTICLE{zbuildermasci2019PASP..131a8003M,
       author = {{Masci}, Frank J. and {Laher}, Russ R. and {Rusholme}, Ben and {Shupe}, David L. and {Groom}, Steven and {Surace}, Jason and {Jackson}, Edward and {Monkewitz}, Serge and {Beck}, Ron and {Flynn}, David and {Terek}, Scott and {Landry}, Walter and {Hacopians}, Eugean and {Desai}, Vandana and {Howell}, Justin and {Brooke}, Tim and {Imel}, David and {Wachter}, Stefanie and {Ye}, Quan-Zhi and {Lin}, Hsing-Wen and {Cenko}, S. Bradley and {Cunningham}, Virginia and {Rebbapragada}, Umaa and {Bue}, Brian and {Miller}, Adam A. and {Mahabal}, Ashish and {Bellm}, Eric C. and {Patterson}, Maria T. and {Juri{\'c}}, Mario and {Golkhou}, V. Zach and {Ofek}, Eran O. and {Walters}, Richard and {Graham}, Matthew and {Kasliwal}, Mansi M. and {Dekany}, Richard G. and {Kupfer}, Thomas and {Burdge}, Kevin and {Cannella}, Christopher B. and {Barlow}, Tom and {Van Sistine}, Angela and {Giomi}, Matteo and {Fremling}, Christoffer and {Blagorodnova}, Nadejda and {Levitan}, David and {Riddle}, Reed and {Smith}, Roger M. and {Helou}, George and {Prince}, Thomas A. and {Kulkarni}, Shrinivas R.},
        title = "{The Zwicky Transient Facility: Data Processing, Products, and Archive}",
      journal = {\pasp},
         year = 2019,
        month = jan,
       volume = {131},
       number = {995},
        pages = {018003},
          doi = {10.1088/1538-3873/aae8ac},
archivePrefix = {arXiv},
       eprint = {1902.01872},
 primaryClass = {astro-ph.IM},
       adsurl = {https://ui.adsabs.harvard.edu/abs/2019PASP..131a8003M}
}

@ARTICLE{zbuilderbellm2019PASP..131a8002B,
       author = {{Bellm}, Eric C. and {Kulkarni}, Shrinivas R. and {Graham}, Matthew J. and {Dekany}, Richard and {Smith}, Roger M. and {Riddle}, Reed and {Masci}, Frank J. and {Helou}, George and {Prince}, Thomas A. and {Adams}, Scott M. and {Barbarino}, C. and {Barlow}, Tom and {Bauer}, James and {Beck}, Ron and {Belicki}, Justin and {Biswas}, Rahul and {Blagorodnova}, Nadejda and {Bodewits}, Dennis and {Bolin}, Bryce and {Brinnel}, Valery and {Brooke}, Tim and {Bue}, Brian and {Bulla}, Mattia and {Burruss}, Rick and {Cenko}, S. Bradley and {Chang}, Chan-Kao and {Connolly}, Andrew and {Coughlin}, Michael and {Cromer}, John and {Cunningham}, Virginia and {De}, Kishalay and {Delacroix}, Alex and {Desai}, Vandana and {Duev}, Dmitry A. and {Eadie}, Gwendolyn and {Farnham}, Tony L. and {Feeney}, Michael and {Feindt}, Ulrich and {Flynn}, David and {Franckowiak}, Anna and {Frederick}, S. and {Fremling}, C. and {Gal-Yam}, Avishay and {Gezari}, Suvi and {Giomi}, Matteo and {Goldstein}, Daniel A. and {Golkhou}, V. Zach and {Goobar}, Ariel and {Groom}, Steven and {Hacopians}, Eugean and {Hale}, David and {Henning}, John and {Ho}, Anna Y.~Q. and {Hover}, David and {Howell}, Justin and {Hung}, Tiara and {Huppenkothen}, Daniela and {Imel}, David and {Ip}, Wing-Huen and {Ivezi{\'c}}, {\v{Z}}eljko and {Jackson}, Edward and {Jones}, Lynne and {Juric}, Mario and {Kasliwal}, Mansi M. and {Kaspi}, S. and {Kaye}, Stephen and {Kelley}, Michael S.~P. and {Kowalski}, Marek and {Kramer}, Emily and {Kupfer}, Thomas and {Landry}, Walter and {Laher}, Russ R. and {Lee}, Chien-De and {Lin}, Hsing Wen and {Lin}, Zhong-Yi and {Lunnan}, Ragnhild and {Giomi}, Matteo and {Mahabal}, Ashish and {Mao}, Peter and {Miller}, Adam A. and {Monkewitz}, Serge and {Murphy}, Patrick and {Ngeow}, Chow-Choong and {Nordin}, Jakob and {Nugent}, Peter and {Ofek}, Eran and {Patterson}, Maria T. and {Penprase}, Bryan and {Porter}, Michael and {Rauch}, Ludwig and {Rebbapragada}, Umaa and {Reiley}, Dan and {Rigault}, Mickael and {Rodriguez}, Hector and {van Roestel}, Jan and {Rusholme}, Ben and {van Santen}, Jakob and {Schulze}, S. and {Shupe}, David L. and {Singer}, Leo P. and {Soumagnac}, Maayane T. and {Stein}, Robert and {Surace}, Jason and {Sollerman}, Jesper and {Szkody}, Paula and {Taddia}, F. and {Terek}, Scott and {Van Sistine}, Angela and {van Velzen}, Sjoert and {Vestrand}, W. Thomas and {Walters}, Richard and {Ward}, Charlotte and {Ye}, Quan-Zhi and {Yu}, Po-Chieh and {Yan}, Lin and {Zolkower}, Jeffry},
        title = "{The Zwicky Transient Facility: System Overview, Performance, and First Results}",
      journal = {\pasp},
         year = 2019,
        month = jan,
       volume = {131},
       number = {995},
        pages = {018002},
          doi = {10.1088/1538-3873/aaecbe},
archivePrefix = {arXiv},
       eprint = {1902.01932},
 primaryClass = {astro-ph.IM},
       adsurl = {https://ui.adsabs.harvard.edu/abs/2019PASP..131a8002B}
}

@ARTICLE{sbuilderblagorodnova2018PASP..130c5003B,
       author = {{Blagorodnova}, Nadejda and {Neill}, James D. and {Walters}, Richard and {Kulkarni}, Shrinivas R. and {Fremling}, Christoffer and {Ben-Ami}, Sagi and {Dekany}, Richard G. and {Fucik}, Jason R. and {Konidaris}, Nick and {Nash}, Reston and {Ngeow}, Chow-Choong and {Ofek}, Eran O. and {O' Sullivan}, Donal and {Quimby}, Robert and {Ritter}, Andreas and {Vyhmeister}, Karl E.},
        title = "{The SED Machine: A Robotic Spectrograph for Fast Transient Classification}",
      journal = {\pasp},
         year = 2018,
        month = mar,
       volume = {130},
       number = {985},
        pages = {035003},
          doi = {10.1088/1538-3873/aaa53f},
archivePrefix = {arXiv},
       eprint = {1710.02917},
 primaryClass = {astro-ph.IM},
       adsurl = {https://ui.adsabs.harvard.edu/abs/2018PASP..130c5003B}
}

@ARTICLE{zackay2016ApJ...830...27Z,
       author = {{Zackay}, Barak and {Ofek}, Eran O. and {Gal-Yam}, Avishay},
        title = "{Proper Image Subtraction{\textemdash}Optimal Transient Detection, Photometry, and Hypothesis Testing}",
      journal = {\apj},
         year = 2016,
        month = oct,
       volume = {830},
       number = {1},
          eid = {27},
        pages = {27},
          doi = {10.3847/0004-637X/830/1/27},
archivePrefix = {arXiv},
       eprint = {1601.02655},
 primaryClass = {astro-ph.IM},
       adsurl = {https://ui.adsabs.harvard.edu/abs/2016ApJ...830...27Z}
}

@INPROCEEDINGS{bertin2011ASPC..442..435B,
       author = {{Bertin}, E.},
        title = "{Automated Morphometry with SExtractor and PSFEx}",
    booktitle = {Astronomical Data Analysis Software and Systems XX},
         year = 2011,
       editor = {{Evans}, I.~N. and {Accomazzi}, A. and {Mink}, D.~J. and {Rots}, A.~H.},
       series = {Astronomical Society of the Pacific Conference Series},
       volume = {442},
        month = jul,
        pages = {435},
       adsurl = {https://ui.adsabs.harvard.edu/abs/2011ASPC..442..435B}
}

@MISC{Bertin2010_swarp,
       author = {{Bertin}, Emmanuel},
        title = "{SWarp: Resampling and Co-adding FITS Images Together}",
 howpublished = {Astrophysics Source Code Library, record ascl:1010.068},
         year = 2010,
        month = oct,
          eid = {ascl:1010.068},
        pages = {ascl:1010.068},
archivePrefix = {ascl},
       eprint = {1010.068},
       adsurl = {https://ui.adsabs.harvard.edu/abs/2010ascl.soft10068B}
}

@ARTICLE{Bertin+1996_sourceextractor,
       author = {{Bertin}, E. and {Arnouts}, S.},
        title = "{SExtractor: Software for source extraction.}",
      journal = {\aaps},
         year = 1996,
        month = jun,
       volume = {117},
        pages = {393-404},
          doi = {10.1051/aas:1996164},
       adsurl = {https://ui.adsabs.harvard.edu/abs/1996A&AS..117..393B}
}

@misc{chambers2019panstarrs1surveys,
      title={The Pan-STARRS1 Surveys}, 
      author={K. C. Chambers and E. A. Magnier and N. Metcalfe and H. A. Flewelling and M. E. Huber and C. Z. Waters and L. Denneau and P. W. Draper and D. Farrow and D. P. Finkbeiner and C. Holmberg and J. Koppenhoefer and P. A. Price and A. Rest and R. P. Saglia and E. F. Schlafly and S. J. Smartt and W. Sweeney and R. J. Wainscoat and W. S. Burgett and S. Chastel and T. Grav and J. N. Heasley and K. W. Hodapp and R. Jedicke and N. Kaiser and R. -P. Kudritzki and G. A. Luppino and R. H. Lupton and D. G. Monet and J. S. Morgan and P. M. Onaka and B. Shiao and C. W. Stubbs and J. L. Tonry and R. White and E. Bañados and E. F. Bell and R. Bender and E. J. Bernard and M. Boegner and F. Boffi and M. T. Botticella and A. Calamida and S. Casertano and W. -P. Chen and X. Chen and S. Cole and N. Deacon and C. Frenk and A. Fitzsimmons and S. Gezari and V. Gibbs and C. Goessl and T. Goggia and R. Gourgue and B. Goldman and P. Grant and E. K. Grebel and N. C. Hambly and G. Hasinger and A. F. Heavens and T. M. Heckman and R. Henderson and T. Henning and M. Holman and U. Hopp and W. -H. Ip and S. Isani and M. Jackson and C. D. Keyes and A. M. Koekemoer and R. Kotak and D. Le and D. Liska and K. S. Long and J. R. Lucey and M. Liu and N. F. Martin and G. Masci and B. McLean and E. Mindel and P. Misra and E. Morganson and D. N. A. Murphy and A. Obaika and G. Narayan and M. A. Nieto-Santisteban and P. Norberg and J. A. Peacock and E. A. Pier and M. Postman and N. Primak and C. Rae and A. Rai and A. Riess and A. Riffeser and H. W. Rix and S. Röser and R. Russel and L. Rutz and E. Schilbach and A. S. B. Schultz and D. Scolnic and L. Strolger and A. Szalay and S. Seitz and E. Small and K. W. Smith and D. R. Soderblom and P. Taylor and R. Thomson and A. N. Taylor and A. R. Thakar and J. Thiel and D. Thilker and D. Unger and Y. Urata and J. Valenti and J. Wagner and T. Walder and F. Walter and S. P. Watters and S. Werner and W. M. Wood-Vasey and R. Wyse},
      year={2019},
      eprint={1612.05560},
      archivePrefix={arXiv},
      primaryClass={astro-ph.IM},
      url={https://arxiv.org/abs/1612.05560}, 
}

@ARTICLE{dey_legacysurveys_2019AJ....157..168D,
       author = {{Dey}, Arjun and {Schlegel}, David J. and {Lang}, Dustin and {Blum}, Robert and {Burleigh}, Kaylan and {Fan}, Xiaohui and {Findlay}, Joseph R. and {Finkbeiner}, Doug and {Herrera}, David and {Juneau}, St{\'e}phanie and {Landriau}, Martin and {Levi}, Michael and {McGreer}, Ian and {Meisner}, Aaron and {Myers}, Adam D. and {Moustakas}, John and {Nugent}, Peter and {Patej}, Anna and {Schlafly}, Edward F. and {Walker}, Alistair R. and {Valdes}, Francisco and {Weaver}, Benjamin A. and {Y{\`e}che}, Christophe and {Zou}, Hu and {Zhou}, Xu and {Abareshi}, Behzad and {Abbott}, T.~M.~C. and {Abolfathi}, Bela and {Aguilera}, C. and {Alam}, Shadab and {Allen}, Lori and {Alvarez}, A. and {Annis}, James and {Ansarinejad}, Behzad and {Aubert}, Marie and {Beechert}, Jacqueline and {Bell}, Eric F. and {BenZvi}, Segev Y. and {Beutler}, Florian and {Bielby}, Richard M. and {Bolton}, Adam S. and {Brice{\~n}o}, C{\'e}sar and {Buckley-Geer}, Elizabeth J. and {Butler}, Karen and {Calamida}, Annalisa and {Carlberg}, Raymond G. and {Carter}, Paul and {Casas}, Ricard and {Castander}, Francisco J. and {Choi}, Yumi and {Comparat}, Johan and {Cukanovaite}, Elena and {Delubac}, Timoth{\'e}e and {DeVries}, Kaitlin and {Dey}, Sharmila and {Dhungana}, Govinda and {Dickinson}, Mark and {Ding}, Zhejie and {Donaldson}, John B. and {Duan}, Yutong and {Duckworth}, Christopher J. and {Eftekharzadeh}, Sarah and {Eisenstein}, Daniel J. and {Etourneau}, Thomas and {Fagrelius}, Parker A. and {Farihi}, Jay and {Fitzpatrick}, Mike and {Font-Ribera}, Andreu and {Fulmer}, Leah and {G{\"a}nsicke}, Boris T. and {Gaztanaga}, Enrique and {George}, Koshy and {Gerdes}, David W. and {Gontcho}, Satya Gontcho A. and {Gorgoni}, Claudio and {Green}, Gregory and {Guy}, Julien and {Harmer}, Diane and {Hernandez}, M. and {Honscheid}, Klaus and {Huang}, Lijuan Wendy and {James}, David J. and {Jannuzi}, Buell T. and {Jiang}, Linhua and {Joyce}, Richard and {Karcher}, Armin and {Karkar}, Sonia and {Kehoe}, Robert and {Kneib}, Jean-Paul and {Kueter-Young}, Andrea and {Lan}, Ting-Wen and {Lauer}, Tod R. and {Le Guillou}, Laurent and {Le Van Suu}, Auguste and {Lee}, Jae Hyeon and {Lesser}, Michael and {Perreault Levasseur}, Laurence and {Li}, Ting S. and {Mann}, Justin L. and {Marshall}, Robert and {Mart{\'\i}nez-V{\'a}zquez}, C.~E. and {Martini}, Paul and {du Mas des Bourboux}, H{\'e}lion and {McManus}, Sean and {Meier}, Tobias Gabriel and {M{\'e}nard}, Brice and {Metcalfe}, Nigel and {Mu{\~n}oz-Guti{\'e}rrez}, Andrea and {Najita}, Joan and {Napier}, Kevin and {Narayan}, Gautham and {Newman}, Jeffrey A. and {Nie}, Jundan and {Nord}, Brian and {Norman}, Dara J. and {Olsen}, Knut A.~G. and {Paat}, Anthony and {Palanque-Delabrouille}, Nathalie and {Peng}, Xiyan and {Poppett}, Claire L. and {Poremba}, Megan R. and {Prakash}, Abhishek and {Rabinowitz}, David and {Raichoor}, Anand and {Rezaie}, Mehdi and {Robertson}, A.~N. and {Roe}, Natalie A. and {Ross}, Ashley J. and {Ross}, Nicholas P. and {Rudnick}, Gregory and {Safonova}, Sasha and {Saha}, Abhijit and {S{\'a}nchez}, F. Javier and {Savary}, Elodie and {Schweiker}, Heidi and {Scott}, Adam and {Seo}, Hee-Jong and {Shan}, Huanyuan and {Silva}, David R. and {Slepian}, Zachary and {Soto}, Christian and {Sprayberry}, David and {Staten}, Ryan and {Stillman}, Coley M. and {Stupak}, Robert J. and {Summers}, David L. and {Sien Tie}, Suk and {Tirado}, H. and {Vargas-Maga{\~n}a}, Mariana and {Vivas}, A. Katherina and {Wechsler}, Risa H. and {Williams}, Doug and {Yang}, Jinyi and {Yang}, Qian and {Yapici}, Tolga and {Zaritsky}, Dennis and {Zenteno}, A. and {Zhang}, Kai and {Zhang}, Tianmeng and {Zhou}, Rongpu and {Zhou}, Zhimin},
        title = "{Overview of the DESI Legacy Imaging Surveys}",
      journal = {\aj},
         year = 2019,
        month = may,
       volume = {157},
       number = {5},
          eid = {168},
        pages = {168},
          doi = {10.3847/1538-3881/ab089d},
archivePrefix = {arXiv},
       eprint = {1804.08657},
 primaryClass = {astro-ph.IM},
       adsurl = {https://ui.adsabs.harvard.edu/abs/2019AJ....157..168D}
}

@ARTICLE{galyam2008ApJ...686..408G,
       author = {{Gal-Yam}, Avishay and {Nakar}, Ehud and {Ofek}, Eran O. and {Cenko}, S.~B. and {Kulkarni}, S.~R. and {Soderberg}, A.~M. and {Harrison}, F. and {Fox}, D.~B. and {Price}, P.~A. and {Penprase}, B.~E. and {Frail}, Dale A. and {Atteia}, J.~L. and {Berger}, E. and {Gladders}, M. and {Mulchaey}, J.},
        title = "{New Imaging and Spectroscopy of the Locations of Several Short-Hard Gamma-Ray Bursts}",
      journal = {\apj},
         year = 2008,
        month = oct,
       volume = {686},
       number = {1},
        pages = {408-416},
          doi = {10.1086/590947},
archivePrefix = {arXiv},
       eprint = {astro-ph/0509891},
 primaryClass = {astro-ph},
       adsurl = {https://ui.adsabs.harvard.edu/abs/2008ApJ...686..408G}
}

@ARTICLE{fremling2016A&A...593A..68F,
       author = {{Fremling}, C. and {Sollerman}, J. and {Taddia}, F. and {Ergon}, M. and {Fraser}, M. and {Karamehmetoglu}, E. and {Valenti}, S. and {Jerkstrand}, A. and {Arcavi}, I. and {Bufano}, F. and {Elias Rosa}, N. and {Filippenko}, A.~V. and {Fox}, D. and {Gal-Yam}, A. and {Howell}, D.~A. and {Kotak}, R. and {Mazzali}, P. and {Milisavljevic}, D. and {Nugent}, P.~E. and {Nyholm}, A. and {Pian}, E. and {Smartt}, S.},
        title = "{PTF12os and iPTF13bvn. Two stripped-envelope supernovae from low-mass progenitors in NGC 5806}",
      journal = {\aap},
         year = 2016,
        month = sep,
       volume = {593},
          eid = {A68},
        pages = {A68},
          doi = {10.1051/0004-6361/201628275},
archivePrefix = {arXiv},
       eprint = {1606.03074},
 primaryClass = {astro-ph.HE},
       adsurl = {https://ui.adsabs.harvard.edu/abs/2016A&A...593A..68F}
}

@ARTICLE{hi4pi_maps,
       author = {{HI4PI Collaboration} and {Ben Bekhti}, N. and {Fl{\"o}er}, L. and {Keller}, R. and {Kerp}, J. and {Lenz}, D. and {Winkel}, B. and {Bailin}, J. and {Calabretta}, M.~R. and {Dedes}, L. and {Ford}, H.~A. and {Gibson}, B.~K. and {Haud}, U. and {Janowiecki}, S. and {Kalberla}, P.~M.~W. and {Lockman}, F.~J. and {McClure-Griffiths}, N.~M. and {Murphy}, T. and {Nakanishi}, H. and {Pisano}, D.~J. and {Staveley-Smith}, L.},
        title = "{HI4PI: A full-sky H I survey based on EBHIS and GASS}",
      journal = {\aap},
         year = 2016,
        month = oct,
       volume = {594},
          eid = {A116},
        pages = {A116},
          doi = {10.1051/0004-6361/201629178},
archivePrefix = {arXiv},
       eprint = {1610.06175},
 primaryClass = {astro-ph.GA},
       adsurl = {https://ui.adsabs.harvard.edu/abs/2016A&A...594A.116H}
}

@ARTICLE{wilms_abund,
       author = {{Wilms}, J. and {Allen}, A. and {McCray}, R.},
        title = "{On the Absorption of X-Rays in the Interstellar Medium}",
      journal = {\apj},
         year = 2000,
        month = oct,
       volume = {542},
       number = {2},
        pages = {914-924},
          doi = {10.1086/317016},
archivePrefix = {arXiv},
       eprint = {astro-ph/0008425},
 primaryClass = {astro-ph},
       adsurl = {https://ui.adsabs.harvard.edu/abs/2000ApJ...542..914W}
}

@ARTICLE{richards2011ApJS..194...29R,
       author = {{Richards}, Joseph L. and {Max-Moerbeck}, Walter and {Pavlidou}, Vasiliki and {King}, Oliver G. and {Pearson}, Timothy J. and {Readhead}, Anthony C.~S. and {Reeves}, Rodrigo and {Shepherd}, Martin C. and {Stevenson}, Matthew A. and {Weintraub}, Lawrence C. and {Fuhrmann}, Lars and {Angelakis}, Emmanouil and {Zensus}, J. Anton and {Healey}, Stephen E. and {Romani}, Roger W. and {Shaw}, Michael S. and {Grainge}, Keith and {Birkinshaw}, Mark and {Lancaster}, Katy and {Worrall}, Diana M. and {Taylor}, Gregory B. and {Cotter}, Garret and {Bustos}, Ricardo},
        title = "{Blazars in the Fermi Era: The OVRO 40 m Telescope Monitoring Program}",
      journal = {\apjs},
         year = 2011,
        month = jun,
       volume = {194},
       number = {2},
          eid = {29},
        pages = {29},
          doi = {10.1088/0067-0049/194/2/29},
archivePrefix = {arXiv},
       eprint = {1011.3111},
 primaryClass = {astro-ph.CO},
       adsurl = {https://ui.adsabs.harvard.edu/abs/2011ApJS..194...29R}
}

@INPROCEEDINGS{jiang2018SPIE10702E..2LJ,
       author = {{Jiang}, Haijiao and {Hu}, Zhongwen and {Xu}, Mingming and {Dai}, Songxin and {Zhang}, Huatao and {Wang}, Lei and {Chen}, Yi},
        title = "{The preliminary design of the next generation Palomar spectrograph for 200-inch Hale telescope}",
    booktitle = {Ground-based and Airborne Instrumentation for Astronomy VII},
         year = 2018,
       editor = {{Evans}, Christopher J. and {Simard}, Luc and {Takami}, Hideki},
       series = {Society of Photo-Optical Instrumentation Engineers (SPIE) Conference Series},
       volume = {10702},
        month = jul,
          eid = {107022L},
        pages = {107022L},
          doi = {10.1117/12.2312550},
       adsurl = {https://ui.adsabs.harvard.edu/abs/2018SPIE10702E..2LJ}
}

@ARTICLE{kasliwal2024TNSAN.340....1K,
       author = {{Kasliwal}, M.~M. and {Fremling}, C. and {Yan}, L. and {Das}, K. and {Verdi}, F. and {Zmudzinas}, J. and {Martin}, C. and {Kirby}, E. and {Xue}, S. and {Ho}, L. and {Herczeg}, G. and {Wu}, X. and {Hu}, Z. and {Ji}, H. and {Matuszewski}, M. and {Bertz}, R. and {Hale}, D. and {Rodriguez}, H. and {Boden}, A. and {Dekany}, R. and {Smith}, R. and {Reiley}, D. and {Nash}, R. and {Milburn}, J. and {Neill}, D. and {Brugger}, J. and {Zarzaca}, R. and {Weber}, B. and {Shapiro}, C.},
        title = "{The First Science Spectrum with the Next Generation Palomar Spectrograph (NGPS): Classification of ZTF24abrfcqd as a Type Ia Supernova}",
      journal = {Transient Name Server AstroNote},
         year = 2024,
        month = nov,
       volume = {340},
        pages = {1},
       adsurl = {https://ui.adsabs.harvard.edu/abs/2024TNSAN.340....1K}
}

@ARTICLE{dragons2023RNAAS...7..214L,
       author = {{Labrie}, K. and {Simpson}, C. and {Cardenes}, R. and {Turner}, J. and {Soraisam}, M. and {Quint}, B. and {Oberdorf}, O. and {Placco}, V.~M. and {Berke}, D. and {Smirnova}, O. and {Conseil}, S. and {Vacca}, W.~D. and {Thomas-Osip}, J.},
        title = "{DRAGONS-A Quick Overview}",
      journal = {Research Notes of the American Astronomical Society},
         year = 2023,
        month = oct,
       volume = {7},
       number = {10},
          eid = {214},
        pages = {214},
          doi = {10.3847/2515-5172/ad0044},
archivePrefix = {arXiv},
       eprint = {2310.03048},
 primaryClass = {astro-ph.IM},
       adsurl = {https://ui.adsabs.harvard.edu/abs/2023RNAAS...7..214L}
}

@software{simpson_2026_19055103,
  author       = {Simpson, Chris and
                  Labrie, Kathleen and
                  Hirst, Paul and
                  Turner, James and
                  Smirnova, Olesja and
                  Rawlings, Mark and
                  Vacca, William and
                  Berke, Daniel},
  title        = {DRAGONS},
  month        = apr,
  year         = 2026,
  publisher    = {Zenodo},
  version      = {4.2.0},
  doi          = {10.5281/zenodo.19055103},
  url          = {https://doi.org/10.5281/zenodo.19055103},
}

@ARTICLE{frostig_winter_2026,
       author = {{Frostig}, Danielle and {Lourie}, Nathan and {Karambelkar}, Viraj and {Kasliwal}, Mansi M. and {Malonis}, Andrew and {Simcoe}, Robert A. and {Stein}, Robert and {Baker}, John W. and {Burdge}, Kevin and {Burruss}, Rick and {Corcoran}, Curt and {De}, Kishalay and {Furesz}, Gabor and {Ganciu}, Nicolae and {Haworth}, Kari and {Heffner}, Carolyn M. and {Hinrichsen}, Erik and {Juneau}, Jill and {Mo}, Geoffrey and {Purdum}, Josiah and {Rose}, Sam and {Soto}, Cruz and {Zolkower}, Jeffry},
        title = "{The WINTER Observatory: A One-Degree InGaAs Survey Camera to study the Transient Infrared Sky}",
      journal = {arXiv e-prints},
         year = 2025,
        month = dec,
          eid = {arXiv:2512.16753},
        pages = {arXiv:2512.16753},
          doi = {10.48550/arXiv.2512.16753},
archivePrefix = {arXiv},
       eprint = {2512.16753},
 primaryClass = {astro-ph.IM},
       adsurl = {https://ui.adsabs.harvard.edu/abs/2025arXiv251216753F}
}

@article{mirar,
  author       = {Viraj Karambelkar and Robert David Stein and others},
  year         = "in-prep",
}

@software{mirar_zenodo,
  author       = {Robert David Stein and
                  Viraj Karambelkar and
                  Sulekha Kishore and
                  Saarah Hall and
                  Benjamin Roulston and
                  Jamie Soon and
                  Danielle Frostig},
  title        = {winter-telescope/mirar: v1.1.0},
  month        = jan,
  year         = 2026,
  publisher    = {Zenodo},
  version      = {v1.1.0},
  doi          = {10.5281/zenodo.18391243},
  url          = {https://doi.org/10.5281/zenodo.18391243},
  swhid        = {swh:1:dir:7e9be464eb9bba4476a5f305782b49b5276ba3f0
                   ;origin=https://doi.org/10.5281/zenodo.10888436;vi
                   sit=swh:1:snp:4ce3e5d7d62d1d36233e42e9cb2c1ff393c2
                   1a55;anchor=swh:1:rel:018a35606502ab6eb0844a46acda
                   750b6674babf;path=winter-telescope-mirar-ecebb5d
                  },
}

@ARTICLE{Gaia2021,
       author = {{Gaia Collaboration}},
        key = {Gaia},
    collaboration = {Gaia Collaboration},
        title = "{Gaia Early Data Release 3. Summary of the contents and survey properties}",
      journal = {\aap},
         year = 2021,
        month = may,
       volume = {649},
          eid = {A1},
        pages = {A1},
          doi = {10.1051/0004-6361/202039657},
archivePrefix = {arXiv},
       eprint = {2012.01533},
 primaryClass = {astro-ph.GA},
       adsurl = {https://ui.adsabs.harvard.edu/abs/2021A&A...649A...1G}
}

@ARTICLE{2mass,
       author = {{Skrutskie}, M.~F. and {Cutri}, R.~M. and {Stiening}, R. and {Weinberg}, M.~D. and {Schneider}, S. and {Carpenter}, J.~M. and {Beichman}, C. and {Capps}, R. and {Chester}, T. and {Elias}, J. and {Huchra}, J. and {Liebert}, J. and {Lonsdale}, C. and {Monet}, D.~G. and {Price}, S. and {Seitzer}, P. and {Jarrett}, T. and {Kirkpatrick}, J.~D. and {Gizis}, J.~E. and {Howard}, E. and {Evans}, T. and {Fowler}, J. and {Fullmer}, L. and {Hurt}, R. and {Light}, R. and {Kopan}, E.~L. and {Marsh}, K.~A. and {McCallon}, H.~L. and {Tam}, R. and {Van Dyk}, S. and {Wheelock}, S.},
        title = "{The Two Micron All Sky Survey (2MASS)}",
      journal = {\aj},
         year = 2006,
        month = feb,
       volume = {131},
       number = {2},
        pages = {1163-1183},
          doi = {10.1086/498708},
       adsurl = {https://ui.adsabs.harvard.edu/abs/2006AJ....131.1163S}
}

@ARTICLE{pypeit:joss_arXiv,
       author = {{Prochaska}, J. Xavier and {Hennawi}, Joseph F. and {Westfall}, Kyle B. and
         {Cooke}, Ryan J. and {Wang}, Feige and {Hsyu}, Tiffany and
         {Davies}, Frederick B. and {Farina}, Emanuele Paolo},
        title = "{PypeIt: The Python Spectroscopic Data Reduction Pipeline}",
      journal = {arXiv e-prints},
         year = 2020,
        month = may,
          eid = {arXiv:2005.06505},
        pages = {arXiv:2005.06505},
archivePrefix = {arXiv},
       eprint = {2005.06505},
 primaryClass = {astro-ph.IM},
       adsurl = {https://ui.adsabs.harvard.edu/abs/2020arXiv200506505P}
}

@MISC{pypeit:zenodo,
       author = {{Prochaska}, J. Xavier and {Hennawi}, Joseph and {Cooke}, Ryan and
         {Westfall}, Kyle and {Wang}, Feige and {EmAstro} and {Tiffanyhsyu} and
         {Wasserman}, Asher and {Villaume}, Alexa and {Marijana777} and
         {Schindler}, JT and {Young}, David and {Simha}, Sunil and
         {Wilde}, Matt and {Tejos}, Nicolas and {Isbell}, Jacob and
         {Fl{\"o}rs}, Andreas and {Sandford}, Nathan and {Vasovi{\'c}}, Zlatan and
         {Betts}, Edward and {Holden}, Brad},
        title = "{pypeit/PypeIt: Release 1.0.0}",
         year = 2020,
        month = apr,
          eid = {10.5281/zenodo.3743493},
          doi = {10.5281/zenodo.3743493},
      version = {v1.0.0},
    publisher = {Zenodo},
       adsurl = {https://ui.adsabs.harvard.edu/abs/2020zndo...3743493P}
}

@INPROCEEDINGS{ldt_12,
       author = {{Levine}, Stephen E. and {Bida}, Thomas A. and {Chylek}, Tomas and {Collins}, Peter L. and {DeGroff}, William T. and {Dunham}, Edward W. and {Lotz}, Paul J. and {Venetiou}, Alexander J. and {Zoonemat Kermani}, Saeid},
        title = "{Status and performance of the Discovery Channel Telescope during commissioning}",
    booktitle = {Ground-based and Airborne Telescopes IV},
         year = 2012,
       editor = {{Stepp}, Larry M. and {Gilmozzi}, Roberto and {Hall}, Helen J.},
       series = {Society of Photo-Optical Instrumentation Engineers (SPIE) Conference Series},
       volume = {8444},
        month = sep,
          eid = {844419},
        pages = {844419},
          doi = {10.1117/12.926415},
       adsurl = {https://ui.adsabs.harvard.edu/abs/2012SPIE.8444E..19L}
}

@INPROCEEDINGS{ldt_14,
       author = {{Bida}, Thomas A. and {Dunham}, Edward W. and {Massey}, Philip and {Roe}, Henry G.},
        title = "{First-generation instrumentation for the Discovery Channel Telescope}",
    booktitle = {Ground-based and Airborne Instrumentation for Astronomy V},
         year = 2014,
       editor = {{Ramsay}, Suzanne K. and {McLean}, Ian S. and {Takami}, Hideki},
       series = {Society of Photo-Optical Instrumentation Engineers (SPIE) Conference Series},
       volume = {9147},
        month = jul,
          eid = {91472N},
        pages = {91472N},
          doi = {10.1117/12.2056872},
       adsurl = {https://ui.adsabs.harvard.edu/abs/2014SPIE.9147E..2NB}
}

@ARTICLE{kong_171205_magnetar,
       author = {{Kong}, De-Feng and {Wang}, Xiang-Gao and {Zheng}, WeiKang and {Filippenko}, Alexei V. and {Wang}, Shan-Qin and {Xin}, L.~P. and {Wei}, J.~Y. and {Han}, X.~H. and {Wang}, J. and {Liang}, En-Wei and {Brink}, Thomas G. and {Reddy}, Naveen A. and {Liu}, Hong-Bang and {Xie}, Fei},
        title = "{GRB 171205A/SN 2017iuk: Is the Central Engine a Magnetar?}",
      journal = {\apj},
         year = 2026,
        month = feb,
       volume = {997},
       number = {2},
          eid = {338},
        pages = {338},
          doi = {10.3847/1538-4357/ae27c0},
       adsurl = {https://ui.adsabs.harvard.edu/abs/2026ApJ...997..338K}
}

@ARTICLE{perley2026arXiv260103337P,
       author = {{Perley}, Daniel A. and {Ho}, Anna Y.~Q. and {McGrath}, Zo{\"e} and {Camilo}, Michael and {Sevilla}, Cassie and {Chen}, Ping and {Schroeder}, Genevieve and {Govreen-Segal}, Taya and {Bochenek}, Aleksandra and {Qin}, Yu-Jing and {Gillanders}, James H. and {Amend}, Benjamin and {Anderson}, Joseph P. and {Andreoni}, Igor and {Aryan}, Amar and {Bellm}, Eric C. and {Bloom}, Joshua S. and {de Boer}, Thomas and {Carney}, Jonathan and {Caiazzo}, Ilaria and {Chambers}, Ken C. and {Charalampopoulos}, Panos and {Chen}, Ting-Wan and {Chen}, Tracy X. and {Coughlin}, Eric R. and {Coughlin}, Michael and {Dennefeld}, Michel and {Dimitriadis}, Georgios and {Fremling}, Christoffer and {Frostig}, Danielle and {Gal-Yam}, Avishay and {Galbany}, Llu{\'\i}s and {Gangopadhyay}, Anjashay and {Ghendrih}, Melzie and {Graham}, Matthew J. and {Gromadzki}, Mariusz and {Groom}, Steven L. and {Guti{\'e}rrez}, Claudia P. and {Hinds}, K.-Ryan and {Huber}, Mark E. and {Inserra}, Cosimo and {Kaiser}, Benjamin C. and {Kasliwal}, Mansi M. and {Koivisto}, Niilo E. and {Lin}, Chien-Cheng and {Liu}, Chang and {Lowe}, Thomas B. and {Magnier}, Eugene and {Mahabal}, Ashish A. and {Milligan}, Andrew and {Minguez}, Paloma and {Mo}, Geoffrey and {M{\"u}ller-Bravo}, Tom{\'a}s E. and {Nicholl}, Matt and {Pessi}, Priscila J. and {Pignata}, Giuliano and {Purdum}, Josiah and {Rehemtulla}, Nabeel and {Rich}, R. Michael and {Sahu}, Anwesha and {Singh}, Avinash and {Smartt}, Stephen J. and {Sollerman}, Jesper and {Srinivasaragavan}, Gokul and {Srivastav}, Shubham and {Stein}, Robert D. and {Schulze}, Steve and {Tweddle}, Jack W. and {Wainscoat}, Richard and {Wise}, Jacob L. and {Yan}, Lin and {Young}, David R.},
        title = "{AT2024wpp: An Extremely Luminous Fast Ultraviolet Transient Powered by Accretion onto a Black Hole}",
      journal = {arXiv e-prints},
         year = 2026,
        month = jan,
          eid = {arXiv:2601.03337},
        pages = {arXiv:2601.03337},
          doi = {10.48550/arXiv.2601.03337},
archivePrefix = {arXiv},
       eprint = {2601.03337},
 primaryClass = {astro-ph.HE},
       adsurl = {https://ui.adsabs.harvard.edu/abs/2026arXiv260103337P}
}

@ARTICLE{perley2020ApJ...904...35P,
       author = {{Perley}, Daniel A. and {Fremling}, Christoffer and {Sollerman}, Jesper and {Miller}, Adam A. and {Dahiwale}, Aishwarya S. and {Sharma}, Yashvi and {Bellm}, Eric C. and {Biswas}, Rahul and {Brink}, Thomas G. and {Bruch}, Rachel J. and {De}, Kishalay and {Dekany}, Richard and {Drake}, Andrew J. and {Duev}, Dmitry A. and {Filippenko}, Alexei V. and {Gal-Yam}, Avishay and {Goobar}, Ariel and {Graham}, Matthew J. and {Graham}, Melissa L. and {Ho}, Anna Y.~Q. and {Irani}, Ido and {Kasliwal}, Mansi M. and {Kim}, Young-Lo and {Kulkarni}, S.~R. and {Mahabal}, Ashish and {Masci}, Frank J. and {Modak}, Shaunak and {Neill}, James D. and {Nordin}, Jakob and {Riddle}, Reed L. and {Soumagnac}, Maayane T. and {Strotjohann}, Nora L. and {Schulze}, Steve and {Taggart}, Kirsty and {Tzanidakis}, Anastasios and {Walters}, Richard S. and {Yan}, Lin},
        title = "{The Zwicky Transient Facility Bright Transient Survey. II. A Public Statistical Sample for Exploring Supernova Demographics}",
      journal = {\apj},
         year = 2020,
        month = nov,
       volume = {904},
       number = {1},
          eid = {35},
        pages = {35},
          doi = {10.3847/1538-4357/abbd98},
archivePrefix = {arXiv},
       eprint = {2009.01242},
 primaryClass = {astro-ph.HE},
       adsurl = {https://ui.adsabs.harvard.edu/abs/2020ApJ...904...35P}
}

@ARTICLE{grupe_2007_late_afterglow,
       author = {{Grupe}, Dirk and {Gronwall}, Caryl and {Wang}, Xiang-Yu and {Roming}, Peter W.~A. and {Cummings}, Jay and {Zhang}, Bing and {M{\'e}sz{\'a}ros}, Peter and {Diaz Trigo}, Maria and {O'Brien}, Paul T. and {Page}, Kim L. and {Beardmore}, Andy and {Godet}, Olivier and {vanden Berk}, Daniel E. and {Brown}, Peter J. and {Koch}, Scott and {Morris}, David and {Stroh}, Michael and {Burrows}, David N. and {Nousek}, John A. and {McMath Chester}, Margaret and {Immler}, Stefan and {Mangano}, Vanessa and {Romano}, Patrizia and {Chincarini}, Guido and {Osborne}, Julian and {Sakamoto}, Takanori and {Gehrels}, Neil},
        title = "{Swift and XMM-Newton Observations of the Extraordinary Gamma-Ray Burst 060729: More than 125 Days of X-Ray Afterglow}",
      journal = {\apj},
         year = 2007,
        month = jun,
       volume = {662},
       number = {1},
        pages = {443-458},
          doi = {10.1086/517868},
archivePrefix = {arXiv},
       eprint = {astro-ph/0611240},
 primaryClass = {astro-ph},
       adsurl = {https://ui.adsabs.harvard.edu/abs/2007ApJ...662..443G}
}

@ARTICLE{grupe_2010_late_afterglow,
       author = {{Grupe}, Dirk and {Burrows}, David N. and {Wu}, Xue-Feng and {Wang}, Xiang-Yu and {Zhang}, Bing and {Liang}, En-Wei and {Garmire}, Gordon and {Nousek}, John A. and {Gehrels}, Neil and {Ricker}, George R. and {Bautz}, Marshall W.},
        title = "{Late-Time Detections of the X-Ray Afterglow of GRB 060729 with Chandra{\textemdash}The Latest Detections Ever of an X-Ray Afterglow}",
      journal = {\apj},
         year = 2010,
        month = mar,
       volume = {711},
       number = {2},
        pages = {1008-1016},
          doi = {10.1088/0004-637X/711/2/1008},
archivePrefix = {arXiv},
       eprint = {0903.1258},
 primaryClass = {astro-ph.CO},
       adsurl = {https://ui.adsabs.harvard.edu/abs/2010ApJ...711.1008G}
}

@ARTICLE{kann2010ApJ...720.1513K,
       author = {{Kann}, D.~A. and {Klose}, S. and {Zhang}, B. and {Malesani}, D. and {Nakar}, E. and {Pozanenko}, A. and {Wilson}, A.~C. and {Butler}, N.~R. and {Jakobsson}, P. and {Schulze}, S. and {Andreev}, M. and {Antonelli}, L.~A. and {Bikmaev}, I.~F. and {Biryukov}, V. and {B{\"o}ttcher}, M. and {Burenin}, R.~A. and {Castro Cer{\'o}n}, J.~M. and {Castro-Tirado}, A.~J. and {Chincarini}, G. and {Cobb}, B.~E. and {Covino}, S. and {D'Avanzo}, P. and {D'Elia}, V. and {Della Valle}, M. and {de Ugarte Postigo}, A. and {Efimov}, Yu. and {Ferrero}, P. and {Fugazza}, D. and {Fynbo}, J.~P.~U. and {G{\r{a}}lfalk}, M. and {Grundahl}, F. and {Gorosabel}, J. and {Gupta}, S. and {Guziy}, S. and {Hafizov}, B. and {Hjorth}, J. and {Holhjem}, K. and {Ibrahimov}, M. and {Im}, M. and {Israel}, G.~L. and {Je{\'l}inek}, M. and {Jensen}, B.~L. and {Karimov}, R. and {Khamitov}, I.~M. and {Kizilo{\v{g}}lu}, {\"U}. and {Klunko}, E. and {Kub{\'a}nek}, P. and {Kutyrev}, A.~S. and {Laursen}, P. and {Levan}, A.~J. and {Mannucci}, F. and {Martin}, C.~M. and {Mescheryakov}, A. and {Mirabal}, N. and {Norris}, J.~P. and {Ovaldsen}, J.-E. and {Paraficz}, D. and {Pavlenko}, E. and {Piranomonte}, S. and {Rossi}, A. and {Rumyantsev}, V. and {Salinas}, R. and {Sergeev}, A. and {Sharapov}, D. and {Sollerman}, J. and {Stecklum}, B. and {Stella}, L. and {Tagliaferri}, G. and {Tanvir}, N.~R. and {Telting}, J. and {Testa}, V. and {Updike}, A.~C. and {Volnova}, A. and {Watson}, D. and {Wiersema}, K. and {Xu}, D.},
        title = "{The Afterglows of Swift-era Gamma-ray Bursts. I. Comparing pre-Swift and Swift-era Long/Soft (Type II) GRB Optical Afterglows}",
      journal = {\apj},
         year = 2010,
        month = sep,
       volume = {720},
       number = {2},
        pages = {1513-1558},
          doi = {10.1088/0004-637X/720/2/1513},
archivePrefix = {arXiv},
       eprint = {0712.2186},
 primaryClass = {astro-ph},
       adsurl = {https://ui.adsabs.harvard.edu/abs/2010ApJ...720.1513K}
}

@ARTICLE{wanderman2010MNRAS.406.1944W,
       author = {{Wanderman}, David and {Piran}, Tsvi},
        title = "{The luminosity function and the rate of Swift's gamma-ray bursts}",
      journal = {\mnras},
         year = 2010,
        month = aug,
       volume = {406},
       number = {3},
        pages = {1944-1958},
          doi = {10.1111/j.1365-2966.2010.16787.x},
archivePrefix = {arXiv},
       eprint = {0912.0709},
 primaryClass = {astro-ph.HE},
       adsurl = {https://ui.adsabs.harvard.edu/abs/2010MNRAS.406.1944W}
}

@ARTICLE{zeh2006ApJ...637..889Z,
       author = {{Zeh}, A. and {Klose}, S. and {Kann}, D.~A.},
        title = "{Gamma-Ray Burst Afterglow Light Curves in the Pre-Swift Era: A Statistical Study}",
      journal = {\apj},
         year = 2006,
        month = feb,
       volume = {637},
       number = {2},
        pages = {889-900},
          doi = {10.1086/498442},
archivePrefix = {arXiv},
       eprint = {astro-ph/0509299},
 primaryClass = {astro-ph},
       adsurl = {https://ui.adsabs.harvard.edu/abs/2006ApJ...637..889Z}
}

@ARTICLE{virgili2009MNRAS.392...91V,
       author = {{Virgili}, Francisco J. and {Liang}, En-Wei and {Zhang}, Bing},
        title = "{Low-luminosity gamma-ray bursts as a distinct GRB population: a firmer case from multiple criteria constraints}",
      journal = {\mnras},
         year = 2009,
        month = jan,
       volume = {392},
       number = {1},
        pages = {91-103},
          doi = {10.1111/j.1365-2966.2008.14063.x},
archivePrefix = {arXiv},
       eprint = {0801.4751},
 primaryClass = {astro-ph},
       adsurl = {https://ui.adsabs.harvard.edu/abs/2009MNRAS.392...91V}
}

@ARTICLE{bromberg2011ApJ...739L..55B,
       author = {{Bromberg}, Omer and {Nakar}, Ehud and {Piran}, Tsvi},
        title = "{Are Low-luminosity Gamma-Ray Bursts Generated by Relativistic Jets?}",
      journal = {\apjl},
         year = 2011,
        month = oct,
       volume = {739},
       number = {2},
          eid = {L55},
        pages = {L55},
          doi = {10.1088/2041-8205/739/2/L55},
archivePrefix = {arXiv},
       eprint = {1107.1346},
 primaryClass = {astro-ph.HE},
       adsurl = {https://ui.adsabs.harvard.edu/abs/2011ApJ...739L..55B}
}

@ARTICLE{nakar2015ApJ...807..172N,
       author = {{Nakar}, Ehud},
        title = "{A Unified Picture for Low-luminosity and Long Gamma-Ray Bursts Based on the Extended Progenitor of llGRB 060218/SN 2006aj}",
      journal = {\apj},
         year = 2015,
        month = jul,
       volume = {807},
       number = {2},
          eid = {172},
        pages = {172},
          doi = {10.1088/0004-637X/807/2/172},
archivePrefix = {arXiv},
       eprint = {1503.00441},
 primaryClass = {astro-ph.HE},
       adsurl = {https://ui.adsabs.harvard.edu/abs/2015ApJ...807..172N}
}

@ARTICLE{nakar2012ApJ...747...88N,
       author = {{Nakar}, Ehud and {Sari}, Re'em},
        title = "{Relativistic Shock Breakouts{\textemdash}A Variety of Gamma-Ray Flares: From Low-luminosity Gamma-Ray Bursts to Type Ia Supernovae}",
      journal = {\apj},
         year = 2012,
        month = mar,
       volume = {747},
       number = {2},
          eid = {88},
        pages = {88},
          doi = {10.1088/0004-637X/747/2/88},
archivePrefix = {arXiv},
       eprint = {1106.2556},
 primaryClass = {astro-ph.HE},
       adsurl = {https://ui.adsabs.harvard.edu/abs/2012ApJ...747...88N}
}

@ARTICLE{soderberg2006Natur.442.1014S,
       author = {{Soderberg}, A.~M. and {Kulkarni}, S.~R. and {Nakar}, E. and {Berger}, E. and {Cameron}, P.~B. and {Fox}, D.~B. and {Frail}, D. and {Gal-Yam}, A. and {Sari}, R. and {Cenko}, S.~B. and {Kasliwal}, M. and {Chevalier}, R.~A. and {Piran}, T. and {Price}, P.~A. and {Schmidt}, B.~P. and {Pooley}, G. and {Moon}, D.-S. and {Penprase}, B.~E. and {Ofek}, E. and {Rau}, A. and {Gehrels}, N. and {Nousek}, J.~A. and {Burrows}, D.~N. and {Persson}, S.~E. and {McCarthy}, P.~J.},
        title = "{Relativistic ejecta from X-ray flash XRF 060218 and the rate of cosmic explosions}",
      journal = {\nat},
         year = 2006,
        month = aug,
       volume = {442},
       number = {7106},
        pages = {1014-1017},
          doi = {10.1038/nature05087},
archivePrefix = {arXiv},
       eprint = {astro-ph/0604389},
 primaryClass = {astro-ph},
       adsurl = {https://ui.adsabs.harvard.edu/abs/2006Natur.442.1014S}
}

@ARTICLE{cano2011ApJ...740...41C,
       author = {{Cano}, Z. and {Bersier}, D. and {Guidorzi}, C. and {Kobayashi}, S. and {Levan}, A.~J. and {Tanvir}, N.~R. and {Wiersema}, K. and {D'Avanzo}, P. and {Fruchter}, A.~S. and {Garnavich}, P. and {Gomboc}, A. and {Gorosabel}, J. and {Kasen}, D. and {Kopa{\v{c}}}, D. and {Margutti}, R. and {Mazzali}, P.~A. and {Melandri}, A. and {Mundell}, C.~G. and {Nugent}, P.~E. and {Pian}, E. and {Smith}, R.~J. and {Steele}, I. and {Wijers}, R.~A.~M.~J. and {Woosley}, S.~E.},
        title = "{XRF 100316D/SN 2010bh and the Nature of Gamma-Ray Burst Supernovae}",
      journal = {\apj},
         year = 2011,
        month = oct,
       volume = {740},
       number = {1},
          eid = {41},
        pages = {41},
          doi = {10.1088/0004-637X/740/1/41},
archivePrefix = {arXiv},
       eprint = {1104.5141},
 primaryClass = {astro-ph.SR},
       adsurl = {https://ui.adsabs.harvard.edu/abs/2011ApJ...740...41C}
}

@ARTICLE{sollerman2006AA...454..503S,
       author = {{Sollerman}, J. and {Jaunsen}, A.~O. and {Fynbo}, J.~P.~U. and {Hjorth}, J. and {Jakobsson}, P. and {Stritzinger}, M. and {F{\'e}ron}, C. and {Laursen}, P. and {Ovaldsen}, J.-E. and {Selj}, J. and {Th{\"o}ne}, C.~C. and {Xu}, D. and {Davis}, T. and {Gorosabel}, J. and {Watson}, D. and {Duro}, R. and {Ilyin}, I. and {Jensen}, B.~L. and {Lysfjord}, N. and {Marquart}, T. and {Nielsen}, T.~B. and {N{\"a}r{\"a}nen}, J. and {Schwarz}, H.~E. and {Walch}, S. and {Wold}, M. and {{\"O}stlin}, G.},
        title = "{Supernova 2006aj and the associated X-Ray Flash 060218}",
      journal = {\aap},
         year = 2006,
        month = aug,
       volume = {454},
       number = {2},
        pages = {503-509},
          doi = {10.1051/0004-6361:20065226},
archivePrefix = {arXiv},
       eprint = {astro-ph/0603495},
 primaryClass = {astro-ph},
       adsurl = {https://ui.adsabs.harvard.edu/abs/2006A&A...454..503S}
}

@ARTICLE{ferror2006AA...457..857F,
       author = {{Ferrero}, P. and {Kann}, D.~A. and {Zeh}, A. and {Klose}, S. and {Pian}, E. and {Palazzi}, E. and {Masetti}, N. and {Hartmann}, D.~H. and {Sollerman}, J. and {Deng}, J. and {Filippenko}, A.~V. and {Greiner}, J. and {Hughes}, M.~A. and {Mazzali}, P. and {Li}, W. and {Rol}, E. and {Smith}, R.~J. and {Tanvir}, N.~R.},
        title = "{The GRB 060218/SN 2006aj event in the context of other gamma-ray burst supernovae}",
      journal = {\aap},
         year = 2006,
        month = oct,
       volume = {457},
       number = {3},
        pages = {857-864},
          doi = {10.1051/0004-6361:20065530},
archivePrefix = {arXiv},
       eprint = {astro-ph/0605058},
 primaryClass = {astro-ph},
       adsurl = {https://ui.adsabs.harvard.edu/abs/2006A&A...457..857F}
}

@ARTICLE{campana2006Natur.442.1008C,
       author = {{Campana}, S. and {Mangano}, V. and {Blustin}, A.~J. and {Brown}, P. and {Burrows}, D.~N. and {Chincarini}, G. and {Cummings}, J.~R. and {Cusumano}, G. and {Della Valle}, M. and {Malesani}, D. and {M{\'e}sz{\'a}ros}, P. and {Nousek}, J.~A. and {Page}, M. and {Sakamoto}, T. and {Waxman}, E. and {Zhang}, B. and {Dai}, Z.~G. and {Gehrels}, N. and {Immler}, S. and {Marshall}, F.~E. and {Mason}, K.~O. and {Moretti}, A. and {O'Brien}, P.~T. and {Osborne}, J.~P. and {Page}, K.~L. and {Romano}, P. and {Roming}, P.~W.~A. and {Tagliaferri}, G. and {Cominsky}, L.~R. and {Giommi}, P. and {Godet}, O. and {Kennea}, J.~A. and {Krimm}, H. and {Angelini}, L. and {Barthelmy}, S.~D. and {Boyd}, P.~T. and {Palmer}, D.~M. and {Wells}, A.~A. and {White}, N.~E.},
        title = "{The association of GRB 060218 with a supernova and the evolution of the shock wave}",
      journal = {\nat},
         year = 2006,
        month = aug,
       volume = {442},
       number = {7106},
        pages = {1008-1010},
          doi = {10.1038/nature04892},
archivePrefix = {arXiv},
       eprint = {astro-ph/0603279},
 primaryClass = {astro-ph},
       adsurl = {https://ui.adsabs.harvard.edu/abs/2006Natur.442.1008C}
}

@INPROCEEDINGS{2007ASPC..376..127M,
       author = {{McMullin}, J.~P. and {Waters}, B. and {Schiebel}, D. and {Young}, W. and {Golap}, K.},
        title = "{CASA Architecture and Applications}",
    booktitle = {Astronomical Data Analysis Software and Systems XVI},
         year = 2007,
       editor = {{Shaw}, R.~A. and {Hill}, F. and {Bell}, D.~J.},
       series = {Astronomical Society of the Pacific Conference Series},
       volume = {376},
        month = oct,
        pages = {127},
       adsurl = {https://ui.adsabs.harvard.edu/abs/2007ASPC..376..127M}
}

@ARTICLE{2022PASP..134k4501C,
       author = {{CASA Team} and {Bean}, Ben and {Bhatnagar}, Sanjay and {Castro}, Sandra and {Donovan Meyer}, Jennifer and {Emonts}, Bjorn and {Garcia}, Enrique and {Garwood}, Robert and {Golap}, Kumar and {Gonzalez Villalba}, Justo and {Harris}, Pamela and {Hayashi}, Yohei and {Hoskins}, Josh and {Hsieh}, Mingyu and {Jagannathan}, Preshanth and {Kawasaki}, Wataru and {Keimpema}, Aard and {Kettenis}, Mark and {Lopez}, Jorge and {Marvil}, Joshua and {Masters}, Joseph and {McNichols}, Andrew and {Mehringer}, David and {Miel}, Renaud and {Moellenbrock}, George and {Montesino}, Federico and {Nakazato}, Takeshi and {Ott}, Juergen and {Petry}, Dirk and {Pokorny}, Martin and {Raba}, Ryan and {Rau}, Urvashi and {Schiebel}, Darrell and {Schweighart}, Neal and {Sekhar}, Srikrishna and {Shimada}, Kazuhiko and {Small}, Des and {Steeb}, Jan-Willem and {Sugimoto}, Kanako and {Suoranta}, Ville and {Tsutsumi}, Takahiro and {van Bemmel}, Ilse M. and {Verkouter}, Marjolein and {Wells}, Akeem and {Xiong}, Wei and {Szomoru}, Arpad and {Griffith}, Morgan and {Glendenning}, Brian and {Kern}, Jeff},
        title = "{CASA, the Common Astronomy Software Applications for Radio Astronomy}",
      journal = {\pasp},
         year = 2022,
        month = nov,
       volume = {134},
       number = {1041},
          eid = {114501},
        pages = {114501},
          doi = {10.1088/1538-3873/ac9642},
archivePrefix = {arXiv},
       eprint = {2210.02276},
 primaryClass = {astro-ph.IM},
       adsurl = {https://ui.adsabs.harvard.edu/abs/2022PASP..134k4501C}
}

@ARTICLE{2026GCN.44160....1P,
       author = {{Perley}, D.~A. and {Schroeder}, G. and {Laskar}, T.},
        title = "{GRB 260310A: VLA Multi-frequency Radio Observations}",
      journal = {GRB Coordinates Network},
         year = 2026,
        month = mar,
       volume = {44160},
        pages = {1},
       adsurl = {https://ui.adsabs.harvard.edu/abs/2026GCN.44160....1P}
}

@ARTICLE{2026GCN.44235....1S,
       author = {{Schroeder}, G. and {Perley}, D.~A. and {Laskar}, T.},
        title = "{GRB 260310A: Continued VLA Multi-frequency Radio Observations}",
      journal = {GRB Coordinates Network},
         year = 2026,
        month = apr,
       volume = {44235},
        pages = {1},
       adsurl = {https://ui.adsabs.harvard.edu/abs/2026GCN.44235....1S}
}

@software{2017ascl.soft04001W,
       author = {{Williams}, Peter K.~G. and {Clavel}, Ma{\"\i}ca and {Newton}, Elisabeth and {Ryzhkov}, Denis},
        title = "{pwkit: Astronomical utilities in Python}",
 howpublished = {Astrophysics Source Code Library, record ascl:1704.001},
         year = 2017,
        month = apr,
          eid = {ascl:1704.001},
archivePrefix = {ascl},
       eprint = {1704.001},
       adsurl = {https://ui.adsabs.harvard.edu/abs/2017ascl.soft04001W}
}

@ARTICLE{lyman2026arXiv260302330L,
       author = {{Lyman}, J.~D. and {O'Neill}, D. and {Killestein}, T. and {Jarvis}, D. and {Kumar}, A. and {Ulaczyk}, K. and {Ackley}, K. and {Chote}, P. and {Dyer}, M.~J. and {Pursiainen}, M. and {Steeghs}, D. and {Godson}, B. and {Magee}, M. and {Mullaney}, J.~R. and {Warwick}, B. and {Belkin}, S. and {Galloway}, D.~K. and {Ramsay}, G. and {Dhillon}, V.~S. and {O'Brien}, P. and {Noysena}, K. and {Kotak}, R. and {Breton}, R.~P. and {Nuttall}, L.~K. and {Gompertz}, B. and {Pollacco}, D. and {Casares}, J. and {Coppejans}, D.~L. and {Eyles-Ferris}, R.~A.~J. and {Graur}, O. and {Kelsey}, L. and {Kennedy}, M.~R. and {Levan}, A. and {Littlefair}, S. and {Mandhai}, S. and {Mata S{\'a}nchez}, D. and {Mattila}, S. and {McCormac}, J. and {Moran}, S. and {Phillips}, C. and {Pu}, K. and {Sahu}, A. and {Shrestha}, M. and {Stanway}, E. and {Starling}, R.~L.~C. and {Vincetti}, L. and {Wickens}, E. and {Wiersema}, K.},
        title = "{The Gravitational-wave Optical Transient Observer (GOTO) data pipeline and workflow for transient discovery}",
      journal = {arXiv e-prints},
         year = 2026,
        month = mar,
          eid = {arXiv:2603.02330},
        pages = {arXiv:2603.02330},
          doi = {10.48550/arXiv.2603.02330},
archivePrefix = {arXiv},
       eprint = {2603.02330},
 primaryClass = {astro-ph.IM},
       adsurl = {https://ui.adsabs.harvard.edu/abs/2026arXiv260302330L}
}

@INPROCEEDINGS{dryer2024SPIE13094E..1XD,
       author = {{Dyer}, Martin J. and {Ackley}, Kendall and {Jim{\'e}nez-Ibarra}, Felipe and {Lyman}, Joseph and {Ulaczyk}, Krzysztof and {Steeghs}, Danny and {Galloway}, Duncan K. and {Dhillon}, Vik S. and {O'Brien}, Paul and {Ramsay}, Gavin and {Noysena}, Kanthanakorn and {Kotak}, Rubina and {Breton}, Rene and {Nuttall}, Laura and {Pall{\'e}}, Enric and {Pollacco}, Don and {Killestein}, Tom and {Kumar}, Amit and {O'Neill}, David and {Kelsey}, Lisa and {Godson}, Ben and {Jarvis}, Dan},
        title = "{The Gravitational-wave Optical Transient Observer (GOTO)}",
    booktitle = {Ground-based and Airborne Telescopes X},
         year = 2024,
       editor = {{Marshall}, Heather K. and {Spyromilio}, Jason and {Usuda}, Tomonori},
       series = {Society of Photo-Optical Instrumentation Engineers (SPIE) Conference Series},
       volume = {13094},
        month = aug,
          eid = {130941X},
        pages = {130941X},
          doi = {10.1117/12.3018305},
archivePrefix = {arXiv},
       eprint = {2407.17176},
 primaryClass = {astro-ph.IM},
       adsurl = {https://ui.adsabs.harvard.edu/abs/2024SPIE13094E..1XD}
}

@software{becker2015ascl.soft04004B,
       author = {{Becker}, Andrew},
        title = "{HOTPANTS: High Order Transform of PSF ANd Template Subtraction}",
 howpublished = {Astrophysics Source Code Library, record ascl:1504.004},
         year = 2015,
        month = apr,
          eid = {ascl:1504.004},
archivePrefix = {ascl},
       eprint = {1504.004},
       adsurl = {https://ui.adsabs.harvard.edu/abs/2015ascl.soft04004B}
}

@ARTICLE{cano2017AdAst2017E...5C,
       author = {{Cano}, Zach and {Wang}, Shan-Qin and {Dai}, Zi-Gao and {Wu}, Xue-Feng},
        title = "{The Observer's Guide to the Gamma-Ray Burst Supernova Connection}",
      journal = {Advances in Astronomy},
         year = 2017,
        month = jan,
       volume = {2017},
          eid = {8929054},
        pages = {8929054},
          doi = {10.1155/2017/8929054},
archivePrefix = {arXiv},
       eprint = {1604.03549},
 primaryClass = {astro-ph.HE},
       adsurl = {https://ui.adsabs.harvard.edu/abs/2017AdAst2017E...5C}
}

@ARTICLE{frail2001ApJ...562L..55F,
       author = {{Frail}, D.~A. and {Kulkarni}, S.~R. and {Sari}, R. and {Djorgovski}, S.~G. and {Bloom}, J.~S. and {Galama}, T.~J. and {Reichart}, D.~E. and {Berger}, E. and {Harrison}, F.~A. and {Price}, P.~A. and {Yost}, S.~A. and {Diercks}, A. and {Goodrich}, R.~W. and {Chaffee}, F.},
        title = "{Beaming in Gamma-Ray Bursts: Evidence for a Standard Energy Reservoir}",
      journal = {\apjl},
         year = 2001,
        month = nov,
       volume = {562},
       number = {1},
        pages = {L55-L58},
          doi = {10.1086/338119},
archivePrefix = {arXiv},
       eprint = {astro-ph/0102282},
 primaryClass = {astro-ph},
       adsurl = {https://ui.adsabs.harvard.edu/abs/2001ApJ...562L..55F}
}

@ARTICLE{irwin2025MNRAS.542.1269I,
       author = {{Irwin}, Christopher M. and {Hotokezaka}, Kenta},
        title = "{Revisiting GRB 060218: new insights into low-luminosity gamma-ray bursts from a revised shock breakout model}",
      journal = {\mnras},
         year = 2025,
        month = sep,
       volume = {542},
       number = {2},
        pages = {1269-1286},
          doi = {10.1093/mnras/staf1309},
archivePrefix = {arXiv},
       eprint = {2412.06736},
 primaryClass = {astro-ph.HE},
       adsurl = {https://ui.adsabs.harvard.edu/abs/2025MNRAS.542.1269I}
}

@ARTICLE{irwin2016MNRAS.460.1680I,
       author = {{Irwin}, Christopher M. and {Chevalier}, Roger A.},
        title = "{Jet or shock breakout? The low-luminosity GRB 060218}",
      journal = {\mnras},
         year = 2016,
        month = aug,
       volume = {460},
       number = {2},
        pages = {1680-1704},
          doi = {10.1093/mnras/stw1058},
archivePrefix = {arXiv},
       eprint = {1511.00336},
 primaryClass = {astro-ph.HE},
       adsurl = {https://ui.adsabs.harvard.edu/abs/2016MNRAS.460.1680I}
}

@ARTICLE{galama1998Natur.395..670G,
       author = {{Galama}, T.~J. and {Vreeswijk}, P.~M. and {van Paradijs}, J. and {Kouveliotou}, C. and {Augusteijn}, T. and {B{\"o}hnhardt}, H. and {Brewer}, J.~P. and {Doublier}, V. and {Gonzalez}, J.-F. and {Leibundgut}, B. and {Lidman}, C. and {Hainaut}, O.~R. and {Patat}, F. and {Heise}, J. and {in't Zand}, J. and {Hurley}, K. and {Groot}, P.~J. and {Strom}, R.~G. and {Mazzali}, P.~A. and {Iwamoto}, K. and {Nomoto}, K. and {Umeda}, H. and {Nakamura}, T. and {Young}, T.~R. and {Suzuki}, T. and {Shigeyama}, T. and {Koshut}, T. and {Kippen}, M. and {Robinson}, C. and {de Wildt}, P. and {Wijers}, R.~A.~M.~J. and {Tanvir}, N. and {Greiner}, J. and {Pian}, E. and {Palazzi}, E. and {Frontera}, F. and {Masetti}, N. and {Nicastro}, L. and {Feroci}, M. and {Costa}, E. and {Piro}, L. and {Peterson}, B.~A. and {Tinney}, C. and {Boyle}, B. and {Cannon}, R. and {Stathakis}, R. and {Sadler}, E. and {Begam}, M.~C. and {Ianna}, P.},
        title = "{An unusual supernova in the error box of the {\ensuremath{\gamma}}-ray burst of 25 April 1998}",
      journal = {\nat},
         year = 1998,
        month = oct,
       volume = {395},
       number = {6703},
        pages = {670-672},
          doi = {10.1038/27150},
archivePrefix = {arXiv},
       eprint = {astro-ph/9806175},
 primaryClass = {astro-ph},
       adsurl = {https://ui.adsabs.harvard.edu/abs/1998Natur.395..670G}
}

@ARTICLE{iwamoto1998Natur.395..672I,
       author = {{Iwamoto}, K. and {Mazzali}, P.~A. and {Nomoto}, K. and {Umeda}, H. and {Nakamura}, T. and {Patat}, F. and {Danziger}, I.~J. and {Young}, T.~R. and {Suzuki}, T. and {Shigeyama}, T. and {Augusteijn}, T. and {Doublier}, V. and {Gonzalez}, J.-F. and {Boehnhardt}, H. and {Brewer}, J. and {Hainaut}, O.~R. and {Lidman}, C. and {Leibundgut}, B. and {Cappellaro}, E. and {Turatto}, M. and {Galama}, T.~J. and {Vreeswijk}, P.~M. and {Kouveliotou}, C. and {van Paradijs}, J. and {Pian}, E. and {Palazzi}, E. and {Frontera}, F.},
        title = "{A hypernova model for the supernova associated with the {\ensuremath{\gamma}}-ray burst of 25 April 1998}",
      journal = {\nat},
         year = 1998,
        month = oct,
       volume = {395},
       number = {6703},
        pages = {672-674},
          doi = {10.1038/27155},
archivePrefix = {arXiv},
       eprint = {astro-ph/9806382},
 primaryClass = {astro-ph},
       adsurl = {https://ui.adsabs.harvard.edu/abs/1998Natur.395..672I}
}

@ARTICLE{kulkarni1998Natur.395..663K,
       author = {{Kulkarni}, S.~R. and {Frail}, D.~A. and {Wieringa}, M.~H. and {Ekers}, R.~D. and {Sadler}, E.~M. and {Wark}, R.~M. and {Higdon}, J.~L. and {Phinney}, E.~S. and {Bloom}, J.~S.},
        title = "{Radio emission from the unusual supernova 1998bw and its association with the {\ensuremath{\gamma}}-ray burst of 25 April 1998}",
      journal = {\nat},
         year = 1998,
        month = oct,
       volume = {395},
       number = {6703},
        pages = {663-669},
          doi = {10.1038/27139},
       adsurl = {https://ui.adsabs.harvard.edu/abs/1998Natur.395..663K}
}

@ARTICLE{stanek2003ApJ...591L..17S,
       author = {{Stanek}, K.~Z. and {Matheson}, T. and {Garnavich}, P.~M. and {Martini}, P. and {Berlind}, P. and {Caldwell}, N. and {Challis}, P. and {Brown}, W.~R. and {Schild}, R. and {Krisciunas}, K. and {Calkins}, M.~L. and {Lee}, J.~C. and {Hathi}, N. and {Jansen}, R.~A. and {Windhorst}, R. and {Echevarria}, L. and {Eisenstein}, D.~J. and {Pindor}, B. and {Olszewski}, E.~W. and {Harding}, P. and {Holland}, S.~T. and {Bersier}, D.},
        title = "{Spectroscopic Discovery of the Supernova 2003dh Associated with GRB 030329}",
      journal = {\apjl},
         year = 2003,
        month = jul,
       volume = {591},
       number = {1},
        pages = {L17-L20},
          doi = {10.1086/376976},
archivePrefix = {arXiv},
       eprint = {astro-ph/0304173},
 primaryClass = {astro-ph},
       adsurl = {https://ui.adsabs.harvard.edu/abs/2003ApJ...591L..17S}
}

@ARTICLE{hjorth2003Natur.423..847H,
       author = {{Hjorth}, Jens and {Sollerman}, Jesper and {M{\o}ller}, Palle and {Fynbo}, Johan P.~U. and {Woosley}, Stan E. and {Kouveliotou}, Chryssa and {Tanvir}, Nial R. and {Greiner}, Jochen and {Andersen}, Michael I. and {Castro-Tirado}, Alberto J. and {Castro Cer{\'o}n}, Jos{\'e} Mar{\'\i}a and {Fruchter}, Andrew S. and {Gorosabel}, Javier and {Jakobsson}, P{\'a}ll and {Kaper}, Lex and {Klose}, Sylvio and {Masetti}, Nicola and {Pedersen}, Holger and {Pedersen}, Kristian and {Pian}, Elena and {Palazzi}, Eliana and {Rhoads}, James E. and {Rol}, Evert and {van den Heuvel}, Edward P.~J. and {Vreeswijk}, Paul M. and {Watson}, Darach and {Wijers}, Ralph A.~M.~J.},
        title = "{A very energetic supernova associated with the {\ensuremath{\gamma}}-ray burst of 29 March 2003}",
      journal = {\nat},
         year = 2003,
        month = jun,
       volume = {423},
       number = {6942},
        pages = {847-850},
          doi = {10.1038/nature01750},
archivePrefix = {arXiv},
       eprint = {astro-ph/0306347},
 primaryClass = {astro-ph},
       adsurl = {https://ui.adsabs.harvard.edu/abs/2003Natur.423..847H}
}

@ARTICLE{woosley1993ApJ...405..273W,
       author = {{Woosley}, S.~E.},
        title = "{Gamma-Ray Bursts from Stellar Mass Accretion Disks around Black Holes}",
      journal = {\apj},
         year = 1993,
        month = mar,
       volume = {405},
        pages = {273},
          doi = {10.1086/172359},
       adsurl = {https://ui.adsabs.harvard.edu/abs/1993ApJ...405..273W}
}

@ARTICLE{pian2000ApJ...536..778P,
       author = {{Pian}, E. and {Amati}, L. and {Antonelli}, L.~A. and {Butler}, R.~C. and {Costa}, E. and {Cusumano}, G. and {Danziger}, J. and {Feroci}, M. and {Fiore}, F. and {Frontera}, F. and {Giommi}, P. and {Masetti}, N. and {Muller}, J.~M. and {Nicastro}, L. and {Oosterbroek}, T. and {Orlandini}, M. and {Owens}, A. and {Palazzi}, E. and {Parmar}, A. and {Piro}, L. and {in't Zand}, J.~J.~M. and {Castro-Tirado}, A. and {Coletta}, A. and {Dal Fiume}, D. and {Del Sordo}, S. and {Heise}, J. and {Soffitta}, P. and {Torroni}, V.},
        title = "{BEPPOSAX Observations of GRB 980425: Detection of the Prompt Event and Monitoring of the Error Box}",
      journal = {\apj},
         year = 2000,
        month = jun,
       volume = {536},
       number = {2},
        pages = {778-787},
          doi = {10.1086/308978},
archivePrefix = {arXiv},
       eprint = {astro-ph/9910235},
 primaryClass = {astro-ph},
       adsurl = {https://ui.adsabs.harvard.edu/abs/2000ApJ...536..778P}
}

@ARTICLE{patat2001ApJ...555..900P,
       author = {{Patat}, Ferdinando and {Cappellaro}, Enrico and {Danziger}, John and {Mazzali}, Paolo A. and {Sollerman}, Jesper and {Augusteijn}, Thomas and {Brewer}, James and {Doublier}, Vanessa and {Gonzalez}, Jean Fran{\c{c}}ois and {Hainaut}, Olivier and {Lidman}, Chris and {Leibundgut}, Bruno and {Nomoto}, Ken'ichi and {Nakamura}, Takayoshi and {Spyromilio}, Jason and {Rizzi}, Luca and {Turatto}, Massimo and {Walsh}, Jeremy and {Galama}, Titus J. and {van Paradijs}, Jan and {Kouveliotou}, Chryssa and {Vreeswijk}, Paul M. and {Frontera}, Filippo and {Masetti}, Nicola and {Palazzi}, Eliana and {Pian}, Elena},
        title = "{The Metamorphosis of SN 1998bw}",
      journal = {\apj},
         year = 2001,
        month = jul,
       volume = {555},
       number = {2},
        pages = {900-917},
          doi = {10.1086/321526},
archivePrefix = {arXiv},
       eprint = {astro-ph/0103111},
 primaryClass = {astro-ph},
       adsurl = {https://ui.adsabs.harvard.edu/abs/2001ApJ...555..900P}
}

@ARTICLE{mazzali2006ApJ...645.1323M,
       author = {{Mazzali}, Paolo A. and {Deng}, Jinsong and {Pian}, Elena and {Malesani}, Daniele and {Tominaga}, Nozomu and {Maeda}, Keiichi and {Nomoto}, Ken'ichi and {Chincarini}, Guido and {Covino}, Stefano and {Della Valle}, Massimo and {Fugazza}, Dino and {Tagliaferri}, Gianpiero and {Gal-Yam}, Avishay},
        title = "{Models for the Type Ic Hypernova SN 2003lw associated with GRB 031203}",
      journal = {\apj},
         year = 2006,
        month = jul,
       volume = {645},
       number = {2},
        pages = {1323-1330},
          doi = {10.1086/504415},
archivePrefix = {arXiv},
       eprint = {astro-ph/0603516},
 primaryClass = {astro-ph},
       adsurl = {https://ui.adsabs.harvard.edu/abs/2006ApJ...645.1323M}
}

@ARTICLE{malesani2004ApJ...609L...5M,
       author = {{Malesani}, D. and {Tagliaferri}, G. and {Chincarini}, G. and {Covino}, S. and {Della Valle}, M. and {Fugazza}, D. and {Mazzali}, P.~A. and {Zerbi}, F.~M. and {D'Avanzo}, P. and {Kalogerakos}, S. and {Simoncelli}, A. and {Antonelli}, L.~A. and {Burderi}, L. and {Campana}, S. and {Cucchiara}, A. and {Fiore}, F. and {Ghirlanda}, G. and {Goldoni}, P. and {G{\"o}tz}, D. and {Mereghetti}, S. and {Mirabel}, I.~F. and {Romano}, P. and {Stella}, L. and {Minezaki}, T. and {Yoshii}, Y. and {Nomoto}, K.},
        title = "{SN 2003lw and GRB 031203: A Bright Supernova for a Faint Gamma-Ray Burst}",
      journal = {\apjl},
         year = 2004,
        month = jul,
       volume = {609},
       number = {1},
        pages = {L5-L8},
          doi = {10.1086/422684},
archivePrefix = {arXiv},
       eprint = {astro-ph/0405449},
 primaryClass = {astro-ph},
       adsurl = {https://ui.adsabs.harvard.edu/abs/2004ApJ...609L...5M}
}

@ARTICLE{soderberg2004Natur.430..648S,
       author = {{Soderberg}, A.~M. and {Kulkarni}, S.~R. and {Berger}, E. and {Fox}, D.~W. and {Sako}, M. and {Frail}, D.~A. and {Gal-Yam}, A. and {Moon}, D.~S. and {Cenko}, S.~B. and {Yost}, S.~A. and {Phillips}, M.~M. and {Persson}, S.~E. and {Freedman}, W.~L. and {Wyatt}, P. and {Jayawardhana}, R. and {Paulson}, D.},
        title = "{The sub-energetic {\ensuremath{\gamma}}-ray burst GRB 031203 as a cosmic analogue to the nearby GRB 980425}",
      journal = {\nat},
         year = 2004,
        month = aug,
       volume = {430},
       number = {7000},
        pages = {648-650},
          doi = {10.1038/nature02757},
archivePrefix = {arXiv},
       eprint = {astro-ph/0408096},
 primaryClass = {astro-ph},
       adsurl = {https://ui.adsabs.harvard.edu/abs/2004Natur.430..648S}
}

@ARTICLE{galyam2004ApJ...609L..59G,
       author = {{Gal-Yam}, A. and {Moon}, D.-S. and {Fox}, D.~B. and {Soderberg}, A.~M. and {Kulkarni}, S.~R. and {Berger}, E. and {Cenko}, S.~B. and {Yost}, S. and {Frail}, D.~A. and {Sako}, M. and {Freedman}, W.~L. and {Persson}, S.~E. and {Wyatt}, P. and {Murphy}, D.~C. and {Phillips}, M.~M. and {Suntzeff}, N.~B. and {Mazzali}, P.~A. and {Nomoto}, K.},
        title = "{The J-Band Light Curve of SN 2003lw, Associated with GRB 031203}",
      journal = {\apjl},
         year = 2004,
        month = jul,
       volume = {609},
       number = {2},
        pages = {L59-L62},
          doi = {10.1086/422841},
archivePrefix = {arXiv},
       eprint = {astro-ph/0403608},
 primaryClass = {astro-ph},
       adsurl = {https://ui.adsabs.harvard.edu/abs/2004ApJ...609L..59G}
}

@ARTICLE{modjaz2006ApJ...645L..21M,
       author = {{Modjaz}, M. and {Stanek}, K.~Z. and {Garnavich}, P.~M. and {Berlind}, P. and {Blondin}, S. and {Brown}, W. and {Calkins}, M. and {Challis}, P. and {Diamond-Stanic}, A.~M. and {Hao}, H. and {Hicken}, M. and {Kirshner}, R.~P. and {Prieto}, J.~L.},
        title = "{Early-Time Photometry and Spectroscopy of the Fast Evolving SN 2006aj Associated with GRB 060218}",
      journal = {\apjl},
         year = 2006,
        month = jul,
       volume = {645},
       number = {1},
        pages = {L21-L24},
          doi = {10.1086/505906},
archivePrefix = {arXiv},
       eprint = {astro-ph/0603377},
 primaryClass = {astro-ph},
       adsurl = {https://ui.adsabs.harvard.edu/abs/2006ApJ...645L..21M}
}

@ARTICLE{cobb2006ApJ...645L.113C,
       author = {{Cobb}, B.~E. and {Bailyn}, C.~D. and {van Dokkum}, P.~G. and {Natarajan}, P.},
        title = "{SN 2006aj and the Nature of Low-Luminosity Gamma-Ray Bursts}",
      journal = {\apjl},
         year = 2006,
        month = jul,
       volume = {645},
       number = {2},
        pages = {L113-L116},
          doi = {10.1086/506271},
archivePrefix = {arXiv},
       eprint = {astro-ph/0603832},
 primaryClass = {astro-ph},
       adsurl = {https://ui.adsabs.harvard.edu/abs/2006ApJ...645L.113C}
}

@ARTICLE{pian2006Natur.442.1011P,
       author = {{Pian}, E. and {Mazzali}, P.~A. and {Masetti}, N. and {Ferrero}, P. and {Klose}, S. and {Palazzi}, E. and {Ramirez-Ruiz}, E. and {Woosley}, S.~E. and {Kouveliotou}, C. and {Deng}, J. and {Filippenko}, A.~V. and {Foley}, R.~J. and {Fynbo}, J.~P.~U. and {Kann}, D.~A. and {Li}, W. and {Hjorth}, J. and {Nomoto}, K. and {Patat}, F. and {Sauer}, D.~N. and {Sollerman}, J. and {Vreeswijk}, P.~M. and {Guenther}, E.~W. and {Levan}, A. and {O'Brien}, P. and {Tanvir}, N.~R. and {Wijers}, R.~A.~M.~J. and {Dumas}, C. and {Hainaut}, O. and {Wong}, D.~S. and {Baade}, D. and {Wang}, L. and {Amati}, L. and {Cappellaro}, E. and {Castro-Tirado}, A.~J. and {Ellison}, S. and {Frontera}, F. and {Fruchter}, A.~S. and {Greiner}, J. and {Kawabata}, K. and {Ledoux}, C. and {Maeda}, K. and {M{\o}ller}, P. and {Nicastro}, L. and {Rol}, E. and {Starling}, R.},
        title = "{An optical supernova associated with the X-ray flash XRF 060218}",
      journal = {\nat},
         year = 2006,
        month = aug,
       volume = {442},
       number = {7106},
        pages = {1011-1013},
          doi = {10.1038/nature05082},
archivePrefix = {arXiv},
       eprint = {astro-ph/0603530},
 primaryClass = {astro-ph},
       adsurl = {https://ui.adsabs.harvard.edu/abs/2006Natur.442.1011P}
}

@ARTICLE{mazzali2006Natur.442.1018M,
       author = {{Mazzali}, Paolo A. and {Deng}, Jinsong and {Nomoto}, Ken'ichi and {Sauer}, Daniel N. and {Pian}, Elena and {Tominaga}, Nozomu and {Tanaka}, Masaomi and {Maeda}, Keiichi and {Filippenko}, Alexei V.},
        title = "{A neutron-star-driven X-ray flash associated with supernova SN 2006aj}",
      journal = {\nat},
         year = 2006,
        month = aug,
       volume = {442},
       number = {7106},
        pages = {1018-1020},
          doi = {10.1038/nature05081},
archivePrefix = {arXiv},
       eprint = {astro-ph/0603567},
 primaryClass = {astro-ph},
       adsurl = {https://ui.adsabs.harvard.edu/abs/2006Natur.442.1018M}
}

@ARTICLE{starling2011MNRAS.411.2792S,
       author = {{Starling}, R.~L.~C. and {Wiersema}, K. and {Levan}, A.~J. and {Sakamoto}, T. and {Bersier}, D. and {Goldoni}, P. and {Oates}, S.~R. and {Rowlinson}, A. and {Campana}, S. and {Sollerman}, J. and {Tanvir}, N.~R. and {Malesani}, D. and {Fynbo}, J.~P.~U. and {Covino}, S. and {D'Avanzo}, P. and {O'Brien}, P.~T. and {Page}, K.~L. and {Osborne}, J.~P. and {Vergani}, S.~D. and {Barthelmy}, S. and {Burrows}, D.~N. and {Cano}, Z. and {Curran}, P.~A. and {de Pasquale}, M. and {D'Elia}, V. and {Evans}, P.~A. and {Flores}, H. and {Fruchter}, A.~S. and {Garnavich}, P. and {Gehrels}, N. and {Gorosabel}, J. and {Hjorth}, J. and {Holland}, S.~T. and {van der Horst}, A.~J. and {Hurkett}, C.~P. and {Jakobsson}, P. and {Kamble}, A.~P. and {Kouveliotou}, C. and {Kuin}, N.~P.~M. and {Kaper}, L. and {Mazzali}, P.~A. and {Nugent}, P.~E. and {Pian}, E. and {Stamatikos}, M. and {Th{\"o}ne}, C.~C. and {Woosley}, S.~E.},
        title = "{Discovery of the nearby long, soft GRB 100316D with an associated supernova}",
      journal = {\mnras},
         year = 2011,
        month = mar,
       volume = {411},
       number = {4},
        pages = {2792-2803},
          doi = {10.1111/j.1365-2966.2010.17879.x},
archivePrefix = {arXiv},
       eprint = {1004.2919},
 primaryClass = {astro-ph.CO},
       adsurl = {https://ui.adsabs.harvard.edu/abs/2011MNRAS.411.2792S}
}

@ARTICLE{bufano2012ApJ...753...67B,
       author = {{Bufano}, Filomena and {Pian}, Elena and {Sollerman}, Jesper and {Benetti}, Stefano and {Pignata}, Giuliano and {Valenti}, Stefano and {Covino}, Stefano and {D'Avanzo}, Paolo and {Malesani}, Daniele and {Cappellaro}, Enrico and {Della Valle}, Massimo and {Fynbo}, Johan and {Hjorth}, Jens and {Mazzali}, Paolo A. and {Reichart}, Daniel E. and {Starling}, Rhaana L.~C. and {Turatto}, Massimo and {Vergani}, Susanna D. and {Wiersema}, Klass and {Amati}, Lorenzo and {Bersier}, David and {Campana}, Sergio and {Cano}, Zach and {Castro-Tirado}, Alberto J. and {Chincarini}, Guido and {D'Elia}, Valerio and {de Ugarte Postigo}, Antonio and {Deng}, Jinsong and {Ferrero}, Patrizia and {Filippenko}, Alexei V. and {Goldoni}, Paolo and {Gorosabel}, Javier and {Greiner}, Jochen and {Hammer}, Francois and {Jakobsson}, Pall and {Kaper}, Lex and {Kawabata}, Koji S. and {Klose}, Sylvio and {Levan}, Andrew J. and {Maeda}, Keiichi and {Masetti}, Nicola and {Milvang-Jensen}, Bo and {Mirabel}, Felix I. and {M{\o}ller}, Palle and {Nomoto}, Ken'ichi and {Palazzi}, Eliana and {Piranomonte}, Silvia and {Salvaterra}, Ruben and {Stratta}, Giulia and {Tagliaferri}, Gianpiero and {Tanaka}, Masaomi and {Tanvir}, Nial R. and {Wijers}, Ralph A.~M.~J.},
        title = "{The Highly Energetic Expansion of SN 2010bh Associated with GRB 100316D}",
      journal = {\apj},
         year = 2012,
        month = jul,
       volume = {753},
       number = {1},
          eid = {67},
        pages = {67},
          doi = {10.1088/0004-637X/753/1/67},
archivePrefix = {arXiv},
       eprint = {1111.4527},
 primaryClass = {astro-ph.HE},
       adsurl = {https://ui.adsabs.harvard.edu/abs/2012ApJ...753...67B}
}

@ARTICLE{margutti2013ApJ...778...18M,
       author = {{Margutti}, R. and {Soderberg}, A.~M. and {Wieringa}, M.~H. and {Edwards}, P.~G. and {Chevalier}, R.~A. and {Morsony}, B.~J. and {Barniol Duran}, R. and {Sironi}, L. and {Zauderer}, B.~A. and {Milisavljevic}, D. and {Kamble}, A. and {Pian}, E.},
        title = "{The Signature of the Central Engine in the Weakest Relativistic Explosions: GRB 100316D}",
      journal = {\apj},
         year = 2013,
        month = nov,
       volume = {778},
       number = {1},
          eid = {18},
        pages = {18},
          doi = {10.1088/0004-637X/778/1/18},
archivePrefix = {arXiv},
       eprint = {1308.1687},
 primaryClass = {astro-ph.HE},
       adsurl = {https://ui.adsabs.harvard.edu/abs/2013ApJ...778...18M}
}

@ARTICLE{fan2011ApJ...726...32F,
       author = {{Fan}, Yi-Zhong and {Zhang}, Bib-Bin and {Xu}, Dong and {Liang}, En-Wei and {Zhang}, Bing},
        title = "{XRF 100316D/SN 2010bh: Clue to the Diverse Origin of Nearby Supernova-associated Gamma-ray Bursts}",
      journal = {\apj},
         year = 2011,
        month = jan,
       volume = {726},
       number = {1},
          eid = {32},
        pages = {32},
          doi = {10.1088/0004-637X/726/1/32},
archivePrefix = {arXiv},
       eprint = {1004.5267},
 primaryClass = {astro-ph.HE},
       adsurl = {https://ui.adsabs.harvard.edu/abs/2011ApJ...726...32F}
}

@ARTICLE{izzo2019Natur.565..324I,
       author = {{Izzo}, L. and {de Ugarte Postigo}, A. and {Maeda}, K. and {Th{\"o}ne}, C.~C. and {Kann}, D.~A. and {Della Valle}, M. and {Sagues Carracedo}, A. and {Micha{\l}owski}, M.~J. and {Schady}, P. and {Schmidl}, S. and {Selsing}, J. and {Starling}, R.~L.~C. and {Suzuki}, A. and {Bensch}, K. and {Bolmer}, J. and {Campana}, S. and {Cano}, Z. and {Covino}, S. and {Fynbo}, J.~P.~U. and {Hartmann}, D.~H. and {Heintz}, K.~E. and {Hjorth}, J. and {Japelj}, J. and {Kami{\'n}ski}, K. and {Kaper}, L. and {Kouveliotou}, C. and {Kru{\.Z}y{\'n}ski}, M. and {Kwiatkowski}, T. and {Leloudas}, G. and {Levan}, A.~J. and {Malesani}, D.~B. and {Micha{\l}owski}, T. and {Piranomonte}, S. and {Pugliese}, G. and {Rossi}, A. and {S{\'a}nchez-Ram{\'\i}rez}, R. and {Schulze}, S. and {Steeghs}, D. and {Tanvir}, N.~R. and {Ulaczyk}, K. and {Vergani}, S.~D. and {Wiersema}, K.},
        title = "{Signatures of a jet cocoon in early spectra of a supernova associated with a {\ensuremath{\gamma}}-ray burst}",
      journal = {\nat},
         year = 2019,
        month = jan,
       volume = {565},
       number = {7739},
        pages = {324-327},
          doi = {10.1038/s41586-018-0826-3},
archivePrefix = {arXiv},
       eprint = {1901.05500},
 primaryClass = {astro-ph.HE},
       adsurl = {https://ui.adsabs.harvard.edu/abs/2019Natur.565..324I}
}

@ARTICLE{wang2018ApJ...867..147W,
       author = {{Wang}, J. and {Zhu}, Z.~P. and {Xu}, D. and {Xin}, L.~P. and {Deng}, J.~S. and {Qiu}, Y.~L. and {Qiu}, P. and {Wang}, H.~J. and {Zhang}, J.~B. and {Wei}, J.~Y.},
        title = "{Spectroscopy of the Type Ic Supernova SN 2017iuk Associated with Low-redshift GRB 171205A}",
      journal = {\apj},
         year = 2018,
        month = nov,
       volume = {867},
       number = {2},
          eid = {147},
        pages = {147},
          doi = {10.3847/1538-4357/aae6c3},
archivePrefix = {arXiv},
       eprint = {1810.03250},
 primaryClass = {astro-ph.HE},
       adsurl = {https://ui.adsabs.harvard.edu/abs/2018ApJ...867..147W}
}

@ARTICLE{urata2019ApJ...884L..58U,
       author = {{Urata}, Yuji and {Toma}, Kenji and {Huang}, Kuiyun and {Asada}, Keiichi and {Nagai}, Hiroshi and {Takahashi}, Satoko and {Petitpas}, Glen and {Tashiro}, Makoto and {Yamaoka}, Kazutaka},
        title = "{First Detection of Radio Linear Polarization in a Gamma-Ray Burst Afterglow}",
      journal = {\apjl},
         year = 2019,
        month = oct,
       volume = {884},
       number = {2},
          eid = {L58},
        pages = {L58},
          doi = {10.3847/2041-8213/ab48f3},
archivePrefix = {arXiv},
       eprint = {1904.08111},
 primaryClass = {astro-ph.HE},
       adsurl = {https://ui.adsabs.harvard.edu/abs/2019ApJ...884L..58U}
}

@ARTICLE{waxman2007ApJ...667..351W,
       author = {{Waxman}, E. and {M{\'e}sz{\'a}ros}, P. and {Campana}, S.},
        title = "{GRB 060218: A Relativistic Supernova Shock Breakout}",
      journal = {\apj},
         year = 2007,
        month = sep,
       volume = {667},
       number = {1},
        pages = {351-357},
          doi = {10.1086/520715},
archivePrefix = {arXiv},
       eprint = {astro-ph/0702450},
 primaryClass = {astro-ph},
       adsurl = {https://ui.adsabs.harvard.edu/abs/2007ApJ...667..351W}
}

@ARTICLE{metzger2011MNRAS.413.2031M,
       author = {{Metzger}, B.~D. and {Giannios}, D. and {Thompson}, T.~A. and {Bucciantini}, N. and {Quataert}, E.},
        title = "{The protomagnetar model for gamma-ray bursts}",
      journal = {\mnras},
         year = 2011,
        month = may,
       volume = {413},
       number = {3},
        pages = {2031-2056},
          doi = {10.1111/j.1365-2966.2011.18280.x},
archivePrefix = {arXiv},
       eprint = {1012.0001},
 primaryClass = {astro-ph.HE},
       adsurl = {https://ui.adsabs.harvard.edu/abs/2011MNRAS.413.2031M}
}

@ARTICLE{rastinejad2022Natur.612..223R,
       author = {{Rastinejad}, Jillian C. and {Gompertz}, Benjamin P. and {Levan}, Andrew J. and {Fong}, Wen-fai and {Nicholl}, Matt and {Lamb}, Gavin P. and {Malesani}, Daniele B. and {Nugent}, Anya E. and {Oates}, Samantha R. and {Tanvir}, Nial R. and {de Ugarte Postigo}, Antonio and {Kilpatrick}, Charles D. and {Moore}, Christopher J. and {Metzger}, Brian D. and {Ravasio}, Maria Edvige and {Rossi}, Andrea and {Schroeder}, Genevieve and {Jencson}, Jacob and {Sand}, David J. and {Smith}, Nathan and {Ag{\"u}{\'\i} Fern{\'a}ndez}, Jos{\'e} Feliciano and {Berger}, Edo and {Blanchard}, Peter K. and {Chornock}, Ryan and {Cobb}, Bethany E. and {De Pasquale}, Massimiliano and {Fynbo}, Johan P.~U. and {Izzo}, Luca and {Kann}, D. Alexander and {Laskar}, Tanmoy and {Marini}, Ester and {Paterson}, Kerry and {Escorial}, Alicia Rouco and {Sears}, Huei M. and {Th{\"o}ne}, Christina C.},
        title = "{A kilonova following a long-duration gamma-ray burst at 350 Mpc}",
      journal = {\nat},
         year = 2022,
        month = dec,
       volume = {612},
       number = {7939},
        pages = {223-227},
          doi = {10.1038/s41586-022-05390-w},
archivePrefix = {arXiv},
       eprint = {2204.10864},
 primaryClass = {astro-ph.HE},
       adsurl = {https://ui.adsabs.harvard.edu/abs/2022Natur.612..223R}
}

@ARTICLE{troja2022Natur.612..228T,
       author = {{Troja}, E. and {Fryer}, C.~L. and {O'Connor}, B. and {Ryan}, G. and {Dichiara}, S. and {Kumar}, A. and {Ito}, N. and {Gupta}, R. and {Wollaeger}, R.~T. and {Norris}, J.~P. and {Kawai}, N. and {Butler}, N.~R. and {Aryan}, A. and {Misra}, K. and {Hosokawa}, R. and {Murata}, K.~L. and {Niwano}, M. and {Pandey}, S.~B. and {Kutyrev}, A. and {van Eerten}, H.~J. and {Chase}, E.~A. and {Hu}, Y.-D. and {Caballero-Garcia}, M.~D. and {Castro-Tirado}, A.~J.},
        title = "{A nearby long gamma-ray burst from a merger of compact objects}",
      journal = {\nat},
         year = 2022,
        month = dec,
       volume = {612},
       number = {7939},
        pages = {228-231},
          doi = {10.1038/s41586-022-05327-3},
archivePrefix = {arXiv},
       eprint = {2209.03363},
 primaryClass = {astro-ph.HE},
       adsurl = {https://ui.adsabs.harvard.edu/abs/2022Natur.612..228T}
}

@ARTICLE{yang2022Natur.612..232Y,
       author = {{Yang}, Jun and {Ai}, Shunke and {Zhang}, Bin-Bin and {Zhang}, Bing and {Liu}, Zi-Ke and {Wang}, Xiangyu Ivy and {Yang}, Yu-Han and {Yin}, Yi-Han and {Li}, Ye and {L{\"u}}, Hou-Jun},
        title = "{A long-duration gamma-ray burst with a peculiar origin}",
      journal = {\nat},
         year = 2022,
        month = dec,
       volume = {612},
       number = {7939},
        pages = {232-235},
          doi = {10.1038/s41586-022-05403-8},
archivePrefix = {arXiv},
       eprint = {2204.12771},
 primaryClass = {astro-ph.HE},
       adsurl = {https://ui.adsabs.harvard.edu/abs/2022Natur.612..232Y}
}

@ARTICLE{gompertz2023NatAs...7...67G,
       author = {{Gompertz}, Benjamin P. and {Ravasio}, Maria Edvige and {Nicholl}, Matt and {Levan}, Andrew J. and {Metzger}, Brian D. and {Oates}, Samantha R. and {Lamb}, Gavin P. and {Fong}, Wen-fai and {Malesani}, Daniele B. and {Rastinejad}, Jillian C. and {Tanvir}, Nial R. and {Evans}, Philip A. and {Jonker}, Peter G. and {Page}, Kim L. and {Pe'er}, Asaf},
        title = "{The case for a minute-long merger-driven gamma-ray burst from fast-cooling synchrotron emission}",
      journal = {Nature Astronomy},
         year = 2023,
        month = jan,
       volume = {7},
        pages = {67-79},
          doi = {10.1038/s41550-022-01819-4},
archivePrefix = {arXiv},
       eprint = {2205.05008},
 primaryClass = {astro-ph.HE},
       adsurl = {https://ui.adsabs.harvard.edu/abs/2023NatAs...7...67G}
}

@ARTICLE{levan2024Natur.626..737L,
       author = {{Levan}, Andrew J. and {Gompertz}, Benjamin P. and {Salafia}, Om Sharan and {Bulla}, Mattia and {Burns}, Eric and {Hotokezaka}, Kenta and {Izzo}, Luca and {Lamb}, Gavin P. and {Malesani}, Daniele B. and {Oates}, Samantha R. and {Ravasio}, Maria Edvige and {Rouco Escorial}, Alicia and {Schneider}, Benjamin and {Sarin}, Nikhil and {Schulze}, Steve and {Tanvir}, Nial R. and {Ackley}, Kendall and {Anderson}, Gemma and {Brammer}, Gabriel B. and {Christensen}, Lise and {Dhillon}, Vikram S. and {Evans}, Phil A. and {Fausnaugh}, Michael and {Fong}, Wen-fai and {Fruchter}, Andrew S. and {Fryer}, Chris and {Fynbo}, Johan P.~U. and {Gaspari}, Nicola and {Heintz}, Kasper E. and {Hjorth}, Jens and {Kennea}, Jamie A. and {Kennedy}, Mark R. and {Laskar}, Tanmoy and {Leloudas}, Giorgos and {Mandel}, Ilya and {Martin-Carrillo}, Antonio and {Metzger}, Brian D. and {Nicholl}, Matt and {Nugent}, Anya and {Palmerio}, Jesse T. and {Pugliese}, Giovanna and {Rastinejad}, Jillian and {Rhodes}, Lauren and {Rossi}, Andrea and {Saccardi}, Andrea and {Smartt}, Stephen J. and {Stevance}, Heloise F. and {Tohuvavohu}, Aaron and {van der Horst}, Alexander and {Vergani}, Susanna D. and {Watson}, Darach and {Barclay}, Thomas and {Bhirombhakdi}, Kornpob and {Breedt}, Elm{\'e} and {Breeveld}, Alice A. and {Brown}, Alexander J. and {Campana}, Sergio and {Chrimes}, Ashley A. and {D'Avanzo}, Paolo and {D'Elia}, Valerio and {De Pasquale}, Massimiliano and {Dyer}, Martin J. and {Galloway}, Duncan K. and {Garbutt}, James A. and {Green}, Matthew J. and {Hartmann}, Dieter H. and {Jakobsson}, P{\'a}ll and {Kerry}, Paul and {Kouveliotou}, Chryssa and {Langeroodi}, Danial and {Le Floc'h}, Emeric and {Leung}, James K. and {Littlefair}, Stuart P. and {Munday}, James and {O'Brien}, Paul and {Parsons}, Steven G. and {Pelisoli}, Ingrid and {Sahman}, David I. and {Salvaterra}, Ruben and {Sbarufatti}, Boris and {Steeghs}, Danny and {Tagliaferri}, Gianpiero and {Th{\"o}ne}, Christina C. and {de Ugarte Postigo}, Antonio and {Kann}, David Alexander},
        title = "{Heavy-element production in a compact object merger observed by JWST}",
      journal = {\nat},
         year = 2024,
        month = feb,
       volume = {626},
       number = {8000},
        pages = {737-741},
          doi = {10.1038/s41586-023-06759-1},
archivePrefix = {arXiv},
       eprint = {2307.02098},
 primaryClass = {astro-ph.HE},
       adsurl = {https://ui.adsabs.harvard.edu/abs/2024Natur.626..737L}
}

@ARTICLE{yang2024Natur.626..742Y,
       author = {{Yang}, Yu-Han and {Troja}, Eleonora and {O'Connor}, Brendan and {Fryer}, Chris L. and {Im}, Myungshin and {Durbak}, Joe and {Paek}, Gregory S.~H. and {Ricci}, Roberto and {Bom}, Cl{\'e}cio R. and {Gillanders}, James H. and {Castro-Tirado}, Alberto J. and {Peng}, Zong-Kai and {Dichiara}, Simone and {Ryan}, Geoffrey and {van Eerten}, Hendrik and {Dai}, Zi-Gao and {Chang}, Seo-Won and {Choi}, Hyeonho and {De}, Kishalay and {Hu}, Youdong and {Kilpatrick}, Charles D. and {Kutyrev}, Alexander and {Jeong}, Mankeun and {Lee}, Chung-Uk and {Makler}, Martin and {Navarete}, Felipe and {P{\'e}rez-Garc{\'\i}a}, Ignacio},
        title = "{A lanthanide-rich kilonova in the aftermath of a long gamma-ray burst}",
      journal = {\nat},
         year = 2024,
        month = feb,
       volume = {626},
       number = {8000},
        pages = {742-745},
          doi = {10.1038/s41586-023-06979-5},
archivePrefix = {arXiv},
       eprint = {2308.00638},
 primaryClass = {astro-ph.HE},
       adsurl = {https://ui.adsabs.harvard.edu/abs/2024Natur.626..742Y}
}

@ARTICLE{blanchard2024NatAs...8..774B,
       author = {{Blanchard}, Peter K. and {Villar}, V. Ashley and {Chornock}, Ryan and {Laskar}, Tanmoy and {Li}, Yijia and {Leja}, Joel and {Pierel}, Justin and {Berger}, Edo and {Margutti}, Raffaella and {Alexander}, Kate D. and {Barnes}, Jennifer and {Cendes}, Yvette and {Eftekhari}, Tarraneh and {Kasen}, Daniel and {LeBaron}, Natalie and {Metzger}, Brian D. and {Muzerolle Page}, James and {Rest}, Armin and {Sears}, Huei and {Siegel}, Daniel M. and {Yadavalli}, S. Karthik},
        title = "{JWST detection of a supernova associated with GRB 221009A without an r-process signature}",
      journal = {Nature Astronomy},
         year = 2024,
        month = jun,
       volume = {8},
        pages = {774-785},
          doi = {10.1038/s41550-024-02237-4},
archivePrefix = {arXiv},
       eprint = {2308.14197},
 primaryClass = {astro-ph.HE},
       adsurl = {https://ui.adsabs.harvard.edu/abs/2024NatAs...8..774B}
}

@ARTICLE{levan2023ApJ...946L..28L,
       author = {{Levan}, A.~J. and {Lamb}, G.~P. and {Schneider}, B. and {Hjorth}, J. and {Zafar}, T. and {de Ugarte Postigo}, A. and {Sargent}, B. and {Mullally}, S.~E. and {Izzo}, L. and {D'Avanzo}, P. and {Burns}, E. and {Ag{\"u}{\'\i} Fern{\'a}ndez}, J.~F. and {Barclay}, T. and {Bernardini}, M.~G. and {Bhirombhakdi}, K. and {Bremer}, M. and {Brivio}, R. and {Campana}, S. and {Chrimes}, A.~A. and {D'Elia}, V. and {Della Valle}, M. and {De Pasquale}, M. and {Ferro}, M. and {Fong}, W. and {Fruchter}, A.~S. and {Fynbo}, J.~P.~U. and {Gaspari}, N. and {Gompertz}, B.~P. and {Hartmann}, D.~H. and {Hedges}, C.~L. and {Heintz}, K.~E. and {Hotokezaka}, K. and {Jakobsson}, P. and {Kann}, D.~A. and {Kennea}, J.~A. and {Laskar}, T. and {Le Floc'h}, E. and {Malesani}, D.~B. and {Melandri}, A. and {Metzger}, B.~D. and {Oates}, S.~R. and {Pian}, E. and {Piranomonte}, S. and {Pugliese}, G. and {Racusin}, J.~L. and {Rastinejad}, J.~C. and {Ravasio}, M.~E. and {Rossi}, A. and {Saccardi}, A. and {Salvaterra}, R. and {Sbarufatti}, B. and {Starling}, R.~L.~C. and {Tanvir}, N.~R. and {Th{\"o}ne}, C.~C. and {van der Horst}, A.~J. and {Vergani}, S.~D. and {Watson}, D. and {Wiersema}, K. and {Wijers}, R.~A.~M.~J. and {Xu}, Dong},
        title = "{The First JWST Spectrum of a GRB Afterglow: No Bright Supernova in Observations of the Brightest GRB of all Time, GRB 221009A}",
      journal = {\apjl},
         year = 2023,
        month = mar,
       volume = {946},
       number = {1},
          eid = {L28},
        pages = {L28},
          doi = {10.3847/2041-8213/acc2c1},
archivePrefix = {arXiv},
       eprint = {2302.07761},
 primaryClass = {astro-ph.HE},
       adsurl = {https://ui.adsabs.harvard.edu/abs/2023ApJ...946L..28L}
}

@ARTICLE{fulton2023ApJ...946L..22F,
       author = {{Fulton}, M.~D. and {Smartt}, S.~J. and {Rhodes}, L. and {Huber}, M.~E. and {Villar}, V.~A. and {Moore}, T. and {Srivastav}, S. and {Schultz}, A.~S.~B. and {Chambers}, K.~C. and {Izzo}, L. and {Hjorth}, J. and {Chen}, T.-W. and {Nicholl}, M. and {Foley}, R.~J. and {Rest}, A. and {Smith}, K.~W. and {Young}, D.~R. and {Sim}, S.~A. and {Bright}, J. and {Zenati}, Y. and {de Boer}, T. and {Bulger}, J. and {Fairlamb}, J. and {Gao}, H. and {Lin}, C.-C. and {Lowe}, T. and {Magnier}, E.~A. and {Smith}, I.~A. and {Wainscoat}, R. and {Coulter}, D.~A. and {Jones}, D.~O. and {Kilpatrick}, C.~D. and {McGill}, P. and {Ramirez-Ruiz}, E. and {Lee}, K.-S. and {Narayan}, G. and {Ramakrishnan}, V. and {Ridden-Harper}, R. and {Singh}, A. and {Wang}, Q. and {Kong}, A.~K.~H. and {Ngeow}, C.-C. and {Pan}, Y.-C. and {Yang}, S. and {Davis}, K.~W. and {Piro}, A.~L. and {Rojas-Bravo}, C. and {Sommer}, J. and {Yadavalli}, S.~K.},
        title = "{The Optical Light Curve of GRB 221009A: The Afterglow and the Emerging Supernova}",
      journal = {\apjl},
         year = 2023,
        month = mar,
       volume = {946},
       number = {1},
          eid = {L22},
        pages = {L22},
          doi = {10.3847/2041-8213/acc101},
archivePrefix = {arXiv},
       eprint = {2301.11170},
 primaryClass = {astro-ph.HE},
       adsurl = {https://ui.adsabs.harvard.edu/abs/2023ApJ...946L..22F}
}

@ARTICLE{rehemtulla2024ApJ...972....7R,
       author = {{Rehemtulla}, Nabeel and {Miller}, Adam A. and {Jegou Du Laz}, Theophile and {Coughlin}, Michael W. and {Fremling}, Christoffer and {Perley}, Daniel A. and {Qin}, Yu-Jing and {Sollerman}, Jesper and {Mahabal}, Ashish A. and {Laher}, Russ R. and {Riddle}, Reed and {Rusholme}, Ben and {Kulkarni}, Shrinivas R.},
        title = "{The Zwicky Transient Facility Bright Transient Survey. III. BTSbot: Automated Identification and Follow-up of Bright Transients with Deep Learning}",
      journal = {\apj},
         year = 2024,
        month = sep,
       volume = {972},
       number = {1},
          eid = {7},
        pages = {7},
          doi = {10.3847/1538-4357/ad5666},
archivePrefix = {arXiv},
       eprint = {2401.15167},
 primaryClass = {astro-ph.IM},
       adsurl = {https://ui.adsabs.harvard.edu/abs/2024ApJ...972....7R}
}

@ARTICLE{clocchiatti2011AJ....141..163C,
       author = {{Clocchiatti}, Alejandro and {Suntzeff}, Nicholas B. and {Covarrubias}, Ricardo and {Candia}, Pablo},
        title = "{The Ultimate Light Curve of SN 1998bw/GRB 980425}",
      journal = {\aj},
         year = 2011,
        month = may,
       volume = {141},
       number = {5},
          eid = {163},
        pages = {163},
          doi = {10.1088/0004-6256/141/5/163},
archivePrefix = {arXiv},
       eprint = {1106.1695},
 primaryClass = {astro-ph.HE},
       adsurl = {https://ui.adsabs.harvard.edu/abs/2011AJ....141..163C}
}
\bibliographystyle{aasjournalv7}

\appendix

\begin{table}
\centering
\caption{Optical spectroscopic observations of GRB\,260310A/AT\,2026fgk. $\Delta t$ is measured from the \emph{Fermi}/GBM trigger (MJD~61109.21) in the rest frame. The first two Gemini/GMOS-N epochs (2026 Apr 05 and Apr 20) were obtained with the slit at PA~$= 47^\circ$ to place both the transient and host simultaneously on the longslit; this PA lay $\approx120^\circ$ from the parallactic angle at the time of each observation (the $q \approx 169^\circ$ and $164^\circ$ respectively), causing atmospheric differential refraction (ADR) to act predominantly perpendicular to the slit and producing significant, wavelength-dependent slit losses that suppress the blue continuum. }
\label{tab:spectra}
\begin{tabular}{cccc}\hline\hline
Date (UT) & MJD & $\Delta t$\,d & Instrument \\\hline
2026 Mar 12 & 61111.25 & $+1.78$ & P60/SEDM  \\
2026 Mar 12 & 61111.40 & $1.90$  & P200/NGPS     \\
2026 Mar 13 & 61112.50 & $2.86$  & P200/NGPS     \\
2026 Mar 14 & 61113.52 & $3.75$  & P200/NGPS     \\
2026 Mar 16 & 61115.47 & $5.43$  & DCT/DeVeny    \\
2026 Apr 05 & 61135.55 & $22.85$ & Gemini/GMOS-N$^{a}$  \\
2026 Apr 20 & 61150.50 & $35.81$ & Gemini/GMOS-N$^{a}$ \\
2026 Apr 23 & 61153.50 & $38.41$ & Gemini/GMOS-N$^{a}$  \\\hline
\end{tabular}

\tablenotetext{a}{Slit oriented at PA~$= 47^\circ$ (for $\Delta t=22.85$\,d, and $\Delta t=35.81$\,d), $\approx\!120\arcdeg$ from parallactic angle. We apply an ADR throughput correction to account for the differential slit loses suffered (Appendix~\ref{asec:gmos_adr_corr}).}
\end{table}

\section{GMOS ADR Correction}
\label{asec:gmos_adr_corr}

To correct for the residual slit losses, we compute a multiplicative throughput correct $C(\lambda) = T(\lambda_{\rm ref})/T(\lambda)$ for each spectrum where $T(\lambda)$ is the fraction of a Gaussian PSF transmitted by a 1\arcsec\ slit when the centroid is placed by the cross slit ADR offset. The wavelength dependent offset is computed from \citet{edlen1966Metro...2...71E} refractive index formula using pressure and temperature measured by the Gemini summit weather station at the time of each observation. The wavelength dependent PSF size is scaled from the median seeing measured by the Mauna Kea Weather Center DIMM monitor during each exposure. The correction is normalized to unity at 6200\AA (the central wavelength), so that the flux at the reference wavelength is preserved and the corrections are largest at the blue end. The corrected spectra are presented in this work, with both the corrected and raw spectra available.

\subsection{Cross-Calibration between X-ray observatories}
\label{asec:xray_calib}
The photon indices and fluxes we measure from EP/FXT, NuSTAR, and Chandra observations show differences that are larger than their statistical uncertainties (Table~\ref{tab:xray}, most strikingly between the simultaneous EP/FXT \& Chandra observations at $\sim$41 days ($\Gamma_{\mathrm{FXT}} = 1.31 \pm 0.13$ versus $\Gamma_{\mathrm{CXO}} = 2.06 \pm 0.24$). While rapid spectral evolution cannot be fully excluded, systematic cross-calibration differences between X-ray observatories are well-documented and likely contribute to these discrepancies. Studies by the International Astronomical Consortium for High-Energy Calibration \citep[IACHEC; ][]{ishida_iachec, madsen_iachec} have shown that photon indices measured by different observatories can disagree by upto $\sim$10\%, and fluxes by $\sim$15\%, even for simultaneous observations of the same source. These systematic differences persist across missions and arise from the uncertainties in the effective area calibration --- particularly at soft energies \citep{nevalainen_calibration}. Furthermore, EP is a recently launched mission \citep{yuan_ep_mission} and its cross-calibration with observatories like Chandra and NuSTAR is currently being characterized (EP collaboration, \textit{in-prep}). We therefore caution that the spectral parameters measured by different instruments should not be directly compared without accounting for these systematic uncertainties. 

\section{SED \& Host Extinction}
\label{asec:sed_host}

Galactic extinction corrected data points pre-break at 7.70\,d were moved to rest frame (correcting for redshift effects following \cite{Hogg2002}). The first data points showing the rise of the afterglow were discarded; so were data points post-break, to minimize supernova contribution. The analysis only uses standard optical and near-infrared broadband filters, discarding broader filter sets. The light curve was fitted with the equation $\log_{10}(L) = \alpha log_{10} t + c$, where $L$ is the luminosity, $\alpha$ the spectral index, and $c$ is the normalization constant for each filter. The optical and NIR spectral energy distribution (SED) of the afterglow was calculated at 3 days, to maximize filter coverage.

\begin{figure}
    \centering
    \includegraphics[width=0.8\linewidth]{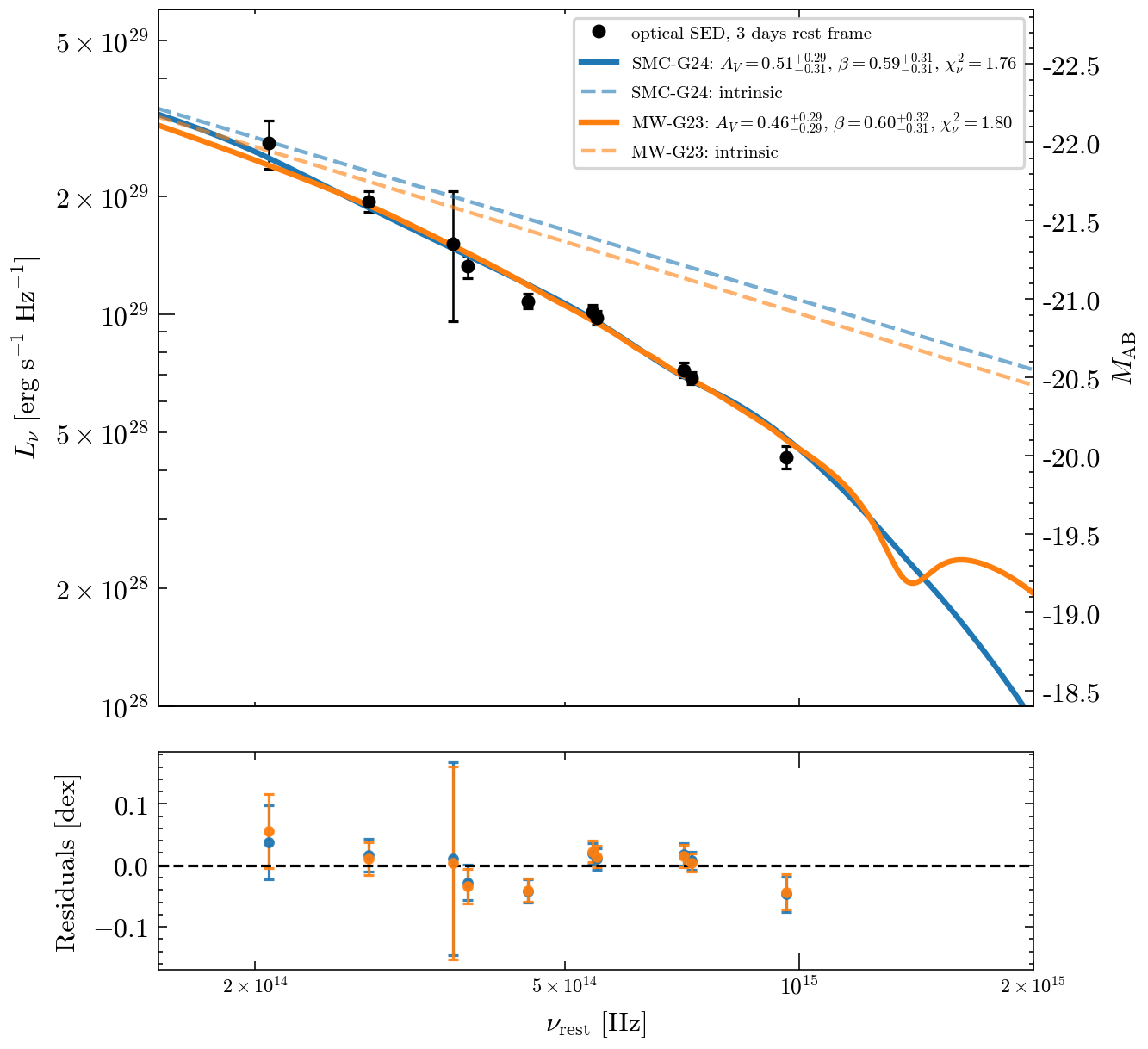}
    \caption{Spectral energy distribution at 3 days post GRB trigger plot fitted with Milky-Way and SMC-like host extinction models from \cite{Gordon23, Gordon_2024_SMC} (solid line), also showing the intrinsic color index $\beta$ for each fit (dashed line).}
    \label{fig:sed_hostext}
\end{figure}

The host extinction models used were the \cite{Gordon23} (MW-G23), a universal, Milky Way-like, total-to-selective extinction ratio $R_V$ dependent model, as well as the \cite{Gordon_2024_SMC} average Small Magellanic Cloud (SMC-G24) extinction model. For the MW-G23 model, we have chosen to fit the data with the total-to-selective extinction ratio $R_V$ set to 3.0, chosen arbitrarily; the SMC-G24 model is not $R_V$ dependent and was calculated at $R_V = 3.02$. The lack of photometric mid and far-UV coverage would make distinguishing between various extinction models and $R_V$ values difficult - the SMC and MW models vary the most in e.g. the far-UV rise or $2175 \AA$ bump - as well as help disentangle the degeneracy in the fit between $A_V$ and $\beta$. The models were accessed via the \texttt{dust\_extinction} package \citep{Gordon2024}. Spectral index $\beta$ and total wavelength-dependent extinction along line-of-sight $A_{\lambda}$ were fitted jointly to the SED data, as shown in Figure~\ref{fig:sed_hostext}.

For the SMC model, $A_V = 0.51^{+0.29}_{-0.30}$, $\beta = 0.59 ^{+0.31}_{-0.31}$, $\chi_R^2 = 1.76$, while the Milky Way-like model produces, for the $R_V = 3.0$ fit gives $A_V = 0.49^{+0.29}_{-0.29}$, $\beta = 0.60^{+0.31}_{-0.31}$, with reduced $\chi^2 = 1.80$. From these results we expect the $A_V$ to be around 0.5 mag, although we warn that it is a value highly degenerate with $\beta$ and $R_V$. We can place a $1\sigma$ upper limit on the extinction of $A_V < 0.80$. The results show that there is non-zero extinction present along the line-of-sight, possibly impacting the supernova search.

\begin{figure}
    \centering
    \includegraphics[width=0.8\linewidth]{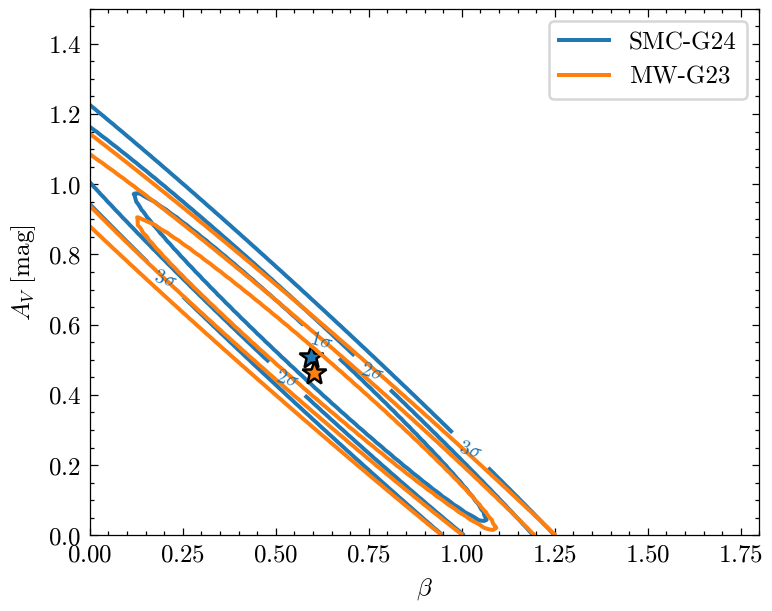}
    \caption{$\chi$ contour plots for MW-G24 and SMC-G23 fits to SED data at three days, illustrating the high degeneracy of $A_V$ and $\beta$ in the fit.}
    \label{fig:sed_hostext_contour}
\end{figure}

\begin{figure}
    \centering
    \includegraphics[width=1\linewidth]{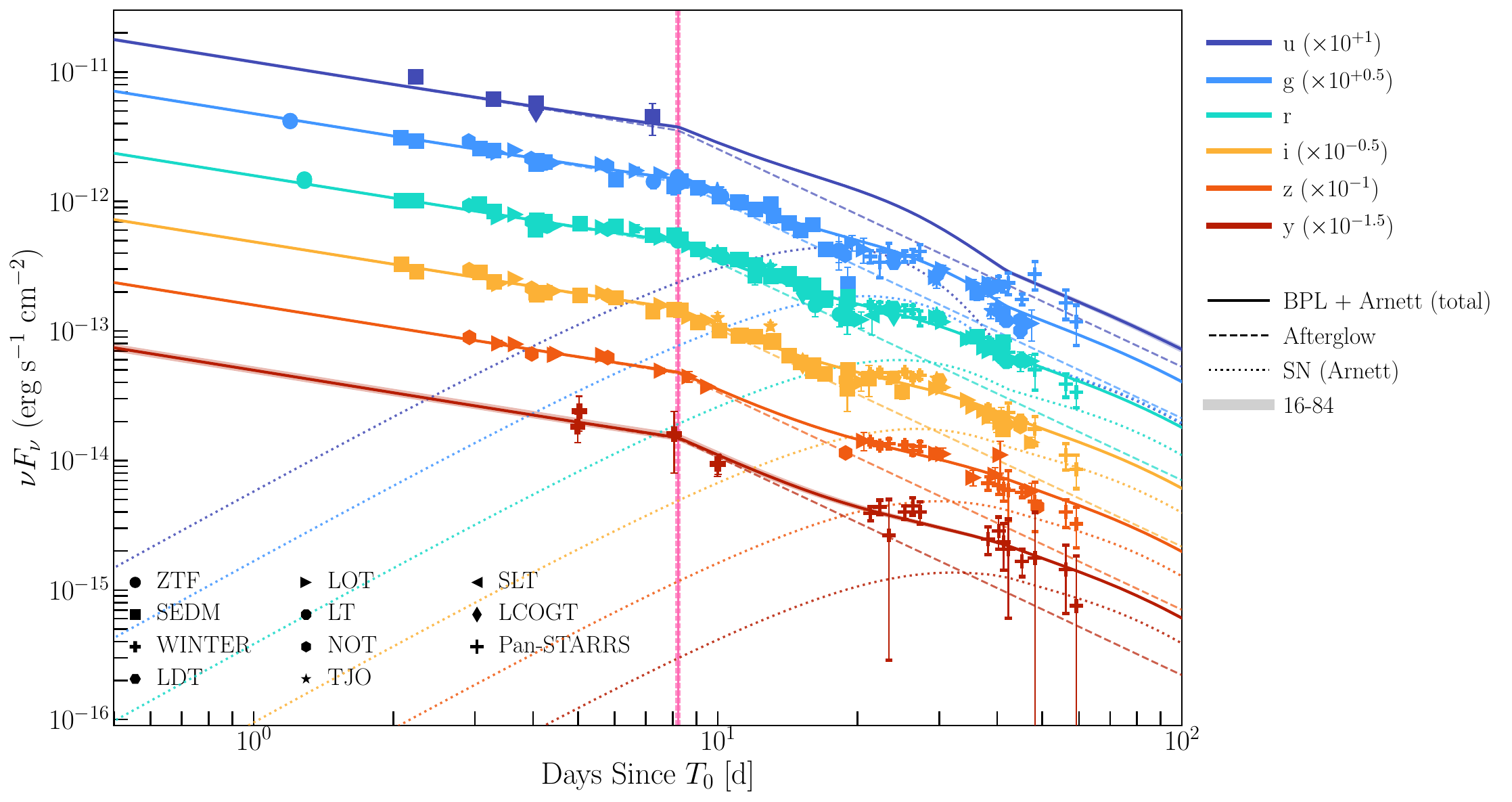}
    \caption{Multi-band light curves of \thisgrb and its associated SN. Each band ($ugrizy$) is shown with a constant offset. Curves are the posterior median of the joint smoothly-broken power-law afterglow + Arnett fit \citep{arnett1982ApJ...253..785A} described in Section~\ref{sec:nimass}. \emph{Solid} line is the total (smoothly-broken power-law + Arnett), \emph{dashed} is the afterglow only, and \emph{dotted} is the SN only---16--84\% posterior credible interval is shown in the shaded envelope. The pink vertical line marks the jet break with its 16--84\% credible interval.}
    \label{fig:bpl_arnett}
\end{figure}

\begin{figure*}
    \centering
    \includegraphics[width=1\linewidth]{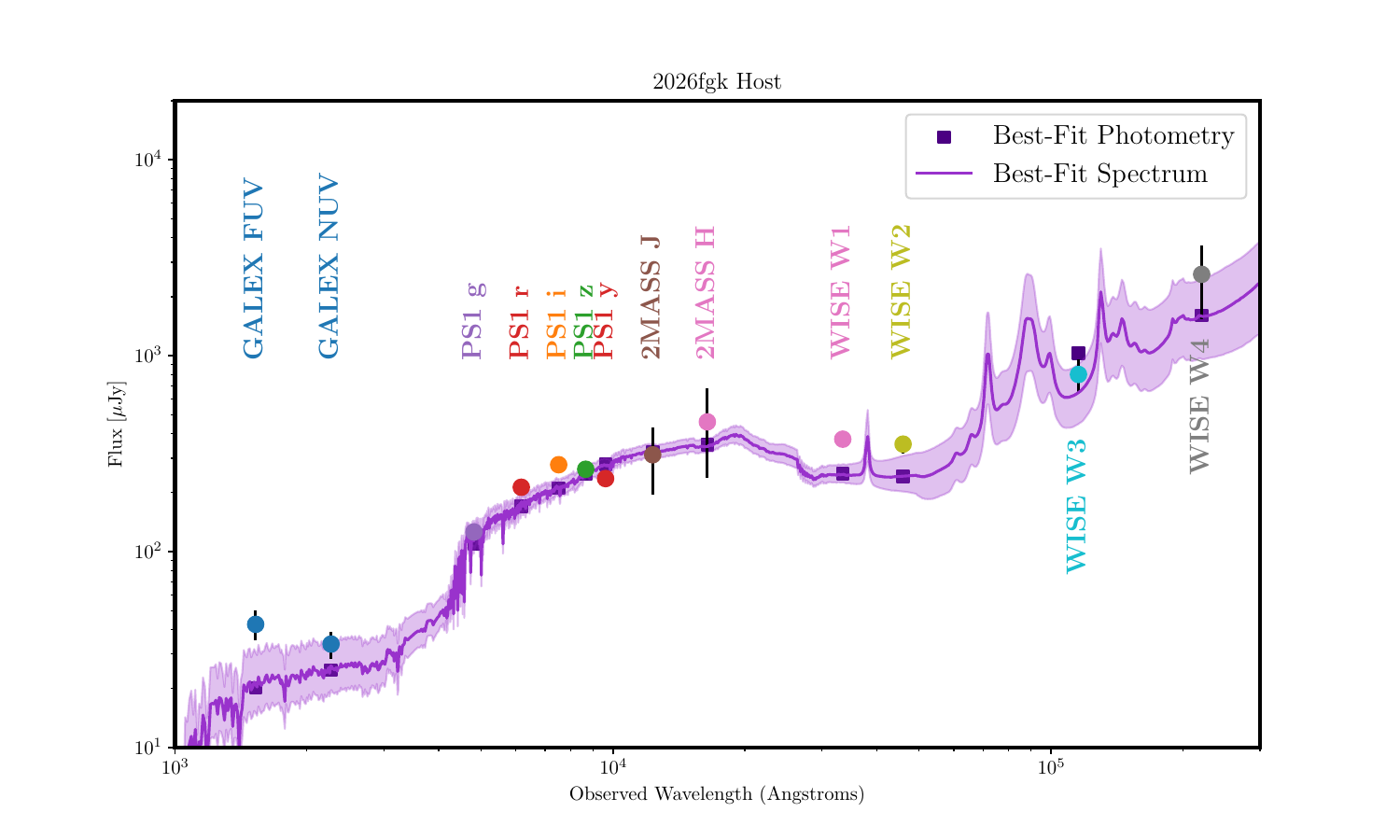}
    \caption{Spectral energy distribution of the \thisgrb host galaxy produced from SED fitting by \textsc{FrankenBlast}. Colored circles show observed photometry from GALEX, PS1, 2MASS, and WISE; filled squares are the best-fit photometry; the solid line and shaded region show the best-fit Prospector spectrum and it's 1$\sigma$ uncertainty.}
    \label{fig:sed1}
\end{figure*}

\begin{figure*}
    \centering
    \includegraphics[width=1\linewidth]{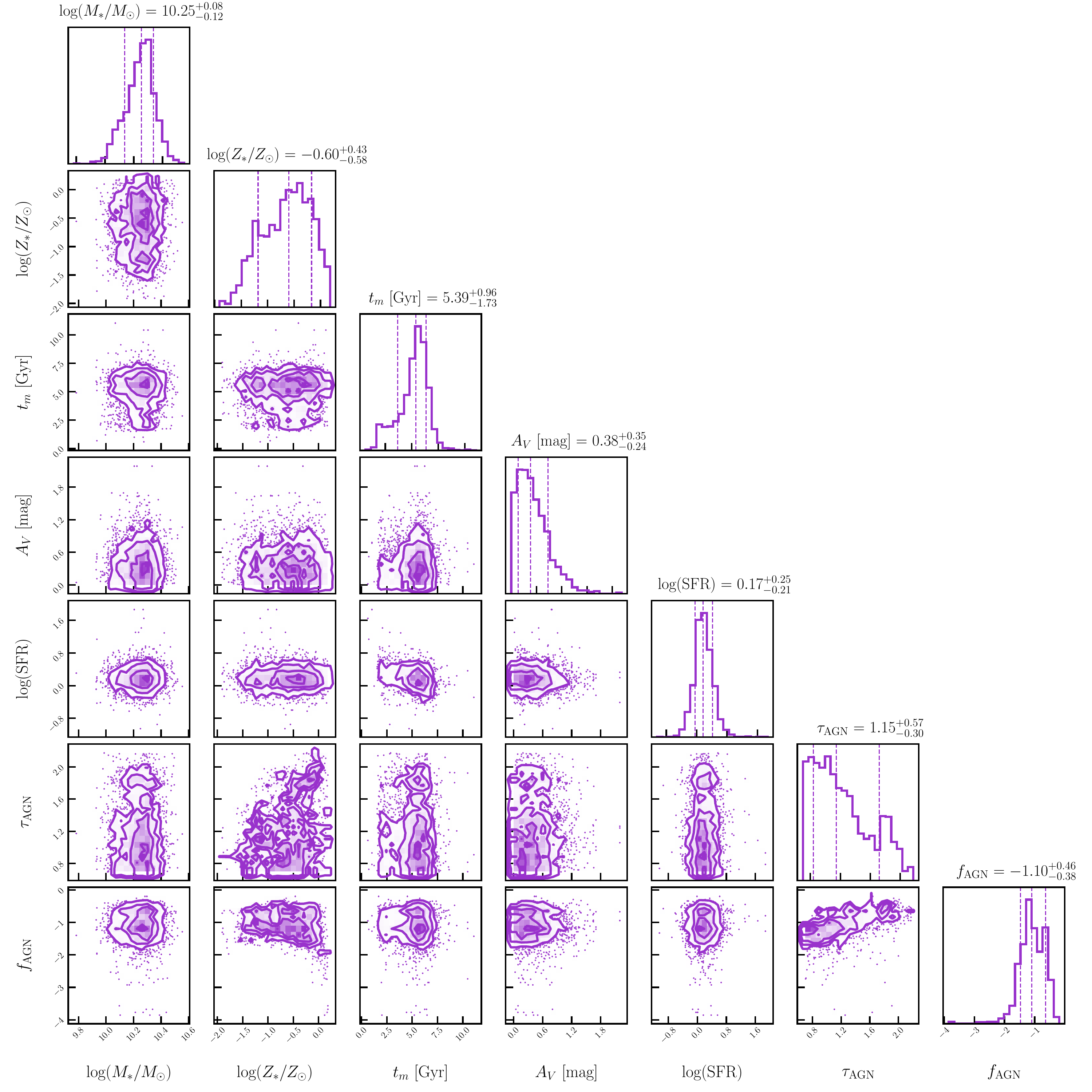}
    \caption{Posterior distributions from \textsc{FrankenBlast} SED fitting of the \thisgrb host galaxy. The fit uses photometry spanning GALEX FUV/NUV through WISE W4---see Section~\ref{sec:host_properties}. See \citet{nugent2026ApJ...997...38N} for further details.}
    \label{fig:sed2}
\end{figure*}

\clearpage
\newpage

\startlongtable
\begin{deluxetable*}{ccccc}
\tablecaption{Radio observations of GRB\,260310A from VLA, NOEMA, OVRO, Effelsberg and SMA. We include the first GCN reported radio measurement from AMI-LA. Phase reported in observer frame. \label{tab:radio}\\ $^{\rm a}$First observation from \citet{GCN.44005} $\cdot$ $^{\rm b}$No 2cm---receiver problems $\cdot$ $^{\rm c}$3 $\sigma$ upper limit.}
\tablehead{
  \colhead{MJD} & \colhead{$\Delta t$ (d)} & \colhead{$\nu$ (GHz)} &
  \colhead{Instrument} & \colhead{$F_\nu$ ($\mu$Jy)} 
}
\startdata
61113.14 & 3.92 & 15.2 & AMI-LA & 9000 $\pm$ 500$^{\rm a}$ \\
61113.18 & 3.97 & 19.0 & VLA & 10804 $\pm$ 174 \\
61113.18 & 3.97 & 21.0 & VLA & 11644 $\pm$ 206 \\
61113.18 & 3.97 & 23.0 & VLA & 12363 $\pm$ 222 \\
61113.18 & 3.97 & 25.0 & VLA & 12956 $\pm$ 264 \\
61113.20 & 3.99 & 13.0 & VLA & 7694 $\pm$ 80 \\
61113.20 & 3.99 & 15.0 & VLA & 8773 $\pm$ 96 \\
61113.20 & 3.99 & 17.0 & VLA & 9729 $\pm$ 131 \\
61113.22 & 4.01 & 9.0 & VLA & 5281 $\pm$ 40 \\
61113.22 & 4.01 & 11.0 & VLA & 6534 $\pm$ 47 \\
61113.23 & 4.02 & 7.0 & VLA & 3922 $\pm$ 29 \\
61113.23 & 4.02 & 5.0 & VLA & 2223 $\pm$ 26 \\
61113.47 & 4.25 & 6.0 & VLA & 3894 $\pm$ 8 \\
61113.47 & 4.25 & 10.0 & VLA & 6368 $\pm$ 13 \\
61113.47 & 4.25 & 15.0 & VLA & 9104 $\pm$ 29 \\
61116.55 & 7.34 & 15.0 & OVRO & 20600 $\pm$ 1500 \\
61116.55 & 7.34 & 15.0 & OVRO & 18100 $\pm$ 1600 \\
61117.19 & 7.98 & 10.45 & Effelsberg & 12000 $\pm$ 1000 \\
61117.24 & 8.03 & 14.25 & Effelsberg & 14000 $\pm$ 1000 \\
61117.24 & 8.03 & 16.75 & Effelsberg & 19000 $\pm$ 2000 \\
61118.68 & 9.47 & 19.25 & Effelsberg & 19000 $\pm$ 3000 \\
61118.68 & 9.47 & 24.75 & Effelsberg & 15000 $\pm$ 3000 \\
61118.95 & 9.74 & 231.744 & NOEMA & 13500 $\pm$ 70 \\
61118.95 & 9.74 & 101.744 & NOEMA & 16630 $\pm$ 20 \\
61118.95 & 9.74 & 216.256 & NOEMA & 13500 $\pm$ 60 \\
61118.95 & 9.74 & 86.256 & NOEMA & 17310 $\pm$ 20 \\
61119.30 & 10.09 & 15.0 & OVRO & 23000 $\pm$ 1500 \\
61119.35 & 10.14 & 151.744 & NOEMA & 14590 $\pm$ 50 \\
61119.35 & 10.14 & 136.256 & NOEMA & 14980 $\pm$ 50 \\
61123.33 & 14.12 & 231.744 & NOEMA & 6540 $\pm$ 90 \\
61123.33 & 14.12 & 86.256 & NOEMA & 11180 $\pm$ 20 \\
61123.33 & 14.12 & 101.744 & NOEMA & 10090 $\pm$ 20 \\
61123.33 & 14.12 & 216.256 & NOEMA & 6730 $\pm$ 70 \\
61123.48 & 14.27 & 136.256 & NOEMA & 8860 $\pm$ 30 \\
61123.48 & 14.27 & 151.744 & NOEMA & 8340 $\pm$ 30 \\
61123.60 & 14.39 & 15.0 & OVRO & 21700 $\pm$ 1600 \\
61125.00 & 15.78 & 230.0 & SMA & 5670 $\pm$ 230 \\
61125.27 & 16.06 & 10.45 & Effelsberg & 14000 $\pm$ 3000$^{\rm b}$ \\
61125.30 & 16.09 & 15.0 & OVRO & 12700 $\pm$ 1700 \\
61126.32 & 17.11 & 43.0 & VLA & 9500 $\pm$ 60 \\
61126.32 & 17.11 & 23.06 & VLA & 11180 $\pm$ 80 \\
61126.32 & 17.11 & 32.0 & VLA & 10210 $\pm$ 110 \\
61126.32 & 17.11 & 19.06 & VLA & 11910 $\pm$ 80 \\
61126.40 & 17.18 & 3.0 & VLA & 2902 $\pm$ 39 \\
61126.40 & 17.18 & 6.0 & VLA & 7530 $\pm$ 53 \\
61126.40 & 17.18 & 10.0 & VLA & 11114 $\pm$ 107 \\
61126.40 & 17.18 & 15.0 & VLA & 11726 $\pm$ 158 \\
61129.35 & 20.14 & 15.0 & OVRO & 13900 $\pm$ 1500 \\
61132.12 & 22.91 & 101.744 & NOEMA & 4430 $\pm$ 20 \\
61132.12 & 22.91 & 86.256 & NOEMA & 4720 $\pm$ 20 \\
61132.28 & 23.07 & 136.256 & NOEMA & 3930 $\pm$ 20 \\
61132.28 & 23.07 & 151.744 & NOEMA & 3880 $\pm$ 30 \\
61132.30 & 23.09 & 15.0 & OVRO & 13300 $\pm$ 1500 \\
61134.36 & 25.16 & 31.87 & VLA & 6740 $\pm$ 30 \\
61134.36 & 25.16 & 41.9 & VLA & 6140 $\pm$ 70 \\
61134.36 & 25.16 & 20.88 & VLA & 7580 $\pm$ 20 \\
61134.44 & 25.22 & 3.0 & VLA & 4997 $\pm$ 40 \\
61134.44 & 25.22 & 6.0 & VLA & 8390 $\pm$ 50 \\
61134.44 & 25.22 & 10.0 & VLA & 8927 $\pm$ 125 \\
61134.44 & 25.22 & 15.0 & VLA & 8163 $\pm$ 215 \\
61134.44 & 25.22 & 1.5 & VLA & 1745 $\pm$ 57 \\
61135.31 & 26.10 & 44.0 & VLA & 3950 $\pm$ 150 \\
61135.33 & 26.12 & 34.0 & VLA & 4807 $\pm$ 67 \\
61135.33 & 26.12 & 32.0 & VLA & 5139 $\pm$ 60 \\
61135.33 & 26.12 & 30.0 & VLA & 5346 $\pm$ 56 \\
61135.33 & 26.12 & 36.0 & VLA & 4603 $\pm$ 62 \\
61135.34 & 26.13 & 25.0 & VLA & 6056 $\pm$ 65 \\
61135.34 & 26.13 & 23.0 & VLA & 6309 $\pm$ 54 \\
61135.34 & 26.13 & 21.0 & VLA & 6670 $\pm$ 74 \\
61135.34 & 26.13 & 19.0 & VLA & 6997 $\pm$ 61 \\
61135.35 & 26.14 & 9.0 & VLA & 9235 $\pm$ 39 \\
61135.35 & 26.14 & 17.0 & VLA & 7275 $\pm$ 140 \\
61135.35 & 26.14 & 11.0 & VLA & 8640 $\pm$ 40 \\
61135.35 & 26.14 & 15.0 & VLA & 7514 $\pm$ 106 \\
61135.35 & 26.14 & 13.0 & VLA & 8126 $\pm$ 52 \\
61135.36 & 26.15 & 4.0 & VLA & 7790 $\pm$ 84 \\
61135.36 & 26.15 & 5.0 & VLA & 9220 $\pm$ 63 \\
61135.36 & 26.15 & 7.0 & VLA & 9670 $\pm$ 68 \\
61135.36 & 26.15 & 3.0 & VLA & 5719 $\pm$ 93 \\
61139.26 & 30.05 & 15.0 & OVRO & 11100 $\pm$ 1300 \\
61139.39 & 30.18 & 151.744 & NOEMA & 3680 $\pm$ 40 \\
61139.39 & 30.18 & 136.256 & NOEMA & 3640 $\pm$ 30 \\
61139.50 & 30.29 & 101.744 & NOEMA & 3910 $\pm$ 40 \\
61139.50 & 30.29 & 86.256 & NOEMA & 4260 $\pm$ 40 \\
61144.79 & 35.58 & 14.25 & Effelsberg & 11000 $\pm$ 3000 \\
61144.81 & 35.60 & 10.45 & Effelsberg & 10000 $\pm$ 2000 \\
61144.81 & 35.62 & 10.45 & Effelsberg & $<$10000$^{\rm c}$ \\
61144.89 & 35.68 & 101.744 & NOEMA & 3310 $\pm$ 20 \\
61144.89 & 35.68 & 86.256 & NOEMA & 3310 $\pm$ 20 \\
61144.89 & 35.68 & 231.744 & NOEMA & 1550 $\pm$ 80 \\
61144.89 & 35.68 & 216.256 & NOEMA & 1820 $\pm$ 70 \\
61145.16 & 35.92 & 3.0 & VLA & 5104 $\pm$ 58 \\
61145.16 & 35.92 & 6.0 & VLA & 6815 $\pm$ 40 \\
61145.16 & 35.92 & 10.0 & VLA & 6472 $\pm$ 143 \\
61145.16 & 35.92 & 1.5 & VLA & 1244 $\pm$ 57 \\
61145.16 & 35.95 & 12.94 & VLA & 6260 $\pm$ 70 \\
61145.16 & 35.95 & 15.0 & VLA & 6110 $\pm$ 20 \\
61145.16 & 35.95 & 16.99 & VLA & 5840 $\pm$ 20 \\
61145.16 & 35.95 & 20.88 & VLA & 5910 $\pm$ 20 \\
61145.16 & 35.95 & 31.87 & VLA & 4820 $\pm$ 30 \\
61145.16 & 35.95 & 41.9 & VLA & 4220 $\pm$ 70 \\
61145.28 & 36.07 & 15.0 & OVRO & 6800 $\pm$ 1500 \\
\enddata
\end{deluxetable*}

\begin{deluxetable*}{ccccccccc}
\tablecaption{X-ray observations of GRB\,260310A from EP, Nu-STAR and Chandra. Photon indices reported here are for the entire full energy bands (EP-FXT 0.5--10\,keV; NuSTAR 2--79\,keV; Chandra 0.3--8\,keV).\label{tab:xray}}
\tablecolumns{7}
\tablewidth{0pt}
\tablehead{
  \colhead{Mid-time} & \colhead{$\Delta t$} & \colhead{Instrument} &
  \colhead{Band} & \colhead{Exp.} & \colhead{$F$} & \colhead{Photon} & \colhead{Fit Statistic} &
  \colhead{Reference} \\ (MJD) & (d) & & (keV) & (s) & (10$^{-12}$ erg s$^{-1}$ cm$^{-2}$) & Index & ({\tt Cstat})
}
\startdata
61111.7330 & 2.53 & EP-FXT & 0.5--10 & 1498 & 5.51 $\pm$ 0.22 & 1.55 $\pm$ 0.09 &111.13/113 & \citet{GCN.43994} \\
61113.5567 & 4.35 & EP-FXT & 0.5--10 & 3628 & 4.48 $\pm$ 0.13 & 1.46 $\pm$ 0.06 & 217.41/198  \\
61116.8642 & 7.66 & EP-FXT & 0.5--10 & 2840 & 3.80 $\pm$ 0.14 & 1.41 $\pm$ 0.07 & 137.30/137 \\
61118.6604 & 9.45 & EP-FXT & 0.5--10 & 2786 & 2.72 $\pm$ 0.12 & 1.52 $\pm$ 0.09 & 105.74/107 \\
61119.7658 & 10.55 & NuSTAR &  2.0--10 & 23380 & 1.15 $\pm$ 0.09 & 1.83 $\pm$ 0.12 & 437.0/370 & \citet{GCN.44063}\\
61120.2583 & 11.05 & EP-FXT & 0.5--10 & 2696 & 2.30 $\pm$ 0.10 & 1.54 $\pm$ 0.10 & 73.37/93 \\
61124.3812 & 15.17 & EP-FXT & 0.5--10 & 2240 & 1.23 $\pm$ 0.09 & 1.50 $\pm$ 0.15 & 38.44/42 \\
61126.5138 & 17.31 & EP-FXT & 0.5--10 & 2921 & 0.90 $\pm$ 0.07 & 1.52 $\pm$ 0.17 & 41.37/42 \\
61132.9008 & 23.69 & EP-FXT & 0.5--10 & 3289 & 0.81 $\pm$ 0.09 & 1.31 $\pm$ 0.15 & 39.74/47 \\
61134.5644 & 25.36 & EP-FXT & 0.5--10 & 3333 & 0.96 $\pm$ 0.07 & 1.25 $\pm$ 0.14 & 48.32/54 \\
61136.1655 & 26.96 & EP-FXT & 0.5--10 & 966$^a$ & 1.21 $\pm$ 0.15 & 1.31 $\pm$ 0.21 & 62.67/221$^a$ \\
61138.7620 & 29.56 & EP-FXT & 0.5-10 & 2659 & 0.80 $\pm$ 0.07 & 1.61 $\pm$ 0.18 & 52.13/304 \\
61141.6334 & 32.43 & EP-FXT & 0.5-10 & 4160 & 0.88 $\pm$ 0.06 & 1.36 $\pm$ 0.13 & 35.93/49 \\
61141.7032 & 32.50 & NuSTAR & 2.0--10 & 12750 & 0.45 $\pm$ 0.07 & 1.64 $\pm$ 0.25 & 398.76/356 & \citet{GCN.44278} \\
61149.7878 & 40.58 & EP-FXT & 0.5-10 & 6306 & 0.68 $\pm$ 0.04 & 1.31 $\pm$ 0.13 & 48.33/59\\
61150.1135 & 40.91 & Chandra & 0.3-10 & 5050 & 0.51 $\pm$ 0.05 & 2.06 $\pm$ 0.24 & 37.08/52 & \citet{yang_chandra_gcn} \\ 
61155.3405 & 46.13 & EP-FXT & 0.5-10 & 3083 & 0.63 $\pm$ 0.07 & 1.17 $\pm$ 0.20 & 28.98/24 \\
\enddata
\tablecomments{(a) This observation was cut short due to passage through the South Atlantic Anomaly, likely affecting the fit.}
\end{deluxetable*}

\clearpage
\newpage

\startlongtable
\begin{deluxetable*}{lcccl}
\tablecaption{IR/Optical/UV photometry of GRB\,260310A.
  $\Delta t$ is measured from the \emph{Fermi}/GBM trigger at
  $T_0 = \mathrm{MJD}~61109.21$ in observer frame.
  $Clear$ = unfiltered/clear; $L$ = GOTO $L$-band (broad optical). Magnitudes are not corrected for Galactic extinction. Upper limits are $3\sigma$.
  \label{tab:opticalphotometry}}
\tablehead{
\colhead{MJD} &
  \colhead{$\Delta t$ (d)} &
  \colhead{Filter} &
  \colhead{Magnitude} &
  \colhead{Telescope/Instrument} 
}
\startdata
61109.22 & 0.01 & 18.84 $\pm$ 0.06 & $L$ & GOTO \\
61109.51 & 0.31 & 16.66 $\pm$ 0.01 & $o$ & ATLAS \\
61109.75 & 0.55 & 17.75 $\pm$ 0.04 & $Clear$ & LAST \\
61109.79 & 0.59 & 17.68 $\pm$ 0.04 & $Clear$ & LAST \\
61109.80 & 0.60 & 17.65 $\pm$ 0.04 & $Clear$ & LAST \\
61109.81 & 0.60 & 17.74 $\pm$ 0.04 & $Clear$ & LAST \\
61109.81 & 0.61 & 17.62 $\pm$ 0.04 & $Clear$ & LAST \\
61109.82 & 0.61 & 17.66 $\pm$ 0.04 & $Clear$ & LAST \\
61109.96 & 0.75 & 17.82 $\pm$ 0.04 & $Clear$ & LAST \\
61110.00 & 0.79 & 17.76 $\pm$ 0.04 & $Clear$ & LAST \\
61110.04 & 0.84 & 17.82 $\pm$ 0.04 & $Clear$ & LAST \\
61110.09 & 0.88 & 17.90 $\pm$ 0.04 & $Clear$ & LAST \\
61110.13 & 0.92 & 17.99 $\pm$ 0.04 & $Clear$ & LAST \\
61110.41 & 1.20 & 18.09 $\pm$ 0.02 & $g$ & ZTF \\
61110.41 & 1.20 & 18.08 $\pm$ 0.06 & $g$ & ZTF \\
61110.41 & 1.20 & 18.08 $\pm$ 0.06 & $g$ & ZTF \\
61110.41 & 1.20 & 18.09 $\pm$ 0.02 & $g$ & ZTF \\
61110.49 & 1.29 & 17.68 $\pm$ 0.02 & $r$ & ZTF \\
61110.49 & 1.29 & 17.64 $\pm$ 0.05 & $r$ & ZTF \\
61110.49 & 1.29 & 17.64 $\pm$ 0.05 & $r$ & ZTF \\
61111.10 & 1.90 & 18.22 $\pm$ 0.02 & $w$ & ATLAS \\
61111.28 & 2.08 & 18.43 $\pm$ 0.05 & $g$ & SEDM \\
61111.29 & 2.08 & 18.10 $\pm$ 0.02 & $r$ & SEDM \\
61111.29 & 2.08 & 17.86 $\pm$ 0.04 & $i$ & SEDM \\
61111.41 & 2.21 & 17.16 $\pm$ 0.09 & $J$ & WINTER \\
61111.44 & 2.23 & 17.28 $\pm$ 0.09 & $J$ & WINTER \\
61111.44 & 2.24 & 18.80 $\pm$ 0.06 & $u$ & SEDM \\
61111.45 & 2.24 & 18.49 $\pm$ 0.05 & $g$ & SEDM \\
61111.45 & 2.24 & 18.09 $\pm$ 0.06 & $r$ & SEDM \\
61111.45 & 2.25 & 18.01 $\pm$ 0.05 & $i$ & SEDM \\
61111.46 & 2.26 & 17.10 $\pm$ 0.21 & $H$ & WINTER \\
61111.95 & 2.74 & 18.60 $\pm$ 0.06 & $Clear$ & LAST \\
61111.98 & 2.77 & 18.18 $\pm$ 0.22 & $L$ & GOTO \\
61112.11 & 2.91 & 18.49 $\pm$ 0.02 & $g$ & NOT \\
61112.12 & 2.91 & 18.19 $\pm$ 0.02 & $r$ & NOT \\
61112.12 & 2.91 & 17.97 $\pm$ 0.04 & $i$ & NOT \\
61112.12 & 2.92 & 17.84 $\pm$ 0.05 & $z$ & NOT \\
61112.20 & 2.99 & 17.34 $\pm$ 0.12 & $J$ & WINTER \\
61112.21 & 3.01 & 16.97 $\pm$ 0.20 & $H$ & WINTER \\
61112.27 & 3.07 & 18.16 $\pm$ 0.02 & $r$ & SEDM \\
61112.28 & 3.07 & 18.64 $\pm$ 0.06 & $g$ & SEDM \\
61112.28 & 3.07 & 18.02 $\pm$ 0.05 & $i$ & SEDM \\
61112.50 & 3.29 & 19.23 $\pm$ 0.10 & $u$ & SEDM \\
61112.50 & 3.29 & 18.67 $\pm$ 0.05 & $g$ & SEDM \\
61112.50 & 3.30 & 18.31 $\pm$ 0.03 & $r$ & SEDM \\
61112.50 & 3.30 & 18.20 $\pm$ 0.03 & $i$ & SEDM \\
61112.57 & 3.37 & 18.71 $\pm$ 0.02 & $g$ & LOT \\
61112.58 & 3.37 & 18.42 $\pm$ 0.02 & $r$ & LOT \\
61112.58 & 3.37 & 18.25 $\pm$ 0.02 & $i$ & LOT \\
61112.58 & 3.38 & 17.96 $\pm$ 0.05 & $z$ & LOT \\
61112.86 & 3.66 & 18.67 $\pm$ 0.02 & $g$ & LOT \\
61112.87 & 3.66 & 18.36 $\pm$ 0.02 & $r$ & LOT \\
61112.87 & 3.66 & 18.14 $\pm$ 0.02 & $i$ & LOT \\
61112.87 & 3.67 & 17.97 $\pm$ 0.04 & $z$ & LOT \\
61113.17 & 3.96 & 18.83 $\pm$ 0.04 & $g$ & NOT \\
61113.17 & 3.97 & 18.50 $\pm$ 0.02 & $r$ & NOT \\
61113.18 & 3.97 & 18.33 $\pm$ 0.03 & $i$ & NOT \\
61113.18 & 3.97 & 18.16 $\pm$ 0.05 & $z$ & NOT \\
61113.23 & 4.02 & 17.91 $\pm$ 0.10 & $J$ & TNG \\
61113.25 & 4.05 & 17.29 $\pm$ 0.10 & $H$ & TNG \\
61113.26 & 4.05 & 18.58 $\pm$ 0.02 & $r$ & LCOGT \\
61113.26 & 4.05 & 18.66 $\pm$ 0.07 & $r$ & SEDM \\
61113.26 & 4.06 & 18.94 $\pm$ 0.05 & $g$ & SEDM \\
61113.27 & 4.06 & 19.47 $\pm$ 0.04 & $u$ & LCOGT \\
61113.27 & 4.06 & 18.42 $\pm$ 0.06 & $i$ & SEDM \\
61113.27 & 4.07 & 19.31 $\pm$ 0.11 & $u$ & SEDM \\
61113.28 & 4.07 & 18.88 $\pm$ 0.04 & $g$ & SEDM \\
61113.28 & 4.08 & 18.48 $\pm$ 0.04 & $r$ & SEDM \\
61113.28 & 4.08 & 18.45 $\pm$ 0.04 & $i$ & SEDM \\
61113.43 & 4.22 & 18.87 $\pm$ 0.03 & $g$ & ZTF \\
61113.43 & 4.22 & 18.90 $\pm$ 0.07 & $g$ & ZTF \\
61113.43 & 4.22 & 18.90 $\pm$ 0.07 & $g$ & ZTF \\
61113.45 & 4.24 & 18.90 $\pm$ 0.05 & $g$ & SEDM \\
61113.45 & 4.24 & 18.50 $\pm$ 0.03 & $r$ & SEDM \\
61113.45 & 4.24 & 18.42 $\pm$ 0.06 & $i$ & SEDM \\
61113.49 & 4.28 & 18.54 $\pm$ 0.07 & $r$ & ZTF \\
61113.50 & 4.29 & 17.86 $\pm$ 0.18 & $J$ & WINTER \\
61113.53 & 4.33 & 17.91 $\pm$ 0.09 & $J$ & WINTER \\
61113.65 & 4.44 & 18.91 $\pm$ 0.03 & $g$ & LOT \\
61113.65 & 4.44 & 18.59 $\pm$ 0.04 & $r$ & LOT \\
61113.65 & 4.45 & 18.38 $\pm$ 0.04 & $i$ & LOT \\
61113.66 & 4.45 & 18.18 $\pm$ 0.05 & $z$ & LOT \\
61113.71 & 4.51 & 18.16 $\pm$ 0.04 & $z$ & LOT \\
61114.10 & 4.90 & 18.81 $\pm$ 0.03 & $w$ & ATLAS \\
61114.20 & 4.99 & 18.19 $\pm$ 0.27 & $y$ & WINTER \\
61114.21 & 5.01 & 17.72 $\pm$ 0.10 & $J$ & WINTER \\
61114.24 & 5.03 & 17.89 $\pm$ 0.33 & $y$ & WINTER \\
61114.25 & 5.04 & 17.77 $\pm$ 0.09 & $J$ & WINTER \\
61114.25 & 5.04 & $>$19.91 & $u$ & SEDM \\
61114.25 & 5.05 & $>$21.22 & $g$ & SEDM \\
61114.26 & 5.05 & 18.55 $\pm$ 0.03 & $r$ & SEDM \\
61114.26 & 5.05 & 18.47 $\pm$ 0.03 & $i$ & SEDM \\
61114.26 & 5.06 & 17.27 $\pm$ 0.22 & $H$ & WINTER \\
61114.29 & 5.08 & 17.86 $\pm$ 0.11 & $J$ & WINTER \\
61114.54 & 5.33 & 17.69 $\pm$ 0.09 & $J$ & WINTER \\
61114.85 & 5.65 & 18.94 $\pm$ 0.03 & $g$ & LOT \\
61114.86 & 5.65 & 18.61 $\pm$ 0.03 & $r$ & LOT \\
61114.86 & 5.65 & 18.41 $\pm$ 0.03 & $i$ & LOT \\
61114.86 & 5.66 & 18.18 $\pm$ 0.04 & $z$ & LOT \\
61114.99 & 5.78 & 18.97 $\pm$ 0.05 & $g$ & NOT \\
61114.99 & 5.79 & 18.65 $\pm$ 0.02 & $r$ & NOT \\
61115.00 & 5.79 & 18.47 $\pm$ 0.04 & $i$ & NOT \\
61115.00 & 5.79 & 18.23 $\pm$ 0.06 & $z$ & NOT \\
61115.21 & 6.01 & 17.81 $\pm$ 0.10 & $J$ & WINTER \\
61115.24 & 6.03 & 19.23 $\pm$ 0.08 & $g$ & SEDM \\
61115.24 & 6.03 & 18.59 $\pm$ 0.03 & $r$ & SEDM \\
61115.24 & 6.04 & 18.51 $\pm$ 0.06 & $i$ & SEDM \\
61115.25 & 6.04 & 17.80 $\pm$ 0.09 & $J$ & WINTER \\
61115.29 & 6.08 & 17.84 $\pm$ 0.12 & $J$ & WINTER \\
61115.50 & 6.29 & 18.01 $\pm$ 0.11 & $J$ & WINTER \\
61115.89 & 6.68 & 19.06 $\pm$ 0.07 & $g$ & LOT \\
61115.89 & 6.69 & 18.64 $\pm$ 0.08 & $r$ & LOT \\
61116.37 & 7.16 & $>$17.38 & $r$ & SEDM \\
61116.37 & 7.17 & $>$18.00 & $i$ & SEDM \\
61116.44 & 7.24 & 19.57 $\pm$ 0.30 & $u$ & SEDM \\
61116.45 & 7.24 & 18.77 $\pm$ 0.10 & $r$ & SEDM \\
61116.45 & 7.25 & 18.78 $\pm$ 0.04 & $i$ & SEDM \\
61116.47 & 7.27 & 19.25 $\pm$ 0.13 & $g$ & ZTF \\
61116.76 & 7.55 & 19.13 $\pm$ 0.03 & $g$ & LOT \\
61116.76 & 7.56 & 18.81 $\pm$ 0.03 & $r$ & LOT \\
61116.77 & 7.56 & 18.65 $\pm$ 0.03 & $i$ & LOT \\
61116.77 & 7.56 & 18.48 $\pm$ 0.05 & $z$ & LOT \\
61117.20 & 7.99 & 18.26 $\pm$ 0.12 & $J$ & WINTER \\
61117.25 & 8.04 & 19.37 $\pm$ 0.14 & $g$ & SEDM \\
61117.25 & 8.05 & 18.76 $\pm$ 0.04 & $r$ & SEDM \\
61117.25 & 8.05 & 18.74 $\pm$ 0.03 & $i$ & SEDM \\
61117.26 & 8.06 & 18.38 $\pm$ 0.54 & $y$ & WINTER \\
61117.27 & 8.07 & 18.22 $\pm$ 0.12 & $J$ & WINTER \\
61117.39 & 8.18 & 19.16 $\pm$ 0.04 & $g$ & ZTF \\
61117.39 & 8.18 & 19.21 $\pm$ 0.10 & $g$ & ZTF \\
61117.39 & 8.18 & 19.21 $\pm$ 0.10 & $g$ & ZTF \\
61117.41 & 8.20 & 18.84 $\pm$ 0.09 & $r$ & ZTF \\
61117.53 & 8.32 & 19.26 $\pm$ 0.06 & $g$ & SEDM \\
61117.53 & 8.33 & 18.84 $\pm$ 0.03 & $r$ & SEDM \\
61117.53 & 8.33 & 18.76 $\pm$ 0.05 & $i$ & SEDM \\
61117.87 & 8.67 & 19.29 $\pm$ 0.04 & $g$ & LOT \\
61117.88 & 8.67 & 18.96 $\pm$ 0.03 & $r$ & LOT \\
61117.88 & 8.68 & 18.88 $\pm$ 0.05 & $i$ & LOT \\
61117.89 & 8.68 & 18.60 $\pm$ 0.10 & $z$ & LOT \\
61118.26 & 9.06 & 19.39 $\pm$ 0.06 & $g$ & SEDM \\
61118.27 & 9.06 & 19.03 $\pm$ 0.04 & $r$ & SEDM \\
61118.27 & 9.06 & 19.00 $\pm$ 0.04 & $i$ & SEDM \\
61118.72 & 9.51 & 19.42 $\pm$ 0.03 & $g$ & LOT \\
61118.72 & 9.51 & 19.09 $\pm$ 0.02 & $r$ & LOT \\
61118.72 & 9.52 & 18.95 $\pm$ 0.02 & $i$ & LOT \\
61118.73 & 9.52 & 18.79 $\pm$ 0.05 & $z$ & LOT \\
61119.19 & 9.98 & 19.42 $\pm$ 0.05 & $g$ & TJO \\
61119.19 & 9.98 & 18.91 $\pm$ 0.21 & $y$ & WINTER \\
61119.19 & 9.99 & 19.08 $\pm$ 0.07 & $r$ & TJO \\
61119.20 & 9.99 & 18.90 $\pm$ 0.10 & $i$ & TJO \\
61119.20 & 10.00 & 18.35 $\pm$ 0.14 & $J$ & WINTER \\
61119.21 & 10.01 & 18.13 $\pm$ 0.34 & $H$ & WINTER \\
61119.23 & 10.02 & 18.90 $\pm$ 0.19 & $y$ & WINTER \\
61119.24 & 10.03 & 18.56 $\pm$ 0.15 & $J$ & WINTER \\
61119.25 & 10.05 & 18.40 $\pm$ 0.36 & $H$ & WINTER \\
61119.28 & 10.07 & 19.57 $\pm$ 0.08 & $g$ & SEDM \\
61119.28 & 10.07 & 19.15 $\pm$ 0.03 & $r$ & SEDM \\
61119.28 & 10.08 & 19.15 $\pm$ 0.05 & $i$ & SEDM \\
61119.40 & 10.19 & 19.54 $\pm$ 0.13 & $g$ & ZTF \\
61119.40 & 10.19 & 19.55 $\pm$ 0.06 & $g$ & ZTF \\
61119.40 & 10.19 & 19.54 $\pm$ 0.13 & $g$ & ZTF \\
61119.47 & 10.27 & 19.16 $\pm$ 0.09 & $r$ & ZTF \\
61120.26 & 11.06 & 19.67 $\pm$ 0.07 & $g$ & SEDM \\
61120.27 & 11.06 & 19.23 $\pm$ 0.06 & $r$ & SEDM \\
61120.27 & 11.06 & 19.25 $\pm$ 0.06 & $i$ & SEDM \\
61120.37 & 11.16 & 18.29 $\pm$ 0.16 & $Hs$ & MIRAGE \\
61120.39 & 11.19 & 18.86 $\pm$ 0.07 & $J$ & MIRAGE \\
61120.44 & 11.24 & 19.68 $\pm$ 0.09 & $g$ & SEDM \\
61120.45 & 11.24 & 19.23 $\pm$ 0.05 & $r$ & SEDM \\
61120.45 & 11.24 & 19.23 $\pm$ 0.06 & $i$ & SEDM \\
61121.23 & 12.03 & 19.55 $\pm$ 0.16 & $r$ & SEDM \\
61121.24 & 12.03 & $>$17.91 & $i$ & SEDM \\
61121.24 & 12.03 & 19.82 $\pm$ 0.10 & $g$ & SEDM \\
61121.24 & 12.04 & 19.33 $\pm$ 0.07 & $r$ & SEDM \\
61121.25 & 12.04 & 19.27 $\pm$ 0.06 & $i$ & SEDM \\
61121.28 & 12.08 & $>$17.62 & $J$ & WINTER \\
61121.34 & 12.14 & 18.93 $\pm$ 0.19 & $Hs$ & MIRAGE \\
61121.37 & 12.16 & 18.90 $\pm$ 0.09 & $J$ & MIRAGE \\
61121.38 & 12.18 & $>$17.10 & $J$ & WINTER \\
61121.47 & 12.26 & 19.36 $\pm$ 0.14 & $r$ & ZTF \\
61122.18 & 12.98 & 19.80 $\pm$ 0.09 & $g$ & TJO \\
61122.19 & 12.98 & 19.37 $\pm$ 0.06 & $r$ & TJO \\
61122.20 & 12.99 & 19.06 $\pm$ 0.08 & $i$ & TJO \\
61122.22 & 13.01 & 19.71 $\pm$ 0.07 & $g$ & SEDM \\
61122.22 & 13.02 & 19.56 $\pm$ 0.05 & $r$ & SEDM \\
61122.22 & 13.02 & 19.37 $\pm$ 0.07 & $i$ & SEDM \\
61122.37 & 13.16 & 18.47 $\pm$ 0.13 & $J$ & WINTER \\
61122.38 & 13.18 & $>$17.92 & $H$ & WINTER \\
61122.41 & 13.20 & 19.93 $\pm$ 0.14 & $g$ & SEDM \\
61122.41 & 13.21 & 19.55 $\pm$ 0.06 & $r$ & SEDM \\
61122.41 & 13.21 & 19.34 $\pm$ 0.07 & $i$ & SEDM \\
61123.38 & 14.18 & $>$18.33 & $J$ & WINTER \\
61123.39 & 14.19 & $>$17.65 & $H$ & WINTER \\
61123.45 & 14.24 & 19.57 $\pm$ 0.14 & $r$ & ZTF \\
61123.46 & 14.26 & 20.07 $\pm$ 0.07 & $g$ & SEDM \\
61123.46 & 14.26 & 19.51 $\pm$ 0.04 & $r$ & SEDM \\
61123.47 & 14.26 & 19.63 $\pm$ 0.07 & $i$ & SEDM \\
61124.28 & 15.08 & 20.21 $\pm$ 0.07 & $g$ & SEDM \\
61124.29 & 15.08 & 19.71 $\pm$ 0.05 & $r$ & SEDM \\
61124.29 & 15.08 & 19.75 $\pm$ 0.07 & $i$ & SEDM \\
61124.37 & 15.17 & 19.22 $\pm$ 0.33 & $J$ & WINTER \\
61124.38 & 15.18 & $>$18.33 & $H$ & WINTER \\
61124.43 & 15.23 & 19.88 $\pm$ 0.04 & $r$ & LCOGT \\
61124.45 & 15.24 & 19.78 $\pm$ 0.05 & $i$ & LCOGT \\
61124.49 & 15.29 & 20.45 $\pm$ 0.26 & $c$ & ATLAS \\
61125.24 & 16.04 & 20.11 $\pm$ 0.12 & $g$ & SEDM \\
61125.25 & 16.04 & 19.86 $\pm$ 0.07 & $r$ & SEDM \\
61125.25 & 16.04 & 19.93 $\pm$ 0.09 & $i$ & SEDM \\
61125.29 & 16.09 & $>$19.88 & $g$ & SEDM \\
61125.30 & 16.09 & 19.73 $\pm$ 0.06 & $r$ & SEDM \\
61125.30 & 16.09 & 19.81 $\pm$ 0.06 & $i$ & SEDM \\
61125.43 & 16.23 & 20.08 $\pm$ 0.20 & $r$ & ZTF \\
61125.50 & 16.29 & 19.95 $\pm$ 0.15 & $o$ & ATLAS \\
61126.10 & 16.89 & 20.41 $\pm$ 0.17 & $w$ & ATLAS \\
61126.22 & 17.01 & $>$21.50 & $g$ & SEDM \\
61126.22 & 17.02 & 20.02 $\pm$ 0.07 & $r$ & SEDM \\
61126.22 & 17.02 & 19.98 $\pm$ 0.09 & $i$ & SEDM \\
61126.31 & 17.10 & 20.58 $\pm$ 0.10 & $g$ & SEDM \\
61126.31 & 17.10 & 19.98 $\pm$ 0.06 & $r$ & SEDM \\
61126.40 & 17.20 & $>$18.68 & $J$ & WINTER \\
61126.42 & 17.21 & $>$18.20 & $H$ & WINTER \\
61127.24 & 18.03 & $>$19.93 & $i$ & SEDM \\
61127.45 & 18.24 & 20.60 $\pm$ 0.20 & $g$ & ZTF \\
61127.45 & 18.24 & 20.58 $\pm$ 0.32 & $g$ & ZTF \\
61127.49 & 18.28 & 20.25 $\pm$ 0.21 & $r$ & ZTF \\
61128.03 & 18.82 & 20.70 $\pm$ 0.06 & $g$ & NOT \\
61128.04 & 18.83 & 20.11 $\pm$ 0.04 & $r$ & NOT \\
61128.05 & 18.84 & 20.10 $\pm$ 0.04 & $i$ & NOT \\
61128.06 & 18.85 & 20.07 $\pm$ 0.06 & $z$ & NOT \\
61128.22 & 19.01 & 20.37 $\pm$ 0.27 & $r$ & SEDM \\
61128.22 & 19.01 & 20.17 $\pm$ 0.06 & $r$ & LCOGT \\
61128.22 & 19.02 & 20.27 $\pm$ 0.36 & $i$ & SEDM \\
61128.23 & 19.03 & $>$18.15 & $H$ & WINTER \\
61128.24 & 19.03 & 20.09 $\pm$ 0.16 & $i$ & LCOGT \\
61128.27 & 19.06 & 21.25 $\pm$ 0.37 & $g$ & SEDM \\
61128.27 & 19.06 & 19.96 $\pm$ 0.19 & $r$ & SEDM \\
61128.27 & 19.07 & 19.90 $\pm$ 0.11 & $i$ & SEDM \\
61128.51 & 19.31 & $>$17.20 & $H$ & WINTER \\
61128.62 & 19.41 & 20.10 $\pm$ 0.09 & $r$ & LOT \\
61128.67 & 19.46 & 20.46 $\pm$ 0.15 & $g$ & LOT \\
61129.39 & 20.18 & $>$18.88 & $J$ & WINTER \\
61129.40 & 20.19 & $>$17.97 & $H$ & WINTER \\
61129.58 & 20.37 & 20.40 $\pm$ 0.15 & $r$ & LOT \\
61129.59 & 20.38 & 20.14 $\pm$ 0.16 & $i$ & LOT \\
61129.82 & 20.62 & 20.58 $\pm$ 0.24 & $g$ & LOT \\
61129.85 & 20.64 & 19.84 $\pm$ 0.18 & $z$ & LOT \\
61130.37 & 21.16 & $>$19.84 & $g$ & SEDM \\
61130.37 & 21.16 & $>$20.10 & $r$ & SEDM \\
61130.37 & 21.17 & 20.08 $\pm$ 0.29 & $i$ & SEDM \\
61130.38 & 21.17 & $>$18.32 & $J$ & WINTER \\
61130.40 & 21.19 & $>$17.56 & $H$ & WINTER \\
61130.53 & 21.33 & 20.69 $\pm$ 0.11 & $g$ & Pan-STARRS \\
61130.54 & 21.33 & 20.10 $\pm$ 0.07 & $r$ & Pan-STARRS \\
61130.54 & 21.34 & 19.98 $\pm$ 0.05 & $i$ & Pan-STARRS \\
61130.55 & 21.34 & 19.89 $\pm$ 0.05 & $z$ & Pan-STARRS \\
61130.55 & 21.34 & 19.90 $\pm$ 0.13 & $y$ & Pan-STARRS \\
61130.70 & 21.49 & 20.31 $\pm$ 0.32 & $r$ & SLT \\
61131.54 & 22.33 & 20.80 $\pm$ 0.26 & $g$ & Pan-STARRS \\
61131.54 & 22.34 & 20.15 $\pm$ 0.10 & $r$ & Pan-STARRS \\
61131.54 & 22.34 & 19.95 $\pm$ 0.06 & $i$ & Pan-STARRS \\
61131.55 & 22.34 & 19.96 $\pm$ 0.07 & $z$ & Pan-STARRS \\
61131.55 & 22.35 & 19.78 $\pm$ 0.14 & $y$ & Pan-STARRS \\
61131.82 & 22.61 & 20.21 $\pm$ 0.07 & $r$ & SLT \\
61132.57 & 23.36 & 20.59 $\pm$ 0.18 & $g$ & Pan-STARRS \\
61132.57 & 23.37 & 20.14 $\pm$ 0.08 & $r$ & Pan-STARRS \\
61132.58 & 23.37 & 20.05 $\pm$ 0.10 & $i$ & Pan-STARRS \\
61132.58 & 23.38 & 19.92 $\pm$ 0.10 & $z$ & Pan-STARRS \\
61132.59 & 23.38 & 20.33 $\pm$ 0.97 & $y$ & Pan-STARRS \\
61133.15 & 23.95 & 20.32 $\pm$ 0.07 & $r$ & LCOGT \\
61133.16 & 23.95 & 20.16 $\pm$ 0.03 & $r$ & LDT \\
61133.16 & 23.96 & 20.05 $\pm$ 0.03 & $i$ & LDT \\
61133.17 & 23.96 & 20.85 $\pm$ 0.09 & $g$ & LDT \\
61133.17 & 23.96 & 20.74 $\pm$ 0.07 & $g$ & LDT \\
61134.20 & 24.99 & $>$20.98 & $r$ & SEDM \\
61134.20 & 25.00 & 20.33 $\pm$ 0.13 & $i$ & SEDM \\
61134.49 & 25.29 & 20.68 $\pm$ 0.10 & $g$ & Pan-STARRS \\
61134.50 & 25.29 & 20.22 $\pm$ 0.12 & $r$ & Pan-STARRS \\
61134.50 & 25.30 & 19.94 $\pm$ 0.05 & $i$ & Pan-STARRS \\
61134.51 & 25.30 & 19.94 $\pm$ 0.06 & $z$ & Pan-STARRS \\
61134.51 & 25.30 & 19.88 $\pm$ 0.13 & $y$ & Pan-STARRS \\
61135.51 & 26.30 & 20.66 $\pm$ 0.12 & $g$ & Pan-STARRS \\
61135.51 & 26.31 & 20.19 $\pm$ 0.08 & $r$ & Pan-STARRS \\
61135.52 & 26.31 & 20.05 $\pm$ 0.06 & $i$ & Pan-STARRS \\
61135.52 & 26.31 & 20.05 $\pm$ 0.08 & $z$ & Pan-STARRS \\
61135.52 & 26.32 & 19.76 $\pm$ 0.17 & $y$ & Pan-STARRS \\
61136.47 & 27.26 & 20.59 $\pm$ 0.13 & $g$ & Pan-STARRS \\
61136.47 & 27.26 & 20.36 $\pm$ 0.14 & $r$ & Pan-STARRS \\
61136.47 & 27.27 & 20.01 $\pm$ 0.07 & $i$ & Pan-STARRS \\
61136.48 & 27.27 & 19.97 $\pm$ 0.10 & $z$ & Pan-STARRS \\
61136.48 & 27.28 & 19.88 $\pm$ 0.21 & $y$ & Pan-STARRS \\
61137.19 & 27.98 & 19.75 $\pm$ 0.14 & $J$ & MIRAGE \\
61137.22 & 28.02 & $>$18.23 & $H$ & WINTER \\
61138.16 & 28.95 & $>$18.26 & $H$ & WINTER \\
61138.69 & 29.49 & 21.12 $\pm$ 0.09 & $g$ & LOT \\
61138.71 & 29.50 & 20.34 $\pm$ 0.04 & $r$ & LOT \\
61138.72 & 29.52 & 20.26 $\pm$ 0.05 & $i$ & LOT \\
61138.73 & 29.52 & 20.08 $\pm$ 0.10 & $z$ & LOT \\
61138.83 & 29.62 & 21.18 $\pm$ 0.14 & $g$ & TJO \\
61138.83 & 29.63 & 20.28 $\pm$ 0.06 & $r$ & TJO \\
61138.84 & 29.63 & 20.07 $\pm$ 0.05 & $i$ & TJO \\
61139.17 & 29.97 & $>$18.26 & $H$ & WINTER \\
61139.24 & 30.03 & 20.36 $\pm$ 0.07 & $r$ & LDT \\
61139.24 & 30.04 & 20.10 $\pm$ 0.07 & $i$ & LDT \\
61139.25 & 30.04 & 21.00 $\pm$ 0.07 & $g$ & LDT \\
61139.39 & 30.18 & 19.74 $\pm$ 0.13 & $J$ & MIRAGE \\
61139.74 & 30.53 & 20.96 $\pm$ 0.07 & $g$ & LOT \\
61139.75 & 30.55 & 20.44 $\pm$ 0.05 & $r$ & LOT \\
61139.76 & 30.56 & 20.24 $\pm$ 0.05 & $i$ & LOT \\
61139.78 & 30.57 & 20.09 $\pm$ 0.12 & $z$ & LOT \\
61143.75 & 34.54 & 20.77 $\pm$ 0.03 & $r$ & LOT \\
61143.77 & 34.56 & 20.49 $\pm$ 0.03 & $i$ & LOT \\
61144.70 & 35.49 & 21.24 $\pm$ 0.09 & $g$ & LOT \\
61144.71 & 35.50 & 20.74 $\pm$ 0.06 & $r$ & LOT \\
61144.72 & 35.51 & 20.62 $\pm$ 0.05 & $i$ & LOT \\
61144.73 & 35.52 & 20.55 $\pm$ 0.15 & $z$ & LOT \\
61145.22 & 36.02 & 20.54 $\pm$ 0.23 & $J$ & MIRAGE \\
61145.37 & 36.17 & 21.42 $\pm$ 0.15 & $g$ & SEDM \\
61145.38 & 36.17 & 20.73 $\pm$ 0.09 & $r$ & SEDM \\
61146.66 & 37.46 & 21.40 $\pm$ 0.08 & $g$ & LOT \\
61146.68 & 37.48 & 20.93 $\pm$ 0.07 & $r$ & LOT \\
61146.69 & 37.48 & 20.69 $\pm$ 0.07 & $i$ & LOT \\
61147.30 & 38.10 & 20.39 $\pm$ 0.28 & $J$ & MIRAGE \\
61147.40 & 38.20 & 21.21 $\pm$ 0.09 & $g$ & Pan-STARRS \\
61147.41 & 38.20 & 20.82 $\pm$ 0.09 & $r$ & Pan-STARRS \\
61147.41 & 38.20 & 20.65 $\pm$ 0.07 & $i$ & Pan-STARRS \\
61147.41 & 38.21 & 20.69 $\pm$ 0.13 & $z$ & Pan-STARRS \\
61147.42 & 38.21 & 20.40 $\pm$ 0.26 & $y$ & Pan-STARRS \\
61147.67 & 38.47 & 21.24 $\pm$ 0.09 & $g$ & LOT \\
61147.68 & 38.48 & 21.00 $\pm$ 0.06 & $r$ & LOT \\
61147.70 & 38.49 & 20.82 $\pm$ 0.10 & $i$ & LOT \\
61147.96 & 38.75 & 21.77 $\pm$ 0.14 & $g$ & TJO \\
61147.96 & 38.76 & 20.84 $\pm$ 0.08 & $r$ & TJO \\
61147.97 & 38.76 & 20.77 $\pm$ 0.09 & $i$ & TJO \\
61148.19 & 38.98 & 20.43 $\pm$ 0.22 & $J$ & MIRAGE \\
61148.77 & 39.56 & 21.37 $\pm$ 0.06 & $g$ & LOT \\
61148.77 & 39.57 & 20.91 $\pm$ 0.05 & $r$ & LOT \\
61148.79 & 39.59 & 20.74 $\pm$ 0.05 & $i$ & LOT \\
61148.80 & 39.60 & 20.49 $\pm$ 0.15 & $z$ & LOT \\
61149.15 & 39.94 & 21.67 $\pm$ 0.16 & $g$ & TJO \\
61149.16 & 39.95 & 20.94 $\pm$ 0.11 & $r$ & TJO \\
61149.16 & 39.96 & 20.90 $\pm$ 0.19 & $i$ & TJO \\
61149.37 & 40.16 & $>$18.90 & $i$ & SEDM \\
61149.49 & 40.28 & 21.17 $\pm$ 0.10 & $g$ & Pan-STARRS \\
61149.49 & 40.29 & 20.98 $\pm$ 0.13 & $r$ & Pan-STARRS \\
61149.49 & 40.29 & 20.86 $\pm$ 0.15 & $i$ & Pan-STARRS \\
61149.50 & 40.29 & 20.73 $\pm$ 0.16 & $z$ & Pan-STARRS \\
61149.50 & 40.30 & 20.24 $\pm$ 0.30 & $y$ & Pan-STARRS \\
61149.78 & 40.57 & 21.32 $\pm$ 0.06 & $g$ & LOT \\
61149.79 & 40.58 & 21.04 $\pm$ 0.04 & $r$ & LOT \\
61149.80 & 40.59 & 20.92 $\pm$ 0.07 & $i$ & LOT \\
61149.82 & 40.61 & 20.11 $\pm$ 0.30 & $z$ & LOT \\
61150.03 & 40.82 & 21.80 $\pm$ 0.06 & $g$ & LT \\
61150.03 & 40.82 & 21.11 $\pm$ 0.08 & $r$ & LT \\
61150.03 & 40.83 & 20.82 $\pm$ 0.21 & $i$ & LT \\
61150.04 & 40.83 & $>$20.76 & $z$ & LT \\
61150.30 & 41.09 & 21.84 $\pm$ 0.17 & $g$ & SEDM \\
61150.30 & 41.09 & 20.83 $\pm$ 0.09 & $r$ & SEDM \\
61150.30 & 41.10 & 21.06 $\pm$ 0.10 & $i$ & SEDM \\
61150.49 & 41.28 & 21.27 $\pm$ 0.10 & $g$ & Pan-STARRS \\
61150.49 & 41.29 & 21.07 $\pm$ 0.12 & $r$ & Pan-STARRS \\
61150.50 & 41.29 & 20.86 $\pm$ 0.11 & $i$ & Pan-STARRS \\
61150.50 & 41.29 & 20.79 $\pm$ 0.21 & $z$ & Pan-STARRS \\
61150.51 & 41.30 & 20.46 $\pm$ 0.42 & $y$ & Pan-STARRS \\
61151.05 & 41.84 & 21.96 $\pm$ 0.08 & $g$ & LT \\
61151.05 & 41.85 & 21.21 $\pm$ 0.07 & $r$ & LT \\
61151.06 & 41.85 & 20.96 $\pm$ 0.08 & $i$ & LT \\
61151.47 & 42.26 & 21.20 $\pm$ 0.20 & $g$ & Pan-STARRS \\
61151.47 & 42.26 & 21.02 $\pm$ 0.24 & $r$ & Pan-STARRS \\
61151.47 & 42.27 & 20.72 $\pm$ 0.21 & $i$ & Pan-STARRS \\
61151.48 & 42.27 & 20.82 $\pm$ 0.46 & $z$ & Pan-STARRS \\
61151.48 & 42.28 & 20.59 $\pm$ 0.77 & $y$ & Pan-STARRS \\
61154.04 & 44.84 & 21.94 $\pm$ 0.06 & $g$ & NOT \\
61154.07 & 44.86 & 21.19 $\pm$ 0.07 & $r$ & NOT \\
61154.08 & 44.88 & 21.00 $\pm$ 0.06 & $i$ & NOT \\
61154.09 & 44.88 & 22.15 $\pm$ 0.14 & $g$ & LT \\
61154.09 & 44.88 & 21.19 $\pm$ 0.19 & $r$ & LT \\
61154.09 & 44.88 & 20.90 $\pm$ 0.16 & $i$ & LT \\
61154.46 & 45.25 & 21.52 $\pm$ 0.10 & $g$ & Pan-STARRS \\
61154.46 & 45.26 & 21.14 $\pm$ 0.08 & $r$ & Pan-STARRS \\
61154.47 & 45.26 & 20.91 $\pm$ 0.07 & $i$ & Pan-STARRS \\
61154.47 & 45.27 & 20.86 $\pm$ 0.10 & $z$ & Pan-STARRS \\
61154.48 & 45.27 & 20.83 $\pm$ 0.26 & $y$ & Pan-STARRS \\
61156.64 & 47.43 & 22.01 $\pm$ 0.29 & $g$ & LOT \\
61156.65 & 47.45 & 21.20 $\pm$ 0.13 & $r$ & LOT \\
61156.67 & 47.46 & 21.30 $\pm$ 0.10 & $i$ & LOT \\
61156.69 & 47.48 & 20.81 $\pm$ 0.17 & $z$ & LOT \\
61157.44 & 48.24 & 21.03 $\pm$ 0.27 & $g$ & Pan-STARRS \\
61157.45 & 48.24 & 21.35 $\pm$ 0.33 & $r$ & Pan-STARRS \\
61157.45 & 48.25 & 21.04 $\pm$ 0.28 & $i$ & Pan-STARRS \\
61157.46 & 48.25 & 21.04 $\pm$ 0.45 & $z$ & Pan-STARRS \\
61157.46 & 48.25 & 20.76 $\pm$ 1.35 & $y$ & Pan-STARRS \\
61158.04 & 48.84 & 21.11 $\pm$ 0.10 & $z$ & NOT \\
\enddata
\end{deluxetable*}

\begin{deluxetable*}{ccc}
\tablecaption{GRB\,260310A host photometry used in SED fitting with \textsc{FrankenBlast}.\label{tab:host_phot}}
\tablecolumns{7}
\tablewidth{0pt}
\tablehead{
  \colhead{Filter} & \colhead{Instrument} & \colhead{Flux $\mu$Jy}
}
\startdata
$NUV$ & GALEX & 33.76 $\pm$ 5.30 \\
$FUV$ & GALEX & 42.57 $\pm$ 7.36 \\
$g$ & PS1 & 126.03 $\pm$ 1.96 \\
$r$ & PS1 & 213.16 $\pm$ 3.54 \\
$i$ & PS1 & 277.81 $\pm$ 3.41 \\
$z$ & PS1 & 263.39 $\pm$ 4.52 \\
$y$ & PS1 & 236.05 $\pm$ 5.65 \\
$J$ & 2MASS & 312.16 $\pm$ 116.45 \\
$H$ & 2MASS & 459.63 $\pm$ 220.56 \\
$W1$ & WISE & 375.08 $\pm$ 26.09 \\
$W2$ & WISE & 353.85 $\pm$ 36.41 \\
$W3$ & WISE & 802.27 $\pm$ 152.57 \\
$W4$ & WISE & 2602.62 $\pm$ 1059.30 \\
\enddata
\end{deluxetable*}

\end{document}